\def\unlockat{\catcode`\@=11}
\def\lockat{\catcode`\@=12}

\input lanlmac.tex
\unlockat
%
%
%
%
%
\newskip\footskip\footskip10pt plus 1pt minus 1pt 
\def\footnotefont{\ninepoint}\def\f@t#1{\footnotefont #1\@foot}
\def\f@@t{\baselineskip\footskip\bgroup\footnotefont\aftergroup\@foot\let\next}
\setbox\strutbox=\hbox{\vrule height9.5pt depth4.5pt width0pt}
\def\listrefs{\footatend\vfill\supereject\immediate\closeout\rfile\writestoppt
\baselineskip=10pt\centerline{{\bf References}}\bigskip{\parindent=20pt%
\frenchspacing\escapechar=` \input \jobname.refs\vfill\eject}\nonfrenchspacing}
\lockat
\input epsf
\input \jobname.defs \def\DefWarn#1{}
\unlockat
\let\lref=\nref
\def\m@ssage{\immediate\write16}  \m@ssage{}

\newif\ifstudvers
\studversfalse

\lref\Ahl{L. V.~ Ahlfors and L.~ Sario, {\it Riemann Surfaces},
Princeton University Press, 1960.}
\lref\Al{ O.~ Alvarez, ``Theory of Strings with Boundaries: Fluctuations, Topology and
Quantum Gravity", Nucl. Phys. {\bf B216} (1983 ) 125.}
\lref\AGIndex{L.~ Alvarez-Gaume, ``Supersymmetry and the Atiyah-Singer
Index Theorem", Comm. Math. Phys. {\bf 90} (1983) 161;\hfil\break
L.~ Alvarez-Gaume, ``Supersymmetry and Index Theory", in {\it NATO Advanced
Study Institute on Supersymmetry}, 1984: Bonn, Germany.}
\lref\lagpg{L.~ Alvarez-Gaum\'e and P.~ Ginsparg, ``The Structure of Gauge
and Gravitational Anomalies", Ann. Phys. {\bf 161} (1985) 423.; Erratum: Ann. Phys.
{\bf 171} (1986) 233.}
\lref\AnGaNaTa{I.~ Antoniadis, E.~ Gava, K.~ Narain, and T.R.~ Taylor, ``Topological
Amplitudes in String Theory", Nucl. Phys. {\bf B413}  (1994) 162.}
\lref\ArCo{E.~ Arbarello and M.~ Cornalba, ``Combinatorial and
Algebro-Geometric Cohomology Classes on the Moduli Space of Curves",
alg-geom/9406008.}
\lref\Ar{See V.~ Arnold, ``Remarks on Enumeration of Plane Purves''; 
``Plane Curves, their Invariants, Perestroikas and Classifications".}
\lref\AsMo{P.~ Aspinwall and D.~ Morrison, ``Topological Field Theory
and Rational Curves", Commun. Math. Phys. {\bf 151} (1993) 245,
hep-th/9110048.}
\lref\atising{M.F. Atiyah and I.M. Singer, 
``The index of elliptic operators I,'' Ann. Math. {\bf 87}(1968)484;
M. F. Atiyah and G.B. Segal, 
``The index of elliptic operators II,'' Ann. Math. {\bf 87}(1968)531;
M.F. Atiyah and I.M. Singer, 
``The index of elliptic operators III,'' Ann. Math. {\bf 87}(1968)546;
``The index of elliptic operators IV,'' Ann. Math. {\bf 93}(1968)119}
\lref\AHS{M.~ Atiyah, N.~ Hitchin and I.~ Singer, ``Self-Duality in Four-Dimensional
Riemannian Geometry", Proc. Royal Soc. (London) {\bf A362} (1978) 425-461.}
\lref\AtBoym{M. F.~ Atiyah and R. ~ Bott, ``The Yang-Mills Equations Over Riemann
Surfaces,'' Phil. Trans. Roy. Soc. London {\bf A308} (1982) 523.}
\lref\AtBomm{M. F.~ Atiyah and R.~ Bott, ``The Moment Map And Equivariant
Cohomology", Topology {\bf 23} (1984) 1.}
\lref\atiyahsinger{M.F.~ Atiyah and I.M.~ Singer,  ``Dirac operators Coupled to
Vector Potentials,''  Proc. Natl. Acad. Sci. {\bf 81} (1984) 2597.}
\lref\atiythrfr{M. F.~ Atiyah, ``New invariants for three and four dimensional mnifolds,''
in the {\it Symposium on the Mathematical Heritage of Hermann Weyl}, R. Wells et.
al. eds. }
\lref\atiyax{M. F.~ Atiyah, ``Topological Field Theory", IHES Publ. Math. {\bf 68} (1989) 175.}
\lref\AtJe{M.F.~ Atiyah and L.~ Jeffrey, ``Topological Lagrangians and
Cohomology,'' Jour. Geom.  Phys. {\bf 7} (1990) 119.}
\lref\BaVi{O.~ Babelon and C.-M.~ Viallet,
``The Geometrical Interpretation of the Faddeev-Popov
Determinant", Phys. Lett. {\bf  85B} (1979) 246.}
\lref\bazrev{J. Baez, Two-dimensional QCD and strings, hep-th/9309067}
\lref\BaTa{J.~ Baez and W.~ Taylor, "Strings and Two-dimensional
QCD for Finite N", MIT-CTP-2266, hep-th/9401041.}
\lref\demeterfi{J.L.F.~ Barbon and K.~ Demetrfi, ``Effective Hamiltonians for
$1/N$ Expansion in Two-Dimensional QCD", hep-th/9406046.}
\lref\Ba{I.~ Bars, ``QCD and Strings in 2d", Talk given at International
Conference on Strings, Berkeley, CA, 24-29 May 1993, hep-th/9312018.}
\lref\BaHa{I.~  Bars and A.~  Hanson, ``Quarks at the Ends of the String",
Phys. Rev. {\bf D13} (1976)1744;
``A Quantum String Theory of Hadrons and its Relation to Quantum Chromodynamics
in Two Dimensions", Nucl. Phys. {\bf B111} (1976) 413.}
\lref\BaSiii{L.~ Baulieu and I.~ Singer, ``Topological Yang-Mills Symmetry",
Nucl. Phys. B. Proc. Suppl.  {\bf 5B} (1988) 12.}
\lref\BaSi{L.~ Baulieu and I.~ Singer, ``The Topological Sigma Model",
Commun. Math. Phys. {\bf 125} (1989) 227-237.}
\lref\BaSii{L.~ Baulieu and I.~ Singer, ``Conformally Invariant Gauge Fixed Actions
for 2-D Topological Gravity",  Commun. Math. Phys. {\bf 135}  (1991) 253-237.  }
\lref\BaTh{L.~ Baulieu and J.~ Thierry-Mieg, ``Algebraic Structure
of Quantum Gravity and Classification of Gravitational
Anomalies", Phys. Lett. {\bf 145B} (1989) 53.}
\lref\becchi{C.M.~ Becchi, R.~ Collina, C.~ Imbimbo, 
``A Functional and Lagrangian Formulation of Two
Dimensional Topological Gravity",  hep-th/9406096.}
\lref\BeGi{A.~ Beilinson and V.~ Ginzburg, ``Infinitesimal structure of
Moduli Spaces of G-Bundles,'' International Mathematics Research
Notices 1992, No. 4. }
\lref\bgv{N.~ Berline, E.~ Getzler, and M.~ Vergne, {\it Heat 
Kernels and Dirac Operators}, Springer 1992} 
\lref\bers{L.~ Bers, ``Finite Dimensional Teichmuller Spaces and 
Generalizations,''  Bull. of 
the AMS, {\bf 5}(1981) 131}
\lref\BeCeOoVa{M.~ Bershadsky, S.~ Cecotti, H.~ Ooguri and C.~ Vafa,
``Kodaira Spencer Theory of Gravity and Exact Results for Quantum String Amplitudes", HUTP-93-A025, hepth/9309140. }
\lref\BeEd{I.~ Berstein and A.L.~ Edmonds, ``On the Classification of
Generic Branched Coverings of Surfaces", Illinois Journal
of Mathematics, Vol {\bf 28}, number {\bf 1}, (1984 ).}
\lref\Bir{J. S.~ Birman, {\it Braids, Links and Mapping Class groups},
pages 11, 12, Princeton University Press, 1975.}
\lref\bbrt{D.~ Birmingham, M.~ Blau, M.~ Rakowski, and 
G.~ Thompson,  ``Topological Field Theories,'' Phys. Rep. {\bf 209} (1991) 129.}
\lref\Bis{J.-M.~ Bismut, ``Localization Formulas, Superconnections,
and The Index Theorem For Families,'' Comm. Math. Phys. {\bf 103} (1986) 127.}
\lref\Bl{M. Blau, ``The Mathai-Quillen Formalism and Topological
Field Theory", Notes of Lectures given at the Karpacz Winter School on
`Infinite Dimensional Geometry in Physics', Karpacz, Poland, Feb 17-29, 1992;
J. Geom and Phys. {\bf 11} (1991) 129.}
\lref\BlThlgt{M.~ Blau and G.~ Thompson, ``Lectures on 2d Gauge
Theories: Topological Aspects and Path Integral Techniques", Presented at the
Summer School in Hogh Energy Physics and Cosmology, Trieste, Italy, 14 Jun - 30 Jul
1993, hep-th/9310144.}
\lref\BlThqym{M.~ Blau and G.~ Thompson, ``Quantum Yang-Mills Theory On
Arbitrary Surfaces",  Int. J. Mod. Phys. {\bf A7} (1992) 3781.}
\lref\blauthom{M.~ Blau and G.~ Thompson, 
``N=2 Topological Gauge Theory, the Euler Characteristic
of Moduli Spaces, and the Casson Invariant", Commun. Math. Phys. {\bf 152} (1993) 41, 
hep-th/9112012.}
\lref\BogSh{ 
N. N.~ Bogoliubov D.V.~ Shirkov, {\it Introduction to the Theory of Quantized Fields}, 
Interscience Publishers, Inc., New York}
\lref\Bo{R.~ Bott, ``On the Chern-Weil Homomorphism and 
the Continuous Cohomology of Lie Groups,'' Adv. in Math.  
{\bf 11} (1973) 289-303.}
\lref\BoTu{R.~ Bott and L.~ Tu, {\it Differential Forms in
Algebraic Topology}, Springer Verlag, New York, 1982.}
\lref\BoHuMaMi{C.P.~ Boyer, J.C.~ Hurtubise, B.M.~ Mann and R.J.~ Milgram,
``The Topology of the Space of Rational Maps into generalized Flag Manifolds",
to appear in Acta. Math.}
\lref\Braam{P.~ Braam, ``Floer Homology Groups for Homology Three-Spheres",
Adv. in Math. {\bf 88} (1991) 131-144.}
\lref\Br{N.~ Bralic, "Exact Computation of Loop Averages in Two-Dimensional
Yang-Mills Theory", Phys. Rev. {\bf D22} (1980) 3090.}
\lref\BRSTrefs{C.~ Becci, A.~ Rouet and R.~ Stora,
``Renormalization of the Abelian Higgs-Kibble Model",
Comm. Math. Phys. {\bf 42} (1975) 127;
``Renormalization of Gauge Theories", \hfill\break
I.V.~ Tyutin, ``Gauge Invariance in Field Theory and in Statistical
Physics in the Operator Formalism", Lebedev preprint FIAN
No. 39 (1975), unpublished\hfill\break
J.~ Fisch, M.~ Henneaux, J.~ Stasheff and C.~ Teitelboim,
``Existence, Uniqueness and Cohomology of the Classical BRST
Charge with Ghosts of Ghosts", Comm. Math. Phys.
{\bf 120} (1989) 379.}
\lref\BrSchSch{L.~ Brink, J.~ Schwarz and J.~ Scherk, ``Supersymmetric Yang-Mills
Theories", Nucl. Phys. {\bf B121} (1977) 77.}
\lref\CaDeGrPa{P.~ Candelas, X.C.~ de la Ossa, P.S.~ Green and L.~ Parkes,
"A Pair of Calabi-Yau Manifolds as an Exactly Solvable Superconformal Theory",
Nucl. Phys. {\bf B359}  (1991) 21.}
\lref\ccd{Y.~ Choquet-Bruhat, C.~ DeWitt-Morette, and M.~ Dillard-Bleick,
{\it Analysis, Manifolds, and Physics}, North Holland, 1977.}
\lref\chen{J.Q.~ Chen,
{\it Group representation Theory for Physicists}, World Scientific, 1987, chapter 7}
\lref\Co{J.H.~ Conway, {\it Functions of One Complex Variable}, Springer.}
\lref\CMROLD{S.~ Cordes, G.~ Moore, and S.~ Ramgoolam,
``Large N 2D Yang-Mills Theory and Topological String Theory",
hep-th/9402107}
\lref\CMRPI{S.~ Cordes, G.~ Moore, and S.~ Ramgoolam,
Proceedings of the 1994 Trieste Spring School on 
Superstrings }
\lref\CMRPII{S.~ Cordes, G.~ Moore, and S.~ Ramgoolam,
Proceedings of the 1994 Les Houches 
school on Fluctuating Geometries. }
\lref\CMRIII{S.~ Cordes, G.~ Moore, and S.~ Ramgoolam,
in Les Houches Session LXII on http://xxx.lanl.gov/lh94.} 
\lref\CreTay{ M. Crescimanno and W. Taylor, 
`` Large N phases of Chiral  $QCD_2$, '' hep-th
9408115}
\lref\danvia{M.~ Daniel and C.M.~ Viallet, 
``The Geometrical Setting of Gauge Theories of the
Yang-Mills Type,'' Rev. Mod. Phys. {\bf 52} (1980) 174}
\lref\wadia{A.~ Dhar, P.~ Lakdawala, G.~ Mandal and S.~ Wadia,
``String Field Theory of Two-Dimensional QCD on a Cylinder: A Realization of
$W ( \infty )$ Current Algebra", Phys. Lett. {\bf B350} (1994); hep-th/9403050.}
\lref\DiInMoPl{R. Dijkgraaf, K. Intrilligator, G. Moore, and R. Plesser, unpublished.}
\lref\DiRu{R.~ Dijkgraaf and R.~ Rudd, unpublished.}
\lref\DiVVTrieste{R.~ Dijkgraaf, H.~ Verlinde and E.~ Verlinde, ``Notes on Topological
String Theory and 2-d  Quantum Gravity", lectures given at {\it Spring School
on Strings and Quantum Gravity}, Trieste, Italy, Apr 24 - May 2, 1990.}
\lref\DiWi{R.~ Dijkgraaf and E.~ Witten, ``Mean Field Theory,
Topological Field Theory, and Multi-Matrix Models", Nucl. Phys. {\bf B342} (1990) 486.}
\lref\Di{J.~ Distler, ``$2-D$ Quantum Gravity, Topological Field Theory,
and the Multicritical Matrix Models,'' Nucl. Phys. {\bf B342} (1990) 523.}
\lref\DiSeWeWi{M.~ Dine, N.~ Seiberg, X.~ Wen and E.~ Witten, ``Nonperturbative
Effects on the STring World Sheet", Nucl. Phys. {\bf B278} (1986) 769,
Nucl. Phys. {\bf B289} (1987) 319.}
\lref\DoApp{S.~ Donaldson, ``An Application of Gauge Theory to Four Dimensional
Topology", J. Diff. Geom. {\bf 18} (1983) 279.}
\lref\Don{S.~ Donaldson, ``Polynomial Invariants for Smooth Four Manifolds", Topology
{\bf 29} (1990) 237.}
\lref\DoKro{S.K.~ Donaldson and P.B.~ Kronheimer, {\it The Geometry of Four-Manifolds},
Clarendon Press, Oxford, 1990.}
\lref\dgcrg{M.R.~ Douglas, ``Conformal Field Theory Techniques for Large $N$
Group Theory,'' hep-th/9303159;
M.R.~ Douglas, ``Conformal Field Theory Techniques in Large $N$ Yang-Mills Theory",
hep-th/9311130,to be published in the proceedings of the May 1993 Carg\`ese
workshop on Strings,  Conformal Models and Topological Field Theories.}
\lref\Dosc{M. R.~ Douglas, ``Some Comments on QCD String,''  RU-94-09,
hep-th/9401073, to appear in Proceedings of the Strings '93 Berkeley conference.} 
\lref\Docft{M.R.~ Douglas, ``Conformal Field Theory Techniques in
Large N Yang-Mills Theory,'' hep-th 9303159. }
\lref\DGLSS{M.R. Douglas, ``Large $N$ Gauge Theory-
Expansions and Transitions,'' hep-th/9409098.}
\lref\DoKa{M.~ Douglas and V.A.~ Kazakov, ``Large $N$ Phase Transition in Continuum
QCD$_2$", Phys. Lett. {\bf B319} (1993) 219, hep-th/9305047 }
\lref\DrZu{ J.M.~ Drouffe and J. B.~ Zuber, ``Strong Coupling and Mean Field Methods
in Lattice Gauge theories", Phys. Rept {\bf 102} (1983) 1. }
\lref\Du{J. J.~ Duistermaat and G. J.~ Heckman, ``On The Variation
In The Cohomology In The Symplectic Form Of The Reduced Phase Space,''
Invent. Math. {\bf 69} (1982) 259.}
\lref\Ed{A. L.~ Edmonds, ``Deformation of Maps to
Branched Coverings in Dimension Two", Annals of Mathematics 
{\bf 110} (1979), 113-125. }
\lref\egh{T.~ Eguchi, P.B.~ Gilkey, and A.J.~ Hanson, 
``Gravitation, Gauge Theories, and  Differential Geometry",
 Phys. Rep. {\bf 66}(1980) 214.}
\lref\echikann{T. Eguchi, H. Kanno, Y. Yamada and S.-K. Yang,
"Topological Strings, Flat Coordinates and Gravitational Descendants"
hep-th/9302048.}
\lref\Ez{C. L.~ Ezell, ``Branch Point Structure of Covering Maps onto
Nonorientable Surfaces", Transactions of the American Mathematical
society, vol {\bf 243}, 1978. }
\lref\Fi{D.~ Fine, ``Quantum Yang-Mills On The Two-Sphere", Commun. Math.
Phys. {\bf 134} (1990) 273;
``Quantum Yang-Mills On A Riemann Surface,''
Commun. Math. Phys. {\bf 140} (1991) 321.}
\lref\FlSympi{A.~ Floer, ``The Unregularized Gradient Flow of the
Symplectic Action", Comm. Pure Appl. Math. {\bf 41} (1988) 775-813.}
\lref\FlSympii{A.~ Floer, ``Symplectic Fixed-Points and Holomorphic Spheres",
Commun. Math. Phys. {\bf 120} (1989) 575-611.}
\lref\FlWitt{A.~ Floer, ``Witten's Complex and Infinite Dimensional
Morse Theory", J. Diff. Geom. {\bf 30} (1989) 207-221.}
\lref\FlThreeD{A.~ Floer, ``An Instanton-Invariant for 3-Manifolds",
Commun. Math. Phys. {\bf 118} (1988) 215-239.}
\lref\Fo{R.~ Forman, ``Small Volume Limits of $2-d$ Yang-Mills,'' preprint,
Rice University (1991).}
\lref\Freed{D.~ Freed and K.~ Uhlenbeck, {\it Instantons and 4-Manifolds}, 
Springer 1984.}
\lref\FrieWind{D.~ Friedan and P.~ Windey, ``Supersymmetric Derivation
of the Atiyah-Singer Index and the Chiral Anomaly", Nucl. Phys. {\bf B235}
(1984) 395.}
\lref\Fu{W.~ Fulton, ``Hurwitz Schemes and Irreducibility
of Moduli of Algebraic Curves", Annals of Math. {\bf 90} (1969) 542.}
\lref\ganor{O. Ganor, J. Sonnenschein and S. Yankielowicz,
"The String Theory Approach to Generalized 2D Yang-Mills Theory",
hep-th/9407114}
\lref\getzler{E.~ Getzler, ``Two-Dimensional Topological Gravity 
and Equivariant Cohomology,'' hep-th/9305013.}
\lref\GlJa{See, e.g., J.~ Glimm and A.~ Jaffe, {\it Quantum Physics},
Springer 1981.}
\lref\Grtalk{D.~ Gross, ``Some New/Old Approaches to QCD,''
Int. Workshop on String Theory, Quantum Gravity and the Unification of Fundamental
Interactions, Rome, Italy, 21-26 Sept. 1992, hep-th/9212148.}
\lref\GrHa{P.~ Griffiths and J.~ Harris, {\it Principles of Algebraic geometry},
p. 445, J.Wiley and Sons, 1978. }
\lref\Gr{M.~ Gromov, ``Pseudo-Holomorphic Curves in Symplectic Manifolds",
Invent. Math. {\bf 82} (1985) 307-347.}
\lref\GrTa{D.~ Gross, ``Two Dimensional QCD as a String Theory",
Nucl. Phys. {\bf B400}  (1993) 161, hep-th/9212149.;
D.~ Gross and W.~ Taylor, ``Two-dimensional QCD is a String Theory",
Nucl. Phys. {\bf  B400}  (1993) 181, hep-th/93011068;
D.~ Gross and W.~ Taylor, ``Twists and Loops in the String
Theory of Two Dimensional QCD", Nucl. Phys. {\bf 403} (1993) 395, hep-th/9303046.}
\lref\grssmatyt{D. Gross and A. Matytsin, ``Instanton Induced Large $N$ Phase
Transitions in Two and Four Dimensional QCD,''
hep-th/9404004}
\lref\GrMat{ D. Gross and A. Matytsin, `` Some properties of Large $N$ Two Dimensional Yang
Mills Theory,'' hep-th 9410054. } 
\lref\GrTatalk{D.J.~ Gross and W.~ Taylor, ``Two-Dimensional QCD
and Strings", Talk presented at  Int. Conf. on Strings '93, Berkeley CA,  24-29 May 1993;
hep-th/9311072.}
\lref\GrWi{D.~ Gross and E.~ Witten, ``A Possible Third Order Phase Transition on the
Large $N$ Lattice Gauge Theory", Phys. Rev. {\bf D21} (1980) 446.}
\lref\GrKiSe{L. ~ Gross, C.~ King and A.~ Sengupta, ``Two-Dimensional Yang-Mills via
Stochastic Differential Equations", Ann. of Phys. {\bf 194} (1989) 65.}
\lref\hamermesh{M.~ Hamermesh, {\it Group Theory and its 
applications to physical problems}, Addison-Wesley 1962}
\lref\Guest{M.A.~ Guest, ``Topology of the Space of Absolute Minima of the Energy
Functional", Amer. J. Math. {\bf 106} (1984) 21-42.}
\lref\harstrom{J.~ Harvey and A.~ Strominger, 
``String Theory and the Donaldson Polynomial",  Commun. Math. Phys. {\bf 151} (1993) 221, hep-th/9108020.}
\lref\HaMo{ J.~ Harris and I.~ Morrison, ``Slopes of Effective Divisors on the
Moduli Space of Stable Curves", Invent. Math. {\bf 99}  (1990) 321-355.} 
\lref\HaMu{ J.~ Harris and D.~ Mumford, ``On the Kodaira Dimension of
the Moduli Space of Curves", Invent. Math. {\bf  67} (1982) 23-86.}
\lref\Hi{F.~ Hirzebruch, {\it Topological Methods in Algebraic Geometry},
Springer Verlag, 1966.}
\lref\Hitchin{N.~ Hitchin, ``The Geometry and Topology of 
Moduli Spaces", in {\it Global Geometry and  Mathematical Physics},
LNM 1451, L.~ Alvarez-Gaum\'e et. al.,  eds.}
\lref\Hora{P.~ Horava, ``Topological Strings and QCD in Two Dimensions",
to appear in Proc. of The Cargese Workshop,1993; hep-th/9311156.}
\lref\Hott{ P.~ Horava, ``Two-Dimensional String Theory and the
Topological Torus", Nucl. Phys. {\bf B386} ( 1992 ) 383-404.}
\lref\Hori{E.~ Horikawa, ``On Deformations of Holomorphic Maps,
I, II, III", J. Math. Soc. Japan, {\bf 25} (1973) 647; ibid {\bf 26}
(1974) 372; Math. Ann. {\bf 222} (1976) 275.}
\lref\Hurt{J.C.~ Hurtubise, ``Holomorphic Maps of a Riemann Surface into a Flag Manifold",
Mc Gill preprint, 1994.}
\lref\husemoller{D.~ Husemoller, {\it Fiber Bundles}, 3$^{{\rm rd}}$ edition, Springer, 1993.}
\lref\Ing{R.E.  Ingram , ``Some Characters of the Symmetric Group",
Proc. Amer. Math. Soc. (1950), 358-369}
\lref\isham{C.J.~ Isham, {\it Modern Differential Geometry  for Physicists},
World Scientific, 1989}
\lref\Kal{J.~ Kalkman, ``BRST Model for Equivariant Cohomology
and Representatives for the Equivariant Thom Class", Commun. Math.
Phys. {\bf 153} (1993) 447;
``BRST Model Applied to Symplectic Geometry", PRINT-93-0637 (Utrecht),
hep-th/9308132.}
\lref\Kan{H.~ Kanno, ``Weil Algebra Structure and Geometrical
Meaning of BRST Transformation in Topological Quantum Field Theory,''
Z. Phys. {\bf C43}  (1989)  477.}
\lref\KaNi{T.~ Karki and A.J.~ Niemi, ``On the Duistermaat-Heckman
Integration Formula and Integrable Models", hep-th/9402041.}
\lref\KaKo{V.A.~ Kazakov and I.~ Kostov, ``Non-linear Strings in
Two-Dimensional $U(\infty)$ Gauge Theory,''  Nucl. Phys. {\bf B176}
(1980) 199-215;
``Computation of the Wilson Loop Functional in Two-Dimensional $U ( \infty )$
Lattice Gauge Theory", Phys. Lett. {\bf B105} (1981) 453;
V.A.~ Kazakov, ``Wilson Loop Average for an Arbitrary Contour in
Two-Dimensional $U(N)$ Gauge Theory", Nuc. Phys. {\bf B179} (1981) 283-292.}
\lref\Kazkos{V.A. Kazakov, ``$U ( \infty )$ Lattice Gauge Theory as a Free Lattice String
Theory", Phys. Lett. {\bf B128} (1983) 316;
V.I.~ Kostov, ``Multicolour QCD in Terms of Random Surfaces",
Phys. Lett. {\bf B138} (1984) 191.}
\lref\Kazwynt{ ``Large $N$ Phase Transition In The Heat Kernel On The $U(N)$ Group,''
V.A. Kazakov and Thomas Wynter, hep-th 9408029. } 
\lref\kazrev{V. Kazakov, Strings, loops, knots and gauge fields,
hep-th/9308135} 
\lref\Ki{F.~ Kirwan, {\it Cohomology of Quotients In Symplectic
And Algebraic Geometry},  Princeton University Press.}
\lref\Kirwan{F.C.~ Kirwan, ``On Spaces of Maps from Riemann Surfaces to
Grassmannians and Applications to the Cohomology of Vector Bundles",Ark. Math.
{\bf 24} (1986) 221-275.}
\lref\jeffkir{L.~ Jeffrey and F.~ Kirwan, ``Localization for Nonabelian 
Group Actions",  alg-geom/9307001.}
\lref\Kn{F.~ Knudsen, ``The Projectivity of the Moduli Space of Stable
Curves",  Math. Scand. {\bf 52} (1983) 161.}
\lref\kobnom{S.~ Kobayashi and K.~ Nomizu, {\it Foundations of 
Differential Geometry I,II}, Interscience Publishers, New York, 1963.}
\lref\Kon{M.~ Kontsevich, ``Intersection Theory On The Moduli Space
Of Curves And The Matrix Airy Function", Commun. Math. Phys. {\bf 147} (1992) 1.}
\lref\kontsevichi{M. Kontsevich and Yu. Manin, 
``Gromov-Witten Classes, Quantum Cohomology, and 
Enumerative Geometry,'' hep-th/9402147; MPI preprint.}
\lref\kontsevichii{M. Kontsevich, 
``Enumeration of rational curves via torus actions,'' hep-th/9405035,
MPI preprint}
\lref\Kos{I.K.~ Kostov, ``Continuum QCD2 in Terms of Discretized Random
Surfaces with Local Weights", Saclay-SPht-93-050; hep-th/9306110.}
\lref\kronmrow{P.B. Kronheimer and T.S. Mrowka, 
``The genus of embedded surfaces in the projective plane,''
preprint.}
\lref\Ku{D.~ Kutasov, ``Geometry on the Space of Conformal Field
Theories and Contact Terms,'' Phys. Lett. {\bf 220B} (1989)153.}
\lref\LaPeWi{ J.M.F.~  Labastida, M.~ Pernici,and E.~ Witten ``Topological
Gravity in Two Dimensions,'' Nucl. Phys. {\bf B310} (1988) 611.}
\lref\lvw{W.~ Lerche, C.~ Vafa , and  N.P.~ Warner , 
``Chiral Rings in $N=2$ Superconformal Theories", Nucl.Phys. {\bf B324} (1989) 427.}
\lref\lz{B.~ Lian and G.~ Zuckerman, ``New Perspectives on the BRST Algebraic STructure of String Theory", Commun. Math. Phys. {\bf 154} (1993) 613; hep-th/9111072.}
\lref\lossev{A. Lossev, ``Descendants constructed from 
matter fields in
topological Landau-Ginzburg theories coupled to 
topological gravity,''
 Theor. Math. Phys,95,N2,(1993),p.317-327, hep-th/9211090;
Structures of K.Saito theory of primitive form in topological theories
coupled to topological gravity, in {\it Integrable models and strings,}
Lecture notes in Physics, 436,
Springer Verlag,1994,
Proc. of the III Baltic RIM Conf.,Helsinki,  September 1993,13-17,
p.172-193; A.Losev and I.Polyubin, 
``On Connection between Topological 
Landau-Ginzburg Gravity and Integrable Systems,'' 
ITEP Preprint PRINT-930412(May 1993), 
submitted to IJMP, hep-th/9305079.}
\lref\manin{Yu. I. Manin, {\it Quantized Fields and Complex 
Geometry}, Springer}
\lref \Ma{ W.S.~ Massey, {\it A Basic Course in Algebraic Topology},
Springer-Verlag, 1991.}
\lref\MaQu{V.~ Mathai and D.~ Quillen,
``Superconnections, Thom Classes, and Equivariant Differential Forms",
Topology {\bf 25} (1986) 85.}
\lref\MDSa{D.~ McDuff and  D.~ Salamon, ``$J$-Holomorphic
Curves and Quantum Cohomology", preprint, 1994.}
\lref\Mig{A.~ Migdal, ``Recursion Relations in Gauge Theories",
Zh. Eksp. Teor. Fiz. {\bf 69} (1975) 810 (Sov. Phys. Jetp. {\bf 42} 413).}
\lref\MitH{A. A.~ Migdal, `` Loop Equations and 1/N Expansion",
Phys. Rep. {\bf 102} (1983)199-290;
G.~ `t Hooft, ``A Planar Diagram Theory for Strong
Interactions, Nucl. Phys. {\bf B72} (1974) 461.}
\lref\Mij{M.~ Mijayima,
``On the Existence of Kuranishi Family for Deformations of Holomorphic
Maps", Science Rep. Kagoshima Univ., {\bf 27} (1978) 43.}
\lref\miller{E.~ Miller, ``The Homology of the Mapping 
Class Group", J. Diff. Geom. {\bf 24} (1986)1}
\lref\milnor{J.W. Milnor and J. D. Stasheff, {\it Characteristic 
Classes}, Princeton, 1974.}
\lref\Min{J. Minahan, ``Summing over Inequivalent Maps in the String
Theory of QCD", Phys. Rev. {\bf D47} (1993) 3430. }
\lref\Minclas{ J.~ Minahan and A.~ Polychronakos, ``Classical Solutions for two
dimensional QCD on the sphere", Nucl. Phys. {\bf B422} (1994) 172; hep-th/9309119 }
\lref\MiPoetd{J.~ Minahan and A.~ Polychronakos, ``Equivalence of Two
Dimensional QCD and the $c=1$ Matrix Model", Phys. Lett. {\bf B312} (1993) 155; hep-th/9303153.}
\lref\MiPoifs{J.~ Minahan and A.~ Polychronakos, "Interacting Fermion
Systems from Two Dimensional QCD", Phys. Lett. {\bf 326} (1994) 288;
hep-th/9309044.}
\lref\MorComp{D.~ Morrison, ``Compactifications of Moduli Spaces Inspired by
Mirror Symmetry", alg-geom/9304007.}
\lref\MoSo{ D.~ Montano and J.~ Sonnenschein, ``Topological
Strings", Nuc. Phys. {\bf B313} (1989) 258}
\lref\Mofad{G.~ Moore, ``Finite in All Directions", hep-th/9305139.}
\lref\mooricm{G. Moore, ``2D Yang-Mills Theory and 
Topological Field Theory,'' hep-th/9409044, to appear in the proceedings of  ICM94}
\lref\mrsb{G.~ Moore and N.~ Seiberg,
 ``Polynomial Equations for Rational Conformal Field
Theories,'' Phys. Lett.  {\bf 212B}(1988)451; 
``Classical and Quantum Conformal Field Theory",
Commun. Math. Phys. {\bf 123}(1989)177;
``Lectures on Rational Conformal Field Theory",
 in {\it Strings 90}, the proceedings
of the 1990 Trieste Spring School on Superstrings.}
\lref\MoSeSt{G.~ Moore, N.~ Seiberg, and M.~ Staudacher,
``From Loops to States in $2D$ Quantum Gravity",  Nucl. Phys.
{\bf B362} (1991) 665}
\lref\Morita{S.~  Morita, ``Characteristic Classes of Surface Bundles", Inv. Math. {\bf 90}
(1987) 551.}\lref\morrpless{D. Morrison and R. Plesser, 
``Quantum Cohomology and Mirror Symmetry 
for a Hypersurface in a Toric Variety,'' to appear.}
\lref\Mu{ D.~ Mumford, ``Towards an Enumerative Geometry
of the Moduli Space of Curves", in {\it Arithmetic and geometry}, M.~ Artin and
J.~ Tate eds., Birkhauser, 1983. }
\lref\MePe{ R.C.~ Myers and V.~ Periwal, ``Topological Gravity and
Moduli Space",  Nucl. Phys. {\bf B333} (1990) 536.}
\lref\nrswils{ S.G.~ Naculich, H.A.~ Riggs, H.J.~ Schnitzer, `The String Calculation of Wilson Loops in Two-Dimensional Yang-Mills Theory, '  hep-th/9406100.}
\lref\NaRiSc{S.G.~ Naculich, H. A.~ Riggs and H.G.~ Schnitzer, ``2D Yang Mills
Theories are String Theories,'' Mod. Phys. Lett. {\bf A8} (1993) 2223; hep-th/9305097.}
\lref\Na{M.~ Namba, ``Families of Meromorphic Functions
on Compact Riemann Surfaces", Lecture Notes in Mathematics,
Number 767, Springer Verlag, New York, 1979.}
\lref\nersessian{A. Nersessian, ``Equivariant Localization: 
BV-geometry and Supersymmetric Dynamics", 
hep-th/9310013}
\lref\niemi{A.J.~ Niemi and O.~ Tirkkonen, ``Equivariance, BRST, 
and Superspace,'' hep-th/9403126}
\lref\obzub{K.H.~ O'Brien and J.-B.~ Zuber, ``Strong Coupling Expansion of Large
$N$ QCD and Surfaces", Nucl. Phys. {\bf B253} (1985) 621.}
\lref\Okubo{ S.~ Okubo,  J. Math. Phys. {\bf 18} (1977)  2382 }
\lref\stora{S.~ Ouvry, R.~ Stora and P.~ Van Baal, 
``On the Algebraic Characterization of Witten's 
Topological Yang-Mills Theory",  Phys. Lett. {\bf 220B} (1989)159}
\lref\ogvf{H. Ooguri and C. Vafa, ``Self-Duality
and $N=2$ String Magic,'' Mod.Phys.Lett. {\bf A5} (1990) 1389-1398; ``Geometry of$N=2$ Strings,'' Nucl.Phys. {\bf B361}  (1991) 469-518.}
\lref\panz{S. Panzeri, ``The c=1 matrix models of two-dimensional Yang Mills Theories, ''
DFTT 36/93, 9307173 }
\lref\park{J.-S. Park, ``Holomorphic Yang-Mills theory on compact Kahler manifolds,'' hep-th/9305095; Nucl. Phys. {\bf B423} (1994) 559;
J.-S.~ Park, ``$N=2$ Topological Yang-Mills Theory on Compact K\"ahler
Surfaces", Commun. Math, Phys. {\bf 163} (1994) 113;
J.-S.~ Park, ``$N=2$ Topological Yang-Mills Theories and Donaldson Polynomials", hep-th/9404009.}
\lref\Part{ A.~ Partensky, J. Math. Phys., {\bf 13}, 1972, 1503 } 
\lref\Pe{ R. C.~ Penner, ``Perturbative Series and the Moduli Space
of Riemann Surfaces, J. Diff. Geo. {\bf 27} (1988) 35-53. }
\lref\PP{A. M.~ Perelomov and V. S.~ Popov, Sov. J. of Nuc. Phys. {\bf 3} (1966) 676}
\lref\Po{J.~ Polchinski, ``Strings and QCD?''
Talk presented at the Symposium on Black Holes,
Wormholes, Membranes and Superstrings,'' Houston,
1992; hep-th/9210045}
\lref\ampi{A.M. Polyakov, ``Gauge Fields as Rings of  Glue,'' Nucl. Phys. {\bf B164} (1980) 
 171.}
\lref\ampii{A.M. Polyakov, ``Quantum Geometry of 
Bosonic Strings,'' Phys. Lett. {\bf 103B}(1981)207}
\lref\Pofs{A.M~ Polyakov, ``Fine Structure of Strings",
Nucl. Phys. {\bf B268} (1986) 406.}
\lref\Pofp{A.M.~ Polyakov, ``A Few Projects in String Theory", PUPT-1394;
hep-th/9304146.}
\lref\PrSe{A.~ Pressley and G.~ Segal, {\it Loop Groups}, Oxford Math. Monographs,
Clarendon, NY, 1986.}
\lref\Piunikhin{S.~ Piunikhin, ``Quantum and Floer Cohomology have the
Same Ring Structure", hep-th/9401130.}
\lref\zuckerman{The following construction goes back to 
D.~ Quillen, ``Rational Homotopy Theory,'' Ann. of  Math. {\bf 90}  (1969) 205.
The application to the present case was developed with G.~ Zuckerman.}
\lref\rajeev{S.~ Rajeev, ``Quantum Hadrondynamics in Two-Dimensions",
preprint hep-th/9401115.}
\lref\Ra{S.~ Ramgoolam, ``Comment on Two-Dimensional $O(N)$ and $Sp(N)$
Yang Mills Theories as String Theories", Nuc. Phys. {\bf B418}, (1994) 30;
hep-th/9307085. }
\lref\Hcas{S.~ Ramgoolam,  Higher Casimir Perturbations of Two Dimensional
Yang Mills Theories, unpublished.} 
\lref\ramloops{S. Ramgoolam, to appear.} 
\lref\RuTi{Y.~ Ruan and G.~ Tian, ``Mathematical Theory of Quantum Cohomology",
preprint 1994.}
%
%
\lref\Rudd{R.~ Rudd, ``The String Partition Function for QCD on the Torus",
hep-th/9407176.}
\lref\Ru{B.~ Rusakov, ``Loop Averages And Partition Functions in $U(N)$
Gauge Theory On Two-Dimensional Manifolds", Mod. Phys. Lett. {\bf A5}
(1990) 693.}
\lref\Sadov{V.~ Sadov, ``On Equivalence of Floer's and Quantum Cohomology",
hep-th/9310153.}
\lref\Sa{D.~ Salamon, ``Morse Theory, the Conley Index and the Floer Homology",
Bull. AMS {\bf 22} (1990) 13-140.}
\lref\Schw{A.~ Schwarz, ``The Partition Function Of Degenerate Quadratic
Functional And Ray-Singer Invariants", Lett. Math. Phys. {\bf 2} (1978) 247.}
\lref\She{S.~ Shenker, "The Strength of Non-Perturbative Effects in String
Theory", in Carg/`ese Workshop on {\it Random Surfaces, Quantum Gravity
and Strings}, Garg`ese, France, May 28 - Jun 1 1990, eds. O.~ Alvarez,
E.~ Marinari, and P.~ Windey (Plenum, 1991)  }
\lref\Segal{G.~ Segal, ``The Topology of Rational Functions", Acta. Math. {\bf 143}
(1979) 39-72.}
\lref\segax{G. B.~ Segal, ``The Definition of Conformal Field 
Theory'', preprint, 1989.}
\lref\segallct{G.B.~ Segal, Lecture at Yale, March, 1992.}
\lref\Se{A.~ Sengupta, ``The Yang-Mills Measure For $S^2$,'' to appear in
J. Funct. Anal., ``Quantum Gauge Theory On Compact Surfaces,'' preprint.}
\lref\singgrib{I.~ Singer, ``Some Remarks on the Gribov Ambiguity",
Commun. Math Phys. {\bf 60} (1978) 7.}
\lref\sohnius{M.F. Sohnius, ``Introducing Supersymmetry,'' 
Phys. Rep. {\bf 128}(1985)39}
\lref\Sp{E. H.~ Spanier, {\it Algebraic Topology}, Mc. Graw-Hill, Inc. 1966.}
\lref\steenrod{N.~ Steenrod, {\it The topology of  Fiber Bundles}, Princeton Univ. Press} 
\lref\storai{R.~ Stora, ``Equivariant Cohomology and 
Cohomological Field Theories,'' lecture notes}
\lref\storaii{R.~ Stora, F.~ Thuillier, J.C.~ Wallet, 
First Carribean Spring School of Math. and Theor. Phys. May 30 (1993) } 
\lref\St{A.~ Strominger,  ``Loop Space Solution of Two-Dimensional
QCD", Phys. Lett. {\bf 101B} (1981) 271.}
\lref\taubes{C.H. Taubes, ``Seiberg-Witten invariants and 
symplectic forms,''  preprint}
\lref\PhaTay{ W.~ Taylor, ``Counting strings and Phase transitions in 2d QCD",
MIT-CTP-2297; hep-th/9404175}  
\lref\vanbaal{P. van Baal, ``An Introduction to Topological 
Yang-Mills Theory,'' Acta Physica Polonica, {\bf B21}(1990) 73}
\lref\VaTopMirr{C.~ Vafa, ``Topological Mirrors", in 
{\it Essays on Mirror Manifolds},   S-T Yau, ed. International 
Press 1992. hep-th/9111017.}
\lref\VaWi{C.~ Vafa and E.~ Witten, ``A Strong Coupling Test of $S$-Duality",
hep-th/9408074.}
\lref\Ve{G.~ Veneziano, ``Construction of a Crossing-Symmetric, Regge-Behaved
Amplitude for Linearly Rising Trajectories,'' Nuovo Cim. {\bf 57A} 190, (1968).}
\lref\DiVeVe{E.~ Verlinde and H.~ Verlinde , ``A Solution of Two-Dimensional Topological Quantum Gravity", Nucl. Phys. {\bf B348} (1991) 457,
R.~ Dijkgraaf, E.~ Verlinde and H.~ Verlinde, ``Loop Equations and Virasoro Constraints
in Nonperturbative 2d Quantum Gravity",  Nucl. Phys. {\bf B348} (1991) 435;
``Topological Strings in $d < 1$", Nucl. Phys. {\bf B352} (1991) 59;
``Notes on Topological String Theory and 2-d Quantum Gravity", in {\it String Theory
and Quantum Gravity}, Proc. Trieste Spring School,
April 24 - May 2,  1990 (World Scientific, Singapore, 1991).}
\lref\verlinde{E.~ Verlinde, ``Fusion Rules and Modular Transformations
in 2d Conformal Field Theory",  Nucl. Phys. {\bf B300} (1988) 360.}
\lref\Vi{J.~ Vick, {\it Homology Theory}, Academic Press, 1973.}
\lref\We{R.O.~ Wells, {\it Differential Analysis on Complex Manifolds},
Springer 1980}
\lref\WeBa{J.~ Wess and J.~ Bagger, {\it Supersymmetry and Supergravity},
Princeton University Press, Princeton,1983.} 
\lref\Wi{K.G.~ Wilson, ``Confinement of Quarks",
Phys. Rev. {\bf D10} (1974) 2445.}
\lref\dsb{E.~ Witten, ``Dynamical Breaking of Supersymmetry",
Nucl. Phys. {\bf B188} (1981) 513.}
\lref\trminus{E.~ Witten, ``Constraints on Supersymmetry  Breaking",
Nucl. Phys. {\bf B202} (1982) 253.}
\lref\ssymrs{E.~ Witten, ``Supersymmetry and Morse Theory",
J. Diff. Geom. {\bf 17} (1982) 661.}
\lref\donaldson{E.~ Witten, ``Topological Quantum Field Theory",
Commun. Math. Phys. {\bf 117} (1988) 353.}
\lref\Witsm{E.~ Witten, ``Topological sigma models ", Commun.
Math. Phys. {\bf 118}, 411-419 (1988). }
\lref\WitHS{E. Witten, ``The Search for Higher Symmetry in String Theory",
Phil. Trans. Royal Soc. London {\bf B320} (1989) 349-357.}
\lref\Witg{ E.~ Witten, ``Topological Gravity,'' Phys. Lett. {\bf 206B} (1988) 601.}
\lref\jonespoly{E.~ Witten, ``Quantum Field Theory and  the Jones Polynomial",
Commun.  Math. Phys. {\bf 121} (1989) 351.}
\lref\Witdgit{E.~ Witten, ``Two-dimensional Gravity and Intersection Theory on Moduli
Space", in {\it Cambridge 1990, Proceedings, Surveys in Differential geometry}, 243-310.}
\lref\Wiag{E.~ Witten, ``Algebraic Geometry Associated with Matrix Models
of Two-Dimensional Gravity", IASSNS-HEP-91/74.}
\lref\Winm{E.~ Witten, ``The N-Matrix Model and gauged WZW models",
Nucl. Phys. {\bf B371} (1992) 191.}
\lref\Witdgt{ E.~ Witten, ``On Quantum gauge theories in two dimensions,''
Commun. Math. Phys. {\bf  141}  (1991) 153.}
\lref\Witdgtr{E.~ Witten, ``Two Dimensional Gauge Theories Revisited",
J. Geom. Phys. {\bf G9} (1992) 303; hep-th/9204083.}
\lref\Witp{E.~ Witten, ``On the Structure of the Topological Phase
of Two Dimensional Gravity", Nucl. Phys. {\bf B340} (1990) 281}
\lref\Witr{E.~ Witten, ``Introduction to Cohomological Field Theories",
Lectures at Workshop on Topological Methods in Physics, Trieste, Italy,
Jun 11-25, 1990, Int. J. Mod. Phys. {\bf A6} (1991) 2775.}
\lref\wttnmirror{E.~ Witten, ``Mirror Manifolds and Topological Field Theory",
hep-th/9112056, in {\it Essays on Mirror Manifolds} International 
Press 1992.}
\lref\wittcs{E.~ Witten, ``3D Chern-Simons as Topological  Open String"; \- hep-th/9207094.}
\lref\wittphases{E.~ Witten, ``Phases of N=2 Theories in Two Dimensions", 
Nucl. Phys. {\bf B403} (1993) 159; hep-th/9301042.}
\lref\wittsusygt{E.~ Witten, ``Supersymmetric Gauge  Theory on a Four-Manifold",
hep-th/9403193.}
\lref\monfour{E. Witten, ``Monopoles and four-manifolds,'' 
hep-th/9411102}
\lref\Zelo{D. P.~ Zhelobenko, Translations of American Math. Monographs, {\bf 40} }

\ifstudvers
  
  {\obeylines\gdef\stripline#1^^M{\toks0{#1}\edef\next{\the\toks0}%
  \ifx\next\eMark\let\next=\endgroup\else\let\next=\stripline\fi\next}}
  \def\eMark{\endremove}
  \def\endremove{\empty}
  
\else
  
  \def\endremove{\vskip0.1truein\hrule height 2pt\vskip0.1truein}
\fi

%
\def\newsec#1{\global\advance\secno by1\message{(\the\secno. #1)}
\global\subsubsecno=0%
\global\subsecno=0\eqnres@t\let\s@csym\secsym\xdef\secn@m{\the\secno}\noindent
{\bf\hyperdef\hypernoname{section}{\the\secno}{\the\secno.} #1}%
\writetoca{{\string\hyperref{}{section}{\the\secno}{\the\secno.}} {#1}}%
\par\nobreak\medskip\nobreak}
\def\eqnres@t{\xdef\secsym{\the\secno.}\global\meqno=1\bigbreak\bigskip}
\def\sequentialequations{\def\eqnres@t{\bigbreak}}\xdef\secsym{}
\global\newcount\subsecno \global\subsecno=0
\def\subsec#1{\global\advance\subsecno by1\message{(\s@csym\the\subsecno. #1)}
\ifnum\lastpenalty>9000\else\bigbreak\fi
\global\subsubsecno=0%
\noindent{{\it\hyperdef\hypernoname{subsection}{\secn@m.\the\subsecno}%
{\secn@m.\the\subsecno.}} #1}\writetoca{\string\quad
{\string\hyperref{}{subsection}{\secn@m.\the\subsecno}{\secn@m.\the\subsecno.}}
{#1}}\par\nobreak\medskip\nobreak}
\global\newcount\subsubsecno \global\subsubsecno=0
\def\subsubsec#1{\global\advance\subsubsecno
by1\message{(\s@csym\the\subsecno.\the\subsubsecno. #1)}
\ifnum\lastpenalty>9000\else\bigbreak\fi
\noindent\quad{\it\hyperdef\hypernoname{subsubsection}%
{\secn@m.\the\subsecno.\the\subsubsecno}%
{\secn@m.\the\subsecno.\the\subsubsecno.} {#1}}\writetoca{\string\qquad 
{\string\hyperref{}{subsubsection}{\secn@m.\the\subsecno.\the\subsubsecno}
{\secn@m.\the\subsecno.\the\subsubsecno.}} {#1}}%
\par\nobreak\medskip\nobreak}

\def\subsubseclab#1{\DefWarn#1\xdef #1{\noexpand\hyperref{}{subsubsection}%
{\secn@m.\the\subsecno.\the\subsubsecno}%
{\secn@m.\the\subsecno.\the\subsubsecno}}%
\writedef{#1\leftbracket#1}\wrlabeL{#1=#1}}


\def\boxit#1{\vbox{\hrule\hbox{\vrule\kern8pt
\vbox{\hbox{\kern8pt}\hbox{\vbox{#1}}\hbox{\kern8pt}}
\kern8pt\vrule}\hrule}}
\def\mathboxit#1{\vbox{\hrule\hbox{\vrule\kern8pt\vbox{\kern8pt
\hbox{$\displaystyle #1$}\kern8pt}\kern8pt\vrule}\hrule}}



\def\Ad{{\mathop{\rm Ad}}}
\def\Aut{{\mathop {\rm Aut}}}
\def\bA{{\bf A}}

\def\bD{{\ID}}
\def\bF{{\IF}}
\def\bG{{\IG}}

\def\bO{{\IO}}

\def\cC{{\cal C}}

\def\cD{{\cal D}}
\def\CD {{\cal D}}

\def\CE {{\cal E}}

\def\CF {{\cal F}}

\def\cG{{\cal G}}
\def\cgp {c_\Gamma^l}
\def\cgm {c_\Gamma^r }
\def\cH{{\cal H}}
\def\CI  {{\cal I}}
\def\cL{{\cal L}}
\def\CL {{\cal L}}

\def\CM {{\cal M}}

\def\cO{{\cal O}}
\def\CO {{\cal O}}
\def\codim{{\mathop{\rm codim}}}
\def\cok{{\rm cok}}
\def\coker{{\mathop {\rm coker}}}

\def\CP {{\cal P }}
\def\CQ {{\cal Q }}
\def\CR {{\cal R}}
\def\CW{{\cal W}}
\def\CY{{\cal Y}}

\def\CS {{\cal S}}
\def\cT{{\cal T}}

\def\CV {{\cal V}}
\def\cV{{\cal V}}

\def\cW{{\cal W}}

\def\cZ {{\cal Z}}
\def\CZ {{\cal Z}}
\def\deg{{\mathop{\rm deg}}}
\def\diagonal{{\mathop{\rm Diag}}}
\def\diff{{\rm diff}}
\def\Diff{{\rm Diff}}
\def\dim{{\mathop{\rm dim}}}
\def\dual{{{}^\ast}}
\def\End{{\mathop{\rm End}}}
\def\eqdef{{\buildrel{\rm def}\over =}}
\def\equil{{\buildrel\sim\over=}}
\def\etwo { {e^2 \over 2} } 

\def\FRAME{{\rm FRAME}}
\def\G {\Gamma}
\def\O{\Omega}

\def\half{{\textstyle{1\over 2}}}
\def\hol{{\HOL}}
\def\HOL{{\cal H}}

\def\Hom{{\mathop {\rm Hom}}}

\def\IB{\relax{\rm I\kern-.18em B}}
\def\IC{\relax\hbox{$\inbar\kern-.3em{\rm C}$}}
\def\ID{\relax{\rm I\kern-.18em D}}
\def\IE{\relax{\rm I\kern-.18em E}}
\def\IF{\relax{\rm I\kern-.18em F}}
\def\IG{\relax\hbox{$\inbar\kern-.3em{\rm G}$}}
\def\IGa{\relax\hbox{${\rm I}\kern-.18em\Gamma$}}
\def\IH{\relax{\rm I\kern-.18em H}}
\def\II{\relax{\rm I\kern-.18em I}}
\def\IK{\relax{\rm I\kern-.18em K}}
\def\IL{\relax{\rm I\kern-.18em L}}
\def\IM{\relax{\rm I\kern-.18em M}}
\def\Im{{\mathop{\rm Im}}}
\def\ker{{\mathop{\rm ker}}}
\def\IN{\relax{\rm I\kern-.18em N}}
\def\inclusionmap#1{{\smash{
        \mathop{\hookrightarrow}\limits^{#1}}}}
\def\ind{{\mathop{\rm ind}}}
\def\Index{{\mathop{\rm ind}}}
\def\IO{\relax\hbox{$\inbar\kern-.3em{\rm O}$}}
\def\Iom{{\inbar\kern-3.00pt\Omega}}
\def\IOm{\relax\hbox{$\inbar\kern-3.00pt\Omega$}}

\def\IP{\relax{\rm I\kern-.18em P}}
\def\IPi{\relax\hbox{${\rm I}\kern-.18em\Pi$}}

\def\IQ{\relax\hbox{$\inbar\kern-.3em{\rm Q}$}}
\def\IR{\relax{\rm I\kern-.18em R}}

\def\ITh{\relax\hbox{$\inbar\kern-.3em\Theta$}}
\def\inbar{\,\vrule height1.5ex width.4pt depth0pt}

\font\cmss=cmss10 \font\cmsss=cmss10 at 7pt
\def\IZ{\relax\ifmmode\mathchoice
{\hbox{\cmss Z\kern-.4em Z}}{\hbox{\cmss Z\kern-.4em Z}}
{\lower.9pt\hbox{\cmsss Z\kern-.4em Z}}
{\lower1.2pt\hbox{\cmsss Z\kern-.4em Z}}\else{\cmss Z\kern-.4em
Z}\fi}
\def \jb{{\bar j}}
\def\ker{{\mathop{\rm ker}}}

\def\liebg{{{\rm Lie}~ \CG}}
\def\lieg{{\underline{\bf g}}}

\def\loclor{{\rm local\ lorentz}}

\def\log {{\rm log}}
\def\MAP{{\rm MAP}}
\def\mapdown#1{\Big\downarrow
        \rlap{$\vcenter{\hbox{$\scriptstyle#1$}}$}}
\def\mapleft#1{\smash{
        \mathop{\longleftarrow}\limits^{#1}}}

\def\mapright#1{\smash{
        \mathop{\longrightarrow}\limits^{#1}}}
\def\mapse#1{\searrow
        \rlap{$\vcenter{\hbox{$\scriptstyle#1$}}$}}
\def\mapsw#1{\swarrow
        \rlap{$\vcenter{\hbox{$\scriptstyle#1$}}$}}
\def\mapup#1{\Big\uparrow
        \rlap{$\vcenter{\hbox{$\scriptstyle#1$}}$}}
\def\MET{{\rm MET}}
\def\Met{{\rm MET}}
\def\min{{\mathop{\rm min}}}
\def\Odagger{
\pmatrix{- ( \delta_\mu{}^\nu \nabla_\gamma + J_\mu{}^\nu \nabla_\beta \epsilon^\beta{}_\gamma ) &
- G_{\mu\nu} \partial_\gamma f^\nu\cr
- \delta^\alpha{}_\gamma J^\nu{}_\mu \partial_\delta f^\mu \epsilon^{\delta\beta} &
{1\over2} ( \delta_\gamma{}^\alpha D^\beta + \delta_\gamma{}^\beta D^\alpha ) \cr}}
\def\ofp{{1\over{4\pi}}}

\def\p {\partial}
\def\pb{\bar{\partial}}
\def\Pfaff{{\rm Pfaff}}
\def\Pf{{\rm Pfaff}}

\def\psm{\psi_-}
\def\psp{\psi_+}

\def\Ram{{\mathop{\rm Ram}}}
\def\rank{{\mathop{\rm rank}}}
\def\Sc {\Sigma_T^c}
\def\sdtimes{\mathbin{\hbox{\hskip2pt\vrule
height 4.1pt depth -.3pt width .25pt\hskip-2pt$\times$}}}
\def\SG{{\Sigma_T}}
\def\sh{{\sqrt{h}}}
\def\Sh{{\Sigma_W}}
\def\sign{{\mathop{\rm sign}}}
\def\sst{\scriptscriptstyle}
\def\Span{{\mathop{\rm Span}}}

\def\ST{{\SG}}
\def\Sw{{\Sh}}
\def\Sym{{\mathop{\rm Sym}}}

\def\tbA{{\widehat{\bA}}}
\def\tbF{{\widehat{\IF}}}
\def\tbG{{\widehat{\bG}}}
\def\tbm {{\bar{\theta}^-}} 
\def\tbp {{\bar{\theta}^+}} 
\def\tbPi{{\widehat{\IPi}}}

\def\tcM{{\widetilde{\cal M}}}

\def\tcV{{\widetilde{\cal V}}}

\def\tm {\theta^-}

\def\tp {\theta^+} 
\def\tr{{\mathop{\rm Tr}}}

\def\vol{{\rm vol}}
\def\vskipabit{{\vskip0.25truein}}
\def\wb {\bar{w}}
 
\def\weylg{{\rm weyl}}
\def\Weyl{{\rm Weyl}}
\def\ymt{$YM_2$}
\def\zb {{\bar{z}}}
%
%
\def\exercise#1{\bgroup\narrower\footnotefont
\baselineskip\footskip\bigbreak
\hrule\medskip\nobreak\noindent {\bf Exercise}. {\it #1\/}\par\nobreak}
\def\endexercise{\medskip\nobreak\hrule\bigbreak\egroup}
%
%
%
\message{S-Tables Macro v1.0, ACS, TAMU (RANHELP@VENUS.TAMU.EDU)}
%
%
\newhelp\stablestylehelp{You must choose a style between 0 and 3.}%
\newhelp\stablelinehelp{You should not use special hrules when
stretching
a table.}%
\newhelp\stablesmultiplehelp{You have tried to place an S-Table
inside another
S-Table.  I would recommend not going on.}%
%
%
\newdimen\stablesthinline
\stablesthinline=0.4pt
\newdimen\stablesthickline
\stablesthickline=1pt
%
%
\newif\ifstablesborderthin
\stablesborderthinfalse
\newif\ifstablesinternalthin
\stablesinternalthintrue
\newif\ifstablesomit
\newif\ifstablemode
\newif\ifstablesright
\stablesrightfalse
%
%
\newdimen\stablesbaselineskip
\newdimen\stableslineskip
\newdimen\stableslineskiplimit
%
%
\newcount\stablesmode
\newcount\stableslines
\newcount\stablestemp
\stablestemp=3
\newcount\stablescount
\stablescount=0
\newcount\stableslinet
\stableslinet=0
%
%
%
\newcount\stablestyle
\stablestyle=0
%
%
\def\stablesleft{\quad\hfil}%
\def\stablesright{\hfil\quad}%
%
%
\catcode`\|=\active%
%
%
\newcount\stablestrutsize
\newbox\stablestrutbox
\setbox\stablestrutbox=\hbox{\vrule height10pt depth5pt width0pt}
\def\stablestrut{\relax\ifmmode%
                         \copy\stablestrutbox%
                       \else%
                         \unhcopy\stablestrutbox%
                       \fi}%
%
%
\newdimen\stablesborderwidth
\newdimen\stablesinternalwidth
\newdimen\stablesdummy
\newcount\stablesdummyc
\newif\ifstablesin
\stablesinfalse
%
%
\def\begintable{\stablestart%
  \stablemodetrue%
  \stablesadj%
  \halign%
  \stablesdef}%
\def\stablesadj{%
  \ifcase\stablestyle%
    \hbox to \hsize\bgroup\hss\vbox\bgroup%
  \or%
    \hbox to \hsize\bgroup\vbox\bgroup%
  \or%
    \hbox to \hsize\bgroup\hss\vbox\bgroup%
  \or%
    \hbox\bgroup\vbox\bgroup%
  \else%
    \errhelp=\stablestylehelp%
    \errmessage{Invalid style selected, using default}%
    \hbox to \hsize\bgroup\hss\vbox\bgroup%
  \fi}%
\def\stablesend{\egroup%
  \ifcase\stablestyle%
    \hss\egroup%
  \or%
    \hss\egroup%
  \or%
    \egroup%
  \or%
    \egroup%
  \else%
    \hss\egroup%
  \fi}%
\def\stablestart{%
  \ifstablesin%
    \errhelp=\stablesmultiplehelp%
    \errmessage{An S-Table cannot be placed within an S-Table!}%
  \fi
  \global\stablesintrue%
  \global\advance\stablescount by 1%
  \message{<S-Tables Generating Table \number\stablescount}%
  \begingroup%
  \stablestrutsize=\ht\stablestrutbox%
  \advance\stablestrutsize by \dp\stablestrutbox%
  \ifstablesborderthin%
    \stablesborderwidth=\stablesthinline%
  \else%
    \stablesborderwidth=\stablesthickline%
  \fi%
  \ifstablesinternalthin%
    \stablesinternalwidth=\stablesthinline%
  \else%
    \stablesinternalwidth=\stablesthickline%
  \fi%
  \tabskip=0pt%
  \stablesbaselineskip=\baselineskip%
  \stableslineskip=\lineskip%
  \stableslineskiplimit=\lineskiplimit%
  \offinterlineskip%
  \def\borderrule{\vrule width \stablesborderwidth}%
  \def\internalrule{\vrule width \stablesinternalwidth}%
  \def\thinline{\noalign{\hrule height \stablesthinline}}%
  \def\thickline{\noalign{\hrule height \stablesthickline}}%
  \def\trule{\omit\leaders\hrule height \stablesthinline\hfill}%
  \def\ttrule{\omit\leaders\hrule height \stablesthickline\hfill}%
  \def\tttrule##1{\omit\leaders\hrule height ##1\hfill}%
  \def\stablesel{&\omit\global\stablesmode=0%
    \global\advance\stableslines by 1\borderrule\hfil\cr}%
  \def\el{\stablesel&}%
  \def\elt{\stablesel\thinline&}%
  \def\eltt{\stablesel\thickline&}%
  \def\elttt##1{\stablesel\noalign{\hrule height ##1}&}%
  \def\elspec{&\omit\hfil\borderrule\cr\omit\borderrule&%
              \ifstablemode%
              \else%
                \errhelp=\stablelinehelp%
                \errmessage{Special ruling will not display
properly}%
              \fi}%
  \def\stmultispan##1{\mscount=##1 \loop\ifnum\mscount>3
\stspan\repeat}%
  \def\stspan{\span\omit \advance\mscount by -1}%
  \def\multicolumn##1{\omit\multiply\stablestemp by ##1%
     \stmultispan{\stablestemp}%
     \advance\stablesmode by ##1%
     \advance\stablesmode by -1%
     \stablestemp=3}%

\def\multirow##1{\stablesdummyc=##1\parindent=0pt\setbox0\hbox\bgroup%

    \aftergroup\emultirow\let\temp=}
  \def\emultirow{\setbox1\vbox to\stablesdummyc\stablestrutsize%
    {\hsize\wd0\vfil\box0\vfil}%
    \ht1=\ht\stablestrutbox%
    \dp1=\dp\stablestrutbox%
    \box1}%
  \def\stpar##1{\vtop\bgroup\hsize ##1%
     \baselineskip=\stablesbaselineskip%
     \lineskip=\stableslineskip%

\lineskiplimit=\stableslineskiplimit\bgroup\aftergroup\estpar\let\temp
=}%
  \def\estpar{\vskip 6pt\egroup}%
  \def\stparrow##1##2{\stablesdummy=##2%
     \setbox0=\vtop to ##1\stablestrutsize\bgroup%
     \hsize\stablesdummy%
     \baselineskip=\stablesbaselineskip%
     \lineskip=\stableslineskip%
     \lineskiplimit=\stableslineskiplimit%
     \bgroup\vfil\aftergroup\estparrow%
     \let\temp=}%
  \def\estparrow{\vfil\egroup%
     \ht0=\ht\stablestrutbox%
     \dp0=\dp\stablestrutbox%
     \wd0=\stablesdummy%
     \box0}%
  \def|{\global\advance\stablesmode by 1&&&}%
  \def\|{\global\advance\stablesmode by 1&\omit\vrule width 0pt%
         \hfil&&}%
  \def\vt{\global\advance\stablesmode by 1&\omit\vrule width
\stablesthinline%
          \hfil&&}%
  \def\vtt{\global\advance\stablesmode by 1&\omit\vrule width
\stablesthickline%
          \hfil&&}%
  \def\vttt##1{\global\advance\stablesmode by 1&\omit\vrule width
##1%
          \hfil&&}%
  \def\vtr{\global\advance\stablesmode by 1&\omit\hfil\vrule width%
           \stablesthinline&&}%
  \def\vttr{\global\advance\stablesmode by 1&\omit\hfil\vrule width%
            \stablesthickline&&}%
  \def\vtttr##1{\global\advance\stablesmode by 1&\omit\hfil\vrule
width ##1&&}%
  \stableslines=0%
  \stablesomitfalse}
\def\stablesdef{\bgroup\stablestrut\borderrule##\tabskip=0pt plus
1fil%
  &\stablesleft##\stablesright%

&##\ifstablesright\hfill\fi\internalrule\ifstablesright\else\hfill\fi%

  \tabskip 0pt&&##\hfil\tabskip=0pt plus 1fil%
  &\stablesleft##\stablesright%

&##\ifstablesright\hfill\fi\internalrule\ifstablesright\else\hfill\fi%

  \tabskip=0pt\cr%
  \ifstablesborderthin%
    \thinline%
  \else%
    \thickline%
  \fi&%
}%
\def\endtable{\advance\stableslines by 1\advance\stablesmode by 1%
   \message{- Rows: \number\stableslines, Columns:
\number\stablesmode>}%
   \stablesel%
   \ifstablesborderthin%
     \thinline%
   \else%
     \thickline%
   \fi%
   \egroup\stablesend%
\endgroup%
\global\stablesinfalse}
%
%
\lockat

\Title{
\vbox{\baselineskip12pt\hbox{hep-th/9411210}\hbox{YCTP-P11-94}}}
{\vbox{
\centerline{Lectures on 2D Yang-Mills Theory,}
\centerline{Equivariant Cohomology}
\centerline{and Topological Field Theories}}}
\bigskip
\centerline{Stefan Cordes, Gregory Moore, and Sanjaye Ramgoolam}
\bigskip
\centerline{stefan@waldzell.physics.yale.edu}
\centerline{moore@castalia.physics.yale.edu}
\centerline{skr@genesis2.physics.yale.edu}
\smallskip\centerline{Dept.\ of Physics}
\centerline{Yale University}
\centerline{New Haven, CT \ 06511}
\bigskip
\bigskip
\centerline{\bf Abstract}
\noindent
These are expository lectures reviewing (1) recent developments
in two-dimensional Yang-Mills theory and (2) the construction 
of topological field theory Lagrangians.  Topological field 
theory is discussed from the point of view of infinite-dimensional 
differential geometry. We emphasize the unifying role of 
equivariant cohomology both as the underlying principle 
in the formulation of BRST transformation laws and as a 
central concept in the geometrical 
interpretation of topological field theory 
path integrals.

\Date{November 19, 1994}

\listtoc\writetoc

\newsec{Introduction} 
\seclab\sIntro


These lectures focus  on a confluence 
of two themes in recent work on geometrical 
quantum field theory. The first theme is 
the formulation of Yang-Mills theory as a 
theory of strings. The second theme is 
the formulation of a class of topological 
field theories known as ``cohomological 
field theories.''

Accordingly, the lectures are divided into 
two parts. Part I  reviews various issues which arise 
in searching for a string formulation of Yang-Mills 
theory. Certain results of part I
 motivate a thorough study of cohomological 
field theory. That is the subject of part II. 

Part I is published in \CMRPI. Part II is published in 
\CMRPII. An electronic version of the entire set 
of lectures is available in \CMRIII.  In  \CMRPII\ 
a cross-reference 
to chapter two of \CMRIII\  is labelled chapter I.2, etc. 

We now give an  overview of
the contents of part I of the lectures. 
A detailed overview of part II 
may be found in chapter \sGRTFT. 
Some background material on differential 
geometry is collected in appendix A.

\subsec{Part I: Yang-Mills as a string theory}
\subseclab\sPrtone

The very genesis of string theory is in attempts 
to formulate a theory of the strong interactions. 
In this sense, the idea that Yang-Mills theory is a 
string theory is over 20 years old. The idea is 
attractive, but has proven elusive. In the last 
two years there has been some significant progress 
on this problem. 

\subsubsec{General issues}
\subsubseclab\sssGeniss

In chapter \sYMasaST\ we discuss the general pros and cons of formulating YM
as a string theory.
The upshot is:
\item{a.)}
There are good heuristic motivations, but few solid results. 
\item{b.)}
String signatures will be clearest in the $1/N$ expansion. 
\item{c.)} Any YM string will differ 
significantly from the much-studied critical string.

We also describe the program initiated by 
D. Gross: examine the $1/N$ expansion of 
two-dimensional Yang-Mills amplitudes and 
look for stringy signatures. Gross' program 
has been the source of much recent progress. 

\subsubsec{Exact amplitudes}
\subsubseclab\sssExctamps

In order to carry out Gross' program, we describe the very beautiful exact solution of 
\ymt\ in chapter \sESofYMinTD.
The upshot is:
\item{a.)}
The Hilbert space is the space of class functions on the gauge group $G$. 
\item{b.)}
The amplitudes are known exactly.
For example, in equation \exctprt\ we derive the 
famous result which states that if a closed oriented
spacetime $\ST$ has area $a$, then the partition function
of Yang-Mills theory for gauge group $G$
is 
\eqn\introi{
Z= \sum_R (\dim ~ R)^{\chi(\ST)}  e^{-e^2 a C_2(R)}}
where we sum over unitary irreducible representations of $G$,
$\chi ( \ST )$ is the Euler character of $\ST$, and $C_2 ( R )$ is the quadratic Casimir in the
representation $R$.
Similar results hold for all amplitudes. 

\subsubsec{Hilbert Space}
\subsubseclab\sssHilSpce

With the exact amplitudes in hand we begin 
our study of the $1/N$ expansion. The first 
step is to understand the Hilbert space of the 
theory in stringy terms. This is done in 
chapter \sFYMTtoSCA. The consequence is that 
one can reformulate the Hilbert space 
of the theory as a Fock space of string states. 
Wilson loops create rings of glue. The topological 
properties of this ring of glue characterize 
the state. 
We also describe briefly some nice relations 
to conformal field theories of fermions and 
bosons. 

\subsubsec{Covering Spaces}
\subsubseclab\sssCovSpcs

The description of states as rings of 
glue winding around the spatial circle 
leads naturally to the discussion of 
coverings of surfaces by surfaces: 
the worldsheet swept out by a ring of glue 
defines a covering of spacetime. 

We review the mathematics of covering surfaces
 in chapter \sCS. The most important 
points are: 

a.) Branched covers are described in terms of 
symmetric groups, see \sssBrnchCov.

b.) Branched covers may be identified with 
holomorphic maps. 

\subsubsec{$1/N$ Expansions of Amplitudes}
\subsubseclab\sssExpdAmpltd

With our understanding of branched covers we 
are then ready to describe in detail the 
beautiful calculations of Gross and 
Taylor of the $1/N$ expansions 
of amplitudes in \ymt\ \refs{\GrTa}. 
We do this in chapter \sONEofYMA.
The outcome is that 
the $1/N$ expansion may be identified with a 
sum over branched covers - or simple 
generalizations of branched covers. 

\subsubsec{Interpreting the sum}
\subsubseclab\sssIntrptsum

The $1/N$ expansion gives a sum over branched 
covers - but how are the branched covers weighted?
In chapter \sEC\ we give the correct interpretation 
of the weights: they are topological invariants - 
Euler characters - of the moduli space of holomorphic 
maps from Riemann surfaces to Riemann surfaces
\CMROLD. 

%
%

\subsubsec{The challenge}
\subsubseclab\sssChllnge

One of the first goals of Gross' program was to find a string action for \ymt. 
The significance of the results of chapters \sEC\ and \sWilloops\ is that they 
provide the bridge between \ymt\ and cohomological field theory. 

Cohomological field theory, when described 
systematically, provides a machine by which 
one can associate a local quantum field theory 
to the study of intersection numbers in  moduli 
spaces of solutions to differential equations. 

\ifig\prtbdiag{Perturbed diagonal.}
{\epsfxsize3.0in\epsfbox{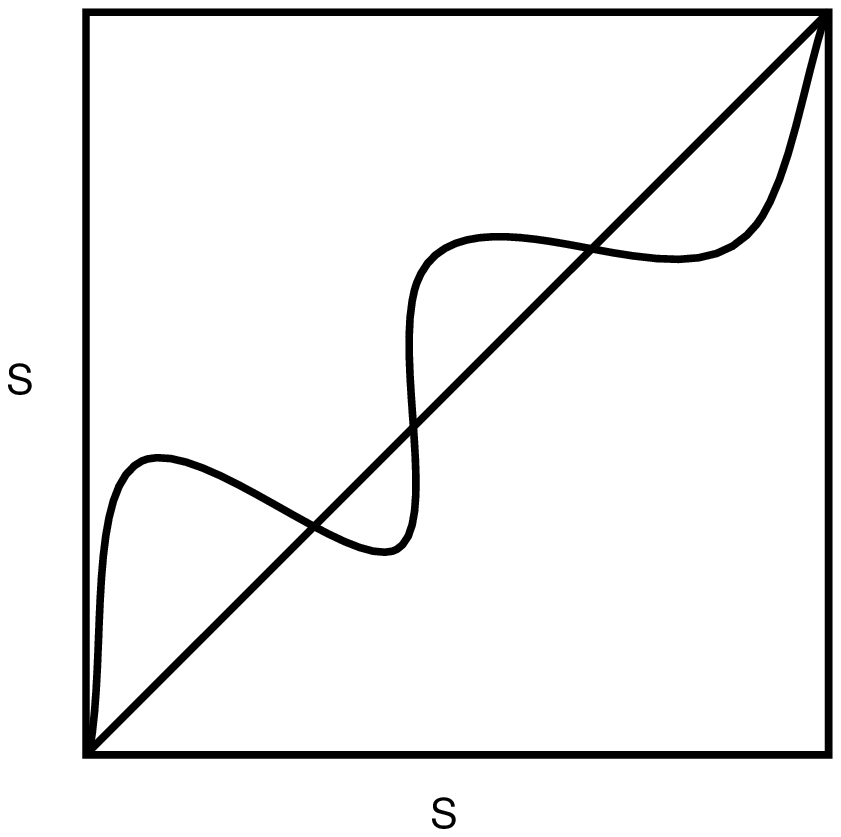}}

The relevance to the Euler characters, discovered in chapter \sEC, can be seen by recalling the
following famous fact. Let $S$ be a space, and consider $S$ sitting along the diagonal of
$S\times S$.
If we perturb the diagonal as in \prtbdiag\ then the intersection number is just $\chi(S)$. 

The machinery is described in  part II and applied in chapter \sYMTasaTST\ to write a string 
action for a string theory equivalent to \ymt. 
This action should be useful for future computations and studies of \ymt.
Much more importantly, it has a natural generalization to {\it four}-dimensional 
target spaces.
The significance of the four-dimensional 
theory is not yet known. Some guesses are described in 
chapter \sFDAC. 
The real challenge is two-fold:
\item{a.)}
Describe the target space physics of the theory in chapter \sFDAC.
\item{b.)}
Far more importantly: can we use the above results to say anything interesting about a
string formulation of $YM_4$? 

\subsec{Related Topics not Discussed}

There are many important topics related to the subject of these lectures which are not
discussed. 
These  include: 

1.  The incorporation of dynamical quarks and the 't Hooft model of D=2 QCD.
Much work on this subject, including a derivation of the 't Hooft equation 
starting from a Nambu-type string action, was carried out by I. Bars and collaborators 
in \refs{\Ba,\BaHa}. 
Recently there has been very interesting progress in relating this topic to a bilocal field
theory with connections to $W_\infty$ symmetry  \refs{\rajeev,\wadia,\demeterfi}. 

2.  Kostov has written an interesting paper inspired by the results of 
Gross-Taylor but applying, in principle, to $D=4$ Yang-Mills.
This carries further an old program initiated by Kazakov, Kostov, O' Brien and Zuber, and  Migdal \refs{\KaKo,\Kos, \obzub , \MitH}.

3. In part II on topological field theory we have emphasized the construction of 
Lagrangians and BRST principles. 
Space does not permit a discussion of the very interesting canonical formulation 
of these theories, which involves Floer's symplectic and 3-manifold homology 
theories. 

4. One issue of great importance in TFT is defining appropriate compactifications 
of moduli spaces.
This is quite necessary for producing explicit, rather than formal,
 results.
Space limitations preclude a discussion of these issues. 

5. In recent years, many very beautiful explicit 
results for topological field theory correlators 
have been obtained, both in topological 
string theory and in topological gauge theory.
For example, there are now explicit formulae 
for quantum cohomology rings and for Donaldson 
invariants. Again, space does not permit a discussion 
of these interesting results.\foot{Quite recently (November, 1994) spectacular 
progress has been made in topological gauge 
theory. See \refs{\kronmrow,\monfour,\taubes} .} 

6. There are many other aspects of topological theories we have not entered upon.
We also have not referenced many authors.
Since 1988 there have been, literally, thousands of papers written on topological field 
theory.
It is not possible to do justice to them all. 
We apologize to authors who feel their work has not been adequately cited. 

\subsec{Some related reviews}

Some aspects of \ymt\ are also reviewed in \refs{\kazrev,\bazrev,\Ba,\GrTatalk,\DGLSS}.
Cohomological field theory 
is also reviewed in \refs{\vanbaal,\Witr,\bbrt\Bl}. 
Finally, the reader should note 
that some parts of this text have appeared in other places. 
Some of the text in chapters 
\sCS, \sEC, and section 
\ssNonchir\ has appeared in \refs{\CMROLD}. 
Parts of  secs. \sstheGCOFGL,\sssHolMpsF, 
and \sFDAC\  have appeared in 
\mooricm.

\newsec{Yang-Mills as a String Theory}
\seclab\sYMasaST

\subsec{Motivations for strings }
\subseclab\ssMforS

The motivations for looking for a Yang-Mills string 
have been extensively discussed elsewhere. See 
\refs{\ampi,\ampii,\MitH,\Po,\Dosc}\ for some 
representative references. 
This is a contentious subject and has been under 
debate for over 25 years. Let us listen in briefly 
to a piece of this interminable argument
between a  visionary physicist/misguided mathematician
and sensible realist/cynical reactionary\foot{We thank many colleagues, especially 
M. Douglas, for participating in such arguments.}.
 
enthusiast: Look, string amplitudes satisfy {\it approximate 
duality!}  This is an experimental result and was, in fact, 
responsible for 
the very genesis of string theories in the late sixties!

skeptic:  
It is, unfortunately, difficult to proceed from these 
experimental results to the correct string version of 
Yang-Mills in anything other than a phenomenological
approach.  

enthusiast:  But there is a second reason.
Consider 't Hooft's 
planar diagrams!  If we hold 
$N e^2$ fixed, then the  diagrams are weighted with 
$N^\chi$ where $\chi$ is the Euler character of a surface 
(in ``index space'') dual to the diagram!
Thus we see the {\it emergence of surfaces 
in weak coupling, large $N$ perturbation theory!}
At large orders of perturbation theory 
we have ``fishnet diagrams!'' This is, of 
course,  the connection between large $N$ matrix 
theories and strings that has been so brilliantly 
exploited in the matrix model 
approach to low-dimensional string theories!

skeptic:
Quite right, but what you see as a strength 
I regard as a weakness: The description of planar diagrams as 
surfaces is only 
clear at large order in perturbation theory. 
Many important physical properties are well-described 
by low orders of perturbation theory, and it is hard to 
see how an effective string description will help. 
How, for example, do you expect to see asymptotic 
freedom with strings, if strings only become effective 
at large orders of perturbation theory?
 
enthusiast: There is a third reason for suspecting 
a deep connection! The {\it natural} variables in 
Yang-Mills are the holonomy 
variables, or Wilson loops: 
$$\Psi(C) = \langle W(C) \rangle$$
Moreover, there is a very geometrical set of 
equations governing these quantities! These are 
the Migdal-Makeenko 
equations:
\eqn\migmak{
\eqalign{
L \Psi &= g^2 \oint dx_\mu \oint dy_\mu \delta^4(x-y) 
\langle \tr~ W(C_{xy})\rangle  \langle \tr~ W(C_{yx})\rangle
(1+ \CO({1\over N})  ) \cr
&\sim 
g^2 \oint dx_\mu \oint dy_\mu \delta^4(x-y) 
\Psi(W(C_{xy})) \Psi(  W(C_{yx})) (1+ \CO({1\over N})  ) \cr}
}
Here $L$ is a kind of Laplacian on functions on 
the space of loops!

skeptic: But the loop variables are 
terribly hard to define. The above equations
involve singular quantities, the Laplacian 
is a very delicate object, and it is not clear 
what kinds of boundary conditions you 
should specify. 
The renormalization of Wilson loops is 
notoriously difficult and any loop 
description of the Hilbert space will be highly 
nontrivial. Moreover, there are nontrivial 
issues of constraints on the wavefunctions 
and independence of variables that have to 
be sorted out. 

enthusiast:  But there is a fourth reason!
Consider the  strong-coupling expansion of 
lattice gauge theory 
$$Z= e^{N^2 F}= \int \prod_{\ell} dU_{\ell}~ 
e^{N \beta  \sum_p \tr (U_p+ U_p^\dagger)} 
= \sum_{n\geq 0}  \beta^n Z_n(N) , $$
where $\ell$ runs over the links and $p$ over the plaquettes. 
When calculating the terms in the expansion, we must 
contract plaquettes and hence we see the 
{\it emergence of surfaces in strong coupling 
perturbation theory!}
For nonintersecting surfaces we literally have a sum over 
surfaces weighted by the Nambu action! 

skeptic:  Well, what you say is true if we 
consider nonintersecting surfaces. Then we may use: 
\eqn\ovlp{
\int dU (U^\dagger)_i^j U^k_l = {1\over N} \delta_i^k \delta_l^j
} 
But in the strong coupling expansion you will have much more 
complicated integrals. When a surface self-intersects you 
will need to weight these in complicated ways. The 
sum over surfaces will be so complicated as to be 
utterly useless.

enthusiast: That is not obvious: the basic integrals of 
$U_{ij}, U^\dagger_{ij}$ have 
$1/N$ expansions so, 
at least formally,  a series expansion exists.  
For example, 
\eqn\ovlpii{\eqalign{
\int dU (U^\dagger)_{i_1}^{j_1}  (U^\dagger)_{i_2}^{j_2} U^{k_1}_{l_1}  U^{k_2}_{l_2}
& = {1\over N^2-1}
[ \delta_{i_1}^{k_1} \delta_{l_1}^{j_1}\delta_{i_2}^{k_2} \delta_{l_2}^{j_2} + PERMS]\cr
&-{1\over N(N^2-1)}
 [ \delta_{i_1}^{k_1} \delta_{j_1}^{l_2}\delta_{i_2}^{k_2} \delta_{j_2}^{l_1} + PERMS]\cr
} 
}
The second term looks strange and contributes at 
leading order in $1/N$.
Nevertheless, one of the 
beautiful achievements of \refs{\Kazkos,\obzub} 
was to find a description of these $1/N$ expansions in 
terms of surfaces. Thus, the large $N$ strong coupling 
expansion of Yang-Mills theory is a string theory! 

skeptic:  But there are more serious objections. You will have 
to connect to the weak coupling phase to see 
continuum physics. Indeed, 
your large $N$ expansion will introduce further 
headaches. After taking the large $N$ expansion 
it is quite possible that there will be a strong-weak 
phase transition which was not present at finite 
$N$. Indeed there are known examples of such 
phase transitions \refs{\GrWi,\DoKa, \GrMat, \Kazwynt }.

enthusiast: But it is not clear how generic these 
phase transitions are. Even if they exist it may be 
possible to calculate some important quantities!

skeptic: It is often said that strings automatically include 
gravity. Don't we expect general 4D strings to have 
gravitons and dilatons - thus ruling them out 
as candidates for QCD? 

enthusiast:  Not at all! There are topological strings 
with no propagating degrees of freedom. There 
are $N=2$ strings \ogvf\ with a single scalar field as 
a propagating degree of freedom! 
 
\bigskip

The debate continues, but it is time to move on. 

\subsec{The case of two dimensions}
\subseclab\ssTheCofTD

The issues debated in the previous section can 
be brought into much sharper focus in the context of 
$D=2$ Yang-Mills theory. Here the loop variables 
advocated by our enthusiast can be defined. Indeed, 
we will do so in chapter \sFYMTtoSCA. Moreover, as we 
will see in chapter \sESofYMinTD,
 exact results for the amplitudes are available. 
D. Gross advocated that one should
 examine these exact results 
in the large $N$ theory and search for  ``stringy'' 
signatures of the coefficients in the $1/N$ expansion 
\refs{\Grtalk}. As we will see, Gross' program has enjoyed 
some degree of success. 

\subsubsec{What to look for} 
\subsubseclab\sssWhtlk

What should we look for to see a string formulation 
of \ymt? There are two basic aspects: 

\vskip0.1truein\noindent
1. Hilbert space

As we learned in the above debate, the natural 
variables in a string description of Yang-Mills are 
the loop variables: 
$$
\prod_j (\tr~ U^j)^{k_j}
\qquad\qquad k_j \in \IZ_+
$$
where $U$ is the holonomy around a loop.
These may be considered as wavefunctions in a Hamiltonian treatment.
We must relate these to a description in terms of $L^2(\CA/\CG )$, where $\CA$ is the
space of spatial gauge fields. 

\vskip0.1truein\noindent
2. Amplitudes

The $1/N$ expansion of the partition
 function should have the general form:
\eqn\expprt{
Z(A,\ST) \sim 
\exp\Biggl[\sum_{h\geq 0} \bigl({1\over N}\bigr)^{2h-2} Z_h(A,\ST) 
\Biggr]}
It is already nontrivial if only even powers of $1/N$ appear and $h\geq 0$. 
In chapter \sONEofYMA\  below we will see that this is indeed true for 
the case of two dimensions. 
The coefficients of the $1/N$ expansion are to be interpreted in terms 
of sums over maps: 
\eqn\smovrmp{
Z_h(A,\ST) = \int_{\CM} d\mu}
where $\CM= \MAP [ \Sw \to \ST ]$ is some space of maps from a {\it worldsheet}
$\Sw$ of genus $h$ to the spacetime $\ST$ of genus $p$. 
We similarly expect other quantities in the theory, e.g. expectation values of  
Wilson loops, to have similar expansions. 

There are two central issues to resolve in making sense of \smovrmp:
\item{$\alpha$.}
What class of maps should we sum over in $\CM$?
The classification of maps $\Sw\to \ST$ depends strongly on what category 
of maps we are working with. 
\item{$\beta$.}
What  is the measure $d\mu$?
(Equivalently, what is the string action?)

The answer to question $\alpha$ is given in  chapter \sEC.
The answer to question $\beta$ is given in chapter \sYMTasaTST.

\newsec{Exact Solution of Yang-Mills in Two Dimensions}
\seclab\sESofYMinTD

\subsec{Special features of two dimensions}
\subseclab\ssSFTD

Exact results in \ymt\ have been developed over the past several years by many
people, beginning with some work of A. Migdal \refs{\Mig}. 
One key feature of two 
dimensions is that there are no propagating degrees of freedom
-- there are no gluons.  
This does not make the theory trivial, but does mean that we must investigate the theory
on spacetimes of nontrivial topology or with Wilson loops to  see degrees of freedom. 
Since there are so few degrees of freedom one might suspect that there is a very large
group of local symmetries.
Indeed,  \ymt\  has a much larger invariance group than just local gauge invariance
${\cal G}$ - it is invariant under area preserving diffeomorphisms,
${\rm SDiff} ( \ST )$. 

The area-preserving diffeomorphism invariance 
may be seen as follows: Once we have chosen 
a metric, $G_{ij}$, on $\ST$ we can map the field strength
$F$, which is a two form, to a Lie-algebra-valued scalar $f$:
\eqn\mapfrm{
f=* F  \longleftrightarrow F= f \mu}
where $\mu$ is an area form. In 
components: 
$$ 
F^a_{ij} = \sqrt{\det G_{ij}}~ \epsilon_{ij}~ f ^a
$$
In terms of $f$ one may write the Yang-Mills action as: 
\eqn\ymactioni{
I_{\rm YM} = {1\over 4 e^2}
\int_{\ST} d^2 x \sqrt {\det G_{ij}} ~\tr f^2 
= {1\over 4 e^2} \int_{\ST} \mu ~\tr f ^2}
The quantity $f $ is a scalar, hence, 
any {\it diffeomorphism} which preserves the volume 
element $d^2 x \sqrt {det G_{ij}}$  is a symmetry of the 
action. 
This makes it ``almost'' generally 
covariant, and this  kills almost all the degrees of freedom.

\subsec{A larger space of theories}
\subseclab\ssLSofT

Let us make several remarks on the 
theory \ymactioni :

\vskip0.1truein\noindent
$\bullet$  \ssLSofT.1 We may rewrite the action as 
\eqn\teek{ I= -\half \int  i \tr (\phi F)  + \half e^2 \mu  \tr \phi^2 }
where $\phi$ is a Lie-algebra valued $0$-form. From this we 
see that the gauge coupling $e^2$ and 
total area $a=\int\mu$ always enter together. 
\vskip0.1truein\noindent
$\bullet$ \ssLSofT.2 
It is natural to generalize the action \teek\ to 
\eqn\invtpoly{
 I=  \int \biggl[ i \tr (\phi F)  + \CU(\phi) \mu\biggr]  }
where $\CU$ is any invariant function on the Lie algebra, $\lieg$ \refs{\Witdgt}. 
Thus, we should regard ordinary \ymt\ as one example of a general 
class of theories parametrized by {\it invariant functions on $\lieg$.}
It is natural to 
restrict to the ring of invariant polynomials 
on $\lieg$.
Explicitly, for $G=SU(N)$, this ring may be described as a polynomial ring 
generated by $\tr~ \phi^k$, so we may describe the 
general theory by coordinates $t_{\vec k}$: 
$$
\CU= \sum t_{\vec k} \prod_j (\tr \phi^j )^{k_j}$$
This is in marked contrast to the 
situation in $D=4$ Yang-Mills
where the only dimension four gauge 
invariant operators are $\tr (F\wedge * F)$ and 
$\tr (F^2)$, the latter being a topological term. 
\vskip0.1truein\noindent
$\bullet$ \ssLSofT.3.
The action
$$
I_{\rm top} = -\half \int  i \tr~ (\phi F)
$$ 
describes 
a topological field theory whose path integral is 
concentrated on flat connections $F=0$.   In the small area 
limit (or the $\CU\to 0$ limit) we must reproduce the 
results of this topological field theory.

\vskip0.1truein\noindent
$\bullet$ \ssLSofT.4.
 The situation in $D=2$ naturally leads to the 
question of whether any information  in $D=4$
Yang-Mills can be extracted from topological 
Yang-Mills. At first this seems absurd given the 
obvious complexity of the physical theory. 
Nevertheless, there are indications that the 
situation is not hopeless. For example we may 
write the $D=4$ action as \bbrt
\eqn\dfract{
\int \tr~ BF + e^2 \tr~ B \wedge * B\qquad .
}
The first term gives a topological theory, although 
it has a much larger gauge invariance than the 
theory with $e^2\not=0$. A second relation 
between a topological and a physical theory 
is given by the twisting procedure described in 
chapter \sTYMT. 

\vskip0.1truein\noindent
$\bullet$ \ssLSofT.5. When comparing results in \ymt\ with 
results from topology, it is important to remember 
that different definitions of the theory can differ 
by the local counterterms
\eqn\stanren{
\Delta I = k_1 {1\over 4 \pi} \int R + k_2~ e^2 \int \mu 
}
leading to an overall ambiguous factor of 
$e^{-k_1(2-2p)  - k_2 e^2 a} $ in the normalization 
of $Z$  on a surface of genus $p$ and area $a$ \Witdgt. 

\ifig\cylquan{Cylinder on which  Yang-Mills is quantized.}
{\epsfxsize3.0in\epsfbox{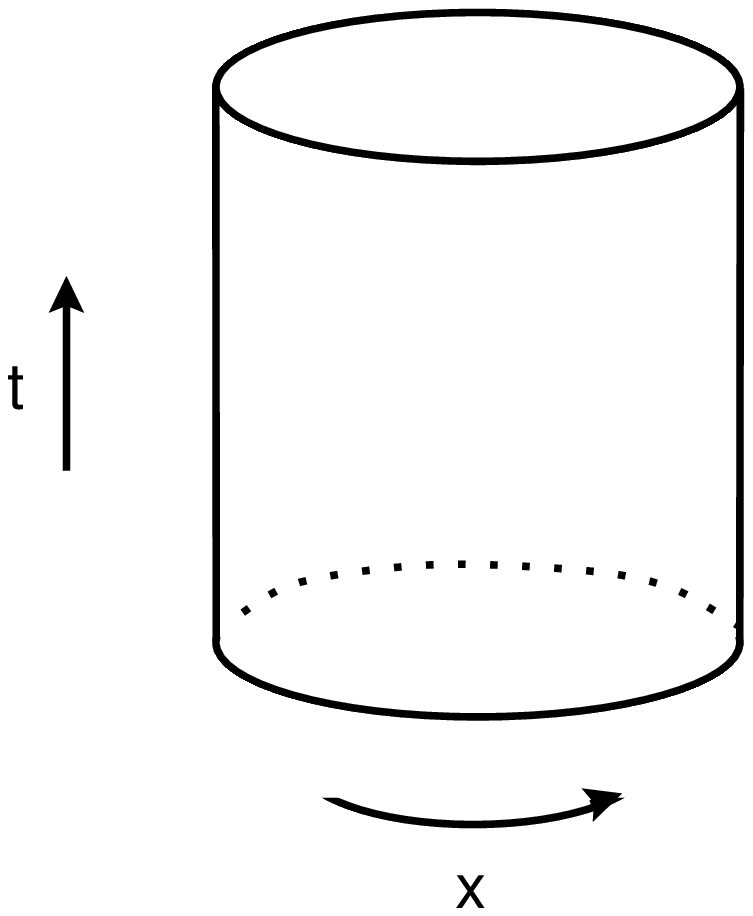}}

\subsec{Hilbert space}
\subseclab\ssHilSP

Consider quantizing the theory  on a cylinder such as  \cylquan\ 
with periodic spatial coordinate $x$ of period $L$.
With the gauge choice 
$A_0=0$,  we may characterize the Hilbert space as 
follows. 
The constraint obtained from varying $A_0$ in 
the Yang-Mills action \ymaction\ is  
\eqn\gauss{ D_1 F_{10}=0.}  
In canonical quantization we must impose the constraint 
\gauss\ on wavefunctions $\Psi[A_1^a(x)]$. By using 
time-independent gauge transformations we can set 
$A_1^a=const$. Then demanding invariance under the 
remaining $x$-independent gauge transformations allows 
us to rotate to the Cartan subalgebra\foot{Recall that a
semi-simple Lie algebra $\lieg$ has a 
unique (up to conjugation) maximal abelian subalgebra 
called the Cartan subalgebra. The commutative group 
obtained by exponentiating the generators is a multi-dimensional 
torus called  the Cartan torus or the maximal torus $T$. The 
Weyl  group, $W$, is the group of outer automorphisms of 
$T\subset G$. $W$ can be thought of as the set of 
$g\in G-T$ such that $gT g^{-1}=T$.  For more details 
see, for example, \refs{\hamermesh,\Zelo}.}.
Finally, we demand 
invariance under the Weyl group. This gives a function 
on $Cartan/Weyl$.  

Alternatively, the constraint \gauss\ becomes an operator 
equation. Let $T_a$ be an orthonormal (ON) basis of $\lieg$, with 
structure constants $[T_a,T_b]=f_{~ab}^c T_c$.  Then the 
Gauss law constraint becomes:
\eqn\gauslw{
\nabla \cdot E^a \Psi =
\Biggl( \p_1 {\delta\over \delta A_1^a(x)} + 
f_{~bc}^a A_1^b(x){\delta\over \delta A_1^c(x)} \Biggr)\Psi = 0 
}
which is solved by wavefunctionals of the form:
\eqn\slvcsnt{
\Psi[A_1^a(x)]= \Psi\Biggl[P \exp[ \int_0^L  dx A_1 ]\Biggr]
}
Demanding invariance under $x$-independent gauge transformations
shows that $\Psi$ only depends on the conjugacy 
class of $U=P \exp  { \int_0^L  dx A_1}$. 
From either point of view we conclude that  \refs {\Witdgt,\MiPoetd}:
\medskip
\boxit{ The Hilbert space of states is the 
space of $L^2$-class functions on $G$. }
The inner product will be 
$$\langle f_1 \mid f_2 \rangle = \int_G dU f_1^*(U) f_2(U) $$
where $dU$ is the Haar measure normalized to give volume 
one. 

We now find a natural basis for the Hilbert 
space. 
By the Peter-Weyl theorem for $G$ compact 
we can decompose 
$L^2(G)$ into the matrix elements of the unitary 
irreducible representations of $G$:
$$L^2(G)=\oplus_{R} R\otimes \bar R$$
Consequently a natural basis for the Hilbert space 
of states - the ``representation basis'' - 
is provided by the characters in the 
irreducible unitary representations. 
The states  $\mid R\rangle$ have wavefunctions
$\chi_R(U)$ defined by 
\eqn\willoop{
\langle U \mid R\rangle \equiv \chi_R(U)\equiv \Tr_R(U)
}

Let us now find the Hamiltonian for this theory. 
In the standard theory the Hamiltonian density 
is $\etwo (E^2+B^2)$, but $B=0$ in $D=2$ so 
$H= \etwo
\int dx 
{\delta\over \delta A^a_1(x)}{\delta\over \delta A^a_1(x)}$.
Acting on functionals of the form \slvcsnt\ we may 
replace
$H \rightarrow  \half e^2L \tr (U{\p\over \p U})^2$.
Now the conjugate momentum  
$\pi_{A_1^a} =  {\delta \over {\delta A_1^a}}$ 
acting on the wavefunctions $\chi_R ( U)$ is given by 
\eqn\piApsi {  \pi_{A_1^a}  \chi_R ( U) =  \chi_R ( T^a U) } 
Since $\sum_a T^a T^a $ evaluated in representation $R$ is  $C_2(R)$, 
the eigenvalue of the quadratic Casimir in 
representation $R$, it follows that
\bigskip

\boxit{The Hamiltonian is diagonalized in the 
representation basis and is the quadratic 
Casimir: $H= \etwo  L C_2(R)$. }

Even in the generalized theories of section \ssLSofT, the 
Hamiltonian is diagonal in the representation basis. The 
action of the invariant polynomials on $\mid R\rangle$ 
is described by polynomials in the Casimir 
operators $C_k(R)$. Explicitly, the Hamiltonian for the 
general theory can be parametrized by 
\eqn\genlham{
H=\sum_{\vec k} \tau_{\vec k} \prod_j (C_j(R) )^{k_j}}
Calculating the change of variables $t\to \tau$ 
involves resolving ordering ambiguities. The coordinates 
carry a degree associated with the degree in the polynomial 
ring. The transformation is upper triangular in terms of 
this degree. For example, a perturbation by $t_k \tr \phi^k$
perturbs the Hamiltonian  by Casimirs of 
degree less than or equal to 
$k$.

\exercise{Research problem}

Consider \ymt\ for the noncompact gauge group 
 $SL(N,\IR)$.  How do the above statements 
generalize?

\endexercise

\ifig\propym{Propagator of Yang-Mills.}
{\epsfxsize3.0in\epsfbox{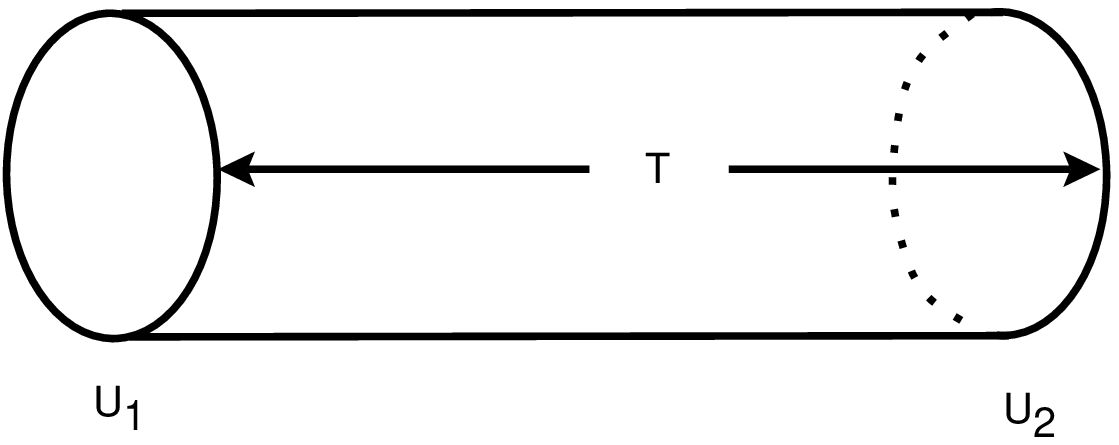}}

\ifig\wvdsk{Wavefunction for disk.}
{\epsfxsize1.5in\epsfbox{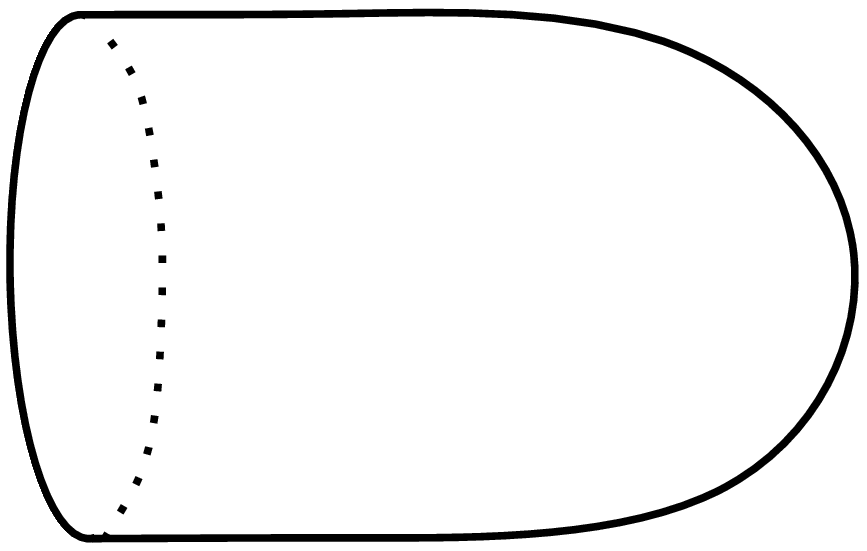}}

\subsec{Exact results for amplitudes}
\subseclab\ssERforA

We review results of \refs{\Mig,\Br,\KaKo,\GrKiSe,\Ru,\Fi,\Se,\BlThqym,\Witdgt,\Fo}.

\subsubsec{Basic Amplitudes}

\vskip0.1truein\noindent
1. Cylinder amplitude. 
Having diagonalized the Hamiltonian, 
we can immediately write the propagator
corresponding to \propym: 
\eqn\heat{
Z(T,U_1,U_2) = \sum_R \chi_R(U_1) \chi_R(U_2^\dagger ) 
e^{-\etwo L T  C_2(R)}}
which is just the heat kernel on the group.  
Note that the combination $\etwo LT = \etwo a$ enters together, as predicted by
${\rm SDiff}(\Sigma)$ invariance. 
In the rest of this chapter we absorb $\etwo$ into $a$. 
For the generalized theories with $H$ given in \genlham\ we simply  replace 
$a C_2(R) \to a \sum \tau_{\vec k} \prod (C_j)^{k_j}$. 

\exercise{Gluing property} 

Using the orthogonality relations of characters prove the 
gluing property:
\eqn\gluepropt{
\int dU~ Z(T_1,U_1,U) Z(T_2,U,U_2)=Z(T_1+T_2,U_1,U_2)
}

\endexercise

2. Cap amplitude, disk amplitude.  
By the gluing property of the propagator it 
suffices to calculate the amplitude for the 
disk in the limit of zero area. At area $=0$ the 
disk amplitude can be calculated in the topological theory, where 
$\CU=0$.   Integrating out $\phi$ sets $F=0$. Now 
the wavefunction $\Psi(U)$, where $U$ is the holonomy 
around  the boundary of the disk,  is supported on 
holonomies of flat connections on the disk, which forces $U=1$:
\eqn\trst{
\Psi(U)=\delta(U,1),}
where $\delta$ is the delta function in the Haar measure. 

Gluing the infinitesimal cap to the cylinder and using \heat\ we find the 
disk amplitude for a disk of area $a$, \wvdsk:
\eqn\disk{
\mathboxit{
Z(a,U)=\sum_R \dim~ R~ \chi_R ( U )~ e^{- a C_2(R)}}}

Using the area-preserving diffeomorphism invariance we may flatten out the disk and regard 
\disk\  as an amplitude for a plaquette.
Indeed this is true for any piece of surface diffeomorphic to the disk. 

\ifig\gllink{Integrating out a link leaves the partition function
invariant.}
{\epsfxsize5.0in\epsfbox{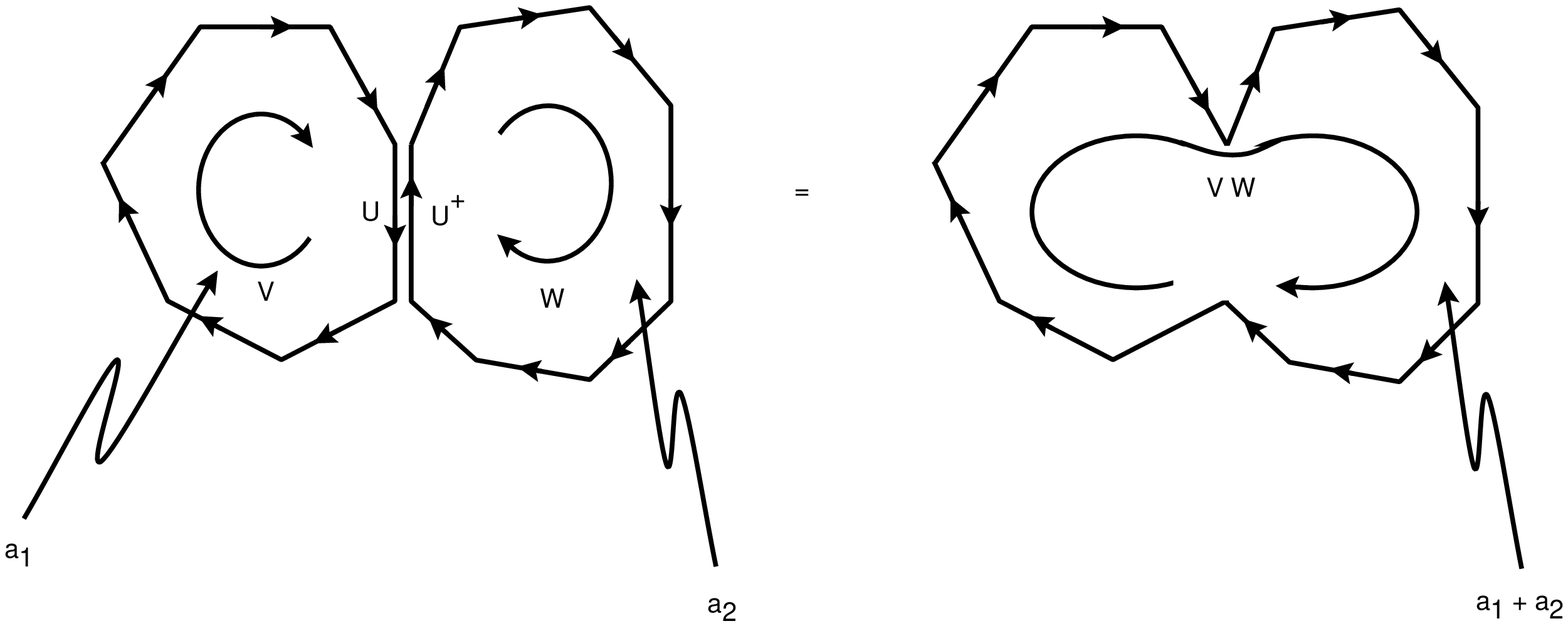}}

\subsubsec{RG Invariance}

Now we can compute the amplitudes for more 
complicated surfaces by standard gluing 
techniques familiar from ``axiomatic topological field 
theory.''

The gluing rules are based on the fundamental 
identity:
\eqn\ehaari{
\int dU~ \chi_{R_1} (VU) \chi_{R_2}(U^\dagger W) =\delta_{R_1,R_2}
 {\chi_{R_1}(VW)\over \dim~R_1}}
which follows from the orthogonality relations for group matrix elements. 
This allows us to glue together disconnected plaquettes.
Suppose that  two regions $\CR_1$ and $\CR_2$ share a common arc $\CI$.
Let $U$ denote $P \exp\int_\CI A$ along $\CI$.
We can glue these two as in \gllink: 

In formulae we have
\eqn\gluet{
\int dU~ Z(a_1, V U ) Z(a_2, U^\dagger W )
=Z(a_1+a_2,  VW)}

We can interpret this as RG invariance of the basic plaquette Boltzman weight. 
If we tried to write the lattice theory with this weight on the plaquettes, we would
produce the exact answer. 
Taking the continuum theory is trivial! 

\exercise{Research problem}

Since the partition function has an exact expression on triangulated surfaces, it 
should be possible to simulate fluctuating geometry on the spacetime $\ST$, using, 
for example, matrix model techniques. 
Investigate the behavior of \ymt\ coupled to  2D quantum gravity.
What are the amplitudes?
Is this theory a string theory?

\endexercise

\ifig\stpsgn{Opening up a genus two surface.}
{\epsfxsize3.0in\epsfbox{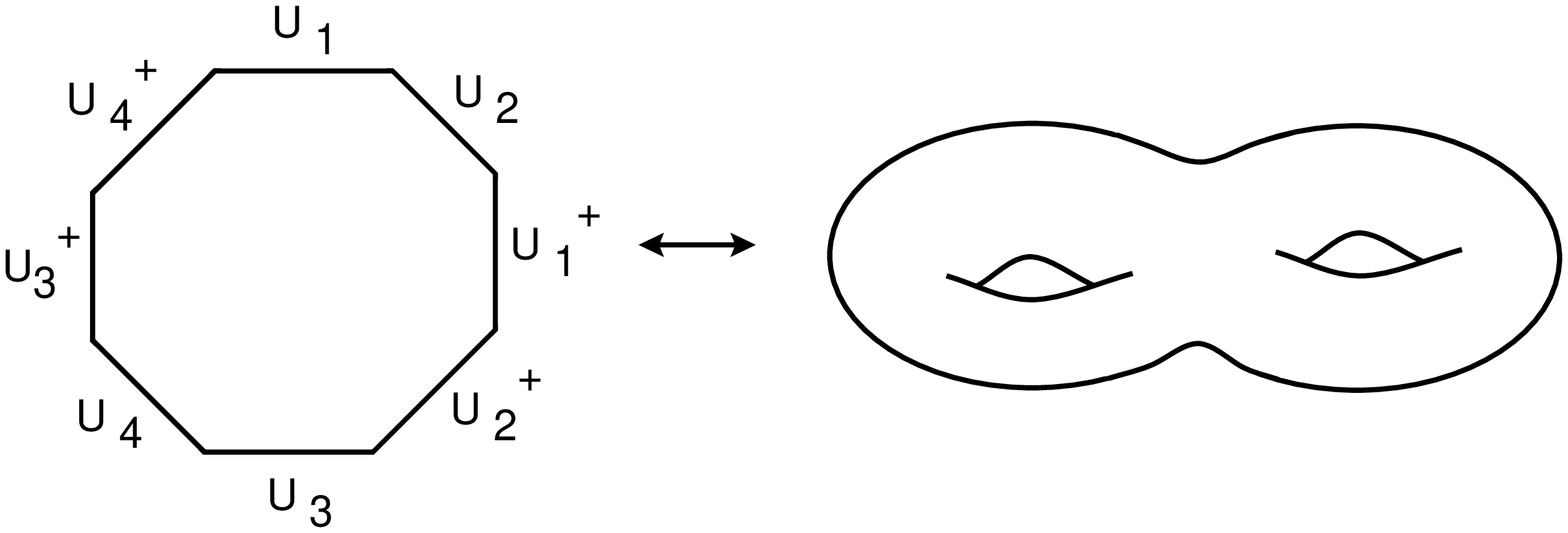}}

\ifig\pntsdg{Decomposition of  a pants diagram.}
{\epsfxsize3.0in\epsfbox{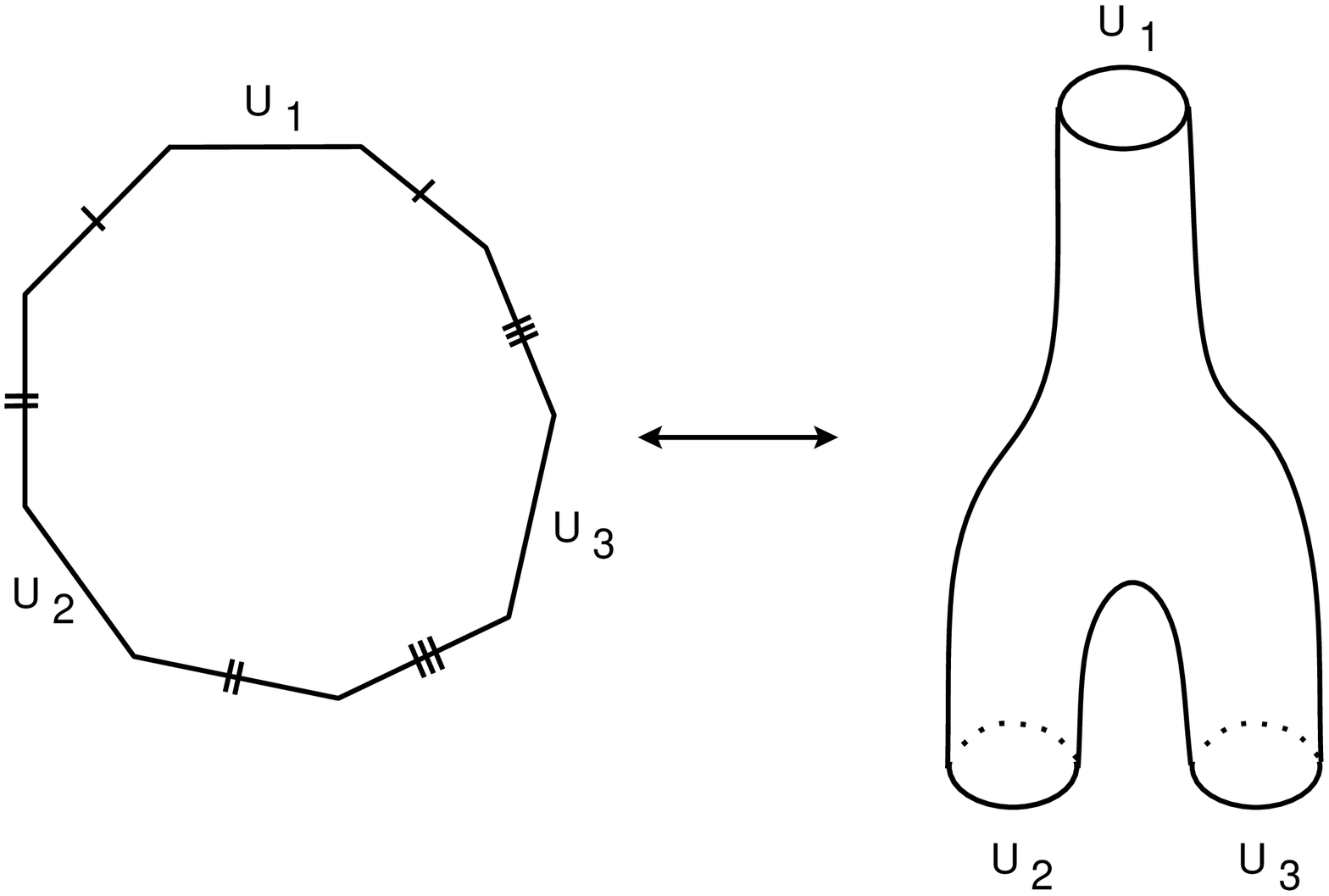}}

\subsubsec{Surfaces with nontrivial topology}

To get more complicated topologies we must glue pieces of  the same boundary to itself.
We can do that using the identity
\eqn\ehaarii{
\int dU~ \chi_R(U V U^\dagger W) = {\chi_R( V)\chi_R(W) \over \dim~ R}}
which follows from \ehaari.
It is a straightforward exercise at this point to derive exact relations for amplitudes:

\exercise{Partition functions}

As an application of this remark let us decompose a  surface $\ST$  of genus $p$ as a
$4 p$-sided polygon
as in \stpsgn. 
By applying \ehaarii\ repeatedly, derive the famous result:
\eqn\exctprt{
Z(\ST, p, a) = \sum_R (\dim~ R)^{2-2p} e^{- a C_2(R) }}
for a closed oriented surface $\ST$ of genus $p$ and area $a$. 

\endexercise

\exercise{3-holed sphere}
  
\item{a.)}
Represent the 3-holed sphere with boundary holonomies $U_1,U_2$ and $U_3$ as a 9-sided
figure with 3 sides pairwise identified as in \pntsdg.
Using \ehaarii\  show that 
\eqn\threehole{
Z(U_1,U_2,U_3;a)
= \sum_R {\chi_{R}(U_1)\chi_{R}(U_2)\chi_{R}(U_3)\over \dim~ R}
e^{-a C_2(R) }}
\item{b.)}
Dividing the surface $\ST$ up into pants, use
\threehole\ to rederive \exctprt. 

\endexercise

\exercise{General surface with boundary}

Generalize the result of the  previous exercise to derive the 
amplitude for a connected surface with $p$ handles and $b$ 
boundaries. Suppose the boundaries carry holonomy 
$U_i$.  Show that
\eqn\Wfuni { Z(\ST,p; U_1, \cdots, U_{b};a ) = \sum_{R} ( \dim~ R)^{2-2p-b } 
e^{ - aC_2(R) } \prod_{i=1}^b  \chi_{R} (U_i )} 

\endexercise

\ifig\nninlps{Two non-intersecting Wilson loops.}
{\epsfxsize3.0in\epsfbox{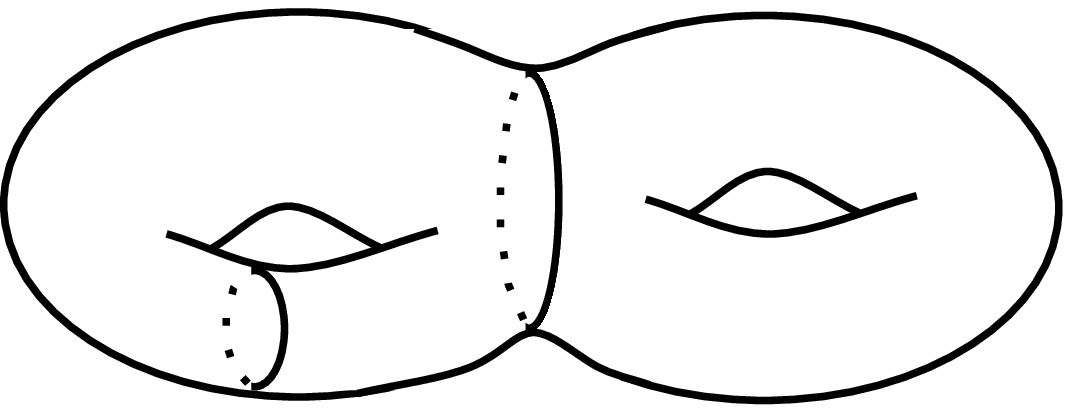}}

\ifig\dftwcrv{Using the orientation of the surface and of the
Wilson line we can define two infinitesimal deformations of
the Wilson line $\Gamma^{l,r}$.}
{\epsfxsize2.0in\epsfbox{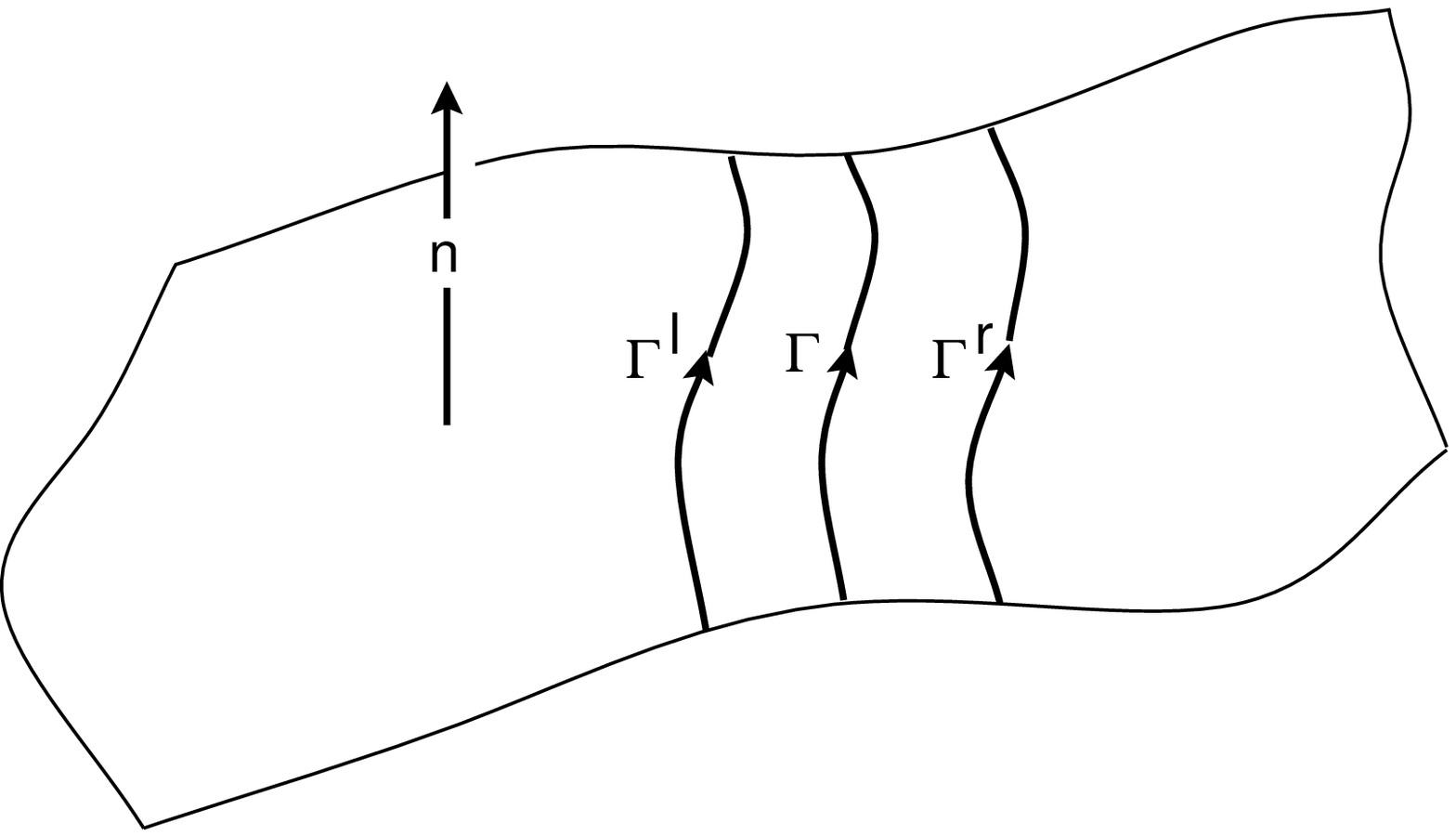}}

\subsec{ Exact  Wilson loop averages}
\subseclab\ssEWLA

\subsubsec{Nonintersecting loops}
\subsubseclab\nonintloops

Suppose that we have a collection $\{ \G \}$
of  curves in $\ST$, as in \nninlps.
Let
\eqn\ccomp{
\ST-\amalg~ \G = \amalg_c~ \Sc}
be the decomposition into disjoint connected components.
Each component $\Sigma_T{}^c$ has $p_c$ handles,  $b_c$ 
boundaries and area $a_c$.
Since $\ST$ and $\G$ are each oriented, each curve $\G$
can be deformed into two curves $\G^{l,r}$ as in \dftwcrv.
We let $c_\G^{l,r}$ denote the label 
of the component $\Sc$ which contains $\G^{l,r}$.
Associated to each curve $\Gamma$ we have a representation 
$R_\G$ and a Wilson loop operator:
\eqn\wlspop{
W(R_\G ,\Gamma)\equiv \tr_{R_\G} P \exp\oint_\G A \qquad\quad .
}
The exact answer for correlation functions of Wilson loops is
given by integrating the amplitude for the surface with boundary 
against the Wilson loop operators:
\eqn\wlext{
\biggl\langle \prod_\G W(R_\G,\G)\biggr\rangle
=\int \prod_\Gamma dU_\Gamma  \prod_c Z(\ST^c; U_{\cgp=c}, U^\dagger_{\cgm=c}) 
\prod_\Gamma W(R_\G,\G)}
where in the second product 
above we include those boundary holonomies
$U_{\cgp}$ with $\Gamma$ such that $\cgp=c$.
Now we use the identity
\eqn\inthol{
\int dU~ \chi_{R_1} (U)  \chi_{R_2} (U)   \chi_{R_3 } (U^\dagger ) 
= N^{R_3}{}_{R_1,  R_2}, }
where
 $N^{R_3}{}_{R_1,R_2}$ are the ``fusion numbers''
defined by the  decomposition of a tensor product into irreducible
representations
\eqn\fusruls{
R_1\otimes R_2=\oplus_{R_3} N^{R_3}{}_{R_1,R_2}~ R_3\qquad  . }
Applying this identity to each boundary
leads to
\eqn\wilavi{\eqalign{
\biggl\langle \prod_\G W(R_\G,\G)\biggr\rangle&=
\sum_{R(c)} \prod_c \bigl(\dim~ R(c)\bigr)^{\chi(\Sc)}
e^{-   a_c C_2(R(c)) }
\prod_\G N^{R(\cgm)}{}_{R(\cgp),R_\G}\cr}}
where we sum over unitary irreps, $R(c)$, for each component $c$,
and $a_c$ denote the areas of the components $\Sc$.
%
%

\subsubsec{Intersecting loops}

Expectation values of insertions of  $\chi_{R_\Gamma}(U_{\Gamma}^\dagger)$ 
for intersecting Wilson loops $\Gamma$, around which the holonomy is $U_{\Gamma}$, 
can be dealt with in a similar way \refs{\Witdgt, \Ru}.
We cut the $\ST$ along the Wilson lines separating it into several regions labeled by
$c$ with areas $a_c$, and $p_c$ handles inside, as in \ccomp.
The Wilson average contains a local factor for each component 
\eqn\Zc{ Z_c = \sum_{R_c} \dim~ R_c~
e^{- a_cC_2(R_c)/2}  \prod_{i=1}^{b_c} \chi_{R_{c}}( U^{(i)}_c )} 
where $U^{(i)}_{c}$ is the holonomy of the gauge field around the $i$'th 
boundary of region $c$. 
The Wilson average can be written as:
\eqn\Wilsav{ \sum_{R_c} \int \prod_{E} 
dU_{E} \prod_{c} Z_c \prod_{\Gamma} \chi_{R_\Gamma}(U_{\Gamma}^\dagger). }
The integral is over all the edge variables $U_E$. 

\ifig\ezfg{ Representations near an intersection of 
Wilson loops.}
{\epsfxsize3.0in\epsfbox{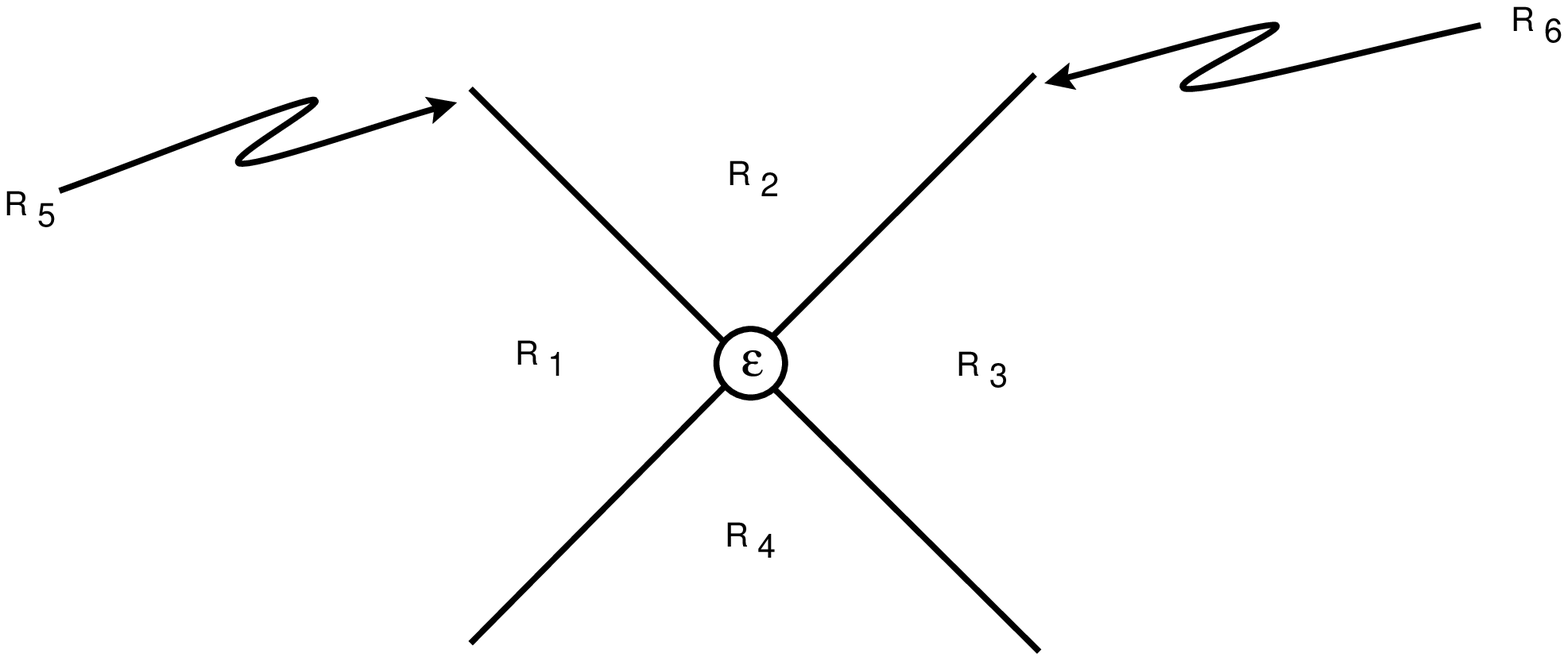}}

The group variable appears three times, from the Wilson line insertion in the representation
$R$, and from the two regions $R_a$ and $R_b$ on 
either side of the edge.
The group integral gives  
a Clebsch-Gordan coefficient for $R_b$ in $R_a
\otimes R$. 
These Clebsch-Gordan coefficients  can be collected 
into factors associated with each vertex.
After summing all the factors associated 
with a given vertex we are left with a $6j$ symbol
where the $6$ representations \foot{We will abbreviate the words
`representation' and `representations'
by `rep' and  `reps.'}  in question belong to 
the two Wilson lines and the four regions neighboring the vertex
\eqn\intwils{
W  = \prod_c \sum_{R_c} \sum_{\epsilon_\gamma}
( \dim~ R_c )^{2-2p_c-b_c}~ e^{ -a_cC_2(R_c)/2 }~
\prod_v G_v (R_c; R_\Gamma; \epsilon_\gamma)}
The index $v$ runs over the vertices and the index $\epsilon_\gamma$ runs over a basis 
for the vector space of intertwiners between $R_a \otimes R $ and $R_b$. 
 
\subsec{\ymt\ and topological field theory}
\subseclab\ssYMTFT

\subsubsec{$e^2=0$ and Flat Connections}
\subsubseclab\sssetzfc

Later on we will discuss at length the relation of \ymt\  to  topological string theory.
Here we sketch briefly some other relations to topological field theories. 

At $e^2=0$, \ymactioni\ becomes a simple ``BF-type'' topological field theory \refs{\bbrt}\ 
with action
\eqn\bftype{
I = {i\over 4 \pi^2} \int_\ST \Tr(\phi F). }
Evidently, $\phi$ acts as a Lagrange multiplier setting $F(A) =0$, so we expect the
theory to be closely connected with the geometry of the moduli space of {\it flat}
connections.
This is defined to be the space
\eqn\fltconn{
\CM(F=0; \ST, P) = \{ A\in\CA(P) \vert F(A) =0\}/\CG(P)}
where $P\to \ST$ is a principal $G$ bundle, and 
the remaining notation is defined in section \ssYMEQA. 
This space is far from trivial.
A flat connection may be characterized by its holonomies around the various 
generators of the homotopy group $\pi_1(\ST)$: 
\eqn\fltcnn{
\CM(F=0; \ST,P) = \Hom(\pi_1(\ST),G) / G}
This description makes clear that it is a manifold
\foot{With singularities, due to the  reducible
connections.
(See section \ssDtwoFLC.)}
of dimension $\dim~ G~ (2 p -2)$.
In order to derive the measure on $\CM$ given by the path integral one must 
gauge fix \'a la Faddeev-Popov.
The resulting path integral - as for all BF theories 
\refs{\Schw,\bbrt}\ reduces to 
a ratio of determinants known as Ray-Singer torsion.
Careful implementation of these considerations \refs{\Witdgt}\ shows 
that
 for gauge group $SU(N)$ we get the symplectic volume of $\CM$
\eqn\sympvol{
Z = {1\over N} \int_\CM {\omega^n \over n!} 
={1\over N} \vol(\CM)}
where $\omega$ is the symplectic form on $\CM$ arising from the symplectic form on $\CA(P)$:
\eqn\sympfrm{
\omega(\delta A_1 , \delta A_2) = {1\over 4 \pi^2} 
\int_\Sigma \Tr ( \delta A_1\wedge \delta A_2)\quad . 
}

Somewhat surprisingly, at $e^2\not=0$ the path integral still contains information 
about the topology of $\CM$.
This requires the ``nonabelian localization theorem" \refs{\Witdgtr,\jeffkir}, and will be
described in chapter \sTYMT\  after we have discussed equivariant cohomology. 

\subsubsec{Chern-Simons theory and RCFT}

The relation of $e^2=0$ \ymt\ to the space 
of flat connections \fltconn\  provides a direct 
link to 3D Chern-Simons theory, and thereby
\refs{\jonespoly}\  to rational conformal field 
theory. 
Indeed, as emphasized in \refs{\Witdgt}, the formula for $Z$, 
equation \exctprt, may be thought of
as the large $k$ limit of the Verlinde formula \verlinde\ for the number of $SU(N)$ level
$k$ conformal blocks on $\Sigma$
\eqn\nmbrblcks{
\sum_{R} \biggl( {S_{00}\over S_{R0} } \biggr)^{2p-2}}
Similarly, \threehole\ is  the limit of Verlinde's formula for the fusion rules
\eqn\frules{
N_{R_1 R_2 R_3} = \sum_{R} 
{S_{RR_1} S_{RR_2} S_{RR_3} \over S_{R0}}}
Here $S_{R_1 R_2}$ is the ``modular transformation matrix''  between representations
$R_1$ and $R_2$ ($0$ refers to the trivial representation). 
The connection arises since, at  $k\to \infty$,  ${S_{R0}\over S_{00} }   \to \dim~ R$
while $S_{R_1 R_2}$ becomes a matrix proportional to $\chi_{R_1} (U_2)$ where
$U_2$ is related to $R_2$ by a map from representations to conjugacy classes of
$G=SU(N)$.
The formula \frules\ was proven, at finite $k$, using methods of conformal field theory,
in \refs{\mrsb}. 

\subsec{Axiomatic Approach}
\subseclab\ssAA

It is possible to give simple axioms, analogous to those given by G. 
Segal and M. Atiyah for
conformal field theory and for topological field theory, respectively, \refs{\segax,\atiyax}\ 
for \ymt. 
The geometric category has as objects collections of oriented circles. 
Morphisms are oriented surfaces with area, cobordant between the circles.
The orientation agrees for the source and disagrees for the target. 

To each circle we assign the Hilbert space of class functions on the gauge group $G$. 
To each surface we have a map of the ingoing Hilbert spaces to the outgoing Hilbert space. 
This map is defined by gluing and by the basic amplitudes

$$
{\rm Tube} = \sum_R \mid R \rangle \langle R \mid e^{-a C_2(R)}
$$

$$
{\rm Pants} = \sum_R \mid R\rangle \otimes \mid R\rangle
\otimes \mid R\rangle {e^{-a C_2(R)}\over \dim~ R\quad .} 
$$

\newsec{From \ymt\  to Strings: The Canonical Approach}
\seclab\sFYMTtoSCA

In this chapter we relate the Hilbert space of class functions to some simple conformal
field theories. 
The main goal is to show how bosonization leads to a natural interpretation of the Hilbert
space in terms of string states. 

We would like to make some kind of sense of 
\eqn\lglmtq{
\lim_{N\to \infty} \CH_{SU(N)}. }
It turns out that the best way to do this is 
to reformulate the Hilbert space in terms of 
free fermions. 

\subsec{Representation theory and Free Fermions}
\subseclab\ssRTandFF

Many aspects of representation theory
have a 
natural description in terms of quantum field theory 
of free fermions. The utility of this point of view has 
been emphasized by M. Douglas, and by 
J. Minahan and A. Polychronakos \refs{\MiPoetd,\dgcrg}. 
We will be following \refs{\dgcrg}.

We have seen above that the Hamiltonian is the 
quadratic Casimir and that the Hilbert space is the 
space of $L^2$ class functions.   
Let us start with the group $G=U(N)$. 
Class functions are determined by their values on the 
maximal torus. We parametrize elements of 
the maximal torus $\Lambda\in T$
by 
\eqn\mxtorus{
\Lambda
= {\rm Diag} \{ z_1,\dots z_N \}
= {\rm Diag} \{ e^{i \theta_1} ,\dots 
e^{i \theta_N} \}}
The Weyl group is the permutation 
group $S_N$, and conjugation by the Weyl group  
on $T$ permutes the $z_i$.
Hence class functions $\psi(\vec \theta)$ are symmetric. 

The inner product on class functions is given by the measure
$$
( \psi, \psi )= {1\over {N! (2\pi )^N }} \int \prod d \theta_i~ \tilde\Delta( \vec{z} )^2
\mid \psi(\vec \theta) \mid^2
$$
where 
$\tilde\Delta( \vec{z} )=\prod_{i<j}\sin{\theta_i-\theta_j\over 2}
= {1\over { (2i)^{  {N(N-1)\over 2}    }}  }\Delta( \vec{z} ) / \prod_i z_i^{(N-1)/2}$, and
$\Delta( \vec{z} )=\prod_{i<j} ( z_i - z_j )$.

Moreover, when acting on class functions the Hamiltonian may be 
written, after a nontrivial calculation,  as:
\eqn\simh{H =   {e^2 \over 2}  {1\over\tilde\Delta ( \vec{z}) }\biggl[
\sum_i (- {d^2\over d\theta_i^2}) - N(N^2-1)/12 \biggr] \tilde\Delta ( \vec{z} )}

Both the measure and the Hamiltonian suggest that it is better to work with 
a totally antisymmetric wavefunction 
$\psi \rightarrow \tilde \Delta ( \vec{z} ) \psi$ with a standard measure 
and $H= \sum_{i=1}^{N} p_i^2$. We are simply describing a 
theory of $N$ noninteracting fermions on the circle 
$0\leq \theta \leq 2 \pi$. The one-body wavefunctions 
are $z^n = e^{i n \theta} $ for $n\in \IZ$. The Hamiltonian, 
and all the higher Casimirs, are diagonalized by the 
Slater determinants: 
\eqn\slater{\psi_{\vec n} ( \vec z )= \det_{1 \le i,j \le N} z_i^{n_j}}
so we will identify states
$\mid R\rangle$ with these Slater determinant states.

The state \slater\ has energy
$E = \sum_i n_i^2 - N(N^2-1)/12$. Under the 
$U(1)$ corresponding to matrices proportional to $1$
the state has $U(1)$ charge 
$Q= \sum n_i$.  
Since the wavefunction vanishes unless all $n_i$ are different, 
 we may assume, without loss of generality, that  $n_1 > n_2 > \cdots >
n_N$. 
We may recover the character of the rep 
corresponding to the  
the Fermi wavefunction \slater\  by 
dividing by $\tilde \Delta ( \vec{z} )$. This gives 
the Weyl character formula:  
\eqn\weyl{\chi_{\vec n}(\vec z) =
{\det_{1\le i,j\le N} z_i^{n_j  } \over \det_{1\le i,j\le N}
z_i^{j-1-n_F}}     = {\det_{1\le i,j\le N} z_i^{n_j + n_F } \over \det_{1\le i,j\le N}
z_i^{j-1}}  .}

\ifig\FilledFermi{The filled Fermi sea}
{\epsfxsize0.4in\epsfbox{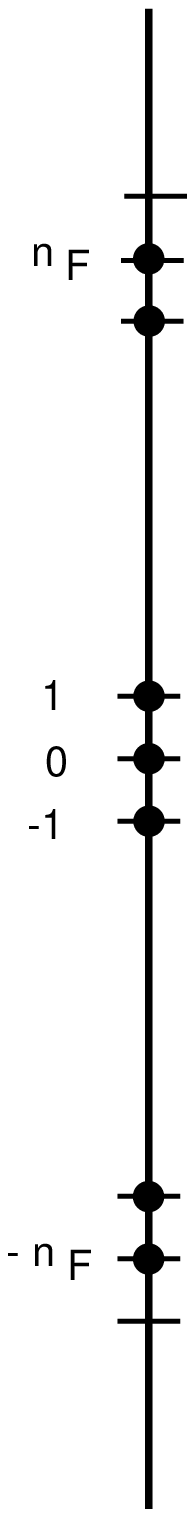}}
The trivial rep corresponds to the 
ground state with $E=Q=0$, with all levels filled 
from $-n_F$ to $n_F$, where  $n_F = (N-1)/2$.
(See \FilledFermi) 
\foot{Assume for simplicity that $N$ is odd.}

The above considerations can also be applied 
to $SU(N)$ rep theory by imposing 
the constraint $\prod z_i =1$. As an $SU(N)$ 
character we are free to shift all the $n_i$ by 
any integer $a$. 

\subsec{$SU(N)$ and $U(N)$} 
\subseclab\ssSUNandUN

$U(N)$ and $SU(N)$ reps
 are easily related by considering 
\eqn\homom{
1\to \IZ_N \to SU(N)\times U(1) \to U(N)\to 1\quad .
}
So 
$$U(N) = {SU(N)\times U(1) \over \IZ_N}. $$
We may  therefore label irreducible $U(N)$ reps by a pair 
$(R,Q)$ where $R$ is an irrep of $SU(N)$. The 
irreps of $SU(N)$ are in turn labeled by Young 
diagrams\foot{See, for example, \refs{\hamermesh,\Zelo}\ for 
a discussion of this.}, $Y$.
Since the kernel of \homom\  is represented trivially, 
the only pairs that occur are 
$(Y,Q=N \ell + n)$ where 
 $Y\in \CY_n^{(N)}$.  We will let 
 $\CY_n$ stand for the set of Young 
diagrams of $n$ boxes, and $\CY_n^{(N)}$
stand for the set of diagrams with 
$\leq N$ rows. When the row lengths are 
specified we write $Y(h_1,\dots , h_r)$.

Given a state $\chi_{\vec n}$, we may describe 
the corresponding Young diagram as follows. 
Using the freedom to shift $n_i\to n_i + a$ 
we may arrange that $n_N=-n_F$. Then 
the Young diagram corresponding to the 
fermion state is $Y=Y(h_1,h_2,\dots h_N)$
where 
\eqn\reprel{ h_j=n_j +j -1-n_F } 
 denotes the number of
boxes in the $j^{th}$ row. Note that $Q=\sum{h_j}$.

The Hamiltonian for the $SU(N)$ theory is 
$$H_{SU(N)} = H_{U(N)} - Q^2/N \qquad .$$

\subsec{Second Quantization} 
\subseclab\ssSQ

It is natural to introduce a second-quantized 
formalism. $B^{+}_{-n}$ creates a mode with wavefunction 
$z^n$, $B_n$ annihilates it, so we introduce: 
\eqn\fermflds{\eqalign{
\Psi(\theta)&= \sum_{n\in \IZ} B_n e^{i n \theta} \cr
\Psi^\dagger(\theta)&= \sum_{n\in \IZ} B^{+}_{-n} e^{-i n \theta}. \cr}
}
The filled Fermi sea satisfies the constraints:
\eqn\fifer{\eqalign{  &B_n \vert 0\rangle
 = 0 ,\qquad  \vert n\vert > n_F \cr
   &B^+_{-n} \vert 0\rangle = 0,\qquad  \vert n\vert \le  n_F .\cr } } 

Translation of operators from first to second quantization 
is standard. An important example for us is the following. 
Class functions act, by multiplication, as operators on 
the Hilbert space. The corresponding second 
quantized operators are given by:
\eqn\upsop{
\Upsilon_n = \sum z_i^n
\qquad\longleftrightarrow\qquad
\int d \theta \Psi^\dagger ( \theta )~ e^{i n \theta} \Psi(\theta). }

\ifig\cplrep{Young diagram for a coupled rep.
Given two reps $R$ and $S$, for $\bar S$ as indicated in the second
diagram, form $R\bar S$ as indicated in the third diagram.}
{\epsfxsize3.0in\epsfbox{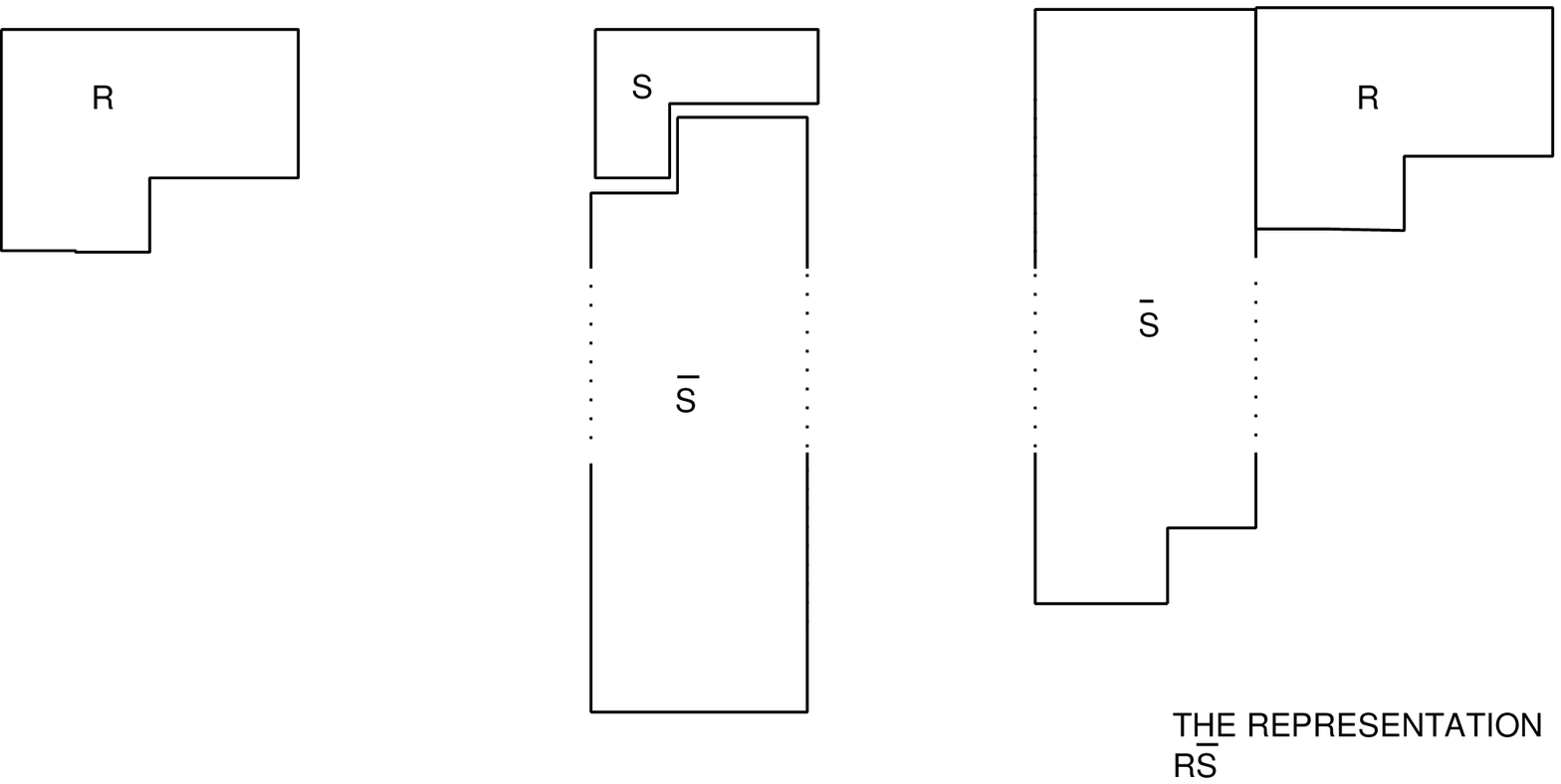}}

\subsec{Large $N$ Limit: Map to CFT}
\subseclab\ssLNL

\subsubsec{Chiral, Antichiral and Coupled Reps}
\subsubseclab\sssCACR

Now we describe a formulation of the large $N$ limit \lglmtq.
The description is based on the idea that in the large $N$ limit, the filled Fermi sea,
corresponding to the trivial rep, has two Fermi levels which are ``far'' from
each other. 

This notion of the levels being far can be made 
a little more
precise. We would like to consider states 
which, in the fermionic language, involve 
excitations around the Fermi sea involving 
changes of level numbers $\Delta n_j$ which 
are small compared to $N$. In terms of 
reps, we characterize these 
states as follows. 
Let $V$ be the fundamental rep, and $\bar V$ its complex conjugate.  
Consider states contained in tensor products of 
small (with respect to $N$) 
numbers of $V$'s and $\bar V$'s. 
Intuitively, we have two decoupled systems 
corresponding to excitations around the two 
Fermi levels. They are decoupled, because 
we only consider small excitations. Thus we 
expect that the large $N$ limit of 
the Hilbert space of class 
functions on $SU(N)$ should naturally be thought of 
as a tensor product: 
\eqn\splthil{
 \CH_{SU(N)} \to \CH_{\rm chiral} \otimes \CH_{\rm antichiral}.
}

In terms of reps, the states we are talking 
about are the following: A collection of particles and 
holes around $n_F$ corresponds to a rep 
$Y(h_1,\dots, h_N)$ described  by \reprel. 
Evidently, we can describe the chiral Hilbert space 
formally as
\eqn\chrlspcee{
\CH_{\rm chiral} = \oplus_{n\geq 0} \oplus_{Y\in\CY_n}
\IC\cdot \mid Y\rangle .}
If we reflect around $n=0$ we obtain a collection 
of particles and holes around $-n_F$. This state will 
correspond to the conjugate rep  $\bar R$. 

A  subtle point arises when we consider states 
corresponding to excitations around both Fermi levels. 
If the excitations around the levels $n_F,-n_F$ 
correspond to reps $R$ and $\bar S$, respectively, 
then the Slater determinant with both excitations present
corresponds to the {\it irreducible} rep 
$R\bar S$, which is defined to be the largest 
irreducible
rep in the decomposition of the 
tensor product $R\otimes \bar S$. When the numbers
of boxes in $Y(R)$ and $Y(S)$ are much smaller than 
$N$, this definition is unambiguous.  The 
reps $R\bar S$ were called 
``coupled reps'' by Gross and Taylor \GrTa.  
The construction is illustrated in terms of 
Young diagrams in \cplrep.

\subsubsec{Mapping to CFT}
\subsubseclab\sssMpCFT

When we have decoupled systems 
it is appropriate 
to define two independent sets of Fermi fields:   
\eqn\bcsys{\eqalign{
\Psi(\theta) &= e^{ i(n_F+\ha) \theta} b(\theta) + 
e^{- i (n_F+\ha) \theta} \bar b(\theta )\cr
\Psi^\dagger(\theta) &= e^{-i(n_F+\ha) \theta} c(\theta) + 
e^{  i(n_F+\ha)\theta } \bar c(\theta). \cr}
}
We introduce complex coordinates $z= e^{i\theta}$,
and define the mode expansions: 
\eqn\bcsysi{\eqalign{
b(z)&= \sum_{n\in \IZ + \ha} b_n z^{n} \cr
c(z)&= \sum_{n\in \IZ + \ha} c_n z^{n} \cr
\bar b( z)&= \sum_{n\in \IZ + \ha} \bar b_n \bar z^{n} \cr
\bar c( z)&= \sum_{n\in \IZ + \ha} \bar c_n \bar z^{n} \cr}
}
$b_n, c_n,\bar b_n, \bar c_n$ are only unambiguously defined for 
$\mid n \mid < <  N$, so \bcsys\ only acts on the states 
in \splthil. The peculiar half-integral moding is 
chosen to agree with standard conventions in CFT. 
In terms of the original nonrelativistic modes we have: 
\eqn\mapmodes{\eqalign{
c_{n} &= B^{+}_{-n_F-\epsilon+n} \cr
b_{n} &= B_{n_F+\epsilon+n} \cr
\bar c_{n} &= B^{+}_{n_F+\epsilon-n}\cr
\bar b_{n} &= B_{-n_F-\epsilon-n}\cr}
}
where $\epsilon=\ha$, so that
\eqn\anticomms{
\{ b_n, c_m\} = \delta_{n+m,0} \qquad \{ \bar b_n, \bar c_m\} = \delta_{n+m,0} 
}
and all other anticommutators equal zero.

We may now reinterpret the fields $b,c,...$. 
Defining $z=e^{i \theta +\tau}$ we see that these 
may be extended to fields in two-dimensions, and 
that they are  (anti-) chiral, that is, they satisfy 
the two-dimensional Dirac equation. We are thus discussing
two relativistic massless Fermi fields  in 
$1+1$ dimensions. 
In this description the trivial rep 
is the product of vacua $\mid 0 \rangle_{bc} \otimes \mid \bar 0 \rangle_{
\bar b\bar c}$
where $b_n \mid 0 \rangle_{bc}= c_n \mid 0 \rangle_{bc}=0 $ for $n>0$. 
The spaces in \splthil\ may be related to the 
CFT statespaces \
\eqn\cftstspc{
\CH_{bc} = {\rm Span}~ \biggl\{ \prod b_{n_i} \prod c_{m_i} \mid 0\rangle_{bc}
\biggr\}\qquad .
}
The space $\CH_{bc}$ has a natural 
grading according to the eigenvalue of
$\sum_{n\in\IZ} :b_n c_{-n} = \oint bc$ (called ``$bc$-number''): 
\eqn\grdspce{
\CH_{bc}=\oplus_{p\in \IZ} \CH_{bc}^{(p)}}
and we may identify:
\eqn\chrlspc{
\CH_{\rm chiral} = \CH_{bc}^{(0)}}
Indeed, in this language the class function corresponding to a rep with a
Young diagram $\mid Y\rangle\in \CH_{\rm chiral}$ may be mapped to a corresponding
fermionic $bc$ state using \reprel\ and then \mapmodes.
The result is: 
\eqn\repfrm{
\mid Y(h_1,\dots, h_N)\rangle 
\quad \leftrightarrow \quad
c_{-h_1+1-\half} \cdots c_{-h_s+s-\half}
b_{-v_1+1-\half} \cdots b_{-v_s+s-\half}
\mid 0\rangle 
}
where $h_i$ are the row-lengths, 
$v_i$ are the column lengths, and $s$ is the 
number of boxes along the leading diagonal. 
Similarly, reps made 
from $\bar n$ tensor products of $\bar V$ would be described 
in terms of $\bar c$ and $\bar b$. As long as $n, \bar n < <  N$ 
there is no ambiguity in this description. 

The following is a quick argument to justify \repfrm. 
For excitations of the free fermion system 
corresponding to particles created at 
levels $a_i$ above the  Fermi level
and holes at levels $-b_i$ below  $n_F$,  
the energy is 
\eqn\ener{ E = \sum_{i=1}^{s} \left [ (n_F + a_i) ^2 - (n_F-b_i)^2 \right ], } 
where $s$ is the number of particles.
Now using the following identity about Young diagrams
(for this identity and generalisations see \Ing)  : 
\eqn\identab{ \sum_{i=1}^{v_1} (h_i-i)^2 
= \sum_{i=1}^{s^\prime} (h_i - i)^2 - (v_i-i+1)^2 + \sum_{i=1}^{v_1} i^2}
where $s^\prime$ is the number of boxes along the leading diagonal, and  $v_i$ is 
the number of  columns in the $i$'th row, we find that the quadratic Casimir (energy)
can be written in the form \ener\ if we make the identification $a_i= h_i-i+1$, 
$b_i= v_i-i$ and $s=s^\prime$ .
Translating this into modes, we get \repfrm.

One very useful aspect of the introduction of 
$b$ and $c$ is that this system is a simple example 
 of a conformal field theory, as 
will be discussed at length by Polchinski at this school. 
We simply note that if we introduce
\eqn\frmvrop{
L_n=   \sum_{m=-\infty}^{\infty}  (n/2 + m) c_{-m} b_{m+n}}
then the $L_n$ satisfy the Virasoro algebra
\eqn\virasoroalg{
[L_n,L_m]=(n-m)L_{n+m} + {c\over 12} n(n^2-1)\delta_{n+m,0}}
with $c=1$. 

Some aspects of large $N$ group theory have very natural CFT fomulations.
For example, the $U(1)$ charge is: 
\eqn\uone{
\eqalign{
Q &=\int d\theta \Psi^{+} \bigl( -i {d \over d \theta} \bigr ) \Psi 
=\sum_n n B^{+}_{-n} B_n\cr
&\rightarrow 
\sum_{n} n c_{-n} b_{n} - \sum_{n}n \bar c_{-n} \bar
b_{n}\cr
&=  L_0 - \bar L_0.  \cr}}
In the second line we have assumed that the operator is acting on states so that the
expression is unambiguous.
 
\vskip0.1truein\noindent
{\bf Remark}:
In the $U(N)$ theory we expect to get 
\eqn\lghlun{
\CH_{U(N)}\to \oplus_p \CH_{bc}^{(p)}\otimes 
\CH_{\bar b\bar c}^{(-p)} . }

\subsec{Bosonization} 
\subseclab\ssBos

The most important application of the CFT 
interpretation is to bosonization of the $b,c$ systems. 
In two-dimensional theories, relativistic bosons and 
fermions can be mapped into each other. Operators
in the Fermi theory have equivalent expressions in 
terms of operators in the Bose theory. 

To motivate this let us consider the CFT version of 
the position space operators: $\Upsilon_n = \tr U^n$
from \upsop.  
We write this in second quantized language and 
substitute \bcsys. Cross terms between barred and 
unbarred fields involve operators that mix the two 
Fermi levels. Since we are only 
interested in the case of the decoupled Fermi level 
excitations we may replace:
\eqn\upsii{\eqalign{
\Upsilon_n &= \tr~ U^n 
\rightarrow  \oint dz~ z^{-1-n} c~ b(z) +
\oint d\bar z~ \bar z^{-1+n} \bar c~  \bar b(\bar z)\cr
&= \sum_m c_{n-m} b_{m} + \bar c_{m-n} \bar b_{-m}\cr
&= \alpha_n + \bar \alpha_{-n} \cr}}
where  we have introduced a field $bc = i \p_z \phi(z)$
which  has expansion
\eqn\bosalg{\eqalign{
\partial_z\phi(z) &=i\sum_{m\in\IZ} \alpha_m z^{m-1}\cr
{}[\alpha_m,\alpha_n]&=
[\bar\alpha_m,\bar\alpha_n]~=~m\delta_{m+n,0}\cr
\left[\alpha_m,\bar\alpha_n\right]&=0.\cr}}

In terms of $\alpha_n$, the Virasoro operators are: 
\eqn\virop{  
L_n = \ha \sum \alpha_{n-m} \alpha_m}
which satisfy \virasoroalg, again with $c=1$. 
Using the $\alpha$ we can define a vacuum 
$\alpha_n \mid 0\rangle =0$ for $n\geq 0$ and a
statespace
\eqn\achalph{
\CH_\alpha = {\rm Span}~ \Biggl\{ 
\mid \vec k \rangle\equiv \prod (\alpha_{-j})^{k_j}\mid 0\rangle
\Biggr\} . }

Bosonization states that there is a natural 
isomorphism: 
\eqn\bsnztin{
\CH_\alpha \cong  \CH_{bc}^{(0)}}
We will not prove this but it can be made very 
plausible as follows. 
The Hilbert space may be graded by 
$L_0$ eigenvalue. The first few levels are: 
\eqn\firstlev{\matrix{
L_0 = 1  & \{b_{-\ha} c_{-\ha} \mid 0 \rangle \} & \{\alpha_{-1} \mid 0 \rangle \}\cr
L_0 = 2  &
\{ b_{-\ha} c_{- {3\over 2}}  \mid 0 \rangle, b_{-{3 \over 2}} c_{- {1\over 2}} \mid 0 \rangle \}  
& \{ \alpha_{-2}\mid 0 \rangle, (\alpha_{-1})^2 \mid 0 \rangle \} \cr}}
At level $L_0=n$, the fermion states are labeled by Young diagrams $Y\in \CY_n$
as in \repfrm.
At level $L_0=n$, the Bose basis elements
are labeled by partitions of $n$.
We will label a partition of $n$ by a 
vector $\vec k=(k_1,k_2,\dots)$ which has almost all
entries zero, such that $\sum_j j k_j =n$.
Bosonization states that the two bases are linearly related: 
\eqn\linrel{
\mid Y \rangle = \sum_{\vec k\in {\rm Partitions} (n)}
\langle \vec k \mid Y\rangle \mid \vec k \rangle}

We have seen that - in terms of the space of class functions - the Fermi basis corresponds
to the rep basis. 
From \upsii\ we see that states in the Bose basis correspond to the class functions: 
\eqn\shuri{
\langle U \mid \vec k \rangle\equiv  \Upsilon(\vec k)  
\equiv \prod_{j=1}^\infty (\Tr~ U^j)^{k_j}}
where $\Tr$ is the trace in the fundamental 
rep.
In fact, a relation such as \linrel\ holds 
for class functions at finite $N$ and is 
a consequence of the Schur-Weyl duality theorem, 
as we explain next. 

\vskip0.1truein\noindent
{\bf Remarks:}
\item{1.}
In the $U(N)$ theory we must consider states $\mid \vec k ,\ell \rangle$ with wavefunctions
\eqn\shurii{
\langle U \mid \vec k,\ell  \rangle\equiv  \Upsilon(\vec k,\ell)  
\equiv (\det U)^{\ell} \prod_{j=1}^\infty (\Tr~ U^j)^{k^j}}
\item{2.}
For other classical gauge groups the connection to free fermions has been studied in
\refs{\panz}.  

\exercise{Commutators}

Check that $\Upsilon_n(U)\leftrightarrow \alpha_{-n} +\bar\alpha_n$
and $\Upsilon_n(U^\dagger)\leftrightarrow \alpha_{n}+\bar\alpha_{-n}$
form a commuting set of operators. 

\endexercise

\subsec{Schur-Weyl duality }
\subseclab\ssWSD

We return to $N<\infty$. 

All unitary irreducible reps of  $SU(N)$ are obtained from reducing tensor 
products of the fundamental rep (for $U(N)$ we need tensor products of both the fundamental and its complex conjugate).  
Consider the action of $SU(N)$ on $V^{\otimes n}$ where $V$ is the fundamental
rep of $SU(N)$. 
Let $\rho(U)$ represent the action of $U$ in $V^{\otimes n}$.    
The symmetric group $S_n$ acts on $V^{\otimes n} $ by permuting the factors.
Let $\tilde{\rho}$ be the map from $S_n$ to $End(V^{\otimes n})$.
It induces a map from the group ring $\IC(S_n)$, also denoted by $\tilde{\rho}$.

\bigskip

\boxit{{\bf Schur-Weyl duality}: The commutant of 
$\rho (SU(N))$ in $V^{\otimes n}$ is $\tilde{\rho}( \IC(S_n) )$, and 
$V^{\otimes n}$ is completely reducible to the 
form:
$V^{\otimes n}\cong \sum_{Y\in\CY_n}  R(Y)\otimes r(Y) $}

In the $SU(N)$ theory irreps can be 
labelled by Young diagrams with fewer than $N$ rows:
$Y\in \CY_n^{(N)}$, where
$\CY_n^{(N)}$ is the set of valid Young diagrams for $SU(N)$.
We denote by $R(Y)$ the corresponding irrep. 
We denote by $r(Y)$ the rep of $S_n$ associated with $Y$.
So if $P_Y$ is the Young projector for the rep of $S_n$ associated with $Y$,
we have \refs{\Zelo}
\eqn\pyeq{   P_Y V^{\otimes n} \cong R(Y)\otimes r(Y) } 

Using Schur-Weyl duality we can relate $\Upsilon_{\vec k}$ to characters of irreducible 
reps: 
\eqn\frobrecip{\eqalign{
\chi_{R(Y)}(U) &=\sum_{\sigma\in S_n} {1\over n!} \chi_{r(Y)}
(\sigma) \Upsilon_{\vec k(\sigma)}(U)\cr 
\Upsilon_{\vec k(\sigma)}(U)&=
\sum_{Y\in \CY_n^{(N)}} \chi_{r(Y)}(C(\vec k)) \chi_{R(Y)}(U). \cr}}

Here we identify conjugacy classes $C(\vec k)$ of $S_n$ with partitions $\vec k$ of $n$,
while $\vec k(\sigma)$ denotes the conjugacy class of an element $\sigma$. 
In the $U(N)$ case we have the same equations, diagonal in $Q$. 
To prove these relations we note that we can write the projector as 
\eqn\projc{
P_Y = {d_{r(Y)}\over n!} \sum_{\sigma\in S_n} \chi_{r(Y)}(\sigma) \sigma , }
where $d_{r(Y)}=\dim~ r(Y)$. 
Then we evaluate $\tr_{V^{\otimes n}} (U P_Y) $ in two 
ways, using \pyeq\ and \projc. Noting that 
\eqn\upsdef{\eqalign{
 \tr_{V^{\otimes n}}\bigl[ U \sigma\bigr]
 &= \sum_{ i_1 .. i_n=1}^{N}
 U_{i_1 i_{\sigma (1) }} 
U_{i_2 i_{\sigma (2) }} \cdots U_{i_n i_{\sigma (n) }} \cr  
&= \Upsilon_{\vec k(\sigma)} (U), \cr 
}}
establishes the first equation. Evaluating
$ \tr_{V^{\otimes n}}\bigl[ U \sigma\bigr]$
in two ways proves the second identity. 
Note that at finite $N$  we 
do not need to add the $\Upsilon$ functions 
constructed from powers of 
$U^\dagger$ since these may be expressed in terms of $U$. 

Comparing \frobrecip\ to the statement of bosonization 
in the $N\to \infty$ theory 
we see that the overlaps of Bose/fermi states are 
simply the characters of symmetric groups: 
\eqn\bosini{
\langle \vec k \mid Y\rangle =
{1\over n!} \chi_{r(Y)} (C(\vec k))}
Using \repfrm\ and the bosonisation formulae \upsii\ we see that this gives
a field theoretic formula for characters of the symmetric groups, as a
vacuum expectation value of free fields. (This fact
goes back to Sato, and  is important in the theory of the 
KP hierarchy.)

\subsec{The string interpretation} 
\subseclab\ssTheSI

Gross and Taylor offered a very elegant string interpretation of the Hilbert space of 
class functions\foot{This interpretation is implicit in some previous works on strings
in the $1/N$ expansion.}.
Consider spacetime to be a cylinder $S^1\times \IR$.
The circle has an orientation. 
Put simply, the one-string Hilbert space is identified with the group algebra 
$\IC[\pi_1(S^1)]$.
The total string Hilbert space is then identified with the Fock space,
${\rm Fock}[\IC[\pi_1(S^1)]]$. Physically,
strings of electric flux are winding around
the circle.

\ifig\strwdng{Strings winding.}
{\epsfxsize3.0in\epsfbox{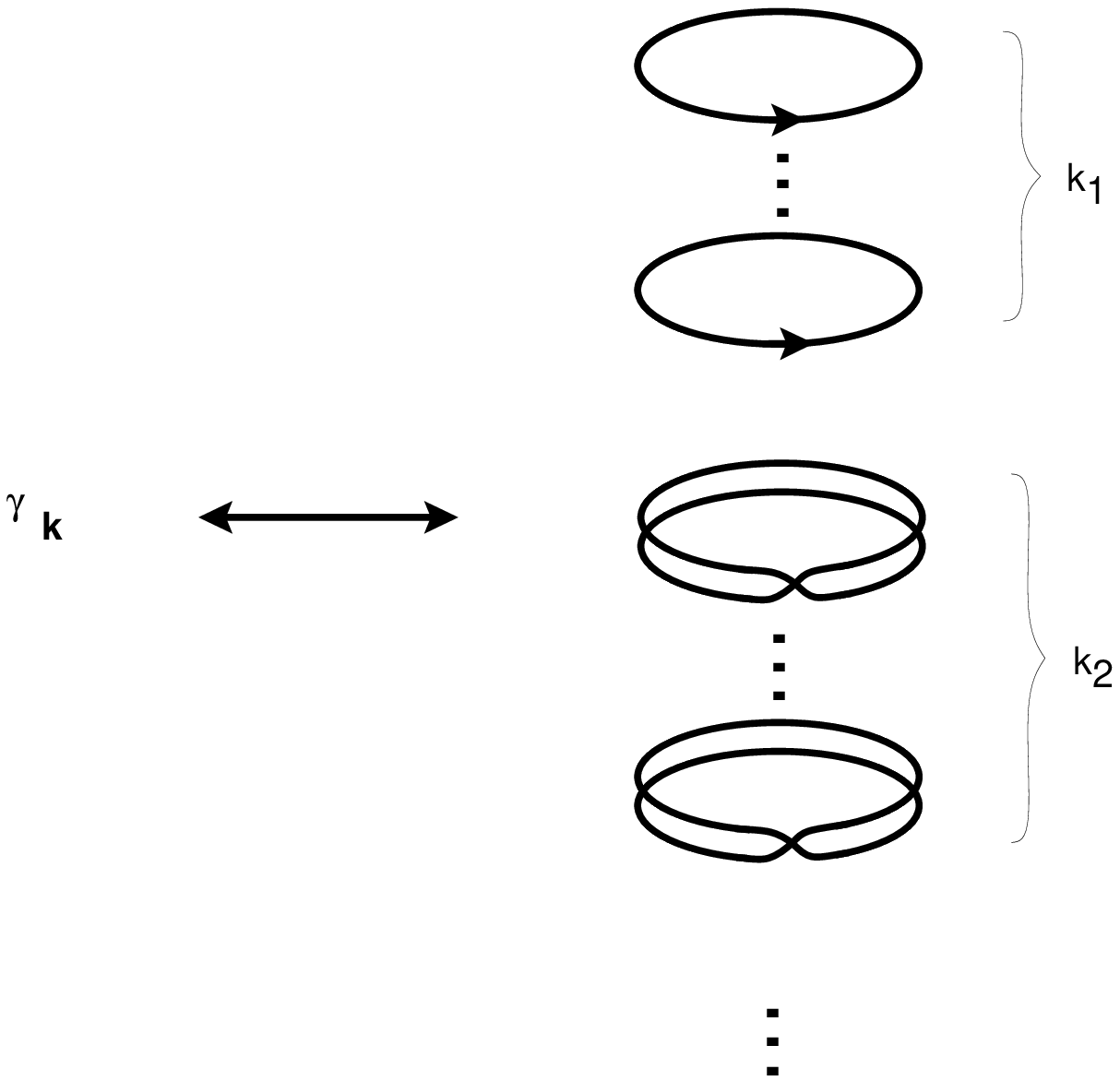}}

For $j> 0$,  $\alpha_{-j}$ creates  a  state of string 
winding $j$ times around the spatial circle in 
the same sense as the orientation. Similarly, 
$\bar \alpha_{-j}$ creates a string with opposite 
orientation.  This is the origin of the names 
``chiral'' and ``antichiral'' above. 
Thus we may picture a chiral state corresponding 
to the partition $\vec k$ as  sets of $k_i$ circles winding around 
a boundary of the target space $i$ times as in \strwdng.
A state corresponding to strings winding in both 
directions would be labelled by 
$\mid \vec k , \vec{\bar k}\rangle$. 
We refer to this basis as the string basis. 

\subsubsec{Back to class functions}
\subsubseclab\sssBtoCF

It is interesting to translate the string states 
back into the language of class functions. 
We have seen that $\mid \vec k\rangle$ corresponds to 
the class function $\Upsilon_{\vec k}$. The story 
for the nonchiral theory is more complicated. 
The state $\mid \vec k_1, \vec k_2\rangle$ 
will correspond to a class function 
$\Upsilon_{\vec k_1,\vec k_2}$. We will also 
use the notation 
$\Upsilon_{v,w}=\Upsilon_{\vec k(v) , \vec k(w)}$ 
for permutations 
$v,w$. These functions will have the form:
\eqn\coupst{ \Upsilon_{\vec k(v) , \vec k(w)} (U,U^\dagger)
 = \Upsilon_{\vec k(v)}(U) 
\Upsilon_{\vec k(w) }(U^\dagger) +\cdots .}
The class functions  in the first term on 
the RHS of \coupst\  are not orthonormal;
the extra terms guarantee that the states satisfy the 
orthogonality relations \refs{\GrTa}, 
\eqn\couport{ 
\int dU~ \Upsilon_{v_1,w_1}(U) \Upsilon_{v_2,w_2} 
(U^\dagger) = \delta ( \vec k_{v_1}, \vec k_{v_2}) 
\delta( \vec k_{w_1}, \vec k_{w_2}) 
\prod_{j}  j^{k^{(j)}_{v_1}}k^{(j)}_{v_1}!
j^{k^{(j)}_{v_2}}k^{(j)}_{v_2}! }
Explicitly,
corresponding  to permutations $v$ and $w$ we have
\eqn\couploop{ 
\Upsilon_{v,w} = \Upsilon_{ \vec k(v), \vec k(w) } 
= \prod_{j}  \sum_{ m=1}^{\min (k_j, l_j) } 
\bigl[ \tr~ (U^j))\bigr] ^{k_j-m}  \bigl[ \tr~ ( {U^\dagger }^j) \bigr]^{l_j-m}
P_{k_j,l_j} (m).}
Here $j$ runs over positive integers; $k_j$ and $l_j$ 
are the numbers of cycles of length $j$ in $v$ and $w$ respectively; and  
$P_{k_j,l_j} (m) ={ k_j \choose m} {l_j \choose m} m! (-1)^m $. 
The functions \couploop\ are called coupled 
loop functions \refs{\GrTa}. 

The class functions $\Upsilon_{\vec k_1,\vec k_2}$ and $\chi_{R\bar S}$ are
related by a generalization of the Frobenius relations \frobrecip.  
The character of the coupled rep is expected to be of the form 
\eqn\crepst{ \chi_{R \bar S} (U) 
= \chi_R( U) \chi_S(U^\dagger) + \sum_{ R^\prime,S^\prime} 
C_{R^\prime,S^\prime} ^{R\bar S} 
\chi_{R^\prime}  (U)\chi_{S^\prime}  (U^\dagger) , } 
where $R^\prime $ and $S^\prime$ are 
Young diagrams  with fewer  
than $n^+$, $n^-$ boxes, respectively, and
the $C$'s  are some constants to be determined. 
For $n^+,n^- < N/2 $, the rep $R\bar S$ 
uniquely determines $R$ and $S$ (this 
becomes clear from \cplrep).  
Together with irreducibility of the coupled reps,  this leads to 
the orthogonality relation : 
\eqn\creporth{ \int dU \chi_{R_1\bar S_1}(U)
\chi_{R_2\bar S_2} (U^\dagger) = \delta_{R_1 R_2} \delta_{S_1S_2}.  }

As a candidate for the generalisation
to the non-chiral case of  the relations
between characters of chiral reps and  
symmetric group characters, which follow 
from Schur-Weyl duality (section \ssWSD), 
 consider  the following sum of  coupled loop functions \GrTa : 
\eqn\crepcan{ 
 \sum_{\sigma^+ \in S_{n^+} , 
\sigma^- \in S_{n^-}  } {\chi_r (\sigma^+ ) 
\over { n^+!} } {\chi_s (\sigma^- ) 
\over { n^-!} }  \Upsilon_{\sigma^+,\sigma^-} (U).  } 
It satisfies the two conditions \crepst\ 
and \creporth. The first is clear by 
using  \coupst. The second follows 
from orthogonality of the coupled loop 
functions and the orthogonality of characters of $S_{n^+}$ and $S_{n^-}$. 
This establishes that 
\eqn\crep{ \chi_{R\bar S} (U) = 
 \sum_{\sigma^+ \in S_{n^+} , \sigma^- \in 
S_{n^-}  } {\chi_r (\sigma^+ ) 
\over { n^+!} } {\chi_s (\sigma^- ) 
\over { n^-!} }  \Upsilon_{\sigma^+,\sigma^-} (U) .}

\subsec{String Interactions} 
\subseclab\ssSI

The group theory Hamiltonian $C_2$ is {\it not} 
diagonal in the string basis of states. Thus, from 
the point of view of the string theory, there are 
nontrivial interactions in the theory. These 
interactions have nice interpretations in 
terms of the splitting and joining of strings \refs{\MiPoetd,\dgcrg}. 

Consider first the $U(N)$ theory. 
In terms of $\alpha$'s, 
$H= \sum n_i^2  - E_0$ translates into 
\eqn\hforal{\eqalign{
H &= e^2 N\biggl[ (L_0 + \bar L_0) + H_{\rm interaction}\biggr] \cr
H_{\rm interaction}& = {1\over N} \Biggl\{
\sum_{n,m>0} \Biggl(\alpha_{-n-m} \alpha_n \alpha_m + 
 \bar \alpha_{-n-m}  \bar\alpha_n  \bar\alpha_m\Biggr) \cr
&\qquad\qquad+ \sum_{n,m<0}  \Biggl( \alpha_n \alpha_m \alpha_{-n-m}+ 
   \bar\alpha_n  \bar\alpha_m\bar \alpha_{-n-m}\Biggr)
\Biggr\} . \cr
}} 

Thus, two strings winding with winding numbers $n,m$ 
around the cylinder will propagate to a third string 
winding $n+m$ times. This is a 3 string interaction. 
For $SU(N)$ we have an additional interaction 
$$\Delta H = {e^2\over N} (L_0 - \bar L_0)^2 \quad . $$

\subsubsec{The Fermi fluid picture} 
\subsubseclab\sssFF

There is another nice way to think about the states in 
the large $N$ limit, which has been employed by 
Douglas. 
In order to have a well-defined momentum operator 
in the large $N$ limit we must scale 
$$P={1\over N} \oint \Psi^\dagger \biggl(-i {d\over d\theta}
\biggr) \Psi
$$
so $\hbar$ in the problem is effectively $1/N$.

\ifig\FermiFluid{Fermi liquid in phase space}
{\epsfxsize1.5in\epsfbox{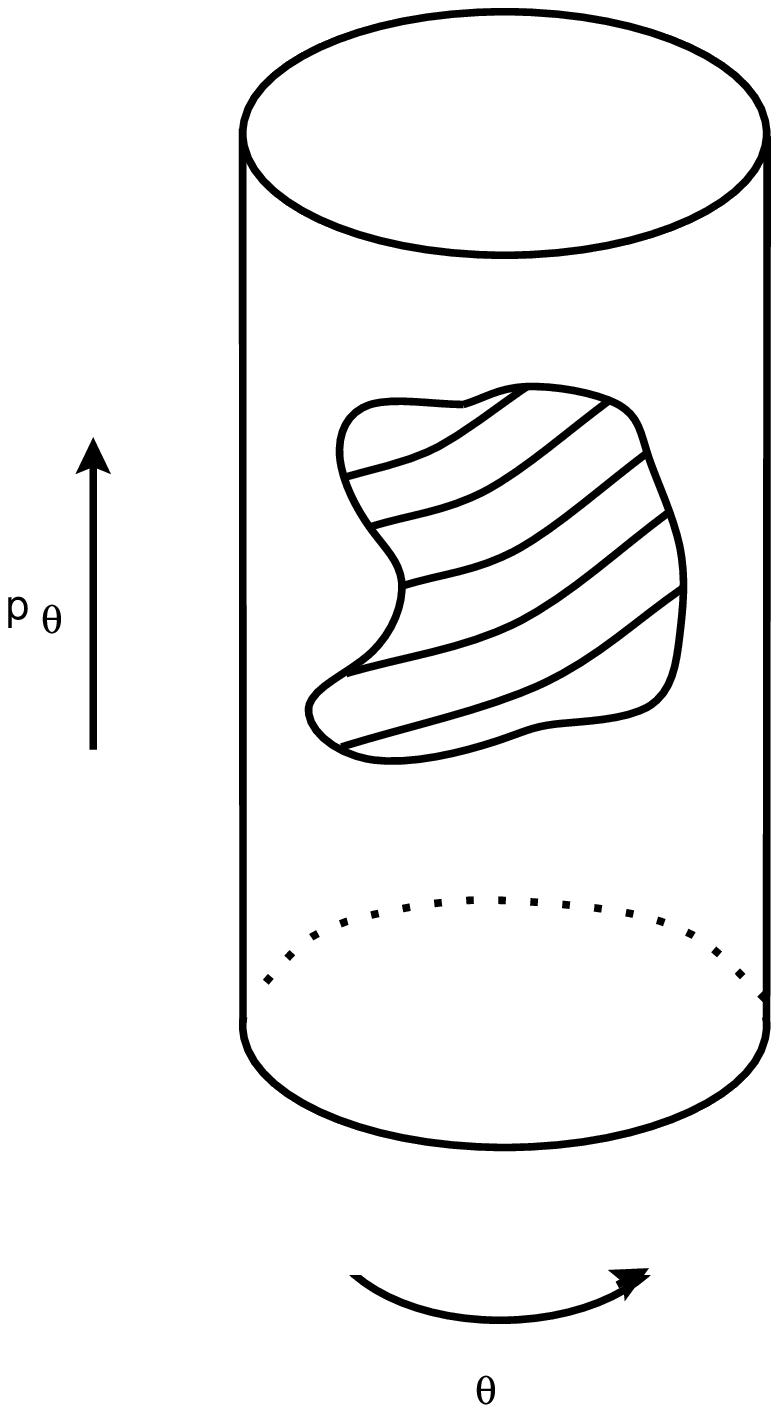}}

We thus have 
$N$ free fermions in a quantum system with 
 $\hbar = 1/N$. The classical limit of such a system 
is best described in terms of phase space. Each 
fermion occupies an area $\Delta p \Delta \theta = \hbar$ 
in phase space, which is, in this case, 
$T^* S^1 = \{ (\theta, p)\}$.
$N$ fermions occupy a region with a volume $=1$.   
Classically, a state is described by density
$\rho(\theta,p)$. Expectation values of operators 
correspond to integrals over phase space: 
\eqn\averg{
\langle \Psi \mid \CO \mid \Psi\rangle 
= \int d \theta dp~ \CO(\theta,p)~ \rho(\theta,p). 
}
Moreover, $\rho$ is a constant density 
times the characteristic function of a region 
of area $=1$, so classically the state may 
be characterized as a region, as illustrated 
in \FermiFluid.

From the Fermi liquid point of view
the energy may be nicely expressed in 
terms of the Fermi momenta $p_{\pm}$: 
\eqn\liqen{
E= \int dp d\theta~ p^2 \rho(p,\theta) 
= \int d \theta {1\over 3} (p_+^3 - p_-^3).  
}

The cubic Hamiltonian arises very 
naturally in the Fermi liquid point of view 
mentioned above. In particular the energy 
\liqen\  becomes \hforal\ under the substitution
\eqn\pplu{\eqalign{ 
&p_+( z) = (N/2 - iz \partial \phi )\cr
& p_- (z) = ( N/2 + i\bar z \bar \partial \phi), \cr  
}}
where we used $\int d\theta \partial_{\tau} 
 \phi =\int d\theta \partial_{\theta}\phi =0$.

\vskip0.1truein
{\bf Remarks:} 
\item{1.} A similar Fermi liquid picture was developed by Polchinski 
for the $c=1$ matrix model. The difference 
here is that the eigenvalues 
live on a circle as opposed to the real 
line, and there is 
no background potential. 
Note that  the decoupling of the two sectors is very 
natural in terms of Fermi levels. 
\item{2.} The Fermi system is completely 
integrable. 
There is an infinite number of conserved Hamiltonians 
corresponding to the center of the universal 
enveloping algebra. These are related to the
infinite parameter family of natural Hamiltonians 
for \ymt. 
For $N=\infty$ the algebra of conserved Hamiltonians 
is 
generated by all the Casimirs $C_k$. 
In terms of the relativistic fermions we have a
$w_\infty \oplus w_\infty$ algebra. This has 
a nice interpretation in terms of area-preserving 
diffeomorphisms of the Fermi sea.

\subsec{Casimirs and Symmetric Groups}
\subseclab\ssCSG

Using Schur-Weyl duality we can express 
$SU(N)$ and $U(N)$ Casimirs in terms of 
quantities for the symmetric group: acting on 
$V^{\otimes n}$ any central element of the universal 
enveloping algebra  must be expressed in 
terms of operators from the symmetric group. 
On the other hand, since
 it comes from $U(N)$, it must be in 
the center of $\IC[S_n]$. 

The relation between $C_k$ and 
characters of symmetric group reps will 
be very important in deriving the $1/N$ 
expansion and finding a string interpretation. 
The most important for our purposes
is 
\eqn\quadcas{\eqalign{
C_2(R(Y)) &= Nn+ 2 {\chi_{r(Y)} (T_{2,n}) \over {d_{r(Y)}}}   \qquad\qquad U(N)\cr
&= Nn+ 2 {\chi_{r(Y)} (T_{2,n}) \over {d_{r(Y)}}}  -n^2/N  \qquad SU(N)\cr}
}
where $Y\in \CY_n^{(N)}$ and 
$T_{2,n}=\sum_{i<j} (ij)\in \IC[S_n]$. The proof of this, and similar 
formulae for the higher Casimirs is rather technical, and 
may be found in the following appendix. 

\subsec{Appendix: Higher Casimirs} 
\subseclab\ssAHC

The relation between Casimirs and symmetric groups
 may be established by writing the value 
of the Casimir on  a rep $R(Y)$ in terms of the row lengths $n_i$ in the Young diagram $Y$.
 This gets  nontrivial for the higher Casimirs ( see for example \PP ). 
Similarly the characters of symmetric groups can be evaluated in terms of the $n_i$, some explicit results are given in  \Ing .   For $C_2$ this method was used in \GrTa , to establish the relation with characters of symmetric groups. Here we will give a more direct construction of the transformation between Casimirs and symmetric group classes,  which generalises easily to the higher Casimir case and which does not require explicitly knowing either the eigenvalues of the Casimirs or the characters of the symmetric groups,  as functions of the $n_i$. 
\foot{Similar (independent) results 
were obtained in \ganor. This appendix is based on \Hcas.}

We start by setting up some notation. 
 A basis for the Lie algebra  $u(N)$ of $U(N)$ is given by the 
matrices $E_{ij}$, $1\le i,j \le N$ which have matrix elements $(E_{ij})_{\alpha \beta}, 
1\le \alpha\le N, 1\le \beta \le N$ : 
\eqn\dfgen{    (E_{ij})_{\alpha \beta} = \delta_{i\alpha} \delta_{j\beta}.} 
They satisfy the relations 
\eqn\reli{  
 E_{ij}E_{kl}  = \delta_{jk} E_{il} }. 
\eqn\relii{                      
 [E_{ij}, E_{kl} ] =  \delta_{jk} E_{il} - \delta_{il} E_{jk} .  }
Let $\phi_{i}, 1\le i \le N $, denote the standard
 basis for the fundamental rep $V$.  
The action of the generators of $u(N)$ is given by 
\eqn\act { E_{ij} \phi_{k} = \delta_{jk} \phi_{i} .}

The following elements of the universal enveloping algebra (UEA ) of $u(N)$ 
\eqn\defcasim{  C_k^{U(N)} =  \sum_{i_1\cdots i_k} E_{i_1i_2} E_{i_2i_3} \cdots E_{i_ki_1}, } 
for $1\le k \le N$, 
generate the centre of the algebra, which we denote by $\Xi_N$ 
 (see for example \Zelo).
 Other sets of generators can be obtained, for example, 
 by changing the order of the $E's$ 
in \defcasim. Any monomial in $\Xi_N$ has a  {\bf natural grading} 
given by the sum of the degrees of all the generating elements in its expression in terms of generators. 
 
Let $\phi^{a}_{i}, 1\le a \le n$, denote the basis   
for the $a'th$ factor $V$  in the tensor product $V^{\otimes n } $ . 
The action of the generators of $u(N)$ on this tensor product  is given by 
\eqn\act { E_{ij} \prod_{a=1}^{n} \phi^{a}_{k} 
= \sum_{a_1=1}^{n}   E^{a_1}_{ij}  \prod_{a=1}^{n} \phi^{a}_{k}.  } 
where 
\eqn\expl{
E^{a_1}_{ij} \phi^{a_2}_{k} = \delta_{jk} \delta_{a_1 a_2} \phi^{a_2}_{ij}. 
}
The operator representing the action 
of the $k'th$ Casimir can therefore be written as
\eqn\castens{ \sum_{a_1, \cdots, a_k \ge 1}^{n}\sum_{i_1\cdots i_k= 1} ^{N} 
 E^{a_1}_{i_1i_2} \cdots E^{a_k}_{i_k i_1}   .}

{}From the Schur-Weyl duality theorem, we expect that there is, in $V^{\otimes n } $, 
  a transformation from operators 
representing the action of $S_n$ to operators representing 
the centre as described by polymomials in the generators of $\Xi_N $. 
The explicit construction of the map uses the generalized 
exchange operators defined in \refs{\chen,\Part}. 

\noindent
{\bf Proposition 1} 

$ \sum_{i_1\cdots i_k} E^{a_1}_{i_1 i_2} E^{a_2}_{i_2i_3} \cdots E^{a_k}_{i_ki_1} 
\equiv P^{a_1a_2\cdots a_k} $  acts in $V^{\otimes n } $ as 
the cyclic permutation $(a_1a_2\cdots a_k)$.

This is checked by using the explicit form of the action. For example, when $k=2$ we have 
\eqn\pf{\eqalign{ 
 \sum_{i_1i_2} E^{a_1}_{i_1i_2}E^{a_2}_{i_2i_1} \phi^{a_1}_{j_1}\phi^{a_2}_{j_2}
=  \sum_{i_1i_2} \delta_{i_2j_1} \delta_{i_1j_2} \phi^{a_1}_{i_1}\phi^{a_2}_{i_2}
= \phi^{a_1}_{j_2}  \phi^{a_2}_{j_1}.  } }
 
This leads to 

\noindent
{\bf Proposition 2}

On $V^{\otimes n }$ we have, for $k \ge 2$, the following equivalence of operators in the universal enveloping algebra (UEA)  of $u(N)$ and the group ring of $S_n$: 
\eqn\cyc{ 
P^{(k)} \equiv \sum_{a_1 \ne  a_2 \ne \cdots \ne  a_k = 1}^{n} E^{a_1}_{i_1i_2}    E^{a_2}_{i_2i_3} 
\cdots E^{a_k}_{i_ki_1}  = k T_k } 
where $T_k$ is the element in the group ring of $S_n$ which is the sum of  all permutations in the conjugacy class characterised by one  cycle of length $k$ and remaining cycles of length $1$. 
We will call the operators $P^{(k)}$ cycle operators.

 The sum on the left  counts cycles containing  fixed numbers $a_1 a_2 \cdots a_k$, once for each permutation of the numbers. Therefore it counts each cycle $k$ times. This explains the factor $k$ multiplying $T_k$. 
More generally we can write UEA operators for any conjugacy class in $S_n$. 

\noindent
{\bf Proposition 3} 

Let the index $j$ run over the 
nontrivial cycles, i.e cycles of length $ z_j \ge 2$
The permutation is represented by the operator : 
\eqn\permuea{   \prod_{j}  \bigl ( \sum_{a_{j1}\ne \cdots \ne a_{jz_j} =1 }^{n} P^{a_{j1}a_{j2}\cdots a_{jz_j} } \bigr ). } 
The sums are  over $a's$ constrained to be all different from each other. 

These   operators can be written in a form where the sums over $a's$ are unrestricted, and 
in this form their relation to the Casimirs $C_k$ will become apparent. We will give 
 a more detailed discussion of the transformation of the cycle operators to Casimirs, and vice versa; the extension to  operators corresponding to arbitrary permutations is straightforward. 
 We will first discuss $P^{(k)}$ operators for $k\le n$. Later we will describe how the transformation is carried out when we relax this restriction.
 
\subsubsec{ From cycle operators to Casimirs} 
We can rewrite the cycle operators: 
\eqn\pop{ \sum_{a_1 \ne  \cdots \ne a_k=1}^{n} P^{a_1,a_2,\cdots a_k} = 
 \sum_{a_1  \cdots  a_k=1}^{n} P^{a_1,a_2,\cdots a_k}  
\prod_{i<j =1}^{ k}  (1- \delta_{a_ia_j}) . }  

The leading term,  obtained by picking $1$ from all the factors in the product over $i,j$
 is the $k'th$ Casimir. 
Terms obtained from the delta functions can be reduced to expressions
in terms of lower order Casimirs,   
by using \reli .  We have then 
 $P_{k} = C_k + \cdots  $ as an operator representing the action of   
$\Xi_N$ in  $V^{\otimes n }$.  For fixed $k$, the equation holds for all 
$n$ and $N$ sufficiently large.

Now we can appeal to the statement of Schur-Weyl 
duality to show that the relation between permutations and Casimirs implies a relation 
between characters of permutations in $r(Y)$ and eigenvalues of Casimir operators in $R(Y)$.  
Using \cyc,  we have 
\eqn\chirel{  \sum_{a_1 \ne a_2\cdots \ne  a_k} P^{a_1\cdots a_k} P_Y V^{\otimes n } 
= (1\otimes kT_k) R(Y) \otimes r(Y) = (P^{(k)}   \otimes 1  )  R(Y) \otimes r(Y),  }
where $P_Y$ is the Young projector associated with the Young diagram $Y$.  
Tracing over the second factor we have

\boxit{ 
\eqn\chireli{ k \chi_{r(Y)} (T_k) = d_{r(Y)}  P^{(k)} (R(Y)) . } 
}

A similar equation can be written for any permutation using the operators defined in 
\permuea. 

Now we discuss how to go   from 
Casimirs to symmetric group characters: 
\eqn\cassym{
{C_k(R(Y)) \over N^{k-1}} 
 =\sum_{\sigma \in S_n} a_k(N,n, \sigma) {\chi_{r(Y)} (\sigma)\over {d_{r(Y)}}} \qquad . 
}
 We will outline  a construction which determines the coefficients
$a_k(N,n, \sigma)$.

Starting from \castens\ we 
separate  the unrestricted sum over $a's$ into  a number of sums,  separated according to the subsets of $a's$ which are equal to each other. 

\vskip0.1truein\noindent
{\bf Examples:} 

We illustrate this procedure by rederiving the formula for $C_2$.
\eqn\ctwo{\eqalign{ 
 &\sum_{i_1 ,i_2=1}^{N} \sum_{a_1,a_2=1}^{n}  E^{a_1}_{i_1i_2}E^{a_2}_{i_2i_1} 
\cr  &= \sum_{i_1 ,i_2=1}^{N} \sum_{a_1\ne a_2=1}^{n}  
E^{a_1}_{i_1i_2}E^{a_2}_{i_2i_1} +  \sum_{i_1 ,i_2=1}^{N} \sum_{a_1=1}^{n}  
E^{a_1}_{i_1i_2}E^{a_1}_{i_2i_1}\cr  
&= 2 {\chi_{r(Y)} (T_2) \over {d_R}} + NC_1 \cr 
&= 2 {\chi_{r(Y)} (T_2) \over {d_R}} + Nn . \cr } }

Applying this procedure for $C_3$ and $C_4$, and collecting terms,  we get 
\eqn\cthree{  C_3(R(Y)) = 3 {\chi_{r(Y)}(T_3) \over {d_R}} + 4N {\chi_{R(Y)}(T_2) \over {d_{r(Y)}}} + N^2n + n(n-1).   } 
\eqn\cfour{ \eqalign{
   C_4(R(Y)) =& 4 {\chi_{r(Y)} (T_4) \over {d_{r(Y)}} } + 9N  { \chi_{r(Y)}(T_3) \over {d_{r(Y)}}} 
\cr 
&+ ( 6N^2 + (6n-10) ) { \chi_{r(Y)} (T_2) \over {d_{r(Y)} } }  + 3N(n)(n-1) + N^3n. \cr}}
 
\subsubsec{ Casimirs for $SU(N)$} 
 Casimirs for $SU(N)$ are closely related to
 those of $U(N)$ ( \refs{\PP,\Okubo}). 
This relation can be used to 
compute the expressions for Casimirs of $SU(N)$
 in terms of characters of symmetric groups. 
Lie algebra elements in  $su(N)$ can be written in terms 
of those of $u(N)$  
\eqn\sugen{  \tilde E_{ij} = E_{ij} - {\delta_{ij} \over N} \sum_{k=1}^{N} E_{kk} . }
They satisfy the same commutation relations as in \relii, have zero trace, and satisfy 
the condition that 
\eqn\surel{ \sum_{i} E_{ii} = 0.}  
Casimirs for $SU(N)$ are defined as follows: 
\eqn\ceeksu{ C_k^{SU(N)} = \sum_{i_1\cdots i_k=1}^{N} 
\tilde E_{i_1i_2} \tilde E_{i_2i_3} \cdots \tilde E_{i_ki_1}.  }
We will sometimes write $C_k^{SU}$, leaving the $N$ label 
implicit. This leads to 
 simple relations between the Casimirs of $SU(N)$ 
and those of $U(N)$:
\eqn\relusu{ C_k^{SU(N)} = \sum_{l=0}^{k} C_{k-l}  {( {-C_1 \over N}) }^{l} 
{k\choose l}, }
where  $C_l = C_l^{(U(N)} $  for $l>0$, and $C_0=N$. 
 These can be used to derive, for example,  the formulae for $C_2$ and $C_3$
\eqn\cthreesu{\eqalign{ 
&C_2^{SU}(R(Y)) = C_2^{(U)} (R(Y)) - 2C_1({C_1\over {N}}) + N (C_1/N)^2\cr
&\qquad \qquad  = C_2^{(U)} (R(Y)) - {n^2\over N}\cr
&C_3^{SU}(R(Y)) = 3 {\chi_{r(Y)}(T_3) \over {d_{r(Y)} }} + 
{(4N -{6n\over N} )}{\chi_{r(Y)}(T_2) \over {d_{r(Y)}}} + N^2n  
-n -2n^2 + {2n^3\over {N^2}}\cr
}}

\ifig\docy{ A string winding twice around space propagates 
to a worldsheet which double covers the cylinder.}
{\epsfxsize3.0in\epsfbox{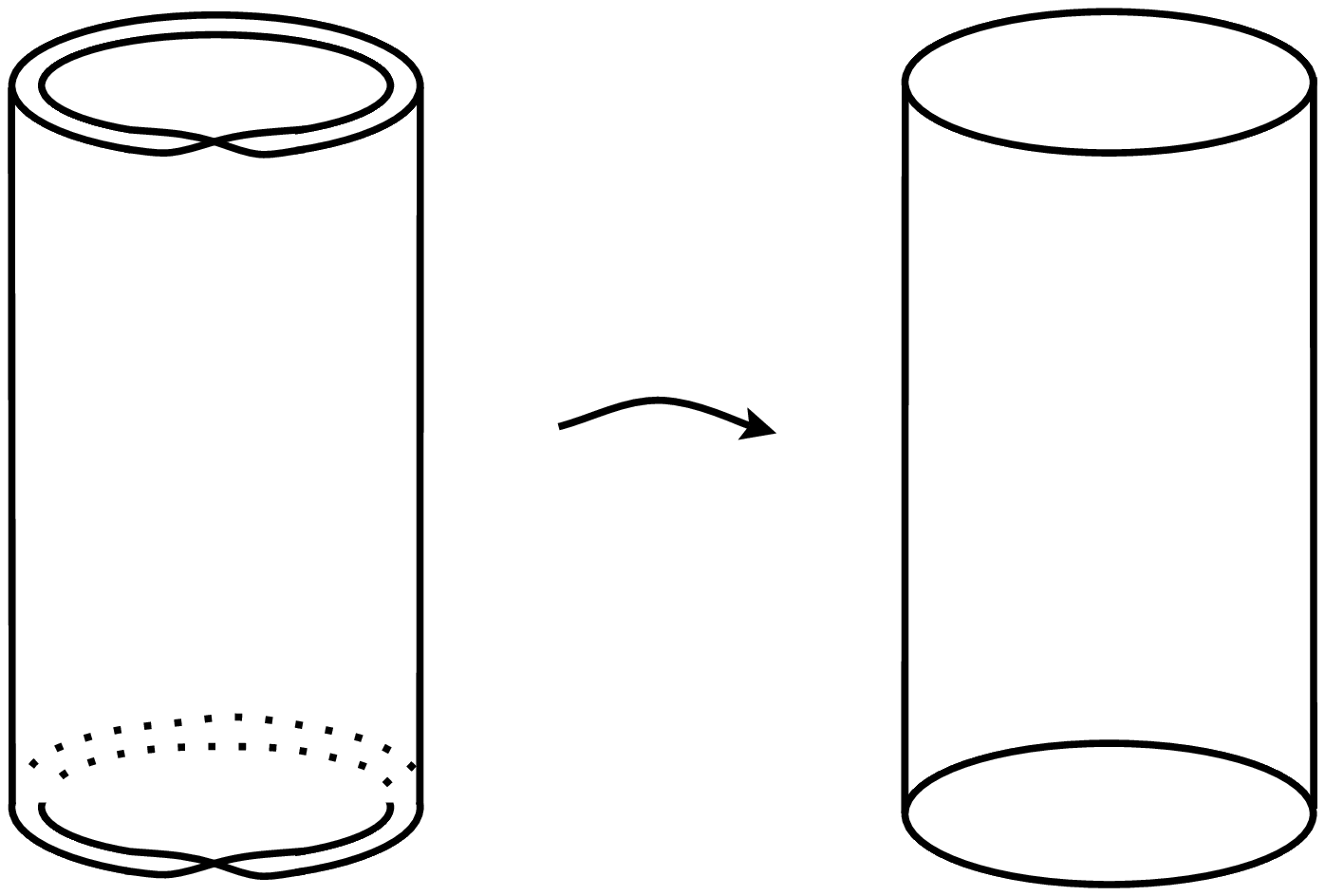}}

\ifig\ezfgii{String interactions described by the Hamiltonian 
of the previous chapter can be interpreted in terms of 
a covering by a pants diagram over the cylinder.}
{\epsfxsize3.2in\epsfbox{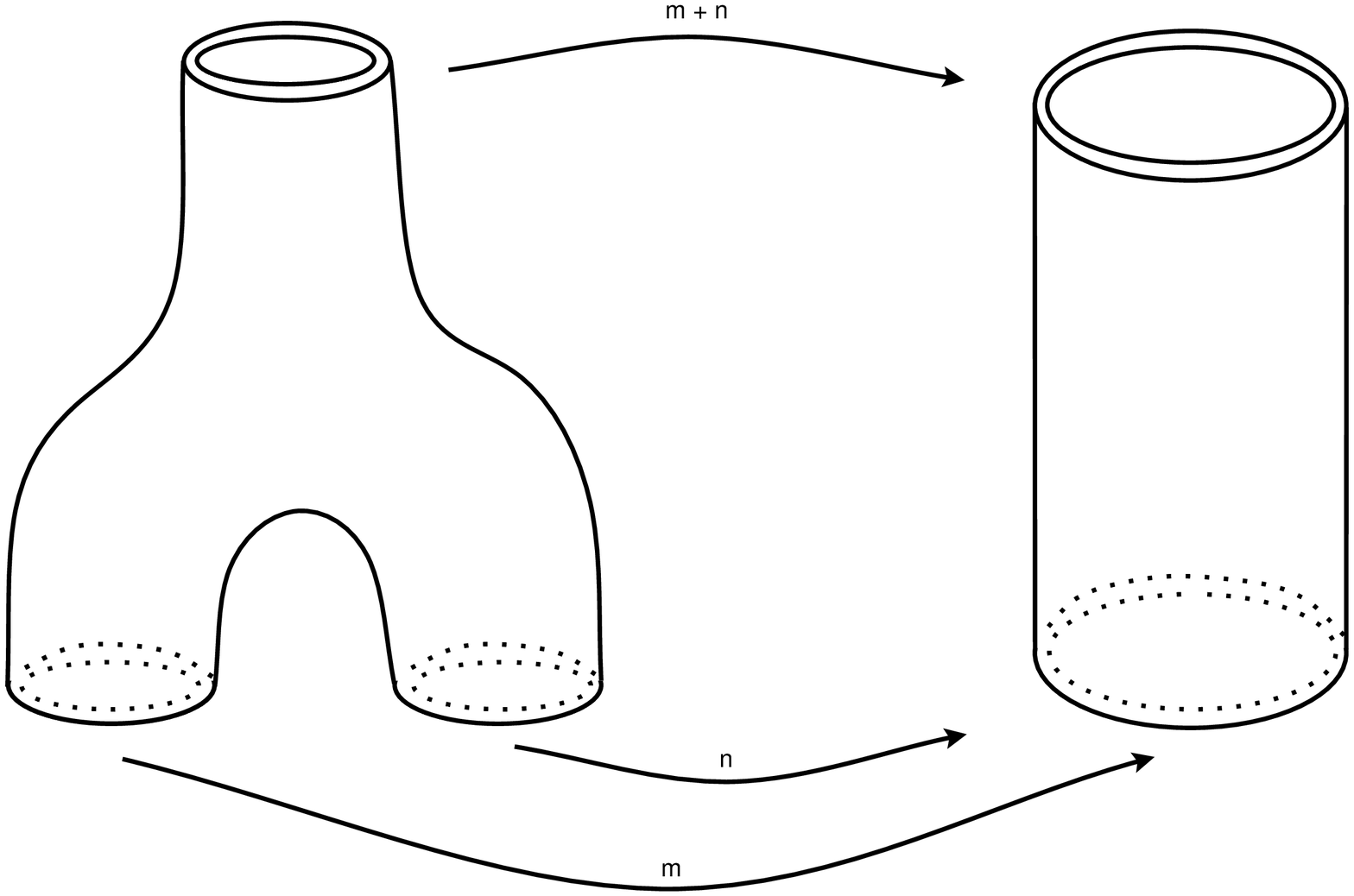}}

\ifig\ezfg{ A choice of generators for the homotopy group
of a punctured surface. The curves $\gamma(P)$ become
trivial if we fill in the puncture $P$.}
{\epsfxsize3.0in\epsfbox{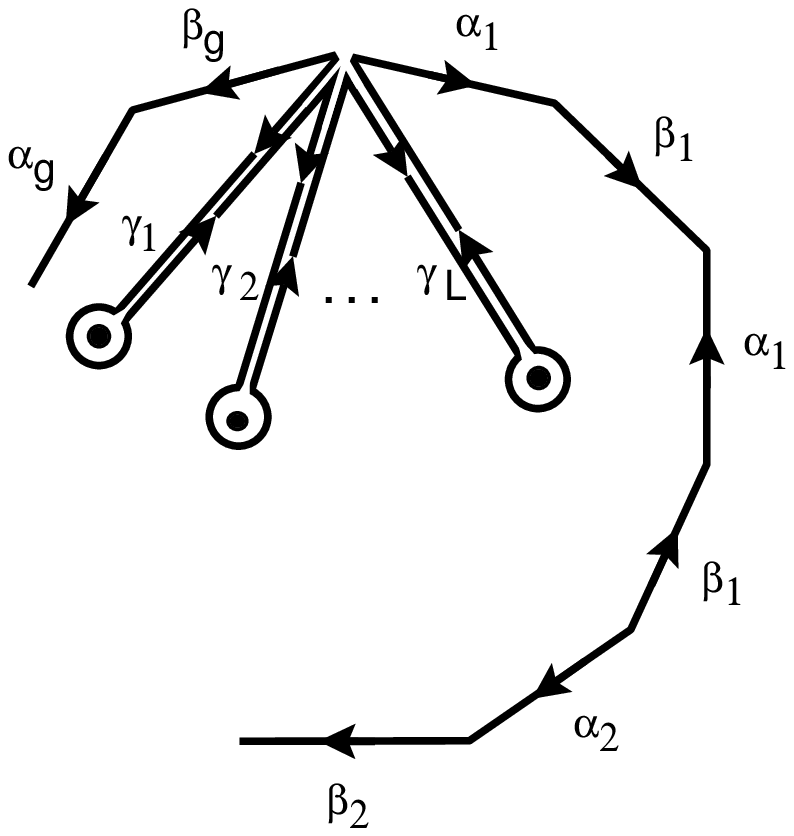}}

\newsec{Covering Spaces} 
\seclab\sCS

\subsec{Motivation}
\subseclab\ssMCS

When we interpret $\tr~ (U^j)$ as an operator creating a string at a fixed 
time, winding $j$ times around the target space direction,
we have introduced the concept of a covering manifold. 
Imagine propagating such a string. Naively we expect that 
propagation of such a string will define a covering 
map with the worldsheet covering the spacetime as in \docy . 

Moreover, if we have two states at an initial 
time they will propagate to a third as governed by 
the interaction Hamiltonian \hforal\ above. 
If we try to interpret this in terms of covering manifolds 
then we are led to an $(n+m)$-fold branched covering of 
the cylinder by the pants diagram as follows:
The lower two circles cover the ingoing circle on the 
cylinder by $n$ and $m$-fold coverings $S^1\to S^1$. 
The top circle on the pants covers the outgoing circle 
on the cylinder $n+m$ times. See \ezfgii.

\subsec{Maps of $\Sw\to\ST$ }
\subseclab\ssMofSwST

\subsubsec{Homotopy groups}

The homotopy groups of punctured Riemann surfaces may be 
presented in terms of generators and relations by
\eqn\fgp{
F_{p,L}\equiv  \biggl \langle \{
\alpha_i,\beta_i\}_{i=1,\ldots,p},\{\gamma_s\}_{s=1,\ldots,L}\vert
\prod_{i=1}^p [\alpha_i,\beta_i]\prod_{s=1}^L \gamma_s=1\biggr \rangle , 
}
where $[\alpha,\beta ]= \alpha \beta \alpha^{-1}\beta^{-1}$. 
Consider a compact orientable surface $\ST$ of genus $p$. If we remove
$L$ distinct points, and choose a basepoint $y_0$,  then there is an
isomorphism
\eqn\fgpiso{
F_{p,L}\cong \pi_1(\Sigma_T-\{P_1,\dots P_L\},y_0).
.}
This isomorphism is not canonical. The choices are
parametrized by the infinite group $\Aut~ (F_{p,L})$.
On several occasions we will make use of a set of generators
$\alpha_i, \beta_i$ and $\gamma_i$ of $\pi_1$ so that, if we cut along curves in the
homotopy class the surface looks like \ezfg.

\subsubsec{Branched covers}
\subsubseclab\sssBrnchCov

\noindent
{\bf Definition \sCS.1}.
\item{a.)}
A continuous map $ f:\Sigma_W \rightarrow \Sigma_T $ is a  {\it branched cover} if
any point $P \in \ST$ has a 
neighborhood $U \subset \Sigma_T$,  such that the inverse
image $f^{-1}(U)$ is a union of disjoint open sets
 on each of which $f$ is topologically
equivalent to the complex map $z\mapsto z^n$ for some $n$.
\item{b.)}
Two branched covers $f_1$ and $f_2$ are
said to be {\it equivalent} if there exists a homeomorphism $\phi :
\Sigma_W \rightarrow \Sigma_W $ such
 that $f_1\circ \phi = f_2 $.

\ifig\brchcvr{Local model of a branched covering. On the 
disk containing $Q$ the map is  $z\to w=z^{n(Q)}$. }
{\epsfxsize3.0in\epsfbox{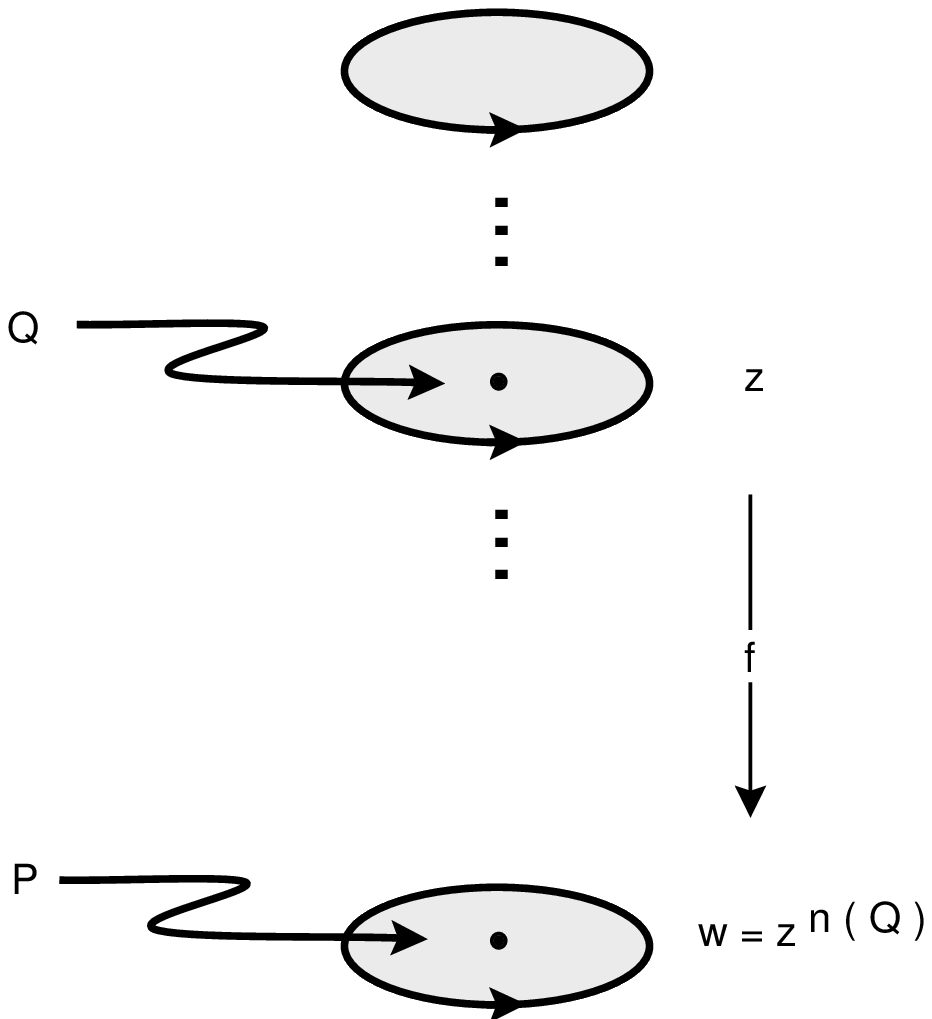}}

\ifig\liftcurv{Lifting of curves in a branched cover 
to produce elements of the symmetric group. }
{\epsfxsize3.5in\epsfbox{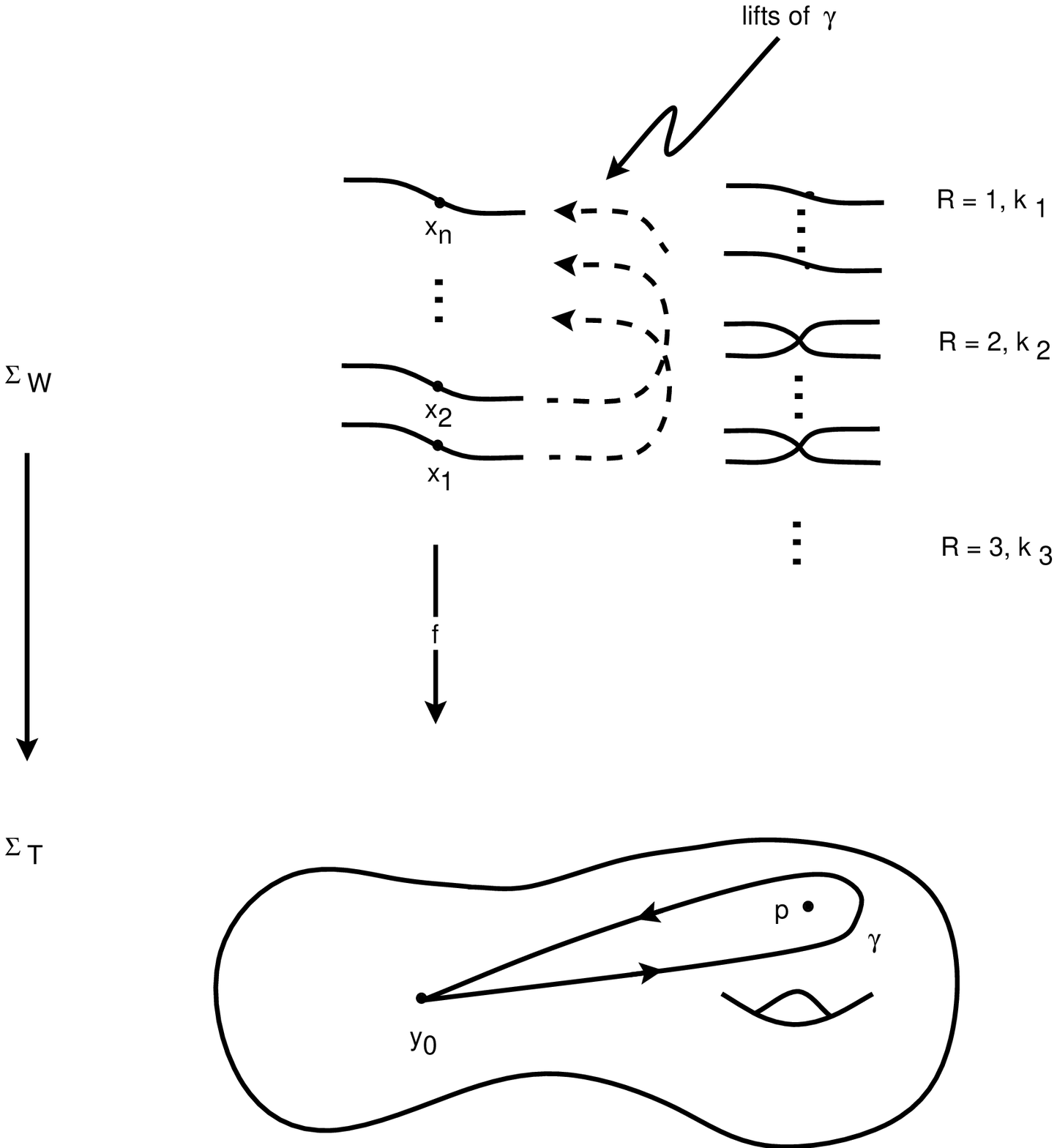}}

Locally, a branched covering looks like \brchcvr. 
For $Q \in \Sw$, the integer $n(Q)$, such that the 
map looks like $z\to w=z^{n(Q)}$, will be called the
{\it ramification index} of $Q$ and will be denoted
$\Ram~(f,Q)$. For any $P \in \ST$ the sum
\eqn\dgreff{
\deg(f)= \sum_{Q\in f^{-1}(P) } \Ram~(f,Q)
}
is independent of $P$ and will be called the {\it index} of $f$ (sometimes the
degree).
Points $Q$ for which the integer $n$ in condition $(a)$  is bigger than $1$ will
be called {\it ramification points}. Points $P\in \ST$ which
are images of ramification points will be called
{\it branch points}.\foot{Unfortunately, 
several authors use these terms
in inequivalent ways.}
  The set of branch points is the branch locus $S(f)$.
The branching number at $P$ is
$$B_P=\sum_{Q\in f^{-1}(P)} [\Ram~(f,Q)-1] $$
The branching number of the map $f$ is $B(f)= \sum_{P\in S(f) } B_P$.
A branch point $P$ for which the branching number is
$1$ will be called a {\it simple branch point}.
Above a simple branch point all the inverse images
have ramification index $=1$, with the exception of
one point $Q$ with index $=2$.

Globally, a branched cover looks like \liftcurv.
We will often use the Riemann-Hurwitz formula:
 If $f:\Sw\rightarrow \ST$
is a branched cover of index $n$ and 
branching number $B$, $\Sw$ has genus $h$, and 
$\ST$ has genus $p$, then :
\eqn\Rhurw{ 
\mathboxit{2h-2 = n(2p-2)+B .}}

\exercise{Riemann-Hurwitz}

Prove the Riemann-Hurwitz formula by relating 
triangulations of $\Sw$ and $\ST$.

\endexercise

Equivalence classes of branched covers may be
related to  group homomorphisms through the
following construction. Choose a point $y_0$ which is not a branch point
and label the inverse images $f^{-1}(y_0)$ by
the ordered set
$\{x_1, \dots x_n \}$.
Following the lift of elements of $\pi_1(\Sigma_T - S ,y_0)$,
the map $f$
induces a homomorphism
$$f_\#:\pi_1(\Sigma_T - S ,y_0)\to S_n\qquad . $$
This construction may be illustrated as in \liftcurv.  

\exercise{}

Suppose $\gamma(P)$ is a curve
surrounding a branch point P as in \ezfg.
There is a close relation between the 
cycle structure of $v_P=f_\# (\gamma(P))$
and the topology of the covering space over a neighbourhood of $P$. 
\item{a.)}
If the cycle decomposition of $v_P$ has $r$ 
distinct cycles then show that $f^{-1}(P)$ has $r$
distinct points.
\item{b.)}
Show that a cycle of length $k$ corresponds 
to a ramification point $Q$ of
index $k$.

\endexercise

With an appropriate notion of equivalence
the  homomorphisms are in 1-1 correspondence
with equivalence classes of branched covers.

\noindent
{\bf Definition \sCS.2}.
Two homomorphisms
$\psi_1,\psi_2:\pi_1(\Sigma_T - S ,y_0)\to S_n$ are said to be
{\it equivalent} if they differ by an inner automorphism of $S_n$,
i.e., if $\exists g$ such that  $\forall x$, $\psi_1(x)=g \psi_2(x)
g^{-1}$.

A crucial theorem for what follows is: 

\noindent
{\bf Theorem \sCS.1}. \refs{\Fu,\Ez}.
Let $S\subset \Sigma_T$ be a finite set and $n$ a positive
integer.
There is a one to one correspondence between equivalence
classes of homomorphisms
$$\psi:\pi_1(\Sigma_T - S ,y_0)\to S_n$$
and equivalence classes of $n$-fold branched coverings of
$\Sigma_T$ with branching locus $S$.

\vskip0.1truein\noindent
{\it Proof:}
We outline the proof which is described in \Ez .
The first step shows that equivalent homomorphisms determine equivalent
branched coverings.  Given a branched cover,
we can delete  the branch points from $\ST$ and the inverse images of the
branch points from $\Sw$ giving  surfaces $\overline {\Sigma}_W$ and
$\overline
{\Sigma}_T$ respectively.
The branched cover restricts to a topological (unbranched)  cover of
$\overline {\Sigma}_T$ by
 $\overline {\Sigma}_W$. One shows that equivalent homomorphisms determine 
equivalent conjugacy classes of subgroups of $\pi_1(\overline \ST)$. 
Now apply a basic theorem  in the theory of covering spaces \Ma\ 
which establishes a one-one correspondence between conjugacy classes of
subgroups of $\pi_1(X)$
 and equivalence classes of topological coverings of  the space $ X$.

Similarly the second step proves that equivalent covers
determine  equivalent homomorphisms. The restriction of $\phi$ in 
definition \sCS.1.b to the inverse
images of $y_0$  determines the permutation which conjugates one homomorphism
into the
other.

Finally, one proves that the map from equivalence classes of homomorphisms to
equivalence classes of branched covers is onto. We cut  $n$ copies of $\ST$
along
chosen generators of $\pi_1(\ST-S)$  (illustrated in \ezfg), and we glue
them
together according to the data of the homomorphism.
$\spadesuit$

\bigskip
This theorem goes back to Riemann.
Since the \ymt\ partition function sums over covering surfaces which
are not necessarily connected we do not restrict to homomorphisms
whose images are transitive subgroups of $S_n$.

\noindent
{\bf Definition \sCS.3}.
An {\it automorphism} of a branched covering $f$ is a homeomorphism $\phi$ such
that
$f\circ \phi = f $.

\vskip0.1truein\noindent
{\bf Examples:} 
\item{1.}
Consider $z\to w=z^n$, a cover of a disk by a disk. 
This has automorphism group $\IZ_n$. 
\item{2.}
The hyperelliptic curve 
$y^2=\prod_{i=1}^{2p+2}(x-e_i)$ is a double 
covering of the plane. For generic $e_i$ 
it has automorphism 
group $\IZ_2$, $(y,x)\to (-y,x)$.

\exercise {Automorphisms of covers}

Let $C(\psi)$ be the subgroup of $S_n$ which fixes 
each element in the image  of  
$\pi_1(\Sigma_T -  S, y_0)$ in $S_n$ under the homomorphism $\psi$. 
\item{a.)}
Show that  ${n!\over \vert C(\psi)\vert}$, the number of cosets of this
subgroup, is the number of distinct homomorphisms related to the
given homomorphism by conjugation in $S_n$.
\item{b.)} 
Show that $\Aut~ f \cong C(\psi)$.

\endexercise

\subsec{Hurwitz spaces}
\subseclab\ssHurSp

The Hurwitz space of branched coverings is 
described in\refs{\Fu,\HaMu}.
Let $H(n,B,p;S)$ be the set
of equivalence classes of branched coverings
of $\Sigma_T$, with degree $n$, branching number $B$, and
branch locus $S$, where $S$ is a set of distinct points on a
surface $\Sigma_T$ of genus $p$.  According to Theorem 
\sCS.1, $H(n,B,p;S)$ is a finite set.
The union of these spaces over sets $S$ with
$L$ elements is the Hurwitz space
 $H(n,B,p,L)$ of equivalence classes of branched coverings of
$\Sigma_T$ with
degree $n$,
branching number $B$ and  $L$ branch points.
Finally let $C_L(\Sigma_T)$ be the configuration space
of ordered $L$-tuples of distinct points on $\Sigma_T$, that is
$$C_L(\Sigma_T)=\{(z_1,\ldots,z_L)\in
\Sigma_T^L\vert z_i \in \Sigma_T, z_i \ne z_j { }\hbox{ for } i \ne j\}.
$$
The permutation group $S_L$ acts naturally on $C_L$ and
we denote the quotient $\CC_L(\Sigma_T)=C_L(\Sigma_T)/S_L$.
There is a map
\eqn\proj{
\pi: H(n,B,p,L)\to \CC_L(\Sigma_T)
}
which assigns to each covering its branching locus.
This map can be made a
topological (unbranched) covering map \Fu\ with
discrete fiber $H(n,B,p;S)$
over $S\in \CC_L$.

The lifting of closed curves in $\CC_L$ will in
general permute different elements of the
fibers $H(n,B,p,S)$. Note however that
$\Aut~ f $ is invariant along any lifted curve
so that $\Aut~ f $ is an invariant of the different
components of $H(n,B,p,L)$.

\subsec{Hurwitz Spaces and Holomorphic maps}
\subseclab\ssHSHM

One great advantage of branched covers is that they allow
us to introduce the powerful methods of complex analysis,
which are crucial to introducing ideas from topological field theory.
Recall that a complex structure $J$ on a manifold $M$ is a 
tensor $J\in \End(TM)$ such that $J^2=-1$. The $\pm i$ 
eigenspaces define the holomorphic and antiholomorphic 
tangent directions. 

In two dimensions a metric on a surface,
$g_{\alpha \beta} dx^\alpha dx^\beta$, determines
a complex structure: $\epsilon(g)\in \Gamma[ \End(T\Sw)]$,
$\epsilon^2=-1$.  
Introduce  the standard antisymmetric
tensor $\hat\epsilon_{\alpha\gamma}$,
\eqn\epsi { \hat\epsilon_{\alpha\gamma} 
= \pmatrix { 0 \qquad 1 \cr -1 \qquad
0}   }
and define
$\epsilon_\alpha^{~ \beta} ( g ) = g^{1/2} \hat\epsilon_{\alpha\gamma}
g^{\gamma\beta}$. Conversely, a complex structure 
determines a conformal class of metrics.

Indeed the moduli space of complex structures is 
just
\eqn\chfvmd{
\CM_{h,0} = \MET (\Sw)/\bigl( \Diff^+ (\Sw)\sdtimes \Weyl(\Sw)\bigr)
}
We can characterize the 
complex coordinate system as the system in which 
$ds^2= e^\phi \mid dz\mid^2$. 

The connection of branched covers and holomorphic 
maps is provided by the following 

\noindent
{\bf Theorem}.
Choose a complex structure $J$ on $\ST$.
Then given a branched cover $f:\Sw\longrightarrow \ST$ there is a
 unique complex structure on $\Sw$ making $ f$ holomorphic \Ahl.

\medskip
\noindent
{\it Proof:} Use the complex structure $f^*(J)$ on $\Sw$.
Equivalently, just pull back a metric on $\ST$, inducing 
$J$, to get a metric $g$ on 
$\Sw$, inducing $\epsilon$. 

Conversely, any  nonconstant
holomorphic map $f:\Sw \longrightarrow \ST$ defines a branched cover.
It follows that we can consider the Hurwitz space $H(n,B,p,L)$ as
 a space of holomorphic maps. Let $H_c(n,B,p,L)$ 
is the space of holomorphic maps from connected worldsheets, 
with degree $n$, branching number $B$ and $L$ branch points. 
The complex structure, $J$, on $\ST$
 induces a complex structure on $H_c(n,B,p,L)$, such that $\pi$
in \proj\  is a holomorphic
fibration. Moreover, the induced complex 
structure on $\Sw$ defines a holomorphic map
$m: H_c(n,B,p,L) \longrightarrow   \CM_{h,0}$ where $\CM_{h,0}$ is
the Riemann moduli space of curves of genus $h$, 
where $h$ can be computed using \Rhurw. 
The image of $H_c$ is a subvariety of $ \CM_{h,0}$.

\subsec{Fiber Bundle approach to Hurwitz space}
\subseclab\ssFBAHS

For comparison with topological field theory we will
need another description of Hurwitz space as the base
space of an
infinite-dimensional fiber bundle. 

Let $\Sw$ be a connected, orientable surface, and suppose
$\ST$ is a Riemann surface with a choice
of K\"ahler metric and complex structure $J$.
Let us begin with the configuration space
\eqn\CurlyM{
\tcM ~=~ \left \{ (f,g) \vert~ f \in \CC^\infty( \Sw, \ST ), g \in \MET~ (\Sw) \right\}
}
where $\CC^\infty( \Sw, \ST )$ is the space of  smooth
($\CC^\infty$) maps,
$f \colon \Sw \to \ST$ and $\MET ( \Sw )$ is the space of smooth
metrics on $\Sw$.

The subspace of
pairs defining a holomorphic map
$\Sw\to\ST$ is then given by
\eqn\bigfam{
\widetilde\CH =\{(f,g): df \epsilon(g) = J df\}\subset \tcM .
}
The defining equation $df \epsilon = J df$ is an equation in
$\Gamma \bigl[ \End (T_x\Sw,T_{f(x)}\ST) \bigr]$.
Indeed in local complex coordinates, where $\epsilon,J$ 
are diagonal with diagonal eigenvalues $\pm i$  we can write: 
\eqn\lcbgfm{
df + J df \epsilon = 2 \pmatrix{0& \pb f\cr \p \bar f & 0\cr} 
}

Let  $\Diff^+(\Sw)\sdtimes \Weyl(\Sw)$ be   the semidirect product of
the group of orientation preserving diffeomorphisms of
$\Sw $ and the group of Weyl transformations on $\Sw$.
There is a natural action of this group on $\tcM$.
Two pairs $(f,g)$  define 
equivalent holomorphic maps iff they are related by 
this group action. Therefore, the quotient space
\eqn\dfofeff{
\HOL (\Sw,\ST)\equiv
 \tilde\CM/\bigl( \Diff^+ (\Sw)\sdtimes \Weyl(\Sw)\bigr)
\qquad
}
parametrizes holomorphic maps $\Sw\longrightarrow \ST$.

\subsec{Riemann and  Hurwitz  Moduli Spaces}
\subseclab\ssCHS

Riemann and Hurwitz moduli spaces have
orbifold singularities 
associated with automorphisms of the surface or 
automorphisms of the cover. It will be quite important 
in chapter \sEC\ that, while $H(n,B,p,L)$ has no 
singularities, the space 
$\HOL~ (\Sw,\ST)$ does have orbifold singularities. 
Roughly speaking $\HOL~ (\Sw,\ST)$ is ``made'' out of the 
spaces  $\amalg_{n(2-2h)-B=2-2p} H(n,B,p,L)$.

These moduli spaces are also noncompact. 
When discussing the intersection theory of 
these moduli spaces it is quite necessary to 
compactify them. The process of compactification 
is very complicated and tricky. In chapter 
\sONEofYMA\ below we describe some of the intuitive 
pictures, associated with collisions of 
branchpoints, that are 
 used in constructing compactifications. 
A compactification of Hurwitz spaces 
has been given 
in \HaMu.

\newsec{$1/N$ Expansions of  \ymt\  amplitudes}
\seclab\sONEofYMA

In this chapter we derive the asymptotic $1/N$ 
expansion of the exact results of \ymt. 

\subsec{Some preliminary identities from group theory}
\subseclab\ssSPIfromGT

\subsubsec{Finite  group identities}

We will need several identities of finite group theory, applied to $S_n$.
First we have the standard orthogonality relations:
\eqn\idenii{\eqalign{
{1\over n!} \sum_\rho \chi_{r_1}(\rho) \chi_{r_2}(\rho^{-1})&=\delta_{r_1,r_2}\cr
\sum_{r\in {\rm Rep}(S_n)}  \chi_r(\rho) \chi_r(\sigma) &=
\delta_{T_\sigma , T_\rho} { n!  \over \vert T_\sigma \vert }  \cr
\Rightarrow {1\over n!} \sum_r d_r \chi_r(\rho) & =\delta(\rho)\cr}
}
where
$\vert T_{\sigma} \vert $ is the order of the conjugacy class containing the permutation 
$\sigma$. 

We will frequently use the fact that:
\eqn\ideni{
\sum_{\sigma\in T} {\chi_r(\sigma)\over d_r}{\chi_r(\rho)\over d_r}
=\sum_{\sigma\in T} {\chi_r(\sigma\rho)\over d_r}
}
where $T$ is any conjugacy class, and $d_r$ is the dimension of the representation $r$ 
of the symmetric group. 

Finally, a simple but important consequence of these identities is that \GrTa: 
\eqn\ideniii{
\bigl({n!\over d_r}\bigr)^2=\sum_{s,t\in S_n} {\chi_r(st s^{-1}t^{-1})\over d_r}}

To prove the last equation,  \ideni\ is used to separate the sum 
of characters into a sum of products of two characters to get
\eqn\prdtwo{ 
n! \sum_{t\in S_n} {\chi_r(t)\over d_r}{\chi_r(t^{-1})\over d_r}\qquad .
}
The orthogonality expressed in \idenii\ 
 is then used to do the sums.  

\subsubsec{Powers of dimensions}

As a final important formula we will  derive a key 
expression for the powers of dimensions of $SU(N)$ representations
in terms of characters  of the symmetric group. 
We begin with 
\eqn\dimfrm{\eqalign{
\dim~ R(Y)  &= {N^n\over n!} \chi_{r(Y)}(\Omega_n) \cr
 \Omega_n  &= \sum_{v\in S_n}
\bigl({1\over N}\bigr)^{n-K_v} v\cr
&=1+{1\over N} T_2 + {1\over N^2}T_3+\cdots \cr}}
obtained by setting $U=1$ in eq. \frobrecip. 
$K_v$ is the number of cycles in the cycle 
decomposition of $v$. 
This generalizes to: 
\eqn\exeri{
(\dim~ R(Y))^m= \biggl({N^n d_{r(Y)}\over n!}\biggr)^m {\chi_{r(Y)}(\Omega_n^m)\over
d_{r(Y)} }
}
where $m$ is any integer, positive or negative. 
To write expressions for inverse powers of the dimension in terms of 
the symmetric group it is convenient to define the inverse of $\Omega_n$ 
in the group algebra. For $N>n$, this inverse always exists, but for $N<n$ it may not. 
In the large $N$  expansion we may always invert it. 
%
%

\subsec{Chiral Gross-Taylor Series} 
\subseclab\ssCGTS

We are now ready to derive the $1/N$ expansion of 
\eqn\crrctsc{
Z=\sum_R (\dim~ R)^{2-2p} e^{-{A\over 2 N} C_2(R)} 
}
where we have put in the correct scaling of the 
coupling constant $e^2 A N \to A$, which is held 
fixed for $N\to \infty$. 

\bigskip
\boxit{
The central idea of the calculation is to use 
Schur-Weyl duality to translate $SU(N)$ representation
theory into $S_n$ representation theory, and 
interpret the latter, geometrically, as data 
defining a branched cover.}

\bigskip

To implement this idea we make the replacement:
\eqn\dfchsmo{
\sum_{R\in {\rm Rep}~ (SU(N))} f(R)  = \sum_{n\geq 0}
 \sum_{Y\in \CY_n^{(N)}} f(R(Y))
}
We first explain the derivation for the ``chiral expansion.''
This keeps only the states in $\CH_{\rm chiral}$ in 
the large $N$ Hilbert space, and is defined by 
dropping the constraint on the number of rows so:
\eqn\dfchsm{
\sum_{R\in {\rm Rep}~ (SU(N))}f(R) \rightarrow \sum_{n\geq 0}
 \sum_{Y \in \CY_n} f(R(Y))
}
Physically this turns out to be the restriction to orientation-preserving 
strings. 

Now we translate the various expressions in 
\crrctsc\ into quantities involving the 
symmetric group. First we have: 
\eqn\dimens{\eqalign{
 (\dim~& R(Y))^{2-2p}\cr 
&=
N^{n(2-2p)} ({d_{r(Y)}\over n!})^2 [({n!\over  d_{r(Y)}})^2]^p 
{\chi_{r(Y)}(\Omega_n^{2-2p})\over d_{r(Y)}} \cr
&= 
N^{n(2-2p)} ({d_{r(Y)}\over n!})^2\sum_{s_i,t_i\in S_n} \prod_i  
{\chi_{r(Y)}(s_i t_i s_i^{-1}t_i^{-1})\over d_{r(Y)}}
{\chi_{r(Y)}(\Omega_n^{2-2p})\over d_{r(Y)}} \cr
&= 
N^{n(2-2p)} ({d_{r(Y)}\over n!})^2\sum_{s_i,t_i\in S_n}  
{\chi_{r(Y)}(\prod_i s_i t_i s_i^{-1}t_i^{-1})\over d_{r(Y)}}
{\chi_{r(Y)}(\Omega_n^{2-2p})\over d_{r(Y)}} \cr
&= 
N^{n(2-2p)} ({d_{r(Y)}\over n!})^2\sum_{s_i,t_i\in S_n}  
{\chi_{r(Y)}(\Omega_n^{2-2p} \prod_i s_i t_i s_i^{-1}t_i^{-1})\over d_{r(Y)}}\cr}
}
Similarly, we can expand the exponential of the 
area using the formula: 
\eqn\quadcas{
C_2(R(Y) ) = n N + 2 {\chi_{r(Y)} (T_2)\over d_{r(Y)}} -{n^2\over N} 
}
%
Using \ideni\ repeatedly we get:
\eqn\dergt{\eqalign{
& (\dim R(Y))^{2-2p} e^{-{A\over 2N}C_2(R(Y))}
 \cr
&= e^{-\half A(n-{n^2\over N^2})}
N^{n(2-2p) } ({d_{r(Y)}\over n!})^2 \cr
  &\sum_{s_i,t_i\in S_n}  
{\chi_{r(Y)}(\prod_i s_i t_i s_i^{-1}t_i^{-1}\Omega_n^{2-2p})\over d_{r(Y)}}
e^{-{A\over N} \chi_{r(Y)}(T_2) } 
\cr
& =\sum_{l\geq 0}{(-A)^l\over {l!}}
e^{-\half A(n-{n^2\over N^2})}
N^{n(2-2p)-l} ({d_{r(Y)}\over n!})^2 \cr 
&\qquad\qquad\qquad \sum_{s_i,t_i\in S_n}  
{\chi_{r(Y)}(\Omega_n^{2-2p} \prod_i s_i t_i s_i^{-1}t_i^{-1})\over d_{r(Y)}}
{\chi_{r(Y)} (T_{2,n}^l)\over {d_{r(Y)} }}
\cr
&=  \sum_{l\geq 0}{(-A)^l\over {l!}}
e^{-\half A(n-{n^2\over N^2})}
N^{n(2-2p)-l } ({d_{r(Y)}\over n!})^2 \cr
& \qquad \qquad\qquad \sum_{s_i,t_i\in S_n}  
{\chi_{r(Y)}(\Omega_n^{2-2p}  T_{2,n}^l \prod_i s_i t_i s_i^{-1}t_i^{-1})\over d_{r(Y)}}
\cr}
}
Now, in the chiral sum we have an unrestricted sum over 
$Y\in\CY_n$ so we use \idenii\ to obtain:
\eqn\ZchiAi{\eqalign{
&Z^+ (A,p,N)\cr
&=1+ \sum_{n\geq 1,\ell\geq 0}^\infty N^{n ( 2 - 2p)-\ell}
e^{-\half A(n-{n^2\over N^2})}{(-A)^\ell\over \ell !}
\sum_{s_i,t_i \in S_n}  {1\over n!} \delta(\Omega_n^{2-2p}
T_{2,n}^\ell \prod_1^p[s_i,t_i])\cr}
}
Acting on an element of the group algebra, 
the delta function 
evaluates the element (regarded  as a function on the group) 
at the identity permutation. 
We now write $n^2=n(n-1)+n$ and 
expand the remaining factor in the exponential 
to get 
the  ``chiral Gross-Taylor series'' (CGTS) as
\eqn\Zchir{
\mathboxit{
\eqalign{
Z^+(A,&p, N)\cr
{}~& =~1+
\sum_{n=1,i,t,h=0}^{\infty} e^{-n A/2} (-1)^i {{ (
A)^{i+t+h}} \over{i!t!h!}} \qquad \qquad \cr
&
 \bigl({1\over N}\bigr)^{n(2p-2)+2h+ i+ 2t} \bigl({n\over 2}\bigr)^h
\bigl({n(n-1)\over 2}\bigr)^t \cr
&
\sum_{p_1,\ldots,p_i \in T_{2,n} }
\sum_{s_1,t_1,\ldots,s_p,t_p\in S_n}
\biggl[ {1\over {n!}}\delta (p_1\cdots p_i \Omega_n^{2-2p}
\prod_{j=1}^p
s_jt_js_j^{-1}t_j^{-1}) \biggr]
.\cr}
 }}

\subsec{Nonchiral Sum}
\subseclab\ssNonCS

The expansion \ZchiAi, based on \dfchsm,
does not give the correct large $N$ asymptotics of
\crrctsc. In \ZchiAi\  the contribution to $Z$ from large representations
with a number of boxes  of order $N$ is separated into infinitely many  terms,  
 each of which is exponentially damped $O(e^{-N})$. 
There are representations $R$ with order $N$ boxes for which ${C_2(R) \over N}$ is 
$O(N^{0})$.  These reps contibute terms to \exctprt\ 
which are not exponentially damped as $N$ goes to infinity. Therefore a  finite
 number of terms in  the expansion \ZchiAi\ will not give an answer differing from 
\exctprt\ by exponentially small amounts. For example,  the representation 
$R$,
with $(N-1)$ rows of length $1$, is conjugate  to the fundamental 
rep and so has the same 
Casimir, $C_2= N-1/N$. In the chiral expansion, however, 
its contribution $(\dim~ R )^{2-2p} e^{-{AC_2(R)\over 2N}}$
is written $(\dim~ R)^{2-2p} e ^{-{A\over 2} \bigl( (N-1) + (1- {1\over{N^2}}-N + 1 )\bigr)}$, the
first term is kept in the exponent and the second is expanded out, giving
 infinitely many terms, each exponentially damped. 

Getting the correct  
asymptotic expansion requires isolating all the representations which have 
${C_2(R)\over N}\sim O(N^0) $,  
 and making sure their contributions appear after
a finite number of terms. Gross and Taylor argued that the most general reps
which satisfy ${C_2(R)\over N} \sim O(N^0)$  are the `coupled reps' 
defined in chapter \sFYMTtoSCA. 
The coupled expansion  is  defined so as to pick up  all the perturbative contributions 
of  the  coupled reps. It is obtained by the replacement 
\eqn\nonchir{  \sum_{{\rm Reps}} f ( {\rm Rep} )  \longrightarrow  \sum_{n^+,n^-} 
\sum_{R \in Y_{n^+} ,S \in Y_{n^-} }  f (R\bar S) .}

\bigskip
The procedure of constructing the coupled expansion may appear strange. 
Indeed at finite $N$, it would overcount reps, which leads in \BaTa\  to a chiral approach 
to the problem of string interpretation at finite $N$.   
 Completing  the  proof that the coupled expansion is the asymptotic expansion  
to  \crrctsc\ requires analyzing the behaviour of ${C_2(R)\over N}$, 
$\dim~ R$ 
and the multiplicity
 of Young diagrams which do not appear after a finite number of terms 
in the coupled expansion, and showing that their contributions are non-perturbative.

In order to find a formula for $\dim~ R\bar S$, 
we can use  \crep:
\eqn\ccarfrm{  
\chi_{R\bar  S } (U) = \sum_{v  \in S_{n^+}, w \in S_{n^-} }
{ \chi_r (v)\over n^+!} { \chi_s (w)\over n^-!}  
\Upsilon_{v,w} (U, U^\dagger) }  
where $R$ corresponds to a Young diagram which
  has $n^{+}$ boxes
 and $S$ corresponds to one with  $n^{-}$ boxes,
 where $r$ and $s$ are the 
reps of $S_{n^+}$ and    $S_{n^-}$ corresponding
 to the same Young diagram. The $\Upsilon_{v,w} $ are the 
coupled loop functions of 
chapter \sFYMTtoSCA.

By setting $U$ to $1$ in \crep, \couploop\  we get  
the expansion in $N$ for $\dim~ R\bar S$. 
\eqn\coup{ \dim~ R{\overline S} = {{N^{n^++ n^-}}\over {n^+! n^-!}}
\chi_{R {\overline S}} (\Omega_{n^+ n^-}), }
where  $\Omega_{n^+, n^-}$ are
certain elements of the group algebra of the symmetric
group $S_{n^+} \times S_{n^-}$ with coefficients in
$\IR ((1/N))$. Explicitly,  $\Omega_{n^+, n^-}$ is an element of the group algebra of
$S_{n^+}\times S_{n^-}$ given by
\eqn\Omegc{ \Omega_{n^+,n^-}
{}~=~ \sum_{v  \in S_{n^+}, w  \in S_{n^-}}
          (v  \otimes w ) P_{v  ,w } ({1\over N^2})
             \bigl({1\over N}\bigr)^{ (n^+ -K_{v }) + (n^- -K_{w })}}
The polynomial $P_{v,w}( {1\over {N^2} })$ can be read off from \couploop :
\eqn\polynvw{  P_{v,w}( {1\over {N^2} })= \prod_j  \sum_m^{min(k_j,l_j) }
P_{k_j,l_j} (m) {1 \over {N^{2m} }}  .} 

\bigskip 

 Manipulations similar to those of the chiral theory (section 
\ssCGTS), 
now  performed for the product group 
$S_{n^+}\times S_{n^-}$ instead of $S_n$,   lead to an expression 
analogous to \Zchir\ 

\eqn\fullgt{
\mathboxit{
\eqalign{
Z&( A,p, N)\cr
&\sim1+ 
\sum_{n^\pm=1,i^\pm=0}^{\infty}
\sum_{p^\pm_1,\ldots,p^\pm_{i^\pm} \in T_2\subset S_{n^\pm}}
\sum_{s^\pm_1,t^\pm_1,\ldots,s^\pm_p,t^\pm_p\in S_{n^\pm}}
\qquad \qquad \cr
&\qquad\bigl({1\over N}\bigr)^{( n^+ + n^- ) (2p-2)+(i^+ + i^- )}
{{(-1)^{(i^++ i^-)}} \over  {i^+! i^-! n^+ ! n^- !}}
( A)^{(i^+ + i^-)}\cr
&\qquad e^{-\half  (n^+ + n^-) A}
e^{\half ((n^+)^2 + ( n^- )^2 - 2 n^+ n^- ) A/N^2}\cr
&\qquad\delta_{{\scriptscriptstyle S_{n^+} \times S_{n^-}}}
\biggl ( p^+_1\cdots p^+_{i^+} p^-_1\cdots p^-_{i^- }
\Omega_{n^+, n^-}^{2-2p}
\prod_{j=1}^p [ s^+_j, t^+_j ] \prod_{k=1}^p [ s^-_k, t^-_k ]
\biggr )
,}}}
where $[\alpha,\beta]= \alpha \beta \alpha^{-1}\beta^{-1}$.
Here $\delta$ is the delta function on the group algebra
of the product of symmetric groups $S_{n^+}\times S_{n^-}$.

\subsec{Area Functions } 
\subseclab\ssArFn

We now want to interpret \fullgt\ and its chiral 
analogue geometrically in terms of coverings. 
For the moment let us put $\Omega\to 1$. We will 
return to the $\Omega$-factor in the next chapter. 

The first salient point is that we have an 
expansion in $1/N^2$ and 
the contribution of connected surfaces to the partition function 
 can be written in the form
\eqn\arpol{\eqalign{   Z^+(A,p,N) = \sum_{h} (1/N)^{ 2p-2} Z^+_{h,p} (A) 
 =\sum_{h} (1/N)^{ 2p-2}\sum_{n} e^{-nA/2}  Z^+_{n, h,p} (A) }}
$Z^+_{h,p} (A)$ is the contribution from surfaces (possibly disconnected) with Euler characteristic $2-2p$.  
$Z^+_{n,h,p} (A) $ is polynomial in $A$, of degree at most $(2p-2)- n(2p-2)=B$.
\item{Case1:}  For $p >  1$, $Z^+_{(h,p)}(A)$ is a finite sum of terms. 
It has a finite, nonvanishing $A\to 0$ limit. 
\item{Case 2:}
For $p=1$, $Z^+_{(1,1)}(A) $ is an 
infinite sum calculated in \GrTa\ to be
 $e^{-A/12}\eta (A)$. The sum converges for $A> 0$. 
$Z_{(2,1)} (A)$ was calculated in 
\dgcrg.  The free energies  for the chiral $U(N)$ theory have  been calculated for $p=1$,
 and $h$ up to eight, 
and its modular properties investigated \refs{\Rudd}. 
\item{Case 3:}
For $p=0$: $ Z^+_{0,0} (A)$  has finite radius of 
convergence and has been investigated in \refs{\PhaTay}.
The range of validity of the coupled Gross Taylor 
expansion for $SU(N)$  is limited to large area in the 
case of a spherical target. Douglas and Kazakov \DoKa\ showed that 
 the leading order (in $1/N$) term in the free 
energy shows a third order phase transition as a function of the area. 
Below the critical area $g^2 A= \pi^2$ the  
large area expansion is not valid. A string 
interpretation of the phase transition has been given in \refs{ \PhaTay , \CreTay}. 
 The detection of any stringy features in the weak 
coupling result would be very interesting.

\subsec{ Geometrical Interpretation of the $A$-dependence} 
\subseclab\ssBTCH

In \Zchir\ we have a sum over 4 positive integers, 
$n,i,t,h$. In this section we will associate geometrical 
pictures with the $A$ dependence coming from these sums. 
These pictures are meant to be heuristic. The interpretation 
can be ambiguous, and sometimes the pictures can be misleading. 
For example, $i,t,h>0$ are the most subtle dependences from the 
topological string point of view.

\subsubsec{$i>0$: Movable branch points}
\subsubseclab\sssMBP

Picking the leading term, $1$,  from the $\Omega_n$ we have 
a sum over $i,t,h$  and the 
the permutations $p_1,\cdots, p_i, s_1, t_1,\cdots , s_p, t_p$.     
First consider the terms with $h=t=0$. We are left with a sum of homomorphisms from $\pi_1( \Sigma_T - \{ i   {\hbox{ punctures}} \}) $  into $S_n$,  where the generators around the $ i$ 
punctures map to the class of simple transpositions. 
 Using theorem \sCS.1, we see that the sum is counting branched
 covers with $i$ points being simple branch points, with weight equal to the inverse 
of the order of the automorphism group of the cover.
 Note that the power of $N$ is the 
Euler character of the worldsheet as given by the Riemann Hurwitz formula \Rhurw. 

\subsubsec{$t>0$: Tubes}

For terms with $h, t > 0 $ the power of $N$ is smaller than that given by \Rhurw, 
by $2t+2h$. This is understood
in terms of  
extra handles and tubes on the worldsheet which map to points \Min.  The $n$ dependence 
in $({n(n-1)\over {2N^2 }})^t$ is understood in terms of maps from  a worldsheet  which has $t$ tubes 
connecting two sheets each.   
The factor ${ n(n-1)\over 2}$
 is the number of ways of choosing which
 pair of sheets is being connnected by the tube.      The ${1\over {t!}}$ is understood as a symmetry factor due to the tubes being indistinguishable.

\vskipabit
\ifig\pintu{ Pinched tube.}
{\epsfxsize3.5in\epsfbox{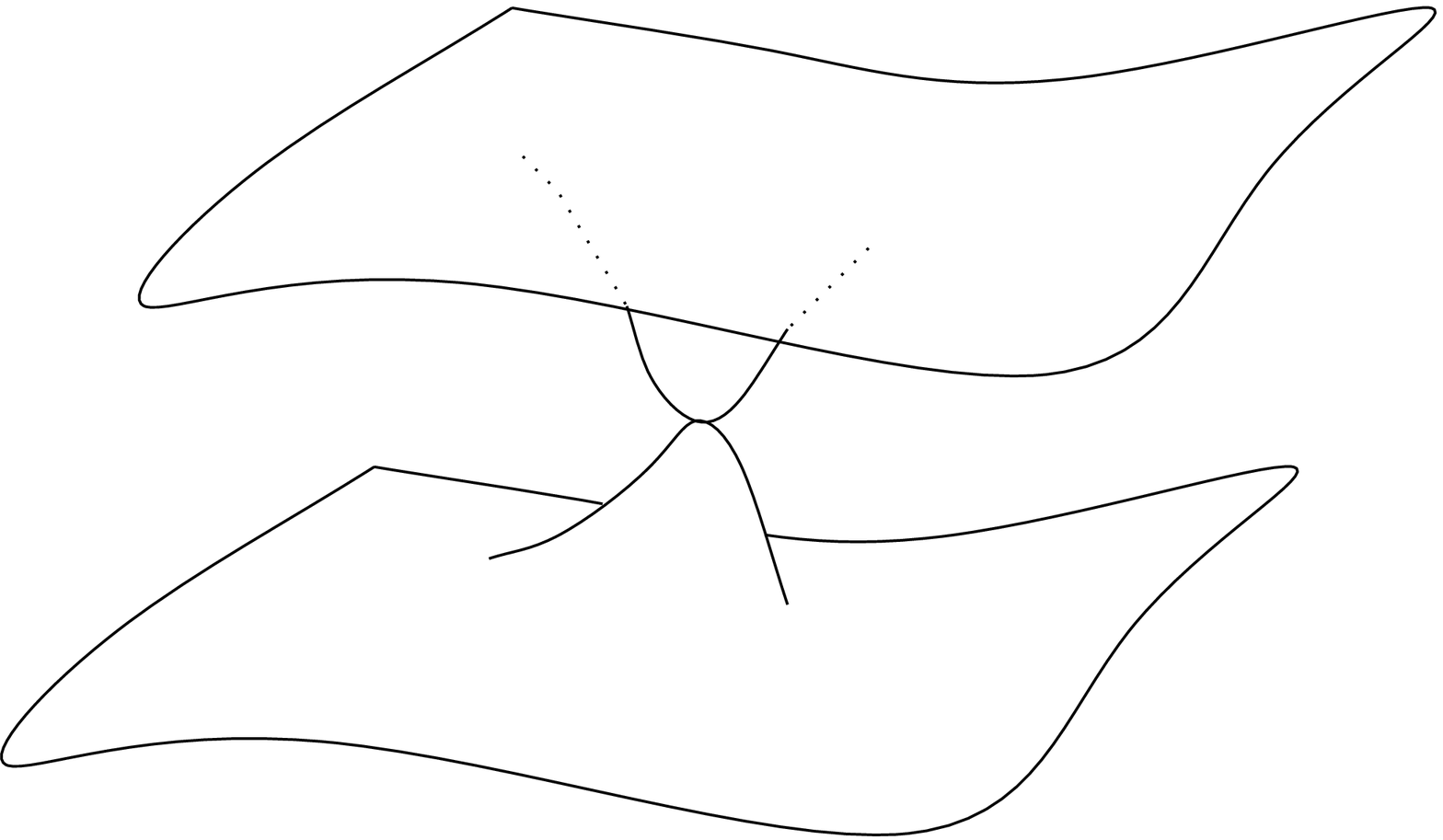}}

The $n$ dependence  can, equivalently,  be 
understood as the weight for the collision 
of two simple branch points to produce a tube. 
\ifig\fico{Three types of collisions of simple ramification points}
{\epsfxsize3.5in\epsfbox{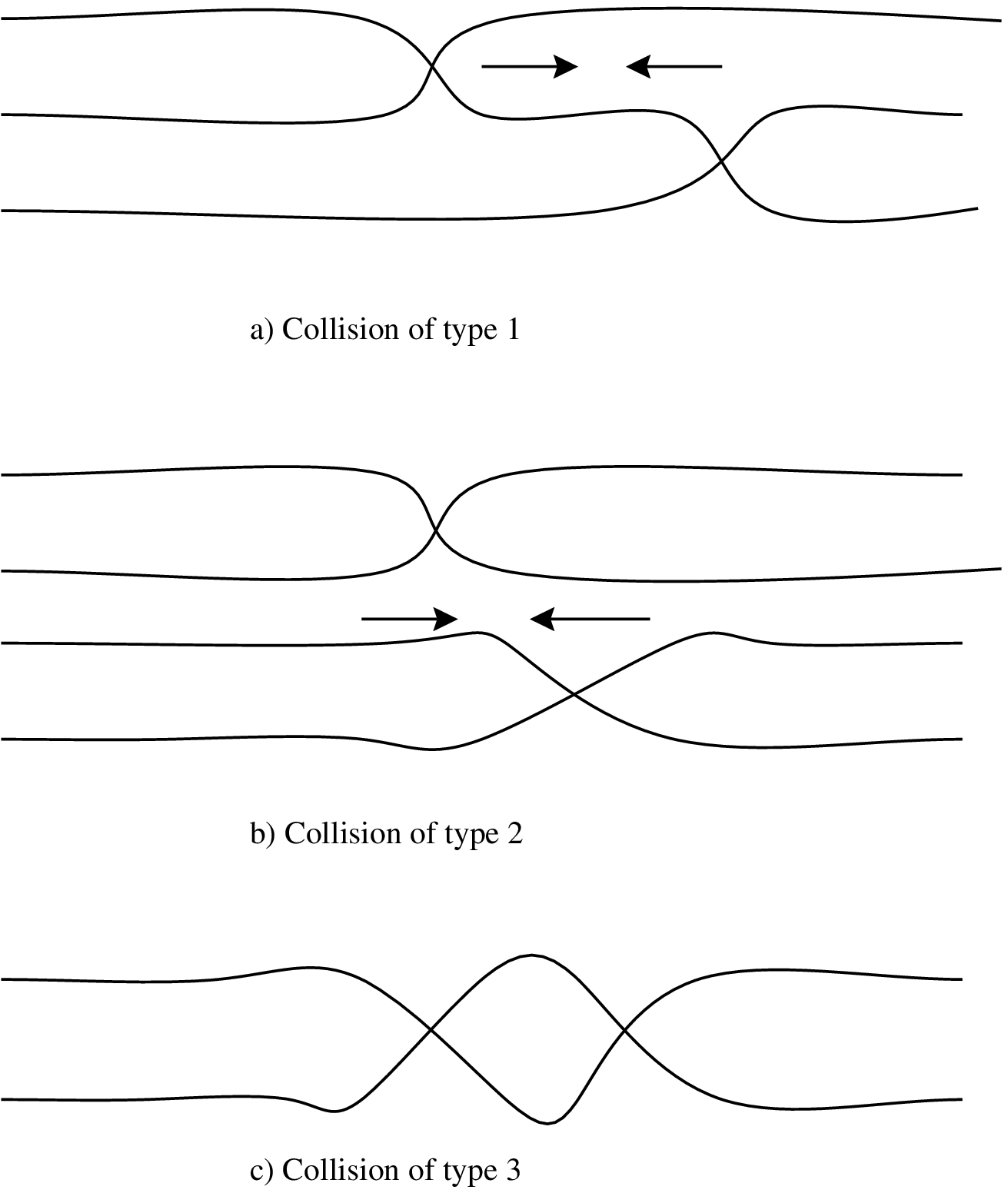 }}
 When two simple branch points approach 
each other, they can produce either a 
ramification point of index $3$, or two 
ramification points of index $2$, or a 
tube with no ramification (see \fico ) .  
This fact can be read off from the following 
multiplication in the class algebra of symmetric groups: 
\eqn\tbcol{ \sum_{q_1,q_2  \in T_2} q_1q_2 = 3 \sum_{q \in T_3} q
+ \sum_{q \in T_{2,2}} q + n(n-1)/2. } 
The coefficient of the identity, corresponding to the
 collisions producing no ramification and a tube,   is exactly 
the number associated with  each of the $t$ tubes.

\ifig\colha{`A'  is a holomorphic map at the boundary of the space of maps ; `B' (a pinch map)  is not.  }
{\epsfxsize5.5in\epsfbox{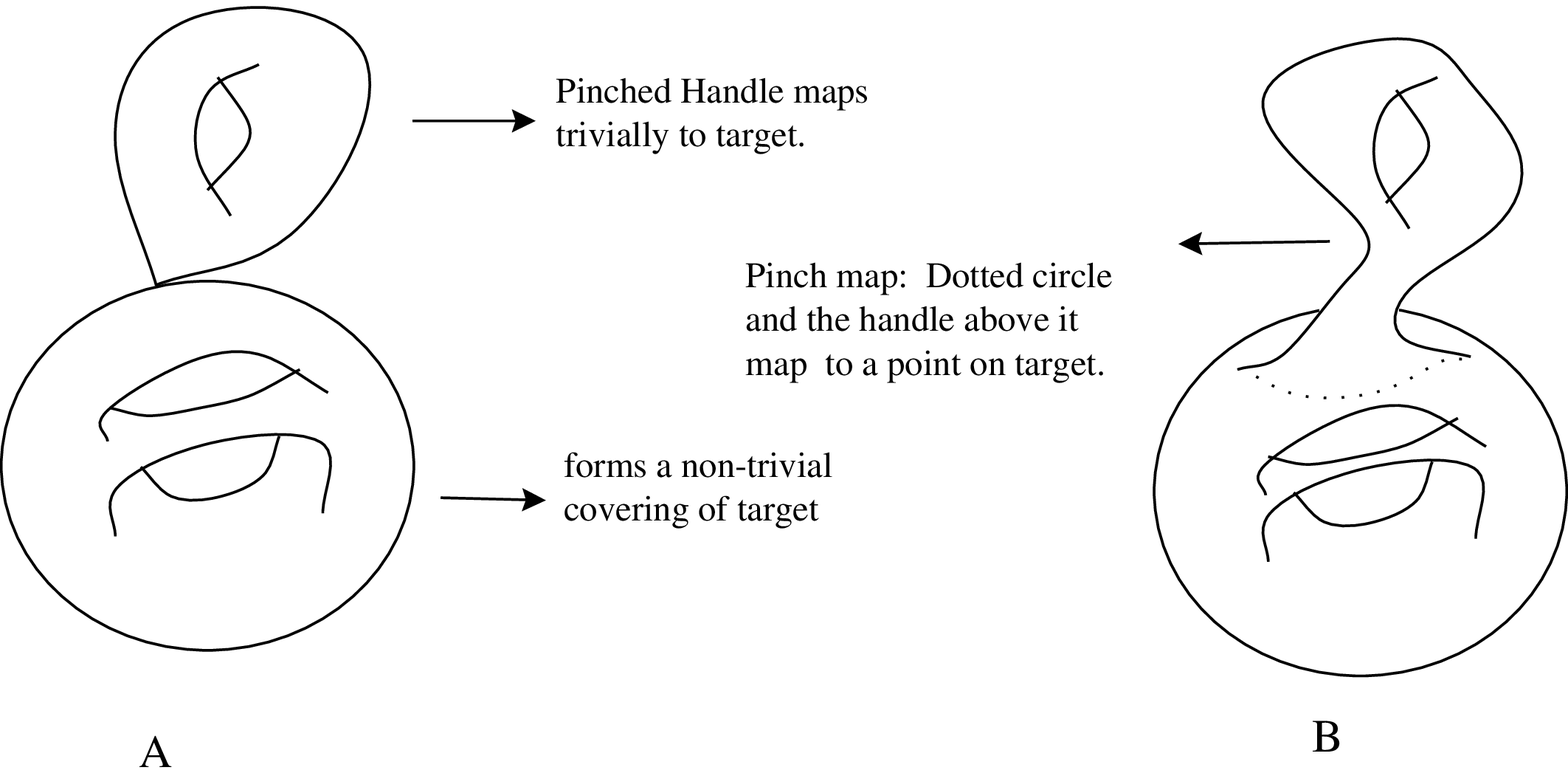}}

\subsubsec{$h>0$: Bubbled handles}

In \Min\  the factor 
 ${(An)^h\over {h! N^{2h}}}$ is 
interpreted in terms of $h$  ``infinitesimally small'' 
handles which map to points on the target.  
The factor $A$ is due to an integration 
over the image of the collapsed handle 
on the target space. There is no 
modulus for the area of the handle 
being pinched off  which is why 
they can be thought to  be infinitesimally small. 
Each handle decreases the Euler character 
by two so the power of $N$ is appropriate. 

The precise nature of the maps contributing the factors 
 ${(An)^h\over {h! N^{2h}}}$ is slightly ambiguous. 
Two different interpretations of these factors
are illustrated in \colha .  If the factors 
were due to maps of type $A$ then  they 
would represent the 
phenomenon of bubbling recently 
discussed in the context of topological 
string theories  \refs{\BeCeOoVa}. 
In  \refs{\BeCeOoVa}\
degenerate instantons  are shown to be required 
in the computation of partition functions  of topological 
field theories. For these degenerate instantons the
 worldsheet has handles, bubbled off from the part 
of the worldsheet mapping nontrivially to the target, which 
are mapped to  points on the target. From this point 
of view there is no dependence on the area of the bubbled off handles because the pull-back
of the K\"ahler  form to these bubbled handles is trivial ( as the map restricted to these handles is constant). 
According to the interpretation of \ymt\ in 
chapter \sYMTasaTST,  collapsed handles relevant for chiral \ymt\
 are  of type A in \colha.

\subsec{Generalizations}

\subsubsec{Higher Casimirs} 

From the string point of view it is natural to ask if we could have higher ramification points 
and higher genus pinched surfaces. These occur when higher Casimir perturbations 
are added. For the fundamental Casimirs,  a simple scaling $t_k C_k (R)={  \lambda_k  C_k (R) \over {N^{k-1} }} $, where $\lambda_k$ is kept fixed as $N \rightarrow \infty$, allows an interpretation in terms of branch points, tubes and handles. Requiring that an interpretation in terms of {\it orientable} worldsheets be possible implies  a symmetry of Casimirs: 
\eqn\cassym{ C_{k} (R, N) = (-1)^{k+1} C_{k} (\tilde R,-N) ,  } 
Here $\tilde R$ is the rep corresponding 
to the Young diagram  which is the transpose of  $R$. The symmetry 
can be proved using a set of diagrammatic rules for the conversion of
Casimirs to characters of symmetric groups described in 
sec. \ssAHC\ and in \ganor. These rules 
can also be used to compute the class algebra
 of symmetric groups, which is related to the counting of covers with inverse 
automorphisms
and to the collision of branch points  \Hcas . 

For products of Casimirs,  a simple string interpretation is possible if 
the coupling to products of Casimirs is scaled by a higher power of ${N^2}$ than 
the product of scalings for  the factors. 
The singular worldsheet  configurations contributing to the partition function of the theory with higher Casimirs involve tubes connecting more than two sheets, and higher genus collapsed 
surfaces (generalising the collapsed handles). 
The string interpretation of higher Casimirs 
has also been studied in \ganor.

\subsubsec{Nonperturbative corrections} 

There are corrections to the coupled expansion 
arising from the fact that we haven't treated the range of Young diagrams being summed 
exactly. The equation \fullgt\ sums over all $Y_{n^+}$ and $Y_{n^-}$ for given $n^\pm$, whereas 
the reps of $SU(N)$ are restricted to have not more than $N -1$ rows. Such corrections were estimated to be $O(e^{-N})$ in \refs{\GrTa,\Rudd}, as expected for 
a string theory where $1/N$ is the coupling constant \She .

\subsubsec{$O(N)$ and $Sp(2N)$ Yang Mills} 

Similar results on $1/N$ expansions have been 
obtained for other gauge groups \refs{\NaRiSc,\Ra}. 
In these cases the worldsheets are not 
necessarily orientable, and there is no 
analogue of the chiral and anti-chiral sectors. 
The maps  involved are again branched covers, 
possibly  with collapsed tubes,  and a new 
ingredient, collapsed cross-caps. The combinatorics 
of the $\Omega$ factors are very similar to that of 
the coupled $\Omega$ factor for the unitary groups. 

\newsec{Euler characters}
\seclab\sEC

\subsec{Chiral GT partition function}
\subseclab\ssCGTPF

In this section\foot{Some of the text of this chapter has been 
cut and pasted from \CMROLD.}
 we make our first connection between the topology of
Hurwitz space and $YM_2$ amplitudes. Consider the CGTS \Zchir.
As in 2D gravity, relations to topological field theory become most transparent
in the
limit $A\rightarrow 0$ where we have a topological theory 
in spacetime. 
Accordingly,  we will study the series
\eqn\Zchiri{
Z^+(0,p, N) = 1+ \sum_{n=1}^{\infty}
N^{n(2-2p)}
\sum_{s_1,t_1,\ldots,s_p,t_p\in S_n}
\biggl[ {1\over {n!}}\delta ( \Omega_n^{2-2p} \prod_{j=1}^p
s_jt_js_j^{-1}t_j^{-1}) \biggr].}

\subsubsec{Recasting the CGTS as a sum over branched coverings}

The first step in rewriting \Zchiri\
is to count the weight of a given power of $1/N$. To this end
we expand the
$\Omega^{-1}$ point as an
element of the free algebra generated by elements of
the symmetric group,
\eqn \Omeginv {
 \Omega_n^{-1} =  1 + \sum_{k=1}^{\infty}
{\sum}^\prime _{v_1\cdots v_k\in S_n}
\biggl({1\over N}\biggr)^{\sum_{j=1}^k n-K_{v_j}  }
 (v_1v_2\cdots v_k ) (-1)^k }
where the primed sum means no $v_i=1$.
We could rewrite \Zchiri\ by  imposing relations
 of the symmetric group  of $S_n$ .  However,
we decline to do this and rather substitute
the expansion \Omeginv\ into \Zchiri\
to obtain
\eqn\Zchirii {\eqalign{Z^+(0,p,N) =~ 1+ 
\sum_{n=1}^{\infty}\sum_{L=0}^{\infty}
&N^{n(2-2p)}
\sum_{s_1,t_1,\ldots,s_p,t_p\in S_n}
\sum_{v_1,v_2, \ldots,v_L \in S_n}^{{}\prime}
\cr
N^{\sum_{j=1}^L (K_{v_j}-n)} &
\biggl[ { d(2-2p,L)\over {n!}}
        \delta(v_1v_2 \cdots v_L \prod_{j=1}^p
s_jt_js_j^{-1}t_j^{-1}) \biggr] \cr}}
 where $d(m,L)$ is defined by
\eqn\genfun {
 (1+x)^m = \sum_{L=0}^{\infty} d(m,L) x^L . }
Explicitly we have
\eqn\dex{ \eqalign{
d(2-2p,L) &= (-1)^L { { (2p+L-3)!}  \over {(2p-3)! L! } } ,\qquad
 \hbox{ for} \qquad p>1 \cr
d(0,L) &= 0 , \hbox {unless}~ L=0 \cr
d(2,L) &= 0 , \hbox {unless}~ L=0,1 ,2. \cr }}
For $p>1$, $\vert d(2-2p,L)\vert $  is the number of ways of  collecting $L$ objects into
$2p-2$ distinct sets. The equation \Zchirii\ correctly gives the partition
function for any $p$ including  zero and one. For example the vanishing of
$d(0,L)$ for $L>0$ means that in the zero area limit only
maps with no branch points contribute to the torus partition function. 
For
genus zero
the vanishing of $d(2,L)$ for $L>2$ means that only maps with no more than two
branch points contribute to the CGTS for the sphere.

To each nonvanishing term in the sum \Zchirii\ we may
associate a homomorphism $\psi:F_{p,L}\to S_ n$,
where $F_{p,L}$ is an $F$ group \fgp,  since  if
the permutations
 $v_1,\dots,v_L, s_1,t_1, \dots, s_p,t_p$
 in $S_n$
satisfy $v_1\cdots v_L \prod_{i=1}^{p}
s_it_is_i^{-1}t_i^{-1} =1$  we may define
\eqn\homomorph{
\psi: \alpha_i\to s_i \qquad \psi:\beta_i \to t_i \qquad
\psi:\gamma_i \to v_i
}
Moreover, if
there exists a
$g \in S_n$ such that
$g \{
v_1, \cdots ,v_L; s_1  , t_1 \cdots s_p ,t_p\} g^{-1}=
 \{
v_1^{'} ,\cdots v_L^{'}; s_1^{'}, t_1^{'} ,
\cdots s_p^{'}, t_p^{'}\}$
as ordered sets
then by definition \sCS.2 the induced homomorphisms are equivalent.
Using the exercise at the end of 
\sssBrnchCov\  the class of $\psi$ will appear in the sum 
 in \Zchirii  $n!/\vert
C(\psi)\vert$
times, where $C ( \psi )$ is the centraliser of $\psi$ in $S_n$.
Therefore, we may  write \Zchirii\ as
\eqn \Zchiii {
Z^+(0,p,N)  =~ 1+ 
\sum_{n=1}^{\infty} \sum_{B=0}^{\infty}
N^{n(2-2p)-B} \sum_{L=0}^{B} d(2-2p,L)
\sum_{\psi \in \Psi (n,B,p,L)}
       {1\over {\vert C(\psi)\vert} } }
where $\Psi (n,B,p,L)$ is the set of  equivalence classes of
homomorphisms  $F_{p,L}\to S_n$, with the condition that the $\gamma_i$
all map
to elements of $S_n$ not equal to the identity.
We have
collected terms with fixed value of:
\eqn\dfb{
B\equiv \sum_{i=1}^{L} (n-K_{v_i}) \qquad .
}

Now we use theorem \sCS.1 to rewrite the sum \Zchiii\ as a sum over
branched coverings.  To do this we must make several
choices.
We choose a point $y_0\in \ST$
and for each $n,B,L,\psi$ we also
make a choice of :
\item{1.}
Some set $S$ of $L$ distinct points on $\ST$.
\item{2.}
An isomorphism \fgpiso.

To each $\psi,S$  we may then associate
a homomorphism
$\pi_1(\ST - S,y_0 )\to S_n$. By theorem \sCS.1
we see that, given a choice of $S$,  to each class $[\psi]$
we associate the equivalence class of a branched covering
$f \in H(n,B,p;S)$, where $f:\Sw\to \ST$.
The genus of the covering surface $h = h ( p, n, B)$ is given by
the Riemann-Hurwitz formula, \Rhurw.
Note that the power of ${1\over N}$ in
\Zchiii\ is simply $2h-2$.
Finally, the centralizer
$C(\psi)\subset S_n$ is isomorphic to
the automorphism group of the associated
branched covering map $f$. The order of
this group, $\vert \Aut~ f \vert$, does not depend on the choice
of points $S$ used to construct $f$.
Accordingly,  we can write $Z^+$ as a sum over
equivalence classes of branched coverings:
\eqn\Zii{
Z^+(0,p,N)  =1+ 
\sum_{n=1}^{\infty} \sum_{B=0}^{\infty}
\sum_{L=0}^{B} \biggl({1\over N}\biggr)^{2p-2}  d(2-2p,L)
 \sum_{f \in H(n,B,p;S) }
{1\over {\vert \Aut~ f\vert} }.  }

\subsubsec{Euler characters}

We have now expressed the CGTS as a sum over
equivalence classes of branched coverings. We
now interpret the weights in terms of the Euler characters of
the Hurwitz space, $\HOL$ of \dfofeff. 

To begin we write
\eqn\euli{
d(2-2p,L) =
{{(\chi_p ) (\chi_p-1) \cdots (\chi_p -L+1)}\over {L!}},}
where $\chi_p = 2-2p$.
The RHS of \euli\ is the Euler character of
the  space $\CC_L(\ST)=C_L(\ST)/S_L$.
This may be
easily proved as follows. Recall that
 it is a general property of fiber bundles $E$
with connected base that the Euler character is
the product of Euler characters of the base and
the fiber \Sp\BoTu :
$$\chi(E) = \chi(F) \chi(B)$$ 
under very general assumptions about the base $B$ and 
the fiber $F$. 

Let $C_{m,n} (\Sigma_T)$ be
the configuration space of $n$ labelled  points on a surface $\Sigma_T$ of
genus $p$ with $m$  {\it fixed} punctures.
There is a fibration
\eqn\fibc{\matrix{
C_{L-1,1} (\Sigma_T) & \mapright{}  &C_{0,L} (\Sigma_T)\cr
                     &              &\mapdown{} \cr
                     &              &C_{0,L-1}(\Sigma_T).  \cr }}
Using the product formula for Euler characters of a fibration we get
\eqn\fiberc{
\chi ( C_{0,L}(\Sigma_T) = (2-2p-(L-1) ) \chi ( C_{0,L-1}(\Sigma_T) ) }
This recursion relation together with $\chi ( C_{0,1}(\Sigma_T))
 =\chi(\Sigma_T) $ gives
$ \chi ( C_{0,L}(\Sigma_T)) =  (\chi_p ) (\chi_p-1) \cdots (\chi_p-L+1)$.
But  $C_{0,L}(\Sigma_T)$ is a topological covering space of $\CC_L(\Sigma_T)$
of degree
$L!$ so this leads to
\eqn\euci { \chi (\CC_L(\ST)) = d(2-2p,L) . }

Using \euci,  we can further rewrite the CGTS as
\eqn\euchar {\eqalign {
Z^+(0,p,N) &=~ 1 +  \sum_{n=1}^{\infty} \sum_{B=0}^{\infty}
N^{n(2-2p)-B} \sum_{L=0}^{B} \chi ( {\CC_L (\ST)) }
     \sum_{ f \in  H( n,B,p,S )  }
{1\over { \vert \Aut~ f\vert } } \cr
}}

Let us now return to the fibration \proj. A straightforward
calculation of the Euler character would give:
\eqn\eulhur{\eqalign{
\chi ( \CC_L (\ST)  )\sum_{ f \in  H( n,B,p,S )  }
 1 &=\chi ( \CC_L (\ST)  )  \vert
H(n,B,p;S) \vert\cr
&=\chi(H(n,B,p,L))\cr}}
where we have again used the fact that the Euler character of a
bundle is the product of that of the base and that of the fiber
\Sp\ (the Euler character of the fiber is
$ {\cal \chi}  (H(n,B,p;S )) = \vert H(n,B,p;S) \vert$).
What actually arises  in the \ymt\ path integral 
is a related sum weighted by $1/\mid \Aut~ f\mid$. 
This is a very encouraging sign.
When $H(n,B,p,L)$ contains coverings with automorphisms
the corresponding space $\hol $ has orbifold singularities.
In the string path integral the moduli space of holomorphic 
maps arises in the form $\HOL$, thus the occurrence of 
orbifold Euler characteristics of $\HOL$ is a clear indication 
of a string path integral reinterpretation of $Z$. 
We introduce the orbifold Euler character
$\chi_{ \rm orb} (H)$ as the Euler character of $\chi( \HOL )$ calculated
by resolving its orbifold
singularities.
The division by the factor $\vert \Aut~ f \vert$ is the correct factor
for calculating the orbifold Euler characteristic of the subvariety
associated to 
$H(n,B,p,L)$ since $\Aut~ f$ is the local orbifold group of the
corresponding points in $\HOL$.
With this understood we naturally define:
\eqn\orbeul{
\chi_{\rm orb} ( (H( n,B,p,L) ))
\equiv \chi ( \CC_L(\ST) )\sum_{ f \in  H( n,B,p,S )  }
{1\over { \vert \Aut~ f\vert } }
}
in the general case.
Thus we finally arrive at our first main result:

\noindent
{\bf Proposition \sEC.1}.
The CGTS is the
generating functional for the orbifold Euler characters of the
Hurwitz spaces:
\eqn\euchario {\eqalign {
Z^+(0,p,N) &=~ 1+ \sum_{n=1}^{\infty} \sum_{B=0}^{\infty}
\biggl({1\over N}\biggr)^{2h-2} \sum_{L=0}^{B}
\chi_{\rm orb} ( H( n,B,p,L) )\cr
}}
where $h$ is determined from $n,p$ and $B$ via the
Riemann-Hurwitz formula.

The $L=B$ contribution in the sum is  the Euler character
of the space of generic branched coverings.
 As described in section \ssCHS, 
compactification of this space involves addition of   boundaries
corresponding to the space of
 maps with higher branch points, i.e.,  where $L < B$.
This leads to an interpretation of 
\eqn\sumchrs{
\sum_{L=0}^{B} \chi_{\rm orb} ( H( n,B,p,L) )
}
as the Euler character of a
partially compactified Hurwitz space $\overline{(H( n,B,p))}$ obtaining by
adding
degenerations of type 1 and 2  (see \fico) 
and their generalizations. Thus, we obtain the Euler 
character of the space of {\it all}  holomorphic maps from 
a smooth worldsheet.

\noindent
{\bf Proposition \sEC.2}: The CGTS is the
generating functional for the orbifold Euler characters of the
analytically compactified Hurwitz spaces:
\eqn\eucharit {\eqalign{
Z^+(0,p,N) &=~ 1+ \sum_{n=1}^{\infty} \sum_{B=0}^{\infty}
\biggl({1\over N}\biggr)^{2h-2}
\chi_{\rm orb} ( \overline{(H( n,B,p))})\cr
&= \exp \biggl[  \sum_{h=0}^{\infty} \biggl({1\over N}\biggr)^{2h-2}
\chi_{\rm orb} (\HOL( \Sw \rightarrow \ST ))\biggr]
 \cr}}
In the second line we have introduced the space of 
holomorphic maps, realized as in section \ssFBAHS, 
from a {\it connected} worldsurface of genus $h$. 
The exponentiation follows from the relation between 
moduli spaces of holomorphic maps from connected 
and disconnected worldsheets. 

\subsubsec{Connected vs. Disconnected surfaces}

In this section we indicate an explicit combinatorical proof 
of the exponentiation in \eucharit. 
In the equation \Zchirii\
$Z^+$ contains contributions from all surfaces, connected and disconnected.
For each $n$, the sum over $S_n$
can be decomposed according to the number of components $c$ of the
worldsheet. For terms in the sum corresponding to
covering surfaces with $n$ sheets and $c$ connected components,
all the permutations in the delta function live
 in a product group $S_{n_1}\times S_{n_2} \times
\cdots S_{n_c}$, where $n_1 + n_2 \cdots + n_c= n$.   Each permutation $\sigma$ in
 the delta function is of the form $\sigma= \sigma^{(1)} \times \cdots \sigma^{(c)} \in S_{n_1}\times S_{n_2} \times
\cdots S_{n_c}$, so the delta function factorises into a product of delta functions
and $K_{(\sigma)}= K_{\sigma^{(1)}}+ K_{\sigma^{(2)} } \cdots + K_{\sigma^{(c)}}$ .   For  each $S_{n_k}$ ($k=1, \cdots
c$) the permutations $s_1^{(k)}, s_2^{(k)}, \cdots s_p^{(k)}, t_1^{(k)}, t_2^{(k)}, \cdots t_p^{(k)}, v_1^{(k)}, \cdots v^{(k)}_L$, generate the whole of $S_{n_k}$ (the condition of transitivity or connectedness). 
 The product group  $S_{n_1}\times S_{n_2} \times
\cdots S_{n_c}$ can be embedded in $S_n$
 in ${n! \over {n_1!n_2!\cdots n_c!}}
{1 \over {c_1!c_2!\cdots c_k!}}$ ways, where $c_1$ of the $n_i$'s are 
equal to one value , $c_2$ are equal    to another etc.
Since the sum in \Zchirii\
counts {\it all} homomorphisms into $S_n$ it sums over all
the embeddings of the subgroups. This guarantees that
sums counting disconnected branched covers 
with fixed branch locus, with
inverse automorphisms, factorise into a  product  over
 connected components. 
 (Without the division by
the automorphism group such  sums do
{\it not} exponentiate \HaMo.)
To complete a direct proof of the  exponentiation, 
we need to write products
like $d(x,L_1)d(x,L_2)$ in terms of  sums over $r$ of expressions
 containing $d(x,L_1+L_2 - r)$.
Such identities can be derived by a simple 
geometrical argument involving a stratification of 
products of configuration spaces  by 
configuration spaces. 

\ifig\fmuep{Local model for a degenerating coupled cover with n=2. The region
between stripes single-covers the annulus. }
{\epsfxsize3.5in\epsfbox{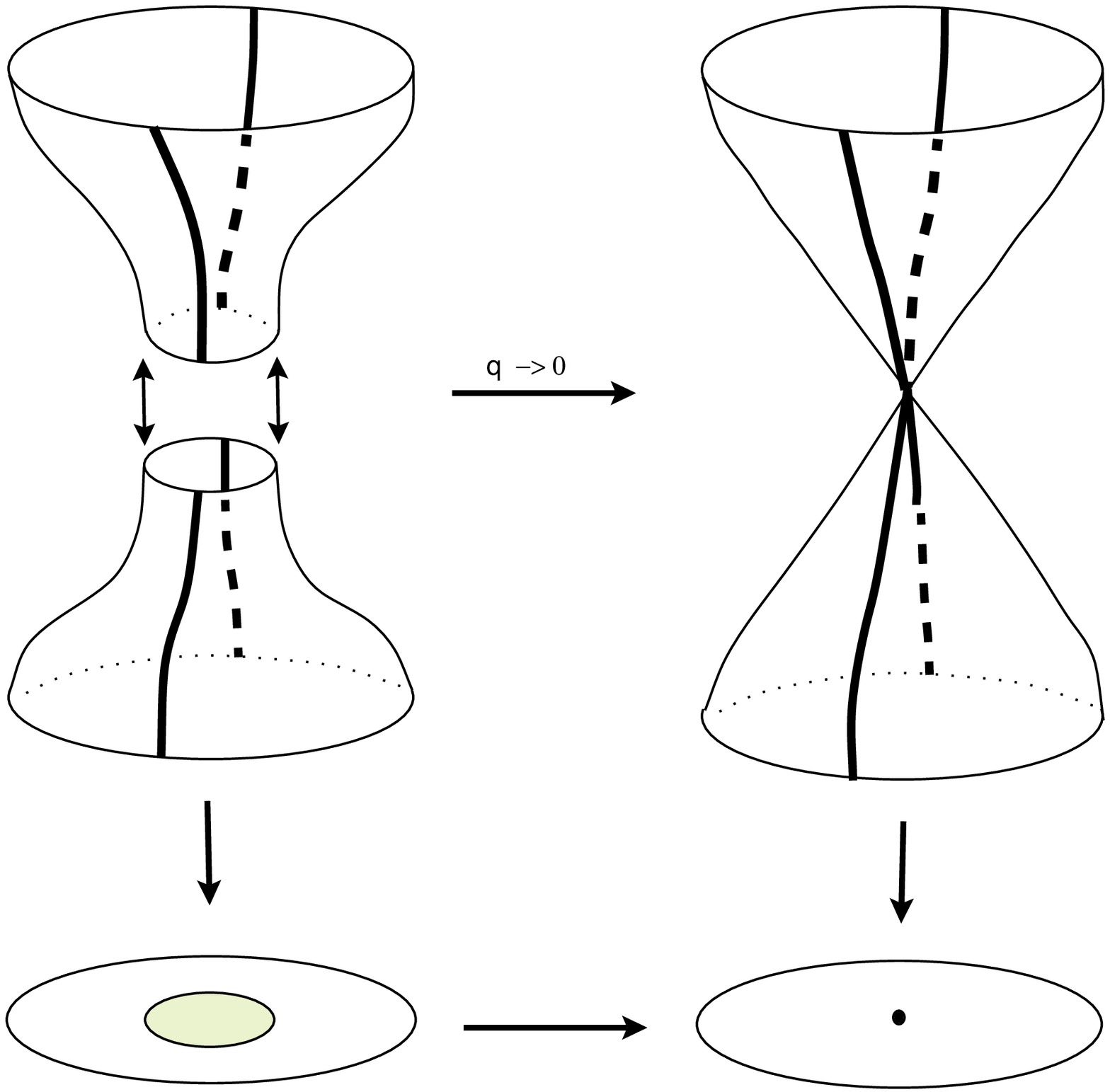}}

\subsec{Nonchiral partition function}

When writing the asymptotics for the full partition function 
similar - but more complicated - manipulations tell the same story \CMROLD .  
Gross and Taylor  showed that the $1/N$ 
expansion of $\Omega_{n^+ n^-} $ admits a 
simple interpretation in terms of maps 
if the worldsheets are allowed to have 
collapsed handles.
One interesting new point  is that some new singular surfaces 
and maps contribute. These maps are neither holomorphic nor 
anti-holomorphic. 
Pictorially, the singularity looks locally like \fmuep .  
Locally such maps are described by considering
a plumbing fixture degenerating to 
 the double point of $\Sw$:
\eqn\plumb{
U_q = \{(z_1,z_2)\vert  z_1 z_2=\eta q,  q \le \vert z_1\vert
,\vert z_2\vert <1\}
}
where $0\leq q <1$ and $\eta$ is an $n^{th}$ root of 
unity for a positive integer $n$. 
On the plumbing fixture we have a family of 
covering maps
\eqn\mdleff{
f^{q,n} (z) ~=~ \cases{z_1^n & for $ q ^{1/2}\le\vert z_1\vert <1$\cr
                         \bar z_2^n & for $ q ^{1/2}\le \vert
z_2\vert <1$. \cr}}

 In the limit $q \rightarrow 0$ we have a map 
which restricts to a  holomorphic  map from one 
disc and to an antiholomorphic  map from  another disc
 which is joined to the first at one point. 

Details
on the combinatorics of such maps
 are given in \CMROLD . The expansion of the coupled $\Omega $
 factors is understood in terms of Euler characters of configuration spaces
 for the motion of both branch points and images of double points \CMROLD. 
The same ideas carry over to gauge groups $O(N)$ and $Sp(2N)$ 
where extra pointlike singularities 
also contribute (double points and collapsed cross-caps). 

\subsec{Inclusion of area}
\subseclab\ssInclar

The same basic reasoning we have used in the
$A=0$ case can be applied to the $A>0$ case.
See \CMROLD.

\newsec{Wilson loops}
\seclab\sWilloops

The ideas of the above sections generalize nicely 
to certain classes of Wilson loop amplitudes 
\GrTa\CMROLD, but space precludes a 
detailed discussion here. Further results on 
the relations between Wilson loop amplitudes 
and covering spaces appear in 
\ramloops. 

\newsec{Introduction to part II: General Remarks on Topological Field Theories}
\seclab\sGRTFT

\subsec{Cohomological Field Theories}

In part II we  review ``cohomological field theory".
Good reviews on the subject already exist 
\refs{\vanbaal, \Witr, \bbrt , \Bl, \DiVVTrieste}, but no previous
review (to our knowledge)
 presents the entire subject systematically from a 
unified point of view. 
We begin with some general remarks on the subject.
Topological field theories, largely introduced by E. Witten, may be grouped 
into two classes: ``Schwarz type" and ``cohomological type".

``Schwarz type" theories \refs{\bbrt}\ have Lagrangians which are metric independent
 and hence - formally- the quantum theory is expected to be topological.
Examples of such theories include the $e^2=0$ \ymt\ theory with Lagrangian 
$$
I = \int \Tr(\phi F)
$$
or 3D Chern-Simons theory
$$
I = \int \Tr(AdA+{2\over 3} A^3)\qquad \qquad .
$$
Mathematically such theories are related to flat connections, knot invariants, Knizhnik-Zamalodchikov equations, etc. 
Physically they are related to rational conformal field theories, the fractional quantum Hall
effect,  anyons etc.
Some aspects of these theories are covered in J. Fr\"ohlich's lectures in this volume. 

In these lectures we will focus on the second type of TFT's - those of ``cohomological type. "
These have a very different flavor.
They are not manifestly metric independent, but have a ``BRST operator," that is, an odd nilpotent operator, $Q$.
Physical observables are $Q$-cohomology classes and amplitudes involving
these observables are metric independent because of decoupling of  BRST trivial degrees
of freedom. 

Cohomological field theory (CohFT) is of very broad significance - each of the examples of 
topological field theories has important applications to mathematics and physics. 
However, the true relevance of topological theories to more traditional problems in physics 
and quantum field theory remains to be seen. 
The main unsolved problem is whether these theories have anything definite and useful to 
say about ``phases" of the theories with propagating degrees of freedom. 
There are many reasons why one might wish to study CohFT. 
Among these we mention the following four: 

\subsubsec{String reformulation of Yang-Mills}

CohFT provides some new tools to attack this old problem.
This is the primary motivation in these lectures.
Having discovered in part I that \ymt\ amplitudes are expressed in terms of 
topological invariants of Hurwitz moduli space we are motivated to search for 
a string theory with the property that
\eqn\eusigi{
\int D [ f, h_{\alpha \beta}, \ldots ]~ e^{-I [ f, h_{\alpha \beta}, \ldots ]}
=\chi_{\rm orb} (\HOL (\Sw,\ST)). }
Here $f:\Sw\to \ST$ and $h_{\alpha\beta}$ is a metric on $\Sw$. 
CohFT provides just the right tools to construct the action $I$, as explained in 
chapter \sYMTasaTST.
In fact, one can go on and, for suitable compact K\"ahler manifolds $X$ derive a theory, 
the ``Euler $\sigma$-model", $\CE \sigma ( X )$, whose partition function is given by
\eqn\eusigii{
Z (\CE \sigma ( X )) = \chi_{\rm orb} ( \HOL ( \Sw, X )). }
This reproduces  the chiral expansion of $QCD_2$. A
 similar construction with a different choice of section
reproduces the orbifold Euler characters for the non-chiral theory. 
 
\subsubsec{Calculation of Physical Quantities}

Some physical quantities can be calculated using methods of CohFT.
This occurs in studies of 4-dimensional string compactification and in studies of 4D
Yang-Mills theory with extended supersymmetries ($N=2$ or $4$). 

In the context of string compactifications, Yukawa couplings
of effective field theories are related to quantum 
cohomology rings which are in turn related to symplectic Floer homology groups. 
Moreover, terms in the effective superpotential of compactified type II strings can be 
calculated from related topological string theories \refs{\BeCeOoVa,\AnGaNaTa}.
Finally, CohFT makes some aspects of the ``mirror symmetry" phenomenon 
transparent \wttnmirror. 

4D supersymmetric Yang-Mills theory has recently begun to undergo a renaissance. 
CohFT has played some role in this development. In a recent paper \VaWi\ the
partition functions of N=4 supersymmetric 
Yang-Mills theories (SYM) were computed
on certain 4-manifolds and shown to satisfy the 
remarkable Montonen-Olive
``$S$-duality conjecture". 
One important part of this story is that 
topologically twisted $N=4$ SYM calculates
the Euler character of the moduli space of  
anti-self-dual instantons \VaWi.
This result is easily understood using the 
formalism we will develop below. 

\subsubsec{Interactions with mathematics}

The subject of topological field theories of cohomological type should be more properly 
called ``enumerative quantum field theory," for it describes intersection theory in moduli 
spaces in the language of local quantum field theory.
The moduli spaces that arise in physics are very canonical and fundamental objects. 

\subsubsec{A Deeper formulation of String theory} 

One compelling reason to study CohFT is that it might prove an essential tool in
formulating a geometrical  basis for string theory. 
Fundamental theories of physics are deeply related to both geometry and symmetry.
This is true of the cornerstones of modern theoretical physics: general relativity
and gauge theory.
It is widely believed that relativity and  gauge theory are encompassed and profoundly
generalized by string theory.
Known results in string theory do indicate that it is a theory combining geometry and
symmetry in a fascinating, new, and not yet understood way.  

The problem of finding a geometrical formulation of string theories is extremely 
difficult, and not terribly well-posed. 
We need to study simpler, but analogous theories.
In standard formulations of  string theory there are infinite numbers of fields/particles, 
and an infinite-dimensional and somewhat mysterious symmetry algebra.
In a sense, most of  these symmetries are ``broken".
Witten has advocated \refs{\Witsm,\WitHS} that topological string theory is another phase of ordinary string theory in which many more symmetries are unbroken and the spectrum
is vastly simpler.
For example, the graviton vertex operator is $Q$-exact, and hence graviton
excitations can be considered to be pure gauge.
Consequently general covariance is unbroken.
One expects that the underlying geometrical significance of topological string
theories will be easier to understand. Indeed, in some examples this has proven to
be the case.
(Chern-Simons as a topological open string \wittcs\ and  ``Kodaira-Spencer
theory"\refs{\BeCeOoVa}.) 

\subsec{Detailed Overview}
\subseclab\ssPRTTwo

We begin our review 
by describing the ``theory of topological field theory".
The essential mathematical elements in constructing cohomological field theories are: 
\vskip0.1truein
\item{a.)}
Equivariant cohomology,
\item{b.)}
(Path) Integral representations of equivariant Thom classes. 
\vskip0.1truein
In chapters \sEQCO, \sINandTIR\ and \sTTwithLS\ we describe in general mathematical
terms the abstract constructions underlying all topological field theories.
The abstract discussion concludes in section \sstheGCOFGL\ where we give a
unified description of an arbitrary cohomological field theory.
The remaining chapters illustrate the principles of  chapters \sEQCO, \sINandTIR\ and
\sTTwithLS\ in several famous examples.

\subsubsec{Equivariant cohomology}
\subsubseclab\sssEvtcoho

The mathematics of equivariant cohomology is described in chapter \sEQCO. 
Equivariant cohomology is the fundamental algebraic structure underlying cohomological
field theory. 
This subject is  related to  constructions in topology involving classifying spaces
$BG$ associated to groups $G$. 
In a sense, cohomological field theory may be regarded as the study of
$H^\bullet (BG)$ and related cohomologies using the language of local QFT.

%

\subsubsec{Mathai-Quillen formalism}
\subsubseclab\sssMQfrmlsm

Topological field theory path integrals are integral representations of Thom classes 
of vector bundles in infinite dimensional spaces. This was 
first pointed out, in the context of Donaldson theory, in an 
important paper of M. Atiyah and L. Jeffrey \refs{\AtJe}.
In chapter \sINandTIR\ we describe the general 
construction of Gaussian 
shaped Thom classes  due to Mathai and Quillen (MQ)\refs{\MaQu}. 
We also describe the important ideas of localization. 
The key result is given in \ssTheLP\ in equation \lociiii. 

\subsubsec{Applications: TFT's without Local Symmetry}
\subsubseclab\sssApplctswoutLS

The results of chapter \sINandTIR\ are applied in chapters \sSQM\ and \sTSM\ to write
topological sigma models. 
In chapter \sSQM\ we show (following \bbrt) how many of the essential ideas of topological
field theory are already contained in supersymmetric quantum mechanics. 

\subsubsec{Local Symmetry}
\subsubseclab\sssLocalsymm

In chapter \sTTwithLS\ we come to grips with the more difficult issue of topological field
theories with local gauge symmetries.
We describe the construction of the ``projection gauge fermion" in section 
\ssABRSTCofPhi.
Using the ideas of this chapter we are able to summarize standard Faddeev-Popov 
gauge-fixing in a way that ties together nicely the two ways ghosts enter in gauge
theories: through the path integral measure and through the purely algebraic BRST
formalism.
This is done in section \ssFPGF. 

\subsubsec{The examples}
\subsubseclab\sssThexmples

The principles of chapter \sTTwithLS\ are illustrated in three standard examples in 
chapters \sTYMT, \sTDTG\ and \sTST, on topological Yang-Mills, 2D gravity, and
topological string theory.
Each theory is presented in a parallel way. 

Finally, as mentioned above, we apply the formalism to the problems described in
part I and derive the underlying string Lagrangian of \ymt. 

\ifig\fbaspic{The basic picture of moduli space. $C$ is 
the space of fields, $M$ above is a slice for moduli 
space. }
{\epsfxsize4.0in\epsfbox{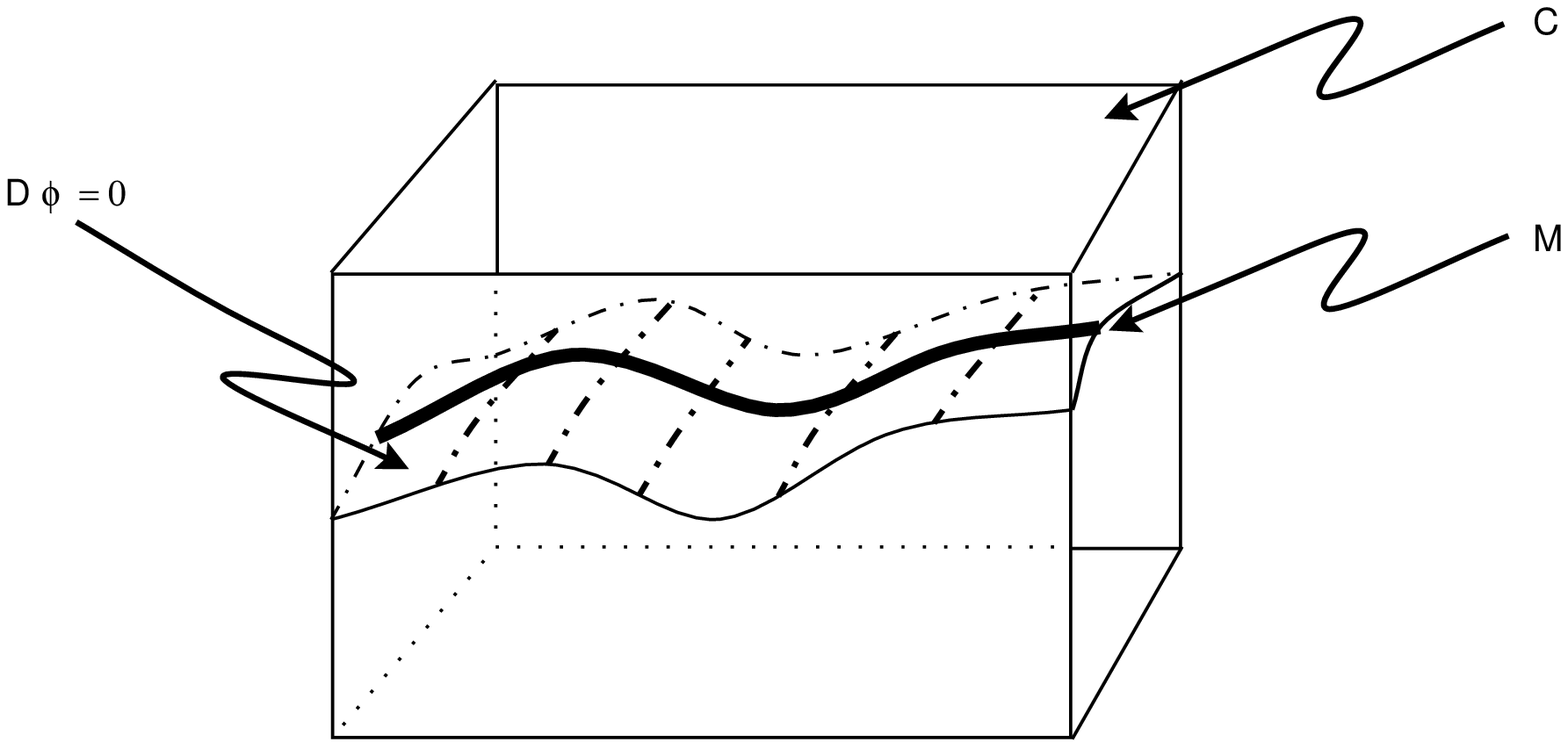}}

\subsec{Paradigm}
\subseclab\ssGenForm

We now describe the basic paradigm for topological 
field theories of cohomological type \refs{\Witr}. 
{}{}From a mathematical point of view, 
topological field/string theory is the study of 
intersection theory on moduli spaces using 
physical methods. In the physical framework these
moduli spaces are presented in the general form: 
\eqn\gnlmdspc{
\CM = \{ \varphi  \in \CC~ \vert~ D \varphi  =0 \} / G}
where $\CM$ is the moduli space, $\CC$ is 
some space of fields, $D$ is some differential 
operator, and $G$ is some group of local 
transformations.  The basic picture for this 
situation is shown in \fbaspic.

Thus we may characterize 
a theory by  three 
basic data:   symmetries, fields, and 
equations \refs{\Witr}. Let us explain the 
role of each of these elements in turn. 
\vskip0.1truein
\item{1.}
{\it Symmetries}: 
The symmetries will generally be characterized by an infinite-dimensional Lie group,
for example a group of gauge transformations, or a diffeomorphism group.
Having chosen the symmetry group $G$ we study the $G$-equivariant cohomology. 
\item{2.}
{\it Fields}:
When we study equivariant cohomology we must choose a model for the 
cohomology.
A choice of differential complex amounts to a choice of fields. 
\item{3.}
{\it Equations}:
Mathematically, the ultimate objects of study are intersection numbers.
The equations are used to isolate interesting subspaces of $\cC$.
One writes Poincar\'e duals to these subspaces using MQ representatives of
Thom forms (thus introducing further fields).
The equations are viewed as the vanishing of a section of a vector bundle: 
$s=0$ where $s(\varphi)=D \varphi$ is the section of a vector bundle we will call the
localization bundle: $s\in \Gamma[ E_{\rm localization}\to \cC ]$. 

\vskip0.1truein\noindent
The above philosophy is given a precise meaning at the end of
chapter \sTTwithLS\ below.  

\subsubsec{General Remarks on Moduli Spaces}

Moduli spaces of solutions to differential equations are pervasive in modern
field theory. 
Examples include moduli spaces of  instantons, monopoles, metrics, 
and holomorphic maps. 
These spaces are all of the form \gnlmdspc\  and share three important 
properties: They are  finite dimensional, generically noncompact, and
generically singular. 

The finite dimensionality means that the intersection theory has a chance of
being well-defined.
Nevertheless, noncompactness is a technical  problem.
Intersection numbers are not defined until one compactifies $\CM$. 
Compactifications of $\CM$ are, in general, difficult to define and 
are also not unique.
Furthermore topological answers depend on the compactification chosen.
Singularities also pose technical problems.
We have already seen orbifold singularities in Hurwitz space above.

\newsec{Equivariant Cohomology}
\seclab\sEQCO

In this chapter we begin developing some of the crucial mathematical background
for topological field theory. 
We begin with some of the algebraic structures in topological field theory.
These, in turn, are related to deep constructions in topology. 
The reason for reviewing this material  is that in all examples
{\it the BRST operator of a topological field theory is the differential for a model of
$\CG$-equivariant cohomology of a space of fields.}
\foot{In some models $\CG$ is the trivial group.}

\subsec{Classifying Spaces}

Let $G$ be a group and $\lieg$ its Lie algebra. 
Some of what follows is only rigorously true when
$G$ is compact, but the formal discussion
 can be applied to any group. In particular, 
in topological field theory, it is applied to infinite dimensional groups.

A $G$-manifold $M$ has an action $x\to g \cdot  x$, for all
$x\in M$ and $g\in G$.
The action of $G$ is said to be {\it free} if, for any $x\in M$,
\eqn\free{
g \cdot x ~=~ x  ~\Longleftrightarrow~ g ~=~ 1}
that is, there are no nontrivial isotropy groups.
If the action of $G$ is free on $M$, then the quotient space $M / G$
forms the base space of a principal $G$ bundle
$$\matrix{
           M             & \mapleft{} & G \cr
\mapdown{\pi} &                      &    \cr
       M / G           &                     &    \cr}
$$
where the quotient space is smooth.
In many cases of interest to physics and mathematics, the group
action is not free.
This leads to the considerations of \sCtoOM\ below. 

\vskip0.1truein\noindent
{\bf Definition}:
To a group $G$ we can associate the {\it universal $G$-bundle}, $EG$, which
is a very special space, satisfying: 
\item{1.}
$G$ acts on $EG$ without fixed points. 
\item{2.}
$EG$ is contractible.

\vskip0.1truein\noindent
{\bf Examples:}
\vskip0.15truein
\begintable
 $G$ | $EG$ | $BG$ \elt
 $\IZ$ | $\IR$ | $S^1$ \elt
 $\IZ^n$ | $\IR^n $ | $S^1 \times \cdots\times S^1$ \elt
 $U(1)=SO(2)$ | $S(H)=\lim_{n\to \infty} S^{2n+1}$ | $\IC P^\infty=\lim_{n\to \infty} \IC P^n$ \elt
 $U(k)$ | $V_k(H)$ | $G_k(H)$  \elt
$\CG$:local gauge transform | $\CA$: Yang-Mills potentials | $\CA/\CG$ \elt
 $\Diff(\Sigma)\times \Weyl(\Sigma)$ | $\MET(\Sigma)$ | $\CM_{h,0}$ \elt
Mod(h,0)  |  Teichm\"uller |  $\CM_{h,0}$
\endtable
\vskip0.15truein

The third and fourth rows are somewhat unfamiliar in physics, 
but the last 3 rows are quite familiar. 
%
Evidently, classifying spaces play an important role in physics.  

\vskip0.1truein\noindent
{\bf Remarks:}
\item{1.}
$EG$ is the ``platonic $G$-bundle."  {\it Any} $G$ bundle is a pullback: 
\eqn\pullback{\matrix{
P\cong f^\ast EG &  & EG\cr
     \mapdown{}     &  & \mapdown{}\cr
              M              & \mapright{f} & BG\cr}}
That is, we can find a copy of any conceivable $G$-bundle 
sitting inside $EG\to BG$ \husemoller.
Moreover, isomorphism classes of bundles 
are in 1-1 correspondence 
with homotopy classes, $[ f\colon M \to BG]$.
\item{2.}
$BG$ is unique up to homotopy type. $EG$ is unique up to 
``equivariant homotopy type."  Recall that two spaces 
$X$ and $Y$ have the same homotopy type if there are maps 
$f\colon X\to Y$ and $g\colon Y\to X$ with $fg$ and $gf$ both 
homotopic to $1$. The maps are said to be 
{\it equivariant} if they commute with the $G$-action.
\item{3.}
There is an explicit combinatorial construction of $EG$ for 
any topological group due to Milnor. See \refs{\husemoller,\steenrod}\
for a description.
\item{4.}
In the last three rows we have ignored an important subtlety, 
namely, that  there are still fixed points.  In the gravity case (last example)
these arise from Riemann surfaces with automorphisms. By 
restricting to diffeomorphisms which preserve $H^\bullet (\Sw,\IZ_3)$ 
one can eliminate all fixed points \bers.
In the Yang-Mills case (third from last example) one must divide the gauge
group by its center (global $\IZ_N$ transformations, for $SU(N)$) and 
cut out the reducible connections described in chapter \sTYMT\ below.
The space of irreducible connections $\CA^{irr}$ is still 
contractible \refs{\AtBoym,\DoKro}.

\exercise{The contractible sphere}

Note that, among other things, the above examples assert that 
the unit sphere in Hilbert space: 
\eqn\unsphhl{
\{(x_1,x_2,\dots )\in \ell_2(\IR)~ \vert~  \sum x_i^2=1 \}}
is contractible! 
Prove this by first showing that the ``Hilbert hotel map"
\eqn\hilhotl{
(x_1,x_2,\dots )\to 
(0,x_1,x_2,\dots )}
is homotopic to 1. Then give a deformation retract of the 
``equator" $\{ \vec x\colon x_1=0\} $ to the ``north pole",
$(1,0,\dots)$.

\endexercise

\subsec{Characteristic classes}
\subseclab\sCC

Let us review briefly some of the theory of  characteristic classes. 
Although $EG$ is contractible, it is a nontrivial 
bundle over $BG$. 
{\it Characteristic classes} are elements of the 
cohomology $H^\bullet (BG)$ which measure the twisting of 
the bundle. By the universal property, characteristic 
classes pull back to all $G$-bundles and measure 
twisting. Indeed, in a sense, all natural ways of measuring
the topology of $P\to M$ are obtained by pullback from $H^\bullet (BG)$.
(Making this statement precise would take us into category theory.)

Characteristic classes of $P\to M$  are formed from the field strengths
$F\in \Omega^2 ( P, \lieg)$ of a connection $A\in \Omega^1 ( P, \lieg)$ on $P$. 
These satisfy:
\eqn\unvrls{\mathboxit{
\eqalign{dA &= F-\half [ A, A]\cr
dF &= -[ A, F]\cr}}}
Let us form 
$$
\widetilde{ch_n}={1\over n!(2 \pi i)^n } \Tr~ F^n\in \O^{2n}(P)
$$
Using \unvrls\ one easily shows that this is a closed form on $P$ (the same is true
for $\CP(F)$, if $\CP$ is any invariant form on the Lie algebra $\lieg$).
Since $EG$ is contractible, its cohomology is trivial.
Indeed $\widetilde{ch_n}$ is exact:
$$\eqalign{
\widetilde{ch_n}
=& d \omega_{0,n}\cr
=& d \left ( {1\over{(n - 1)! (2 \pi i )^n}} \sum_{i=0}^{n-1}
{1\over{(n+i)}} \Tr[( d A )^{n-i-1} ( A )^{2i + 1} ] \right ). \cr}
$$
To get interesting cohomology we must discuss closed forms on $BG$.
This leads to the notion of {\it basic forms}.

\ifig\puonei{Principal $U ( 1 )$ bundle over $M$.}
{\epsfxsize3.5in\epsfbox{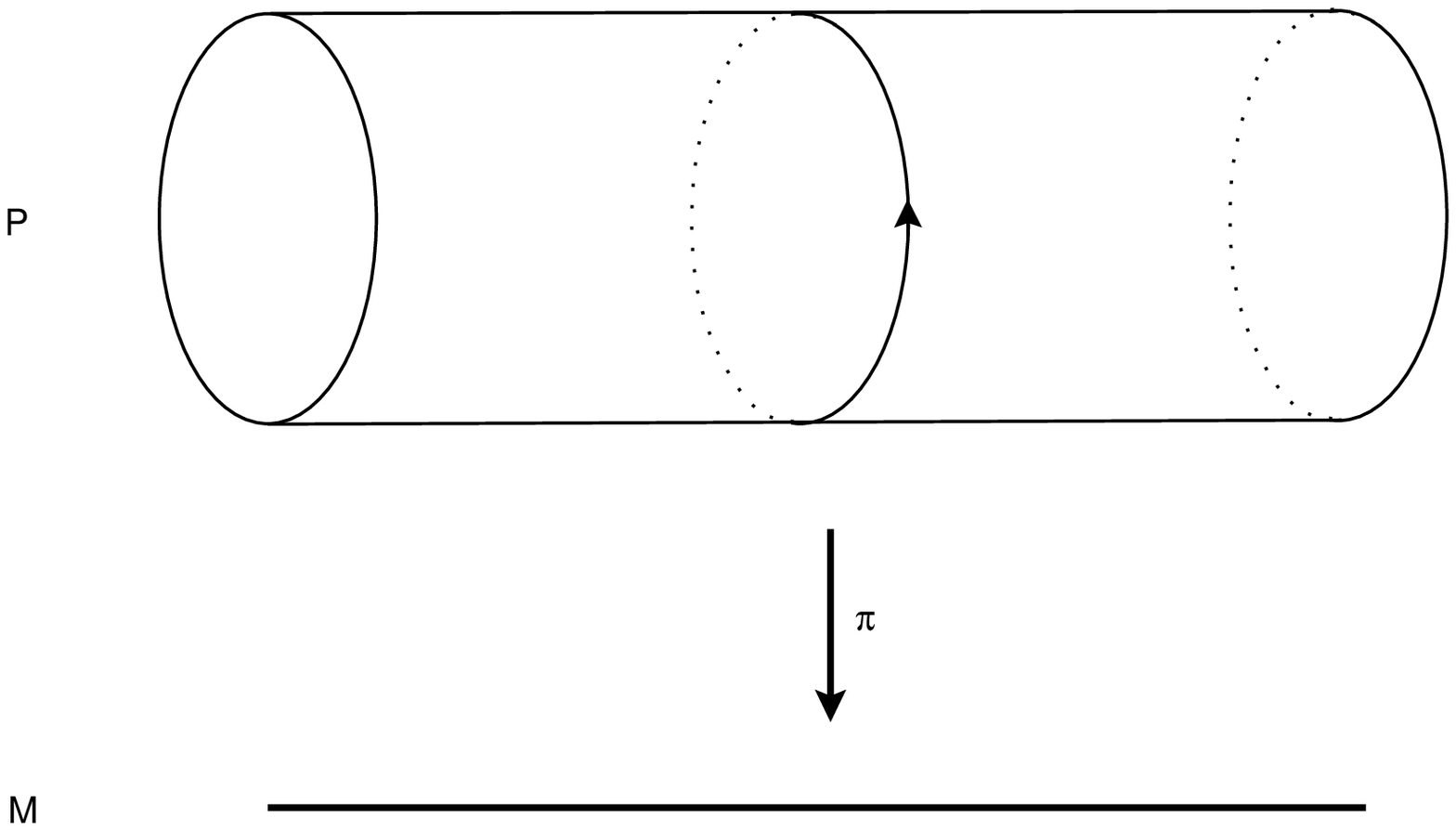}}

\subsubsec{Basic Forms}

Recall from \sssVTV\  that a principal bundle $P$ has an action of $X\in \lieg$ on $P$
via the vertical vector fields $\xi(X)$ (we will drop the $\xi$ in our notation.)
For example, for a principal $U(1)$ bundle the action looks like \puonei.
There are {\it two}  associated actions of $\lieg$ on the differential forms $\O (P)$:
\eqn\twoacts{\eqalign{
\iota (X)\colon \O^{k} ( P ) \to \O^{k-1}  ( P ) & \qquad {\rm contraction} \cr
\CL(X) \equiv [\iota (X),d]_+ :\O^k ( P ) \to \O^k ( P ) & \qquad {\rm Lie\ derivative} \cr}}

Forms $\widetilde \omega$ on $P$ which are of the form $\pi^*(\omega)$ 
for $\omega \in\O (M)$ are
called ``basic".  Such forms are characterized by 
\eqn\bscfrms{\eqalign{
\iota (X) \widetilde \omega & = 0 \qquad\quad {\rm no\ vertical\ components}\cr
\CL(X) \widetilde \omega & = 0 \qquad\quad {\rm no\ vertical\ variation}.  \cr}}

It is easily checked that $\widetilde{ch}_n$ are basic, so that 
$\widetilde{ch}_n = \pi^*(ch_n) $.
The forms $ch_n$ define nontrivial cohomology classes on $BG$. 
It can be shown that the cohomology class is independent of the choice of
connection $A$.  

\vskip0.1truein\noindent
{\bf Remark}:
We have defined these classes using differential forms.
In fact they can be defined purely topologically as elements of the 
{\it integral cohomology} $H^\bullet (BG;\IZ)$ \milnor.

\subsec{Weil Algebra}
\subseclab\ssWA

We now introduce an algebraic analog of $EG$. 
We may view a connection as a map from the 
dual of the Lie algebra, $\lieg\dual$, to a differential form:
\eqn\weilhomi{\eqalign{
A\in \Omega^1(P,\lieg) \qquad & \qquad  
\longleftrightarrow\qquad A\colon\lieg\dual \to \Omega^1(P)\cr
F\in \Omega^1(P,\lieg) \qquad & \qquad  
\longleftrightarrow \qquad  F\colon\lieg\dual \to \Omega^2(P)\cr}}
where $\lieg\dual $ is the dual to $\lieg$. 
Let us return to \unvrls.
The reinterpretation \weilhomi\ motivates the 
\vskip0.1truein\noindent
{\bf Definition}: The Weil algebra of $\lieg$ is 
the differential graded algebra (DGA) 
$$\CW ( \lieg ) = S ( \lieg\dual~ ) \otimes \Lambda ( \lieg\dual~ )$$
where $S( \cdot )$ is the symmetric algebra and
$\Lambda ( \cdot )$ is the exterior algebra.
It may be described in terms of generators by choosing 
a basis $\lieg=\{ T_i \}$ and taking generators
for  $S( \lieg \dual ~)$
given by $\{ \phi^i \}_{i=1,\ldots, \dim~ G}$ of degree 2
and generators for $\Lambda ( \lieg\dual~ )$ given by
$\{ \theta^i \}_{i=1,\dots, \dim~ G}$, of degree 1. 
The Weil algebra becomes a differential algebra upon
introducing  the differential:
\eqn\WeilDeriv{\eqalign{
d_{\CW}  \theta^i ~=&~ \phi^ i -\half f^i_{~ jk}  \theta^j  \theta^k
                  \cr
d_{\CW} \phi^i   ~=&~ - f^i_{~ jk}   \theta^j\phi^k\cr}}
where $f^i_{~ jk}$ are the structure constants of $\lieg$.
It may be seen that $d_{\CW}$ is nilpotent,
$d_{\CW}^2 = 0 $. 
 
\vskip0.1truein\noindent
{\bf Remarks:} 

\item{1.} One could {\it define}  
a connection on $P$ as a homomorphism $\CW(\lieg)\to \Omega(P)$. 

\item{2.} 
The Weil algebra may be thought of ``more 
physically" in terms of ``$b,c,\beta,\gamma$ systems"
as follows. 
The space $S(\lieg\dual~)$ may be identified with 
the space of functions on the Lie algebra. 
We denote generators of the polynomial 
functions  (with respect to the 
basis $T_i$) by $\gamma^i$, of degree two. 
Similarly $\CW(\lieg)$ may be identified with the space of functions on a superspace
\foot{The $\Pi$ indicates that the fiber is considered 
odd. See, for example, \manin.  }
 built
from the tangent bundle of $\lieg$,  $\hat \lieg=\Pi T\lieg$. 
The odd generators of functions on the tangent fibers are denoted by $c^i$.
So a function on the superspace $\hat \lieg$ is a superfield
$$
\Phi(\gamma^i, c^i). 
$$
This space of functions is graded.
The grading is referred to as ``ghost number" in physics. 
Now we introduce an algebra of operators on the space of functions: 
\eqn\comrels{
[\beta_i , \gamma^j] = \delta_i^j \qquad\qquad 
\{ b_i, c^j\} = \delta_i^j}
The action of the operator:
\eqn\bcbetgam{
d_{\cal W} \to Q= - f^i_{jk} c^j \gamma^k \beta_i  - 
\half f^i_{jk} c^j c^k b_i +\gamma^i b_i}
coincides with the Weil differential. Note that $Q$ is very reminiscent of  the form of 
``BRST operators" for supersymmetric gauge principles.
We will comment further on this connection in section \ssECvLAC\ below. 

\subsubsec{Basic subcomplex}

The Weil algebra has  properties analogous to those of $EG$. 
Just as $EG$ is contractible, the cohomology of the Weil algebra 
is trivial:  
$$H^\bullet ( {\cal W} (\lieg),d_\CW) =\delta_{ \bullet, 0} \IR .$$ 

\exercise{Trivial cohomology}

Show that the cohomology is trivial by choosing 
a different set of generators for $W(\lieg)$. (Hint: Use the first 
equation to make one generator exact). 

\endexercise

To get interesting cohomology we take 
our cue from the theory of characteristic classes 
reviewed above and  introduce two differential operators on $\CW (\lieg )$:
the interior derivative, $I_i$, and Lie derivative, $L_i$, defined
by their actions on the generators:
\eqn\DefDeriv{\eqalign{
I_i  \theta^j ~=&~ \delta_i^j\cr
I_i \phi^j ~=&~ 0\cr
L_i ~\colon=&~ [ I_i, d_\CW ]_+ \cr}}
$I_j$ has degree $-1$ while $L_j$ is of degree zero.
\vskip0.1truein\noindent
{\bf Definition}: An element  $\eta \in \CW ( \lieg )$ will be called
\item{a.} {\it Horizontal} if
$\eta \in \cap_{i=1}^{\dim G} \ker ( I_i  )$,
\item{b.} {\it Invariant} if
$\eta \in  \cap_{i=1}^{\dim G} \ker ( L_i )$,
\item{c.} {\it Basic} if $\eta$ is both horizontal and invariant.

\exercise{Action of $L_i$}
Show that $L_i$ are the generators of the 
co-adjoint action of $\lieg$ on $\lieg\dual$. 
\endexercise

We denote the basic subcomplex of $\CW(\lieg)$ by $B\lieg$. 
It is straightforward to calculate $H^\bullet ( B\lieg )$, since 
on the basic subcomplex $d_\CW$ is zero! Hence we 
need only determine explicitly $B\lieg$. Horizontality implies that 
we are in $S( \lieg\dual~ )$, and invariance translates into 
invariance under the coadjoint action of $\lieg$ on $\lieg\dual$. 
We denote the invariant elements under this $G$-action 
by $S(\lieg\dual~)^G$. 

To summarize, we have: 
$$\eqalign{
B\lieg  &= \CW(\lieg)_{\rm basic} \cr
H^\bullet (B\lieg)&= S(\lieg\dual~)^G \cr}
 $$
These are the invariant polynomials on the Lie algebra $\lieg$, 
i.e., $B\lieg$ is the algebra of Casimir invariants. 
The relation between $B\lieg$ and $BG$ is more than just an 
analogy:
\vskip0.1truein\noindent
{\bf Theorem \sEQCO.1}: If $G$ is a compact connected Lie group then 
$H^\bullet ( BG ) = H^\bullet ( B\lieg )=S( \lieg\dual~ )^G$.
\vskip0.1truein\noindent
For a proof see, for example, \AtBomm.
\vskip0.1truein\noindent
{\bf Example:} $G=SO(2)=U(1)$.
As we have already seen, the classifying space of $U ( 1 )$ is $\IC P^\infty$,
whose cohomology is a polynomial algebra on a single generator, $\Omega$,
of degree two.
Further $\lieg = u ( 1 )$ has a single generator, so that
$$
H^\bullet ( \IC P^\infty ) = S (u(1)\dual~ ) = \IC[ \O ]
$$ 
\vskip0.1truein\noindent
{\bf Remark}: 
Already in finite dimensions the theorem is not  true for noncompact groups.
A proper statement involves ``continuous cohomology"\refs{\Bo}. 
In topological field theory the relevant groups are infinite dimensional and certainly
not compact. 
Nevertheless, the cohomologies are remarkably close to those of related compact groups,
although, as we will see, they involve spacetime in an interesting way
(through the ``descent equations"). 

\subsec{Equivariant Cohomology of Manifolds}
\subseclab\sCtoOM

Suppose a manifold $M$ has a $G$-action.  
In general, there are fixed points and $M/G$ is not a manifold.  
It is difficult to discuss the cohomology in such situations.  
A standard trick in algebraic topology is to replace 
$$
M\to  EG\times M
$$
This has a free $G$-action so 
$$
EG\times_G M \equiv {EG\times M\over G},
$$
where $g\cdot (e,x)=(e g^{-1}, gx)$,  
is a manifold. On the other hand, $EG$ is contractible so $EG\times M \cong M$ in
homotopy theory.  
Note, if the $G$ action on $M$ is free then indeed $EG\times_G M\cong M/G$
in homotopy theory.
In general, we can regard $EG\times_G M$ as a bundle over $BG$ with fiber $M$. 

The above observations motivate the topological definition of equivariant cohomology:
\vskip0.1truein\noindent
{\bf Definition.}  The  topological $G$-equivariant cohomology of 
$M$ is
\eqn\tpleqv{
H^\bullet_{G,{\rm topological}} (M) \equiv H^\bullet (EG\times_G M)}
Note that $H^\bullet _G(pt) = H^\bullet (BG)$ is highly nontrivial!

As in our previous discussion, there is a corresponding algebraic description.
Algebraically, the replacement  $M\to EG\times M$ is analogous to 
$$
\O( M ) \to \CW( \lieg ) \otimes \O ( M )
$$
We must define {\it basic forms}.
Let $X_i$ be the vector fields on $M$ corresponding to the action of $T_i\in\lieg$. 

\vskip0.1truein\noindent
{\bf Definition}: Elements of $\eta \in \CW ( \lieg ) \otimes \Omega ( M )$
will be called
\item{a.} {\it Horizontal} if
$\eta \in \cap_{i=1}^{\dim G}\ker ( I_i \otimes 1 + 1 \otimes \iota ( X_i ) )$,
\item{b.} {\it Invariant} if
$\eta \in \cap_{i=1}^{\dim G}\ker ( L_i \otimes 1 + 1 \otimes \cL(X_i) )$,
\item{c.} {\it Basic} if $\eta$ is both horizontal and invariant.
\vskip0.1truein\noindent
%
%
\vskip0.1truein\noindent
{\bf Definition}: 
The algebraic $G$-equivariant cohomology of $M$ is 
$$
H^\bullet_{G,{\rm algebraic}} ( M)
\equiv H^\bullet ( (\CW(\lieg) \otimes \O(M))_{\rm basic}, d_T)
$$
where
\eqn\totdiffl{
d_T = d_\CW\otimes 1 + 1\otimes d}
is the differential. 

Analogous to Theorem \sEQCO.1\ above, we have:
\vskip0.1truein\noindent
{\bf Theorem \sEQCO.2 } For $G$ compact 
$$
H^\bullet_{G,{\rm topological}} (M)= H^\bullet_{G,{\rm algebraic}} ( M)
$$
Again, for further details, see \AtBomm. 
\vskip0.1truein\noindent
{\bf Example}:
One of the most important cases for us is when $G=\Diff(\Sigma)\times \Weyl$, then,
according to our table,  $H^\bullet_G ( pt )=H^\bullet (\CM_{h,0})$.
Clearly, $G$-equivariant cohomology is related to 2D topological gravity.
If we let $M$ be the configuration space of a sigma model, $\MAP ( \Sw, X )$,  then
$G$-equivariant cohomology is related to 2d topological gravity coupled to a sigma model. 
The fact that $H^\bullet_{S^1} ( pt )=\IC[ \Omega ]$ is a polynomial algebra (with elements
of arbitrarily high degree = ghost number) is probably related to the ``special states" of
$D\leq 2$ string theory \lz. 

\subsec{Other formulations of equivariant cohomology}

This section follows \Kal.
It is typical in mathematics that a given object, cohomology 
groups for example, can be characterized or formulated in 
many very different ways. This is true of equivariant 
cohomology. 
There are three algebraic models commonly encountered 
in the literature.  The first model is the Weil model discussed above. 

\subsubsec{Cartan Model}
\subsubseclab\sssCartan

The extreme simplicity of $B\lieg$ suggests that a complex 
much simpler than the Weil complex suffices.
\vskip0.1truein\noindent 
$\bullet$ Complex: 
\eqn\cartcompl{
 S(\lieg\dual~) \otimes \O(M) \qquad .}
\vskip0.1truein\noindent
$\bullet$ Differential:
\eqn\CartanDiff{\eqalign{ 
d_{\sst \cC} \phi^i ~=&~ 0\cr
d_{\sst \cC} \eta   ~=&~ ( 1 \otimes d - \phi^i \otimes \iota_i ) \eta\cr
&= (d-\iota_\phi) \eta\cr}}
for $\eta\in \O(M)$. 
Note that $d_\cC^2 = - \phi^i \otimes \CL_i $ and
in general, $d_\cC$  does not square to zero. On the invariant subcomplex
defined by:
\eqn\crtcplx{
\O_G(M)\equiv \bigl ( S ( \lieg\dual~ ) \otimes \Omega ( M ) \bigr )^G}
it does square to zero:  $d_\CC^2=\phi^i L_i\otimes 1 \to 0$.
Elements of $\O_G(M)$ are called {\it equivariant differential forms}. 

We will see
\eqn\spclcse{
H^\bullet \Bigl ( \bigl ( \CW ( \lieg ) 
\otimes \Omega ( M ) \bigr )_{\rm basic},
d_\cW \Bigr ) ~\equil~
H^\bullet \Bigl ( \bigl ( S ( \lieg\dual~ ) \otimes \Omega ( M ) \bigr )^G,
d_{\sst \cC} \Bigr )}
as a special case of a more general result. 

\subsubsec{BRST Model}  
\subsubseclab\sssBRSTM

In the physical context of topological field theories it turns out that another model of
equivariant cohomology arises naturally.
This is called the ``BRST model,'' or, sometimes, the ``intermediate model.'' 
As a vector space, the complex of the BRST model is identical to that of the Weil model:
$$
\cW ( \lieg ) \otimes \Omega ( M )
$$
but now the differential $d_B$ is:
\eqn\BRSTOp{
d_B ~=~ d_{ \CW} \otimes 1 + 1 \otimes d
    + \theta^i \otimes \cL_i - \phi^i \otimes \iota_i}
with $d_B^2=0$. 
As in the Weil model, the cohomology of this complex is trivial.
One must restrict to a subcomplex in order to calculate the equivariant cohomology. 
The analog of the basic subcomplex of the Weil model is in fact the subcomplex 
$ \bigl ( S ( \lieg\dual~ ) \otimes \Omega ( M ) \bigr )^G$ of the Cartan model.
On this subcomplex we can clearly  identify $d_\cC=d_B$ (and so $d_\cC^2 = 0$.) 

\subsubsec{Equivalences} 

It was shown by Kalkman that the BRST and Weil models of equivariant 
cohomology are related by the algebra automorphism of conjugation by 
$\exp( \theta^i \iota_i )$:
\eqn\algaut{
e^{\theta^i \iota_i} d_T e^{- \theta^i \iota_i} = d_B}
where $d_T$ is given by \totdiffl. 

\exercise{}

Prove \algaut\ by computing separately $e^{\Ad ~\theta^i \iota_i} d_\CW $
and $e^{\Ad ~\theta^i \iota_i} d $. 
Note that it follows immediately that $d_B^2=0$, something which is not 
manifest from \BRSTOp. 

\endexercise

Similar computations to those above show that 
\eqn\mpbsc{\eqalign{
e^{\Ad~ \theta^i \iota_i} [ I_i \otimes 1 + 1 \otimes \iota_i ]
&=  I_i \otimes 1\cr
e^{\Ad ~ \theta^i \iota_i}[ L_i \otimes 1 + 1 \otimes \cL_i ]
&= [ L_i \otimes 1 + 1 \otimes \cL_i ]\cr}}
from which it follows that the basic subcomplex of the Weil model is mapped to the
Cartan subcomplex in the BRST model.
This proves \spclcse. 

\vskip0.1truein\noindent
{\bf Remark:}
A very similar conjugation appears in the papers \refs{\echikann,\getzler}
which relate the ``string picture'' and the ``matter picture'' of topological string 
theory (in the canonical formalism).

\subsubsec{Axiomatic Formulation}

The definition of equivariant cohomology can be axiomatized. 
We are here following \getzler.
It is not difficult to see that the 
equivariant cohomology groups 
satisfy the following three properties: 

\item{1.} {\it Normalisation}: If the $G$-action is free, then
$H_{\sst G}^\bullet ( M ) \equil H^\bullet ( M / G )$.
\item{2.} {\it Homotopy Invariance}: If $f\colon M_1 \to M_2$ is
an equivariant map inducing a homotopy equivalence, then
$f^\ast \colon H_{\sst G}^\bullet ( M_2 )
\to H_{\sst G}^\bullet ( M_1 )$ is an isomorphism.
\item{3.} {\it Mayer-Vietoris}: If $M = U \cup V$, where $U$ and $V$
are invariant open submanifolds of $M$, then there is the long
exact sequence
$$\matrix{
                &                                       &
                &                                       
                                                        &
                &                                       &
                \cr
                & \cdots                                &
\longrightarrow & H_{\sst G}^{\bullet-1} ( U )
           \oplus H_{\sst G}^{\bullet-1} ( V )          &
\longrightarrow & H_{\sst G}^{\bullet-1} ( U \cap V )   &
\longrightarrow \cr
\longrightarrow & H_{\sst G}^\bullet ( M )              &
\longrightarrow & H_{\sst G}^\bullet ( U )
           \oplus H_{\sst G}^\bullet ( V )              &
\longrightarrow & H_{\sst G}^\bullet ( U \cap V )       &
\longrightarrow \cr
\longrightarrow & \cdots                                &
                &                                       &
                \cr}
$$

These three properties are all clear from the topological definition.
In fact, they serve to characterize the cohomology groups uniquely and thus serve as an 
axiomatic definition of equivariant cohomology. Technically, 
equivariant cohomology is a contravariant functor from the category of $G$-manifolds to the category of 
graded vector spaces, $H_{\sst G}^\bullet ( M )$. 

\subsec{Example 1: $S^1$ -Equivariant Cohomology}

\subsubsec{Point}

We have already seen $H_{S^1}^\bullet (pt) = \IC[\O]$, where 
$\O$ is of degree two. 

\ifig\twosph{Figure of $S^2$ with $U ( 1 )$ action.}
{\epsfxsize2.0in\epsfbox{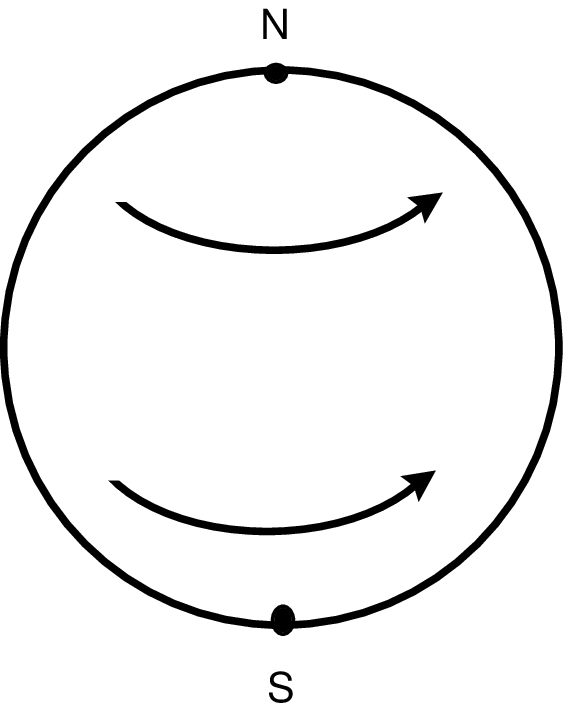}}

\subsubsec{$S^2$}

Consider the standard $U(1)$ action on the two-sphere, \twosph.
It is interesting to compare the different formulations here. 
{}From the axiomatic point of view we calculate the equivariant cohomology
$H_{S^1}^\bullet ( S^2 )$ as follows.
Introduce the standard open covering of $S^2$ by two disks:
$U_1 = S^2 - \{ \infty \}$ and $U_2 = S^2 - \{ 0 \}$.
Then the Mayer-Vietoris long exact sequence for the equivariant
cohomology reads:
\eqn\MVEquiv{\matrix{
   0 \to        &         H_{\sst S^1}^0 ( S^2 )        &
\longrightarrow & H_{\sst S^1}^0 ( U_1 ) \oplus
H_{\sst S^1}^0 ( U_2 )                                  &
\longrightarrow & H_{\sst S^1}^0 ( U_1 \cap U_2 )       &
\longrightarrow \cr
\longrightarrow & H_{\sst S^1}^1 ( S^2 )                &
\longrightarrow & H_{\sst S^1}^1 ( U_1 )
           \oplus H_{\sst S^1}^1 ( U_2 )                &
\longrightarrow & H_{\sst S^1}^1 ( U_1 \cap U_2 )       &
\longrightarrow \cr
\longrightarrow & H_{\sst S^1}^2 ( S^2 )                &
\longrightarrow & H_{\sst S^1}^2 ( U_1 )
           \oplus H_{\sst S^1}^2 ( U_2 )                &
\longrightarrow & H_{\sst S^1}^2 ( U_1 \cap U_2 )       &
\longrightarrow \cr
\longrightarrow & \cdots                                &
                &                                       &
                \cr}}%
$S^1$ does not act freely on $S^2$ since the north and south poles
are fixed points.
$S^1$ does, however, have a free action on $U_1 \cap U_2$, so that
by the normalisation axiom,
$H_{S^1}^i ( U_1 \cap U_2 ) = H^i ( U_1 \cap U_2 / S^1 )$,
which vanishes for $i \ge 1$.
As a result \MVEquiv\ splits into subsequences:
\eqn\MVSub{\matrix{
    0 \to       &         H_{\sst S^1}^0 ( S^2 )        &
\longrightarrow & H_{\sst S^1}^0 ( U_1 ) \oplus
H_{\sst S^1}^0 ( U_2 )                                  &
\longrightarrow & H_{\sst S^1}^0 ( U_1 \cap U_2 )       &
\longrightarrow \cr
\longrightarrow & H_{\sst S^1}^1 ( S^2 )                &
\longrightarrow & H_{\sst S^1}^1 ( U_1 ) \oplus
H_{\sst S^1}^1 ( U_2 )                                  &
\longrightarrow &                      0                &
\longrightarrow \cr
\longrightarrow & H_{\sst S^1}^2 ( S^2 )                &
\longrightarrow & H_{\sst S^1}^2 ( U_1 )
           \oplus H_{\sst S^1}^2 ( U_2 )                &
\longrightarrow &                     0                 &
\longrightarrow \cr
\longrightarrow & \cdots                                &
                &                                       &
                \cr}}
For $i \ge 2$, we simply have
\eqn\ECSoSt{
H_{\sst S^1}^i ( S^2 )
~\equil~ H_{\sst S^1}^i ( U_1 ) \oplus H_{\sst S^1}^i ( U_2 ). }
Using the axiom of homotopy invariance, we see that the maps
$\varphi_i\colon U_i \to pt$ are equivariant maps which induce
homotopy equivalences, so that $H^0=\IC$ and hence:
\eqn\ECSoSt{
H_{\sst S^1}^\bullet ( S^2 )
~\equil~ \{ f ( \Omega ) \oplus g ( \Omega ) \in \IC [ \Omega ]~
\vert~ f  ( 0 ) = g ( 0 ) \}}
where $\IC [ \Omega ]$ is the polynomial algebra in a variable
$\Omega$ of degree $2$.

In terms of the Cartan model we begin by writing the Cartan differential as
$d_\cC = d - \O~ \iota_{\p / \p \phi}$.
Then $h_1( \O ) ( \alpha + \O \cos\theta) + h_2(\O)$ is a cohomology class, 
where $\alpha$ is the solid angle, in local coordinates: $d \phi~ d(\cos\theta)$. 
Remembering that $\O$ has degree 2, we recover the description 
\ECSoSt\  of the cohomology with 
$f(\O)=h_2(\Omega) $ and $g(\O)=h_2(0)+\O h_1(\O)$.

\subsec{Example 2: $\CG$-Equivariant cohomology of $\CA$}

We will discuss this in much greater detail in chapter \sTYMT, but for the present
we note that the space of gauge connections, $\CA$, 
on a principal bundle is the universal bundle 
for $\CG$, the group of gauge transformations. 

The Cartan model can be represented as
\eqn\dondii{\eqalign{
d_\cC A &=\psi\cr
d_\cC \psi & = - D_A \phi \cr
d_\cC \phi  &=0 \cr}}
where $\psi= \widetilde d A$.
In less condensed notation, $A_\mu^a(x)$ are ``coordinates" on $\CA$; 
$\psi_\mu^a(x) = \widetilde d A_\mu^a(x)$ are a basis of 1-forms; and $\phi^a(x)$
are functions on the Lie algebra.

\exercise{}

a.) Prove the second equation in \dondii. 

b.) Show that,
\eqn\rtwoclss{\eqalign{
\CO_2^{(0)}(P) & = {1\over 8 \pi^2} \Tr \phi^2(P)\cr
\CO_2^{(1)}(\gamma) & = {1\over 4 \pi^2} 
\int_\gamma \Tr(\phi \psi)\cr
\CO_2^{(2)}(\Sigma) & = {1\over 4 \pi^2} 
\int_\Sigma \Tr(\phi F- \half \psi\wedge \psi)\cr}}
are closed equivariant forms on $\CA$.
Here $P$ is a point in $X$, $\gamma$ is a curve and $\Sigma$ is a surface.
We will return to these forms in several later sections. 

\endexercise

\subsec{Equivariant Cohomology vs. Lie-Algebra Cohomology}
\subseclab\ssECvLAC

In the BRST quantization of gauge theories one works with Lie-algebra cohomology;
in topological field theory one works with equivariant cohomology.
It is natural to wonder how these two cohomologies are related to one another. 
It turns out that\foot{We would like to thank G. Zuckerman for explaining 
this to us.} {\it equivariant cohomology of a Lie algebra $\lieg$ is the same as a ``supersymmetrized" Lie algebra cohomology of a corresponding graded Lie algebra
$\lieg_{\rm super}$}. 

\subsubsec{Lie Algebra cohomology and BRST quantization}

Let $\lieg$ be the Lie algebra spanned by $T_i$ with $[T_i, T_j] = f^k{}_{ij} T_k$.
The ordinary Lie algebra cohomology is defined using the complex
$\Lambda^\bullet \lieg\dual$. 
We can take generators to be anticommuting elements $c^i$ of degree 1.
The action of the differential is: 
\eqn\brstop{
Q_{\lieg} c^i = -\half f^i{}_{jk} c^j c^k}
$Q_\lieg$ squares to zero by the Jacobi identity. 
Introducing a conjugate operator
$$
\{ b_i, c^j\} = \delta_i^j
$$
we can write $Q_\lieg=-\half f^i{}_{jk} c^j c^kb_i$. 
The cohomology of $Q_\lieg$, $H_Q^\bullet (\lieg)$, is the {\it Lie-algebra cohomology}. 

Now let $V$ be a $\lieg$-module.
We can define the Lie-algebra cohomology with coefficients in $V$,
$H_{Q_\lieg}^\bullet (\lieg,V)$,  by considering the complex
\eqn\brstcplx{
\Lambda^\bullet \lieg\dual \to \Lambda^\bullet \lieg\dual~\otimes V}
and the action of the differential
\eqn\brstdffl{
Q_{\lieg} \to  c^i \otimes \rho(T_i) + Q_{\lieg}}
where $\rho ( T_i )$ is the representation of $T_i$ in $V$.
In more ``physical " notation we have:
\eqn\brstdffi{
Q_\lieg = c^i \rho(T_i) - \half f^i_{~jk} c^j c^k b_i}
\par\noindent
{\bf Example:} One famous example is string theory where $\lieg=Vir(c)$ is the Virasoro
algebra, where 
$V$ is a representation provided by a CFT of central charge $c$. 
It is a well-known fact that $Q_\lieg$ only squares to zero for $c=26$.
In this case (as in the ordinary one) the cohomology defines the space of physical states. 

Note that all the above goes through if we replace a Lie algebra by a super-Lie algebra.
The ghosts $c^i$ carry opposite statistics to the generators $T_i$. 

\subsubsec{Supersymmetrized Lie Algebra cohomology}

To the Lie algebra $\lieg$ we now associate a differential graded Lie algebra (DGLA)
$\lieg[\epsilon]= \lieg \otimes \Lambda^\ast  \epsilon$ where $\deg( \epsilon )=-1$
and $\deg( T_i )=0$;  $\epsilon$ is Grassmann, so $\epsilon^2=0$.
The differential is defined by $\p \epsilon = 1$.  
The DGLA, $\lieg[ \epsilon ]$, is generated by $T_i$ and $\widetilde T_i =T_i \otimes \epsilon$.
It has structure constants: 
\eqn\dglast{
\eqalign{
[ T_i, T_j] &= f^k{}_{ij} T_k\cr
[ T_i, \widetilde T_j ]&=f^k{}_{ij} \widetilde T_k\cr
[ \widetilde T_i, \widetilde T_j ]&=0\cr}}
The graded exterior algebra,  $\Lambda^\bullet \lieg[\epsilon]\dual$, may be identified
with $S( \lieg\dual~ ) \otimes \Lambda( \lieg\dual~ )$. 
Indeed it is generated by $\gamma^i$ and $c^i$ of degrees 2 and 1, respectively.
The differential is defined by:
\eqn\dgondf{
\p\dual~ c^i = \gamma^i \qquad \p~\dual~ \gamma^i =0}
where $\p\dual$ is dual to $\p$. 

This super-Lie algebra has a BRST differential for $\lieg[\epsilon]$-Lie-algebra cohomology.
We introduce $b_i$ and $\beta_i$ in the usual way and get the differentials: 
\eqn\suprlacoh{\eqalign{
Q_{\lieg[\epsilon]} &=-f^i{}_{jk}  c^j \gamma^k \beta_i-
\half f^i{}_{jk} c^j c^k b_i  \cr
\p\dual & = \gamma^i b_i \cr
d_\CW & = Q_{\lieg[\epsilon]} + \p\dual .\cr}}
where the last line makes use of the remark at the end of 
section \ssWA. 

Moreover, let $V$ be a $\lieg$-module.
Then we can promote $V\to \O^\bullet(V)$  to get a $\lieg[\epsilon]$-module
with $X$ and $\widetilde X$ acting by
\eqn\dgmi{
X\to \CL_X \qquad \qquad X\otimes \epsilon \to \iota_X . }
In fact, $( \O^\bullet ( V ), d )$ is a differential graded module (DGM) for the DGLA
$\lieg[\epsilon]$. 
The total differential: 
\eqn\ttldiff{
Q  =c^i \CL_i + \gamma^i \iota_i + Q_{\lieg[\epsilon]}
 +  \p\dual \otimes 1+1\otimes d}
coincides with the ``BRST" model differential of section \sssBRSTM ! 

\subsec{Equivariant Cohomology and Twisted $N=2$ Supersymmetry}
\subseclab\ssECandTNTwoS

Twisted $N=2$ supersymmetry algebras are closely related to equivariant cohomology. 
For example, consider the twisted $N=2$ supersymmetry algebra in two dimensions.
The relations \foot{Note that in this context $L_0$ denotes the zero mode of the
bosonic stress-energy tensor.}: 
\eqn\twsttii{\eqalign{
[J_0, G_0]=-G_0 \qquad & \qquad [ {\cal J}, \iota_\xi ] = - \iota_\xi\cr
[J_0, Q_0]=+ Q_0  \qquad & \qquad [ {\cal J}, d ] = d \cr
[G_0, Q_0]=L_0  \qquad & \qquad [ d, \iota_\xi ] = {\cal L}_\xi\cr}}
show a perfect parallel with equivariant cohomology for an action generated by a single
vector field $\xi$; that is, for  $S^1$-equivariant cohomology.  
Here ${\cal J}$ measures the degree of the form. 

There is a beautiful generalization of this \refs{\segallct,\storai}. 
Consider the loop space of a manifold $X$, which we will denote $LX$. 
Consider differential forms on $LX$, denoted $\O^\bullet (LX)$.
Formally we may think of $\O^\bullet ( LX )$ as a continuous tensor 
product of forms on $X$: 
$\O^\bullet ( LX ) = \otimes_{\theta\in S^1} \O^\bullet ( X )_\theta$.
{}From this point of view we may speak of 
a ``local degree" and a ``local exterior derivative". 
If $f(\theta)$ is a function on the circle we may form
\eqn\geomntwo{\eqalign{
\deg_f \equiv {\cal J}_f  & \equiv \oint d \theta~ f ( \theta )~ \widetilde d X^\mu ( \theta )~
\iota( {\p \over \p X^\mu(\theta)} )\cr
d_f = \oint f(\theta)  d_\theta & = \oint d \theta~ f(\theta)~ \widetilde d X^\mu(\theta)~
{\delta \over \delta X^\mu(\theta)}\cr}}

On the other hand, there is a natural action of $G= \Diff(S^1)$ on
$LX$, and given an element of the Lie algebra,  $v(\theta){\p \over \p \theta}$, there
is a corresponding vector field 
$$V=
\oint v ( \theta ) {{\partial X^\mu ( \theta )}\over{\partial \theta}} {\p \over \p X^\mu(\theta)}
$$
on $LX$.
Accordingly we have operators $L_V$ and $\iota_V$ on $\O^\bullet (LX)$. 
One easily checks that these operators 
satisfy the algebra: 
\eqn\gmntwii{\eqalign{
[ L_{V_1}, L_{V_2}] = L_{[ V_1, V_2]}\qquad &\qquad  [ {\cal J}_f, {\cal J}_g] = 0\cr
[ L_{V_1}, \iota_{V_2}] = \iota_{[ V_1,V_2]} \qquad &\qquad [ {\cal J}_f, \iota_V] = -\iota_{f V}\cr
[ L_{V},d_f] = -d_{f^\prime V } \qquad &\qquad  [ {\cal J}_f, d_g] = d_{f g} \cr
[\iota_V, d_f]_+ = L_{f V} & - {\cal J}_{V f'} \cr
[L_V, {\cal J}_f] &= - {\cal J}_{V f'} \cr}}
Defining Fourier modes via $f_n=e^{i n \theta}, v_n = e^{i n \theta}$, we have: 
\eqn\ntwoalg{\eqalign{
{\cal J}_{f_n}  & \to {\cal J}_n\cr
d_{f_n} & \to G_n^+\cr
\iota_{V_n}  & \to -i G_n^-\cr
L_{V_n} & \to -i L_n\cr}}
in which case \gmntwii\ becomes the topologically twisted $N=2$ superconformal algebra with 
central extension $c=0$.
Recall that the twisted $N=2$ algebra with central charge $c$ is generated by
$L_m,J_m,G_m$ and $Q_m$ with $m\in \IZ$ and relations: 
\eqn\twstntwo{
\eqalign{
[L_m, L_n ] = (m-n)L_{m+n} \qquad &\qquad  [J_m, J_n] = 0\cr
[L_m, G_n] =(m-n)G_{m+n}\qquad &\qquad [J_m, G_n] = -G_{m+n} \cr
[L_m,Q_n] = -n Q_{m+m} \qquad &\qquad  [J_m, Q_n] = Q_{m+n} \cr
[G_m,Q_n]_+ = 2 L_{m+n} & +n J_{m+n}+\half c m(m+1)\delta_{m+n,0} \cr
[L_m,J_n] &= - nJ_{m+n}-\half c m(m+1) \delta_{m+n,0} \cr}}
See, for example, \refs{\lvw, \DiVeVe}.
Putting $f=1$ in \gmntwii\ we see that ${\cal J}_f$, $d_f$, $\iota_v$, and $\CL_v$ satisfy 
the basic relations needed to define $\Diff(S^1)$-equivariant cohomology. 

\subsec{Equivariant Cohomology and Symplectic Group Actions}
\subseclab\ssECSGA

Equivariant cohomology finds a very natural application in the important example of
Hamiltonian actions of Lie groups $G$ on symplectic manifolds $( M, \omega)$, 
where $\omega= \ha \omega_{ij} dx^i dx^j$ is the 
symplectic form. In this case 
there are vector fields $V_a$ acting on $M$:
\eqn\algvees{
[ V_a, V_b] = f_{~ab}^c V_c}
with corresponding Hamiltonians generating the flows: 
\eqn\hamilts{
\iota_{V_a} \omega = - dH_a . }
Equation \hamilts\ 
 has a lovely reinterpretation in 
equivariant cohomology.  Note that  the 
Cartan differential becomes 
\eqn\sympdffl{
D= d - \phi^a \iota_{V_a}. 
}
The definition \hamilts\ is equivalent to the statement 
 that $\omega-\phi^a H_a$
is an equivariantly closed form: 
\eqn\evclsd{
D(\omega-\phi^a H_a) = 0 . 
}
This will be useful in our discussion of localization
below.

\subsubsec{Case of \ymt}
\subsubseclab\CofYMT

This remark becomes particularly interesting in 
the context of \ymt. 
As we saw in subsection \sssetzfc, $\CA$ carries a 
natural symplectic structure in $D=2$. 
Here the moment map for the action of 
the group of gauge transformations is
simply \AtBoym:
$$\mu(A) = -{1\over 4 \pi^2} F $$
so, from \rtwoclss\ we recognize $\CO_2^{(2)}$ as 
the equivariant extension of the symplectic form. 

\ifig\inclusion{Example of an inclusion.}
{\epsfxsize3.0in\epsfbox{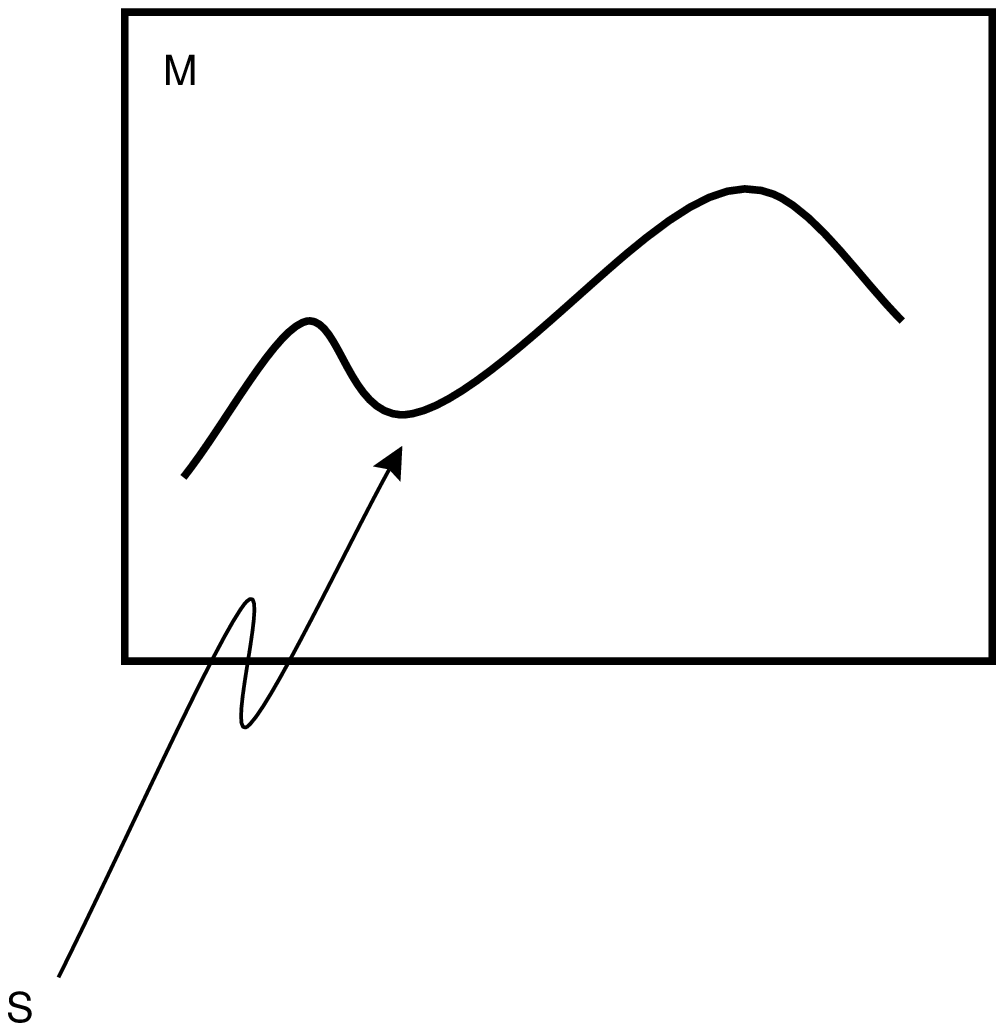}}

\newsec{Intersection Numbers and their Integral Representations}
\seclab\sINandTIR

Let us return to the general picture of 
moduli spaces shown in \fbaspic\ of chapter \sGRTFT.
Relating the path integral over the fields $\CC$ 
to an integral over moduli space involves two steps:  
localization to $D \varphi=0$ and projection by the gauge freedom.
These steps correspond to the two basic ways in which spaces may 
be related to each other: inclusion and fibration, shown
in \inclusion\ and \fibration, respectively. 
In cohomological field theory each of these steps is associated with a BRST principle. 

In this chapter we will illustrate the first of the 
two basic constructions which allow us to
write integrals over small spaces in terms of integrals over large spaces. 
In \inclusion\  the basic problem is: given a cohomology class $\xi\in H^\bullet (M)$ and
a submanifold $i: S\hookrightarrow M$ can we localize an integral over $M$ to an integral
over $S$? 
The solution to this problem involves the construction of the Poincar\'e dual,
$\eta[S\hookrightarrow M]\in H^\bullet (M)$, which satisfies: 
\eqn\pdlprop{
\int_S i^\ast ( \xi ) = \int_M \xi\wedge \eta[S\hookrightarrow M]}
So $\eta$ is just a fancy version of a delta function. 
If $S$ has codimension $k$ then $\eta[ S \hookrightarrow M]\in H^k ( M )$.

Note that as a corollary of this remark if  $S_1,\dots, S_m$ intersect transversally and
$\sum_{i=1}^m \codim~ S_i = \dim~ M$ then 
$$
\int \eta_1 \wedge \cdots \wedge \eta_m 
= \#(S_1\cap \cdots \cap S_m)
$$
is the intersection number of the varieties. 

\subsec{Thom Class and Euler Class}
\subseclab\ssTCandEC

\subsubsec{Thom class}

A central object of study in topological field theory is the Thom class of a 
vector bundle. 
{}From a Thom class we can construct Poincar\'e duals as well as Euler classes. 
We review briefly this notion in this subsection.
This material is standard ( an excellent textbook on the subject is \refs{\BoTu}). 

The cohomology of Euclidean space is $H^j( \IR^n )= \delta_{j,0} \IR$.
It is possible to define another cohomology - cohomology with compact
supports - by restricting to differential forms with 
compact support. For $  \IR^n$ the cohomology groups are: 
\eqn\cptsupp{
H_c^j ( \IR^n ) \cong  \delta_{j,n}\IR}
In physics a more natural notion than compact support is rapid decrease: 
i.e., Gaussian decay at infinity.
Cohomology for forms with rapid decrease is the same as cohomology with compact
supports \MaQu, but the case of rapid decrease generalizes to quantum 
field theory.
An explicit generator of the cohomology is: 
\eqn\bumpform{
\biggl({1\over \pi t}\biggr)^{n/2} e^{-(x,x)/t}~ dx^1\wedge\cdots \wedge dx^n. }
Note that integration gives the explicit  isomorphism in \cptsupp: 
$$
\eta\in H_{RD}^j(\IR^n) \longmapsto \pi_\ast  \eta
=\int_{\IR^n} \eta\in \IR
$$

If $E \to M$ is a vector bundle, with standard fiber $V$ and fiber metric $(\cdot , \cdot )_V$,
then one may also consider cohomology with rapid decrease along the fiber,
$H_{\sst VRD} ( E )$.
Thus, we only consider closed forms with Gaussian decay along the fiber directions and a 
form is only considered exact if it can be written as
$d$ of another form with rapid decrease. 

Given the Gaussian decay along the fiber, we may define  integration along the fiber:
$$
\pi_\ast\colon \Omega^\bullet _{\sst VRD} ( E ) \longrightarrow \Omega^{\bullet-n} ( M )
$$
where $n = \rank~ E$. 
\par
There is a Poincar\'{e} lemma for the vertical rapid decrease cohomology:

\vskip0.2truein\noindent
{\bf Theorem}: (Thom Isomorphism) For an orientable vector bundle
$E \to M$ of rank $n$, integration along the fiber defines an
isomorphism:
$$
\pi_\ast\colon H_{\sst VRD}^\bullet ( E ) ~\equil~ H^{\bullet-n} ( M ).
$$
\vskip0.2truein\noindent
{\bf Proof}: See Bott and Tu \refs{\BoTu}.
\par
The image of ${\bf 1} \in H^0 ( M )$ under $\pi_\ast^{-1}$ determines a cohomology class
\eqn\thomcls{
\pi_\ast^{-1} (1) \equiv \Phi (E) \in H^n_{\sst VRD} ( E )}
called the {\it Thom class} of the oriented vector bundle $E$.
In terms of this class, the Thom isomorphism is given by
\eqn\ThomIso{\eqalign{
\cT\colon H^\bullet ( M ) ~\longrightarrow&~ H^{\bullet+n}_{\sst VRD} ( E )\cr
\cT ( \omega ) ~\longmapsto&~ \pi^\ast ( \omega ) \wedge \Phi(E).\cr}}
In the following sections we will find explicit representatives of the 
Thom class. 

\subsubsec{Euler class} 

For an extensive discussion of the Euler class of a vector bundle see \BoTu.
One quick way to define it is to introduce a connection $\nabla$ compatible with a metric
on $E$.
The curvature $F$ is an antisymmetric matrix of two-forms. 
The Pfaffian of an antisymmetric matrix $A$ satisfies ${\rm \Pf} (A)^2=\det A$ and may be
defined by 
\eqn\pffdff{\eqalign{
{\rm \Pf}(A) &\equiv {1\over m! 2^m} 
\sum_{\sigma\in S_{2m}} \sign(\sigma) A_{\sigma(1) \sigma(2)}\cdots
A_{\sigma(2m-1) \sigma(2m)}\cr
&\equiv \int d\rho~ \exp~ {1\over2} \rho \cdot A \cdot \rho\cr}. }
where the integral is over real Grassmann variables $\rho$.

The Euler class of $E$ may be defined to be 
the cohomology class 
\eqn\dfneulcl{
 e(E \to M) \equiv [{1\over (2 \pi)^m} {\rm \Pf}(F)]. 
}
We will use variously the notations $e(E), \chi(E)$, 
 for the Euler class of a 
vector bundle $E$. When we 
wish to emphasize the base space we will write
$e(E\to M)$ and $\chi(E\to M)$. 

\subsubsec{Two key properties} 
\subsubseclab\sssTwokps

The Thom class will be central to our discussion of  topological field theory because of 
two key facts.
We will derive these facts from the explicit (Mathai-Quillen) representative of $\Phi(E)$
below. 

\item{P1.} {\it Euler Class}.  
Let $s\colon M\to E$ be {\it any} section of $E$, then $s^\ast (\Phi(E))$ is a closed form and
its cohomology class coincides with the {\it Euler class}  $\chi(E\to M) \in H^{2m} (M)$,
where $2m$ is the rank  of the vector bundle $E$. 
Note that if $\rank~ E >\dim~ M$ then the Euler class is necessarily trivial. 

\ifig\ZeroSet{The zero set of a generic section}
{\epsfxsize3.0in\epsfbox{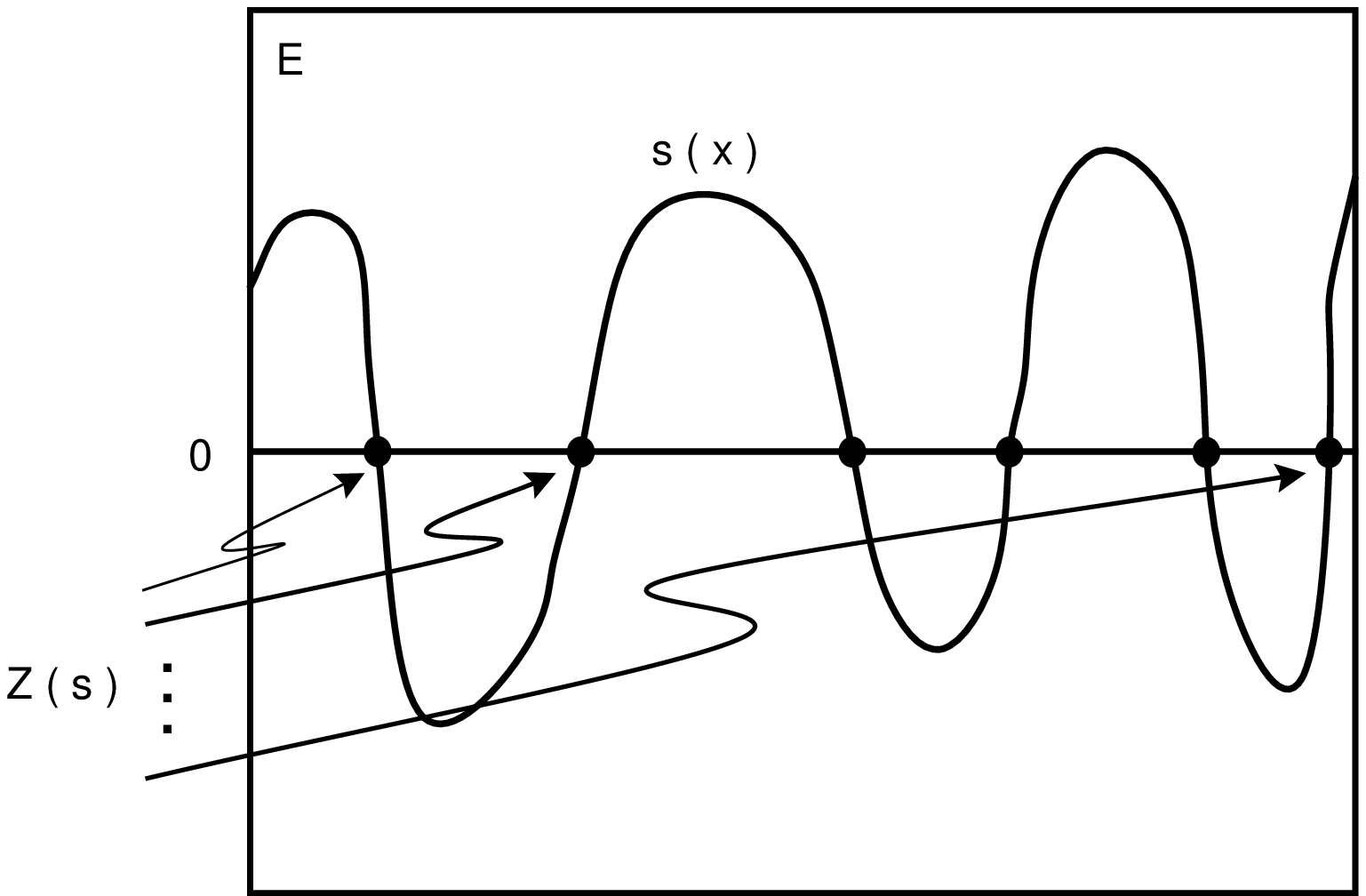}}

\item{P2.} {\it Localization Property: }
Let  $\CZ(s)$ denote the zero locus of a 
section $s$ of $E$ as in \ZeroSet. 
If $s$ is a {\it generic} section
\foot{Technically, if $s$ is transversal to the zero section in $E$.} then 
\eqn\loc{
\mathboxit{
\int_{\CZ(s)}  i^\ast \CO= \int_{M} {s}^\ast (\Phi(E)) \wedge  \CO}}
That is,
$
s^\ast  ( \Phi ( E )) = \eta [ \CZ ( s ) \to M]
$. 
This makes sense since a generic section vanishes on a set of codimension $\rank~ E$. 
\vskip0.1truein\noindent
In the application to topological field theory one interprets $D \varphi$ in section \ssGenForm\
as a section of some vector bundle over the space of  fields: $s(\varphi) = D \varphi$ defines  
$s\in \Gamma[E_{\rm localization} \to \CC]$. 
Then \loc\  is the key property allowing one to localize the integral to the subspace 
$D\varphi=0$ in \fbaspic.
In topological field theory we often use nongeneric sections, in which case this equation
receives an important correction described in section \sssGLF\ below.

\subsec{Universal Thom Class} 
\subseclab\ssUTC

We will now show how to construct a nice explicit representative for $\Phi(E)$ 
by first constructing a ``universal" representative
\MaQu. 
While $E$ might be twisted and difficult to work with, we can replace constructions 
on $E$ by equivariant constructions on a trivial bundle. 
Let $E$ be an orientable real vector  bundle of rank $2m$, with standard fiber $V$.
Let  $(\cdot , \cdot)_V$ be a nondegenerate bilinear form on $V$, defining an 
orthogonal group $G=SO(V)$ with Lie algebra $\lieg_s$.
We can identify $E$ as a bundle associated to a principal $SO(V)$ 
bundle $P\to M$, where $P$ is the $SO(V)$ bundle of all orthonormal (ON)
oriented frames on $E$:
\eqn\trvtrck{
\pi: P\times V \to E= {P\times V\over G}}
It is easier to work with the trivial bundle $P\times V$. 
Moreover, by \tautequi\  sections of $E$ are simply $G$-equivariant functions, 
$P\to V$, and forms on $E$ are identified with basic forms on $P\times V$: 
$$
\O^\bullet ( E ) \cong \O^\bullet ( P\times V)_{\rm basic}. 
$$
The correspondence between the DeRham theory on $P$ and the Weil algebra,
$\cW( \lieg_s )$, which guided us in the previous chapter now suggests the definition:

\vskip0.1truein\noindent
{\bf Definition:} A form $U\in \CW( \lieg_s ) \otimes \Omega(V)_{\rm RD}$ 
will be called a {\it universal Thom form} (in the Weil model) if it satisfies: 
\item{(i)}
U is basic
\item{(ii)}
$Q  U=0 $,  where $Q=d_\CW + d$.  
\item{(iii)}
$\int_V U = 1$

The reason $U$ is useful is that if we choose a connection $\nabla$ on $E$ compatible
with the fiber metric, then we can obtain
 a representative $\Phi(E,\nabla)$ of the Thom
class as follows. 

As we have noted, a connection on $E$, (equivalently, a connection on $P$) is the same
thing as a choice of Weil homomorphism $w\colon \cW ( \lieg_s ) \to \O (P)$. 
We then have a diagram: 
\eqn\diagrm{
\matrix{\cW ( \lieg_s ) \otimes \Omega ( V ) & \mapright{w} & \O ( P \times V )\cr
\mapup{} & & \mapup{} \cr
( \cW ( \lieg_s ) \otimes \Omega ( V ))_{\rm basic}
& \mapright{w} & \O(P\times V)_{\rm basic} \cr
&   &  \cr
& \mapse{\bar w} & \mapup{\pi^\ast}\cr
&  & \O ( E ) \cr}}
Applying the Weil homomorphism, $w$, to
$U \in \cW ( \lieg_s ) \otimes \Omega^\bullet ( V )$ gives 
$w ( U ) \in \Omega^\bullet ( P \times V)$. This form is basic 
and therefore $w(U) = \pi^\ast ( \Phi ( E, \nabla ))$, that is, 
\eqn\gttmcl{
 \Phi ( E, \nabla )= \bar w(U) 
}
for some form $\Phi( E, \nabla) \in H^{2m}_{VRD} ( E )$. 
Using the defining properties of the Thom class of $E$ described in the previous section,
we see that properties $(i)$, $(ii)$ and $(iii)$ suffice to prove  that $\bar w(U)$ represents the
Thom class of $E$. 
\par\noindent
{\bf Remark:} 
We have used here the Weil model of equivariant cohomology. 
One can also construct a universal class in the Cartan or BRST model, and we will do
so below. 

\subsec{Mathai-Quillen representative of the Thom class} 

We first define the Mathai-Quillen (MQ) class.
Then, by writing it in various ways we will show that it manifestly satisfies criteria  $(i)$, $(ii)$
and $(iii)$ above and hence represents the universal Thom class. 

Let $x^a$ be orthonormal (ON) coordinates for $V$ and let
$\theta^A\in \CW^1( \lieg_s )$ and $\phi^A \in \CW^2( \lieg_s )$ be generators of the Weil
algebra of $\lieg_s= so ( V )$. 
We will try to guess a form for an explicit  universal Thom class.
We know that Euler classes are given by Pfaffians of the curvature, so we start with 
$$
U~ {\buildrel ? \over \sim}~ \Pf ( \phi )
$$
but we need a form on $V$ of rapid decrease, so given \bumpform\ we refine our guess to 
$$
U~ {\buildrel ? \over \sim}~ \Pf ( \phi )~ e^{-(x,x)_V} dx^1 \wedge \cdots \wedge dx^{2m}
$$
This doesn't satisfy criterion $(iii)$ above so we again refine 
our guess to: 
$$U~ {\buildrel ? \over \sim}~ {1 \over ({2 \pi})^m} \Pf(\phi)~ 
e^{-(x,x)_V - (dx,\phi^{-1} dx)_V} 
$$
Finally a necessary condition for (i) is that $U$ be horizontal, so introduce the 
notation
 $\nabla x = d x + \theta \cdot x$, where $\theta$ is the connection on $V$,
i.e.
$$
(\nabla x)^a= d x^a + \theta^A (T_A)^a_{~b }\cdot x^b
$$
and, finally, define the MQ representative:
\eqn\frstrep{
\mathboxit{
U={1\over (2\pi )^m}  \Pf ( \phi )~ e^{-(x,x)_V -(\nabla x, \phi^{-1} \nabla x)_V}
\in \CW( \lieg_s ) \otimes \Omega^\bullet ( V )
}
}
The $\phi^{-1}$ is slightly formal, but makes sense if one expands the exponential and
combines with the Pfaffian.

\vskip0.1truein\noindent
{\bf Theorem} \MaQu:  $U$ is a universal Thom class.  

In the next two sections we will prove this theorem by checking properties
$(i)$, $(ii)$ and $(iii)$ of section \ssUTC. 
Clearly, $U$ is $SO(V)$ invariant. 
Moreover, $\nabla x$ is horizontal:
$$
\widetilde I(X) \nabla x = X(x) + I_X(\theta) x =0
$$
so $U$ is horizontal.
\foot{Recall that the action of $T_A$ on $V$ is given by
$X_A=-(T_A)^a_{~b} x^b {\p\over\p x^a}$.}
Thus, $U$ is basic, checking property $(i)$. 
It is less evident that $\int_V U=1$ and that $U$ is closed. 
In order to make these properties manifest we proceed to integral representations of $U$.
These integral representations are at the heart of the connection to TFT. 

\subsec{Integral Representation: Antighosts} 

We begin by introducing anticommuting orthonormal coordinates  $\rho_a$  for $\Pi V^\ast$, 
where the $\Pi$ signifies that the coordinates are to be regarded as anticommuting. 
In the field-theoretic context the $\rho$'s are the ``antighosts." 
When we consider the complex 
$\O^\bullet ( \Pi V^\ast )$ the $\rho$ generate $\O^0 (\Pi V^\ast )$ but
will be considered to be of degree $-1$. 
We now consider the form:
$$
\exp\biggl[ {1\over 4}(\rho,\phi \rho)_{V^\ast} + i \langle \nabla x,\rho\rangle  \biggr]
\in \CW( \lieg_s ) \otimes \Omega^\bullet (V) \otimes \Omega^\bullet ( \Pi V^\ast )
$$
where $\langle \cdot, \cdot \rangle$ is the dual pairing on $V$.
We perform the Berezin integral over the  $\rho$ to get the (fermionic) integral
representation of  $U$:
\eqn\secrep{
\mathboxit{
U= {1\over \pi^m} e^{-(x,x)_V} \int_{\Pi V^\ast} d\rho~
\exp\biggl[{1\over 4}( \rho, \phi \rho)_{V^\ast}
+ i\langle \nabla x, \rho\rangle \biggr]  \in \CW( \lieg_s )\otimes \Omega^\bullet (V)}}

{}From this representation it is obvious that $\int_V U=1$.
The reason is that to get a top form on $V$ we have to take the term in the 
exponential with the top degree of $dx$.
This also pulls off the top form in $\rho$.
Then
\eqn\chkint{\eqalign{
\int_V U  &= {1\over \pi^m}~ \int_V \int_{\Pi V^\ast} d \rho~
e^{-( x, x)_V} {i^{2m}\over (2m)!} \bigl( dx^a \rho_a\bigr)^{2m}\cr
&= {1\over \pi^m}~ \int_V dx^1\wedge -\wedge dx^{2m}~ e^{-( x, x)}  =1\cr}}

It remains to show that $U$ is closed and defines an element of cohomology.
For this we need the BRST integral representations. 

\vskip0.1truein\noindent
{\bf Remarks}:
\item{1.}
A nice check on all this is to pull $\bar w(U)$ back by the zero section of $E$.
Putting $x=0$ in \secrep\  gives a fermionic integral 
representation of a Pfaffian, hence: 
$$
s_0^\ast \bar{w}(U)
= s_0^\ast  {1\over (2\pi )^m} \Pf(F)
=e(E)
$$ 
recovering the  Euler character from the pullback of the zero section. 
\item{2.}
We are using an ON set of coordinates $\rho$ in writing the measure here.
The measure is not invariant under general linear transformations and if in the basis
$\rho_a$ the metric on the fiber is $G_{ab}$, then the proper measure is
$d \rho \sqrt{\det G}$. 

\subsec{Integral Representation: Q-Invariance}
\subseclab\ssIRQI

In order to write a manifestly closed expression for $U$ we enlarge the equivariant
cohomology complex to:
\eqn\seccplx{
\CW( \lieg_s )\otimes \Omega^\bullet ( V ) \otimes \Omega^\bullet ( \Pi V^\ast )}
and consider the following differential
\eqn\seccplxi{
Q_{\CW} =d_\CW\otimes 1\otimes 1 + 1\otimes d\otimes 1 + 1\otimes 1\otimes \delta
}
$d_\CW$ is the Weil differential of 
chapter \sEQCO, while $\delta$ is the de Rham differential in $\Pi V^\ast$.
Explicitly: 
$$
\delta \pmatrix{ \rho_a \cr \pi_a \cr}
=\pmatrix{ 0 & 1\cr  0 & 0\cr}  \pmatrix{ \rho_a \cr \pi_a \cr} $$
The grading, or ghost numbers of $\rho$ and $\pi$ are $-1$ and $0$, respectively. 

Consider the ``gauge fermion" 
\eqn\anthrgf{
\Psi = -i \langle \rho, x\rangle + {1\over 4} ( \rho,\theta  \rho )_{V^\ast}
- {1\over 4} ( \rho, \pi )_{V^\ast}
\quad\in\quad \CW( \lieg_s ) \otimes \Omega^\bullet ( V ) \otimes 
\Omega^\bullet ( \Pi V^\ast )}
which in orthonormal coordinates reads:
\eqn\secgf{
\Psi =   - \rho_a (i x^a - {1\over 4}\theta^{ab} \rho_b + {1\over 4} \pi_a)}
Expanding the action and doing the Gaussian integral on $\pi$ leads to the third
representation: 
\eqn\thrdrep{\mathboxit{
U =  \int_{V^\ast \times \Pi V^\ast} 
\prod_{a=1}^{2m} {d\pi_a\over\sqrt{2\pi}}  {d\rho_a\over\sqrt{2\pi}}~  
e^{Q_\CW ( \Psi )}}}
The advantage of this representation is that 
\eqn\comint{
\int Q_\CW\biggl (\cdots \biggr ) =  ( d + d_ \cW ) 
 \int \biggl(\cdots \biggr)
}
which follows from the simple observation that 
\eqn\drvtfrm{
\delta = \pi_a {\p\over \p \rho_a}}
and hence $\int \delta ( \cdots ) = 0$ by properties of the Berezin integral over $\rho$.
Since the integrand is $Q_\CW$-closed, it immediately follows from \comint\
that $U$ is closed in $\CW(\lieg_s)\otimes \Omega^\bullet (V)$.
Thus, we have finally proven that $U$ satisfies criteria $(i)$, $(ii)$ and $(iii)$ of 
section \ssUTC\ and hence  $U$  is a universal Thom form. 

\subsec{Cartan Model Representative}
\subseclab\ssCartMod

A universal Thom form
$U_\cC\in ( S( \lieg_s\dual~ )\otimes \O( V ))^G$ 
can also be constructed in the Cartan model of equivariant cohomology.
$U_{\cC,t}$
is obtained by a differential on the complex
$
S( \lieg_s\dual~ ) \otimes \O^\bullet ( V ) \otimes \O^\bullet ( \Pi V^\ast )
$
defined by:
\eqn\brsii{ \eqalign{
Q_\cC x  & = (d - \iota_\phi ) x  \cr
Q_\cC \pmatrix{\rho_a \cr \pi_a\cr} &= 
\pmatrix{0 & 1\cr  - \CL_\phi & 0\cr} 
\pmatrix{\rho_a \cr \pi_a\cr}  \cr}}
We then may take the much simpler gauge fermion:
\eqn\mqii{\eqalign{
U_{\cC,t} &= {1\over (2\pi)^{2m}} 
\int_{V^\ast \times \Pi V^\ast} d \pi~ d \rho~
e^{Q_\cC \biggl[ \rho_a (- i x^a - t  \pi_a)\biggr]}\cr
&=({1\over 4 \pi t})^m\int_{ \Pi V^\ast}d \rho~ 
\exp\biggl[ -{1\over 4 t} ( x, x)_V + i  \langle \rho, d x \rangle 
+ t~  ( \rho, \phi \rho)_{V^*} \biggr]\cr}}
It may appear strange that we do {\it not} include the covariant derivative in $x$ in this representation. 
When using this representation we must remember one subtlety in applying the analog
of \diagrm\  to the Cartan model.
The map:
$$
( S ( \lieg_s\dual~ )\otimes \O ( V ), d_\cC) \rightarrow ( \O ( P \times V), d). 
$$
given by naively applying the Weil homomorphism, so that  $\phi\to F$, is {\it not} a
chain map.
For example, $d_\cC \phi=0$,  while $d F= -[A,F]$.
Given a connection on $P$ we can define a horizontal projection of a form
$\omega^{\rm horizontal}
( X_1, \dots , X_n) \equiv \omega(X_1^h , \dots ,  X_n^h )$.
The map taking $\phi\to F$ and 
projecting on the horizontal component 
$$
(( S( \lieg_s\dual~ )\otimes\O ( V ))^G, d_\cC) \rightarrow ( \O(P\times V)_{\rm basic} , d)
$$
is a 
chain map
and thus, in the Cartan model, the pullback of the Thom class becomes
\eqn\crtvrs{
w ( U_{\cC,t} )^{\rm horizontal} = \pi^\ast (\Phi_t ( E, \nabla )) \qquad . }

\vskip0.1truein\noindent
{\bf Remark}:
While the gauge fermion is much simpler in the Cartan model, 
one must, in principle, make a horizontal projection.
This can be very awkward in gauge theories where the connection 
on the space of gauge potentials is nonlocal in spacetime. 
Nevertheless, the Cartan model is still used in topological 
gauge theories and string theories for reasons explained 
in chapter \sTTwithLS\  below. 

\subsec{Other integral representations of $U$} 
\subseclab\ssOIRofU

The definitions of $Q$ and of the gauge fermion are far from unique.  
Two representatives which appear in the literature are, first:

\vskip0.1truein\noindent
$\bullet$ Complex: $\CW( \lieg_s )\otimes \Omega^\bullet (V)
\otimes \Omega^\bullet ( \Pi V^\ast )$.
\par\noindent
$\bullet$ Differential: 
\eqn\brsi{ \eqalign{
Q \theta & = \phi\cr
Q \phi &= 0\cr
Q x^a & = \nabla x^a\cr
Q \rho_a &= \pi_a\cr
Q \pi_a & = \phi_{ab}\rho_b\cr}}
\par\noindent
$\bullet$ Thom form: 
\eqn\mqi{
U= {1\over (4\pi)^{2m}} 
\int_{V\times \Pi V^*} d \rho d \pi~ e^{Q \biggl[ \rho_a (- i x^a-  \pi_a)\biggr]}}
We have $Q^2= \CL_\phi$ so it is 
necessary to restrict  to the $G$-invariant subcomplex
to get a differential. 
This representative occurs, for example,  in \Bl.  A second
example is: 
\vskip0.2truein\noindent
$\bullet$ 
Complex: $\CW( \lieg_s )\otimes \Omega^\bullet (V) \otimes \Omega^\bullet (\Pi V^\ast)$:
\par\noindent
$\bullet$ Differential: 
\eqn\extrabrsii{ \eqalign{
Q x^a & = d x^a \cr
Q \rho_a &= \bar \pi_a+ \theta_a^b \rho_b\cr
Q \bar \pi_a & = -\phi_a^b \rho_b + \theta_a^b \bar \pi_b\cr}}
$\bullet$ Gauge fermion:
\eqn\mqiism{
U = {1\over (4\pi)^m} 
\int_{V\times \Pi V^*} d \rho d \pi~ 
e^{- Q \biggl[ \rho_a (i x^a - \bar \pi_ a)\biggr]}}
This representation is used, e.g., in \Witsm.  
It is also more natural in supersymmetric quantum mechanics. 
It is simply related to the representation of section \ssIRQI\ by 
the shift: $\bar \pi_a = \pi_a - \theta_a^b \rho_b$. 

\subsec{Dependence on choices and ``BRST Decoupling"}
\subseclab\ssDOCandBRSTD

In constructing the form \thrdrep\ we made many choices.
We chose ON coordinates $\rho_a$ to construct the measure $d\rho$; 
we chose a metric $( \cdot, \cdot )_V$ on $V$ and we chose a specific gauge fermion \secgf. 
We now explain that these choices are unimportant. 

The independence of choices follows from the basic principle of ``BRST decoupling."
{}From the derivation of the universal Thom class, it is clear that any change in gauge 
fermion $\Psi\to \Psi+ \Delta \Psi$ will produce a valid representative so long as 1.) 
The resulting $U$ is basic, and  
2.) $U$ can be normalized to have integral 
one, i.e., $\int_V U\in \CW(\lieg_s)$  is a nonvanishing scalar. 

One important aspect must be borne in mind when applying these rules to topological field theories. 
In TFT one is integrating over noncompact infinite-dimensional function spaces.
In order to make precise statements it is necessary to impose boundary conditions.
A change in $\Psi$ can lead to a change in the ``on-shell" value of $\pi$ which in 
turn can bring in new $Q$ fixed-points.
(See section \sssQFPT\ below.)
We will see an
example
 of this in our discussion of supersymmetric quantum mechanics
in section \ssRunvac.  

\subsubsec{Independence of coordinates $\rho_a$}

An important special case of the above remark is that the Thom class is independent
of the metric on $V$. 
First, although we used an ON set of coordinates  $\rho_a$,  the superspace 
measure $d \rho d \pi$ is invariant under changes of coordinates
$\rho\to A \rho, \pi\to A^{-1} \pi$ for all $A \in GL ( V )$. 

\subsubsec{Independence of metric on $V$}

Let us consider a more general gauge fermion: 
$
\Psi=i c_1\langle \rho, x \rangle - c_2 ( \rho, \theta \rho )_{V^*} + c_3 ( \rho, \pi)_{V^*}
$
and write $U=\int \exp[  Q_\CW ( \Psi ) ]$ with $Q_\CW$ defined as in \seccplxi.  
The resulting class $U$ is automatically closed, but is basic iff $c_2= - c_3$.
In this case, letting $c_2= t$ and absorbing $c_1$ into $x$  we obtain the gauge fermion
\eqn\gfunv{\eqalign{
\Psi_t   &= - \rho_a ( i x^a - t \theta^{ab} \rho_b + t \pi_a )\cr
&=-i \langle \rho, x \rangle + t ( \rho, \theta \cdot \rho )_{V^\ast} - t ( \rho, \pi )_{V^\ast} \cr}}
and we  can define the properly normalized universal Thom class as
\eqn\nrmunv{
U_t \equiv ({1\over 4 \pi t})^m \int d \rho~ 
\exp~ \biggl[ -{1\over 4 t} ( x, x )_V + i \langle \rho, \nabla  x \rangle
+ t  ( \rho, \phi \rho )_{V^\ast} \biggr]}
The corresponding Thom class is denoted
\eqn\thomtee{
w( U_t )= \pi^\ast ( \Phi_t ( E, \nabla )). }
The parameter $t$ may be interpreted as the scale of the metric $( \cdot , \cdot )_V$
on $V$. 

\subsec{Superspace and ``physical notation"}
\subseclab\ssSandPN

\subsubsec{Functions on superspace = Differential forms}
\subsubseclab\sssSprsce

Let us introduce the supermanifold $\widehat M$ whose odd coordinates are
generated from the fibers of $T^\ast M$.
In local coordinates we may write $(x^i, \psi^i)$. A {\it function} on $\widehat M$ is the 
same thing as a differential form on $M$.
That is,  we have the basic tautology: 
\eqn\bsctaut{
\mathboxit{
\hat\CF(M)\equiv \CC^\infty(\hat M) \cong \Omega^\bullet (M)}}
the correspondence simply being given by $\psi^i\leftrightarrow d x^i$.
While \bsctaut\ is trivial, it is used repeatedly in the following. 

If $M$ is a manifold we may integrate differential forms of top degree,
$\Omega^n ( M )$, but in general
we cannot integrate functions, since $M$ has no natural measure.
On the other hand $\widehat M$ has a very natural measure: 
\eqn\supmeas{
\hat \mu= d x^1\wedge - \wedge dx^n  
d\psi^1\wedge - \wedge d\psi^n}
Under changes of variables the Bose and Fermi determinants 
cancel\foot{The supermanifold $\hat M$ is ``split" so there is no subtlety here.}.
If $\hat \omega \in C^\infty ( \hat M )$ corresponds to the differential form $\omega$
then we have
\eqn\tautii{
\int_M  \omega  = \int_{\hat M} \hat \mu~ \hat \omega}
Thus, superspace integration is integration of  differential forms. 

\vskip0.1truein\noindent
{\bf Remark}:
Note that when pulled back by a section $s$, the expression $\int_M s^\ast\- \Phi_t(E,\nabla)$ is most
naturally interpreted as an integral over the superspace $\widehat{E^\ast}$
associated to the {\it total space} $E^\ast$. 

\subsubsec{Physical interpretation}
\subsubseclab\sssPI

Now we can outline how the form \secrep\nrmunv\ occurs in physics. 
The term in the exponential is an action. 
In physics the base manifold is a space of  fields ${\varphi } \in\CC$ and the section is typically
of the form $D \varphi $ for some operator $D$. 
The first term in the action in \nrmunv\ is 
$(s,s)_V= \mid D\varphi \mid^2$ and 
gives the purely bosonic terms in the action. 
For examples we have
\eqn\bsctrms{\eqalign{
\parallel s (\varphi) \parallel^2 & = \int \sqrt{g}~ \Tr F^+(A)^2
\qquad\qquad\qquad~{\rm in\ chapter}\  \sTYMT\cr &= \int \sqrt{h}~ \mid df + J df\epsilon(h)\mid^2 \qquad\quad {\rm in\  chapter}\  \sTSM \cr}}

The term $\langle \nabla s,\rho \rangle $ gives the kinetic energy of the fermions.
Recall that given a connection we have
$$
\nabla s  \in \O^1( \CC; E). 
$$
Under the correspondence \bsctaut\ $d  \varphi \in \O^1(\CC)$ corresponds to a 
``ghost field'' $\chi$. 
We may write $\nabla s = \CD \chi$ for some differential operator $\CD$. 
Since $\nabla s$  is valued in $E$,   $\CD$  is a map from ghosts to antighosts. 
Finally, $(\rho,\phi \rho)_{V^\ast}$ is a four-fermi interaction, 
since $\phi$ becomes the curvature of $\nabla$ which, by the correspondence \bsctaut,
is quadratic in $\chi$. 

Now let us interpret the parameter $t$ in \gfunv\ and \nrmunv. 
By a rescaling of $\rho$ we can bring $t$ out in front of the action so that 
\eqn\conhbr{
Q\Psi_t = {1\over t} \Biggl[-{1\over 4} \parallel s( \varphi )\parallel^2 + \cdots
\Biggr]}
Thus we may identify $t$ with $\hbar$ or, alternatively, 
with a coupling constant in the physical model. As we will 
see in section \ssTheLP\  this implies that in these 
theories the semiclassical or weak-coupling expansion 
is exact. 

The independence of cohomology classes from the {\it metric} of the fiber $V$ also 
means that the energy-momentum tensor of the theory is a BRST commutator. 
For example, in \bsctrms\ we see that the spacetime metric is used to define
$( \cdot, \cdot )_V$. 

\subsec{The Localization Principle}
\subseclab\ssTheLP

\subsubsec{Localizing support}

We  have shown many different ways to rewrite the MQ representative of the universal
Thom class \frstrep. 
In particular, we have written a family of representatives 
$ \Phi_t ( E, \nabla )$.
{}From section \ssDOCandBRSTD\ the cohomology class of 
$ \Phi_t ( E, \nabla )$ cannot depend
on  $t$.
On the other hand, if we choose a limit  representative with $t \to 0^+$ then the 
support of  $s^\ast ( \Phi_t ( E, \nabla ))$ will be sharply peaked on the zero-set of $s$.
Consider the integral
\eqn\conintg{
\int_M s^\ast (\Phi_t)~ \CO}
where $\CO \in H^\bullet ( M )$. 
According to the method of steepest descents or stationary phase, the  $t \to 0^+$
asymptotics of the integral are given by an integral along $\CZ ( s )$ and a Gaussian 
integration in the normal directions.
In the limit $t \to 0^+$ the Gaussian approximation becomes arbitrarily good.
On the other hand, from what we just said the integral cannot depend on $t$!
\medskip
\boxit{{\it Therefore, the Gaussian approximation is exact.}}

\subsubsec{Example: {}From Gauss-Bonnet to Poincar\'e-Hopf}

We now work out the above principle for the case of $E=TM$ where $M$ is a Riemannian 
manifold with metric $g_{ij}$.
We take the Levi-Civita connection on $TM$.
Let $V=V^i \p_i$ be a section of $TM$. Then, considered as an element of
$\cC^\infty ( \widehat M )$, 
\eqn\gsbni{\eqalign{
V^\ast& (\Phi_t ( TM, \nabla ))\cr
&= {1\over (4 \pi t)^m} \int d \rho \sqrt{g}~
\exp \biggl[ -{1\over 4 t} g_{ij} V^i V^j + i \rho_j (\widehat{\nabla V})^j
+ tR_{ijkl} \rho^i \cdot \rho^j \psi^k \psi^l
\biggr] \cr}}
where the\ $~\widehat{ }~$\ indicates that we 
are to think of the corresponding differential
as a function on superspace.
Setting $V=0$, or taking $t \to \infty$, we get 
\eqn\getpfaff{
{1\over m! (2 \pi)^m} \sqrt{g}~ \Pfaff [g^{ii^\prime}R_{i^\prime j^\prime} g^{j^\prime j} ]
= {1\over m! (2 \pi)^m}~ \Pfaff ( R^i_{~ j} ) }
giving
the Euler characteristic $\chi ( TM ) = \chi ( M )$, expressed according to the Gauss-Bonnet
formula.

On the other hand, letting $t \to 0^+$ we see that the integral 
\eqn\spscint{\eqalign{
&\int_{\hat M} \hat \mu~ V^\ast ( \Phi_t ( TM, \nabla ))\cr
& = {1\over (4 \pi t)^m}~ \int dx d\psi d \rho \sqrt{g}
\exp \biggl[ -{1\over 4 t} g_{ij} V^i V^j + i \rho_j ( \widehat{\nabla V} )^j
+ t  ~ R_{ijkl} \rho^i \cdot \rho^j \psi^k \psi^l
\biggr]. \cr}}
is concentrated at the zeroes, $P$, of the vector field $V$.
Let $x^i$ be local coordinates in the neighborhood of a zero of $V$ such that:
\eqn\xpdv{
V^i = V^i_{~ j} x^j + \CO( x^2 )}
We will assume that the zero is generic so that $\det V^i_j \not= 0$. 
Then at $x=0$:  
$$
( \widehat{\nabla V} )^i= V^i_{~ j} \psi^j + \CO ( x )
$$
in the superspace language.
Now do the integral \spscint\ in the neighborhood of $x=0$.
As $t \to 0^+$ the Gaussian approximation gives an
integral over $\rho$ and $\psi$ leading to a factor of 
$$
(-1)^m \sqrt{g} \det V^i_{~ j}
$$
(The extra factor of $\sqrt{g}$ arises because we are not using an orthonormal basis
of $\rho_i$. )
The bosonic Gaussian integral yields
$$ 
{1\over \sqrt{\det ( V^{tr} g V )}}
$$
Thus, {\it the boson and fermion determinants cancel up to sign} and the contribution of 
the fixed point is just
\eqn\vfldindx{
\sign \det V^i_{~ j} = \Index(V,P). }
The index of $V$ at $P$ may be thought of more conceptually as the winding number
of the map $S^{n-1} \to S^{n-1}$ given by a first order zero of  $V$. 
Finally, since the Gaussian approximation is exact,
\spscint\ is just the sum over fixed points of the 
quantity \vfldindx. 
The $t$-independence of the integral proves the Poincar\'e-Hopf theorem.
In summary we see that the MQ representative gives a formula for the 
Euler character that interpolates smoothly between the Gauss-Bonnet and
Poincar\'e-Hopf formulae for $\chi(M)$:
\eqn\localization{
\matrix{
 &  & \int \hat \mu V^* (\Phi_t )& & \cr
 & \swarrow &  & \searrow&  \cr
\sum_{\{ P \vert V(P)=0 \}} \Index(V,P)
& & & & 
\int_M e(TM)\cr} }
\vskip0.2truein

\ifig\VanSection{A section with vanishing locus along the $x$-axis.}
{\epsfxsize3.0in\epsfbox{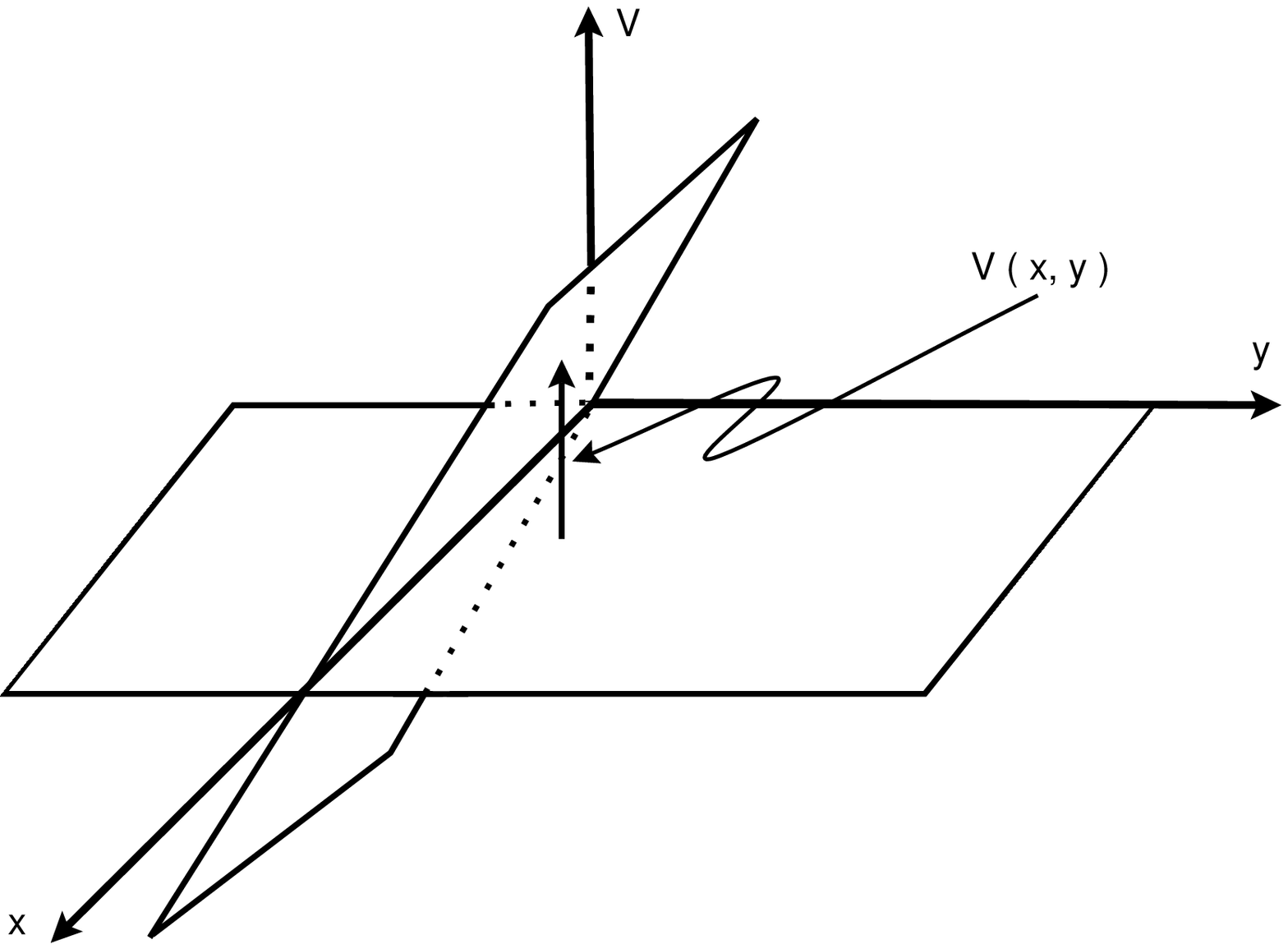}}

\subsubsec{General Localization Formula}
\subsubseclab\sssGLF

Consider the vector bundle
$$\matrix{
E & \longleftarrow & V\cr
\downarrow & & \cr
M & & \cr}
$$
Let us now consider any section $s$ of $E$, not necessarily generic.
The proper statement of the localization formula is given in \refs{\Winm}, sec. 3.3. 
Near the submanifold $\CZ(s)$ we can choose coordinates $(x,y)$ for $M$ so that the
equation $y^i=0$ describes $\CZ(s)$ locally as in \VanSection.

A section $s$ may be written locally as $s(x,y)=( x, y; \vec v ( x, y ))$.
We assume the section is not too special so that $\vec v(x,y)$ vanishes 
linearly at $y=0$: $\vec v(x,y)= y^i \vec v_i(x)+\cdots $. 
Let $E^\prime_{x,y=0} = {\rm Span}~ \{ \vec v_i ( x ) \}$, and consider the sequence of
bundles over $\CZ(s)$: 
\eqn\wittseqi{
0 \rightarrow E^\prime \mapright{} E \mapright{} E/E^\prime \mapright{} 0}
It is shown in \refs{\Winm,\Wiag} that if $P$ is Poincar\'e dual to
$\chi(E/E^\prime \to \cZ ( s ))$ in
$\cZ ( s )$, then $i ( P )$ is Poincar\'e dual to $\chi(E \to M)$ in $M$.
Translating this to a statement about cohomology we have:
\eqn\lociiii{
\mathboxit{
\int_M \chi(E \to M) \wedge \CO = 
\int_{\CZ(s)} i^\ast (\CO) \wedge \chi(E/E^\prime \to \CZ(s))}}
where $i\colon \CZ ( s ) \hookrightarrow M$ is the inclusion.

This fact arises in applications to topological string theories as follows. 
As noted in section \sssPI, the connection allows us to write 
$\nabla s \in \O^1 ( M; E )$.
A 1-form with values in $E$ can be regarded as a linear operator: 
\eqn\dfnop{
\nabla s: T_p M \to E_p}
Over $\CZ ( s )$, we have
$$
\nabla s \vert_{\CZ ( s )} = d s + \theta s \vert_{\CZ( s )} = d s \vert_{\CZ ( s )}
$$
Further given any tangent vector $V \in TM \vert_{\CZ ( s )}$ and section $s$, we may
write in local coordinates:
$$\eqalign{
V =& \xi^i {\partial\over{\partial x^i}} + \zeta^i {\partial\over{\partial y^i}}\cr
s =& \vec{e}_\alpha~ s^\alpha\cr
s^\alpha ( x, y ) =& v_i{}^\alpha y^i + O ( y^2 )\cr}
$$
so that
$$\eqalign{
\nabla s\vert_{\CZ ( s )} ( V ) 
=& ds ( V ) \vert_{\CZ ( s )}\cr
=& \vec{e}_\alpha v_i{}^\alpha \zeta^i\cr}
$$
If $v_i^{\alpha }$ is injective 
 (giving precise meaning to ``not too special'' above \wittseqi ), then
$$
\nabla s \vert_{\CZ ( s )} (V)= 0 \qquad\Longleftrightarrow\qquad \zeta^i = 0
$$
but such $V$ are in $T \CZ ( s )$, so that
\eqn\zrmd{
\ker(\nabla s )\mid_{Z(s)} \cong T\CZ(s). }
That is, ``the ghost zeromodes are tangent to $\CZ(s)$."

We can identify 
\eqn\dfepr{
E^\prime = \Im (\nabla s )}
hence the exact sequence \wittseqi\  becomes: 
\eqn\wittseq{
0 \mapright{} E^\prime \mapright{} E \mapright{} \cok(\nabla s )\mapright{} 0} 
{}From the topological statement  \lociiii, we obtain: 
\eqn\locii{
\mathboxit{
\int_M s^\ast ( \Phi ( E, \nabla )) \wedge \CO = 
\int_{\CZ ( s )} i^\ast ( \CO ) \wedge \chi ( \cok (\nabla s ) \to \CZ(s))}}
\bigskip

\exercise{Ghost number selection rule}

Show that $\CO$ must be a differential form of degree
\eqn\gstcnv{
\deg(\CO) = \Index (\nabla s)}
for \locii\ to be nonzero. Here $\Index (T)$ for a 
linear operator $T$ is the index: 
$$
\Index (T)\equiv \dim ~ \ker~ T - \dim ~ \cok~ T
$$

In applications to TFT, $\nabla s$ is a Fredholm operator, so even though $E$ and
$E^\prime$ are infinite dimensional $E / E^\prime$ is finite dimensional. 

\endexercise

The result \locii,  which will be very important in all the examples below, can also be
proved directly from \nrmunv. 
If  $\cok (\nabla s ) \not=0$, then there will be {\it antighost zeromodes}.
That is, some of the fermionic coordinates $\rho$ cannot be absorbed by bringing down 
the kinetic term from the exponential.
These must be absorbed by the curvature term.
The fermionic integral computes the Pfaffian of the appropriate curvature, producing
the RHS of \locii. 

\vskip0.1truein\noindent
{\bf Remarks}: 
\item{1.}
Note that for a generic section 
$\coker~ \nabla s =\{0\}$ and  \locii\  reduces to \loc\ above. 
In this case $\chi$ is just $\pm 1$, and is determined by whether the isomorphism
$\nabla s: TM/ \ker \nabla s \to E$ is orientation preserving or reversing. 
\item{2.}
$\CZ(s)$ might have many components.
In QFT applications it will typically have infinitely many components. 
\item{3.}
The formula \locii\ can be generalized to the 
case where $s$ vanishes quadratically in $y$. 
In that case there are further determinants. 

\subsubsec{$Q$ Fixed-Points}
\subsubseclab\sssQFPT

The localization can also be understood from another point of view  \Winm.
If the action of $Q$ is free, then it determines an odd fibration of the supersymmetric
field space, $\widehat{E^\ast}$.
Any integral can then be broken up into an integral over the even base space times
an integral over the odd fibre.
Since in our case the argument of the integral is $Q$-invariant, the fibre integral
gives the ``volume" of the fibre.
But this is zero, since
$$
\int d \theta~ 1 = 0
$$
for any Grassmann variable $\theta$ (collective coordinate for $Q$).
Thus if $Q$ acts freely, then integrals over $Q$-invariant observables vanish.

If $Q$ acting on $\widehat{E^\ast}$ has a fixed point set, $\CZ$, then the entire
contribution to the integral comes from the neighborhood of $\CZ$.

Evidently, the fixed points of $Q$ are: 
$$
\pi_a = 0  \qquad \qquad \psi^i {\p \over \p \phi^i} s =0
$$ 
The second equation is solved by $\psi=0$
in the directions normal to $\CZ(s)$. 
Since $\pi$ only enters through a 
Gaussian integral we can eliminate it  by completing the 
square.
Then the equation $\pi_a=0$ becomes the nontrivial equation
\eqn\qfxpt{
\pi^j = -2 \theta^{jk}\rho_k + {i\over t} s^j =0}
hence the fixed points are at $s^j=0$, as expected. 

\subsubsec{Physical Interpretation}
\subsubseclab\sssPI

The $Q$ fixed-point theorem, or localization theorem restricts 
fields $\phi\in \CC$ to the 
classical solution space.
The space $\CZ(s)$ is a space of collective coordinates. 
$Qs=0$ gives the equations for fermion zeromodes which, under the superspace 
correspondence \bsctaut\ correspond to cotangent vectors to $\CZ(s)$.
$Q$-invariance implies that Bose/Fermi determinants cancel up to sign simplifying the integral over collective coordinates.  
The equation \gstcnv\  is related to anomalies
 in the conservation of a ghost number current. 

\subsec{Partition Functions}
\subseclab\ssPF

Suppose $\CO_i$ are closed forms on $\CM$ and that $\omega_i=i^\ast (\CO_i)$ 
are forms on $\CZ(s)$.
Again $i\colon \cZ ( s ) \hookrightarrow \CM$ is the inclusion map.
One elegant way to summarize the cohomology ring, or intersection ring of $\CZ(s)$ is
given by the generating functional:
\foot{
We are looking at a generating 
functional for correlators. Adding the operators to 
the action might change the nature of the integral. 
In such cases one should take the  
$\alpha$ to be nilpotent.}
\eqn\gncohrg{
\langle e^{\alpha^i \omega_i} \rangle_{\CZ(s)}
\equiv \int_{\CZ(s)} e^{\alpha^i \omega_i}~ \chi( E/E^\prime \to \cZ ( s ))}
On the other hand, by \lociiii\ and \locii\ we can represent
this as
\eqn\ghcohrgi{
Z(\alpha^i) = \int_{\hat E^\ast} \hat \mu~ e^{Q(\Psi) + \alpha^i \widehat \CO_i}}

\boxit{
The ``partition function" of the theory perturbed by operators $\hat \CO_i$
is the generating functional of the cohomology ring  of $\CZ(s)$. }

\subsec{Localization and Integration of Equivariant Differential Forms}
\subseclab\ssLandIofEDF

An illustration of the localization ideas used in this chapter is the localization proof of 
the Duistermaat-Heckman (DH) formula, and the generalization of this argument to
the nonabelian localization theorem for integrals of equivariant differential forms \Witdgtr. 

\subsubsec{The Duistermaat-Heckman theorem}
\subsubseclab\ssstheDHT

Let $(M,\omega)$ be a symplectic manifold of dimension $2n$, with a Hamiltonian action of $U(1)$.
Thus there is a vector field $V$ generating the flows with Hamiltonian:
$\iota_V \omega = - dH$. 
The DH theorem expresses 
\eqn\symplint{
\int_M {\omega^n\over n!} e^{\alpha H}}
in terms of the critical points of $H$. 

To prove the DH theorem let us recall from section \CofYMT\  that  $\omega+ H$
is an equivariantly closed form for the equivariant differential $D=d+ \iota_V$. 
(Here we have evaluated $\phi=-1$.)
Let us rewrite \symplint\ as 
\eqn\symplinti{
\bigl({1\over \alpha}\bigr)^n \int_M e^{\alpha(\omega+H)}}
The proof of the localization theorem proceeds by considering the integral 
\eqn\symplintiii{
\bigl({1\over \alpha}\bigr)^n \int_M e^{\alpha(\omega+H) + t D\lambda}}
where $\lambda$ is an invariant one-form. 
Using the basic fact that 
\eqn\ibpii{
\int_M \alpha~ D \beta =0}
for an equivariantly closed $\alpha$, and the fact that $D^2 \lambda = \cL_\phi \lambda=0$
we see that \symplintiii\ is t-independent and, in particular, is equal to \symplinti. 
One way to prove the DH formula is now to choose the $U(1)$-invariant one-form 
$\lambda=-g(V,\cdot)$ where $g$ is a $U(1)$-invariant metric\foot{Any metric can
be made $U(1)$ invariant by averaging.}. 
Then $D\lambda = d\lambda - g(V,V)$, so the $t\to \infty$ limit will localize the integral 
on $g(V,V)=0$, i.e. at $V=0$ \refs{\AtBomm, \bgv}. 

We can put the integral \symplintiii\  in the context  of this chapter by translating to the
superspace $\widehat M$, so $\hat \omega=\ha \omega_{ij} \psi^i \psi^j$. 
We furthermore interpret $D$ as a BRST operator, acting on equivariant differential
forms: $D\rightarrow Q$.
Then $\lambda=\psi^i \pi_i$ is interpreted as a gauge fermion $\Psi$. 
Thus the integral: 
\eqn\symplintii{
\bigl({1\over \alpha}\bigr)^n \int_{\hat M}
\hat \mu~  e^{t D \lambda+ \alpha(\omega+H)} 
=\int \hat \mu~ e^{Q\Psi + \CO}}
is exactly of the form of a ``partition function" for a ``theory" with a $Q$-exact action as
in the previous section! 

We may immediately apply the $Q$-fixed-point theorem to prove the DH formula. 
The action of $Q$ is given by $x^i\to \psi^i$, $\psi^i\to V^i$ and hence the $Q$-fixed
points are $V^i(x)=0$.
In order to obtain the local factors we interpret the differential of $Q$ as an 
operator on the local tangent space to $\hat M$. The result is the famous 
DH formula expressing $\bigl({1\over \alpha}\bigr)^n \int_M e^{\alpha(\omega+H)}$
as a sum over the critical points of $H$. 

For further discussion of interpretations of the DH formula, localization and relations 
to supersymmetric/BV geometry see \refs{\AtBomm,\bgv, \Witdgtr,\jeffkir,\KaNi, \nersessian, \niemi}
and references therein. 

\subsubsec{Nonabelian Localization}
\subsubseclab\sssNL

The localization proof of the Duistermaat-Heckman formula has 
a broad generalization \refs{\Witdgtr,\jeffkir}. 
Suppose $X$ is a manifold (not necessarily symplectic) with 
an action of a group $G$.
Consider an equivariant differential form $\alpha\in \Omega_G(X)$
(recall the definition in sec. \sssCartan.)
Nonabelian localization simplifies integrals of the form
\eqn\inteqfrm{
\int_X \alpha \in S( \lieg\dual~ ). }

The integral \inteqfrm\ is a polynomial function on $\lieg$.
One would like to integrate this over $\lieg$.
Therefore a convergence factor is required:
\eqn\equiint{
\oint_{\epsilon,X} \alpha \equiv {1\over \vol G} 
\int_{\lieg} d \phi \int_X \alpha e^{-{\epsilon\over 2} (\phi,\phi)}}
Here $( \cdot , \cdot )$ is an invariant  metric on  $\lieg$, and $d \phi$ is the Euclidean
measure on $\lieg$.
This integration was introduced in \Witdgtr\ and is called integration of equivariant forms.
It will be useful to us in chapter \sTTwithLS. 

\equiint\ satisfies the basic integration by parts property: 
\eqn\ibpiii{
\oint_{\epsilon,M} \alpha~ D \beta =0}
if $\alpha$ is equivariantly closed. 
Therefore  by standard arguments: 
\eqn\nadh{
\oint_{\epsilon,M} \alpha~ e^{t D\lambda}}
is independent of $t$.
Once again we introduce superspace notation and consider  
\eqn\nadhi{
\int d \phi~ \hat \mu~  \hat \alpha~
e^{t D \lambda  - \epsilon \ha (\phi,\phi)}}
We shall view this as a partition function in a cohomological field theory with 
$Q\longrightarrow D$ and gauge fermion $\Psi= \lambda$.
The theory has been perturbed by the BRST invariant observable $(\phi,\phi)$.
Localization follows in the standard way by comparing the $t\to 0$ and $t\to \infty$ 
limits.
Choosing a gauge fermion $\lambda= \psi^i \pi_i(x)$ one can do the Gaussian
$\phi$-integral resulting in a suppression of the integrand by 
\eqn\lclizii{
\exp\biggl[ - \ha t^2 \sum_a (V_a^i \pi_i)^2 \biggr]
}
where $V_a^i\p_i$ are the vector fields corresponding 
to the action of generators of $\lieg$. It follows from 
\lclizii\ that the integral localizes at
$$
\lambda(V_a) = V_a^i \pi_i =0
$$
This is the nonabelian localization theorem. 

\subsubsec{Application: Weak-coupling limit of \ymt}
\subsubseclab\sssWkcpl

One application of the nonabelian 
localization theorem is a formula for the 
partition function of \ymt\ \Witdgtr
\foot{An alternative approach to this result  is discussed
in \BlThlgt.}. 
The key observation is that we can write 
the \ymt\ partition function as
\eqn\ymtss{\eqalign{
Z_{\rm physical}
&= {1 \over \vol~ \CG} \int_{\liebg \times \widehat \CA } d \phi \hat \mu_\CA~
\exp\Biggl\{
- \biggl[ {i \over 4 \pi^2} \int_\Sigma \Tr~ ( \phi F - \half \psi \wedge \psi )
\biggr]\cr
&\qquad\qquad\qquad\qquad\qquad\qquad\qquad
-\biggl [ \epsilon \int_\Sigma \mu~ {1\over 8 \pi^2} \Tr~ \phi^2 ( P ) \biggr] \Biggr\}\cr}}
where $\epsilon$ is related to the gauge coupling of chapter 
\sFYMTtoSCA\  by 
\eqn\rltcplings{
e^2= 2 \pi^2 \epsilon}
and $\widehat \mu_\CA=dA d \psi$ is the usual superspace measure.
\ymtss\ coincides with the partition function of the theory because the integral over $\psi$ is just such as
to give the symplectic volume element on $\CA$:
\eqn\sympvlele{
\int_{\hat \CA } \widehat \mu_\CA~ \exp~ [ {i\over 4 \pi^2} 
\int_\Sigma \Tr~ ( \half \psi\wedge \psi)]~
\leftrightarrow  ~ {\omega^n\over n!}}
The remaining action then coincides with \teek. 

Note that  \ymtss\ is a special case of integration of equivariant
differential forms for $\O_\CG(\CA)$.
Thus we can apply the nonabelian localization theorem.
To do this we must introduce a $Q$-exact action from a $\CG$-invariant one-form $\lambda$. 
In \Witdgtr\  the choice 
\eqn\ymtwgf{
\lambda=t \int \Tr~ \psi \ast D f =t \int \mu\Tr~ \psi^\alpha D_\alpha f
}
is made where $f$ is defined in \mapfrm. 
Localization to $\lambda(V)=0$ is then identical to localization to the solutions of the
classical Yang-Mills equations.
These consist of both flat and nonflat connections.
For gauge group $SU(N)$ the 
latter have a field strength which can be put in the form
$F= 2 \pi \omega \diagonal~ \{ n_1, n_2, \dots , n_N \} $, where $\sum n_j =0$ and
$\omega$ generates $H^2(\ST,\IZ)$.
The action is $I ={1\over \epsilon a} \sum n_i^2$.
Thus nonabelian localization implies the 
important result that the $e^2\to 0$ limit of $Z$ is
of the form
\eqn\wkcpl{
Z(e^2 a, \ST, SU(N))
{\buildrel e^2\to 0 \over \longrightarrow}
 \int_{\CM} \rho + \CO(e^{-c/(\epsilon a)} )}
where $\CM$ is the moduli space of flat connections discussed in sections \ssYMTFT\
and  \fltconn. 
We will discuss $\rho$, which turns out to have {\it polynomial } dependence on $\epsilon$,
further in section \ssRelphysym. 

\newsec{Supersymmetric Quantum Mechanics}
\seclab\sSQM

The  subject of topological field theory may be traced back to Witten's fundamental work
on dynamical supersymmetry breaking \refs{\dsb, \trminus}.
This work led naturally to a reformulation of DeRham theory and Morse theory in terms
of supersymmetric  quantum mechanics \ssymrs.
In turn Witten's formulation of Morse theory led to Floer's development of 
the homology theory that bears his name. 
Finally Floer homology of three-manifolds directly led to the formulation of Donaldson theory 
\refs{\atiythrfr, \donaldson}. 

In this section we will present one of the simplest 
topological theories: Supersymmetric
Quantum Mechanics (SQM)\refs{\dsb, \trminus,\AGIndex}.
Supersymmetric quantum mechanics is discussed as an example of a topological field 
theory in \refs{\bbrt}.
The relation to  the Mathai-Quillen formalism was pointed out  in \refs{\blauthom, \Bl}. 

\subsec{Action and Supersymmetry}
\subseclab\ssRTZO

Let $X$ be a Riemannian manifold
 with  metric $G_{\mu\nu}$. 
The degrees of freedom of supersymmetric quantum mechanics on $X$, $SQM(X)$, 
consist of a coordinate $\phi^\mu(t)$ on $X$ and fermionic coordinates
$\Psi^\mu(t)$ and $\bar \Psi^\mu(t)$, which together may be thought of as components of 
a superfield: 
\eqn\sclrspfld{
\Phi^\mu = \phi^\mu + \bar \theta \Psi^\mu + \theta\bar \Psi^\mu  + \bar\theta \theta F^\mu}
Here $F^\mu$ is an auxiliary field. 

The action of $SQM(X)$ is given by:
\eqn\sqmact{
I_{SQM} = \int d t \Bigl ( G_{\mu\nu} s^\mu s^\nu - i \bar \Psi_\mu D_t \Psi^\mu
- {1 \over 4} R^\mu{}_{\nu\rho\sigma} 
\bar \Psi_\mu G^{\nu \nu^\prime} \bar \Psi_{\nu^\prime} \Psi^\rho \Psi^\sigma \Bigr )}
where
\eqn\sqmdf{\eqalign{
s^\mu & = \dot \phi^\mu + G^{\mu \nu}\partial_\nu W\cr
D_t \Psi^\mu &=\nabla_t \Psi^\mu
+ G^{\mu\nu}  (\nabla_\nu \nabla_\lambda W ) \Psi^\lambda\cr
\nabla_t \Psi^\mu & = {d\over dt} \Psi^\mu
+ \Gamma^\mu_{\nu \rho} \dot \phi^\nu \Psi^\rho\cr}}
and $W$ is a {\it real}-valued function on $X$.
Note the close resemblance to the MQ form of  the previous section.
Below we will make this correspondence exact. 

The theory is, of course, supersymmetric and has a standard superspace
construction which we will sketch in \sssDRSQM.
If we make the field redefinition
\eqn\NuF{
\bar F_\mu = 2 i G_{\mu\nu} ( F^\nu + \dot\phi^\nu )
+ \partial_\lambda G_{\mu\nu} \Psi^\lambda\bar\Psi^\nu}
the usual supersymmetry transformations take the simple form:
\eqn\intrdque{\eqalign{
{\bf Q} \phi^\mu =& \Psi^\mu\qquad\qquad\qquad\qquad\qquad
{\bf Q} \Psi^\mu = 0\cr 
{\bf Q} \bar \Psi_\mu =& \bar F_\mu \qquad\qquad\qquad\qquad\qquad
{\bf Q} \bar F_\mu = 0. \cr}}
Evidently ${\bf Q}^2=0$. 

As emphasized in  \bbrt,  SQM  provides a simple example of a topological field theory. 
The nilpotent fermionic symmetry ${\bf Q}$ can  be interpreted as a BRST operator.
(The proper  physical interpretation of the BRST cohomology will appear 
below.)
Moreover, \sqmact , is derived from a ${\bf Q}$-exact action:
\eqn\qxtact{
I_{SQM} = \int dt~ \Biggl \{ {\bf Q}, \bar \Psi_\mu \left( i s^\mu
+ {1 \over 4} G^{\lambda \mu} \bar \Psi_\kappa \Gamma^\kappa_{\lambda\nu} \Psi^\nu
+ {1 \over 4} G^{\mu \nu} \bar F_\nu\right)\Biggr \}}
After integrating out $\bar F_\mu$, to get
\eqn\Feom{
\bar F_\mu
= -2 ( i G_{\mu\nu} s^\nu - \half \Gamma^\nu_{\lambda\mu} \Psi^\lambda \bar\Psi_\nu )}
we recover \sqmact. 

Topological invariance implies that  the partition function 
\eqn\sqmpfi{
Z_{SQM} = \int e^{-I_{SQM}}}
is a topological invariant, i.e. independent of any einbein one puts on the one-dimensional
space, and, if $X$ is compact, is independent of $W$.
Since the action is ${\bf Q}$-exact, the reasoning of  \sssQFPT\ shows that $Z_{SQM}$
localizes on the set of ${\bf Q}$-fixed points. 
From \Feom\ we have
$$
\bar F_\mu = -2 i G_{\mu\nu} \dot\phi^\nu
+ \Gamma^\nu{}_{\lambda\mu} \Psi^\lambda \bar \Psi_\nu -  2 i \partial_\mu W . 
$$
It is apparent from \intrdque\ that the ${\bf Q}$-fixed points correspond
to $\Psi^\mu =  \bar F^\mu = 0$.
This is the space of instantons given by
\eqn\instans{
\CM = \cases{
\{ \phi^\mu \in {\rm Map}~ ( S^1, X )~ \vert~ \phi^\mu(t)={\rm constant} \} \cong X &
if $W=0$\cr
& \cr
\{ P \in X~ \vert~  dW(P)=0 \} &
if $W \not=0$\cr}}

\vskip0.1truein\noindent
{\bf Remark}:
The above formulae might appear unmotivated. 
They arise naturally from dimensional reduction $1+1\to 0+1$ of the 2D nonlinear sigma model 
reviewed in section \sssDRSQM\  below. 

\subsec{MQ interpretation of SQM}

Many aspects of the general discussion of chapter \sINandTIR\ are 
explicitly realized in SQM, which may be viewed as the application
of the MQ formalism to the infinite-dimensional geometry of the space
of unbased (differentiable) loops in  $X$:
\eqn\loopsp{
LX\equiv \MAP ( S^1, X )\qquad .}
Heuristically, we may think of coordinates on this manifold as
$\phi^\mu(t)$, $\mu=1,\dots \dim X$,  and $t\in S^1$. 
The tangent space, corresponding to the spaces of small deformations 
$\delta \phi^\mu(t)$ of a  loop, is identified with 
$$
T_\phi(LX) \cong \Gamma(\phi^*(TX) ). 
$$ 
We will apply the MQ formalism to the bundle $E=T~ LX$.
This bundle has a natural metric
\eqn\lpmet{
(\gamma_1, \gamma_2)_{\phi}
= \oint_0^1 dt~ G_{\mu \nu}(\phi(t)) \gamma_1^\mu(t) 
\gamma_2^\nu(t)\qquad . 
}
In local coordinates, the metric is $G_{\mu \nu}(\phi(t_1)) \delta(t_1-t_2)$.
The Riemannian geometry of $LX$ is therefore almost identical to that 
of $X$, with some extra delta functions entering expressions when written
in terms of local coordinates.
For example, the  Levi-Civita connection, $\nabla$, on $LX$ is just 
the pullback connection from $X$.
It acts on a vector field 
$$
V= \oint dt~ V^\mu(\phi,t) {\partial \over \partial \phi^\mu(t)}
$$
to produce 
$$
\nabla V = \oint dt_1 dt_2 \Biggl[ {\delta V^\mu (\phi,t_1)\over \delta \phi^\nu(t_2)}
+ \Gamma^\mu_{\nu \lambda}(\phi(t_2 ))V^\lambda ( \phi, t_1 )  \delta(t_1 -t_2) \Biggr ]
{\partial \over \partial \phi^\mu(t_1)} \otimes \widetilde d \phi^\nu(t_2) 
$$
where $\{ {\partial \over {\partial \phi^\mu ( t )}} \}$ and $\{ \widetilde d \phi^\mu ( t ) \}$
are to be viewed as bases of $T~ LX$ and $T^\ast LX$, respectively.  
In local coordinates the curvature is just the curvature of  $X$ multiplied by
$\delta$-functions. 

Having specified our connection we choose a section of $E$ to be: 
\eqn\sqmsect{
[ s ( \phi ) ]^\mu(t) = \dot \phi^\mu ( t ) + G^{\mu \nu}\partial_\nu W ( t )}
An easy calculation, using the basic tautology \bsctaut\ 
to replace
$\widetilde d \phi^\mu(t)\to \Psi^\mu(t)$ shows that 
\eqn\nbess{
\nabla s = \oint dt \Biggl[ \nabla_t \Psi^\mu + 
(\nabla^\mu\nabla_\lambda W) \Psi^\lambda\Biggr]
{\partial \over \partial \phi^\mu(t)}}
Identifying coordinates for the dual bundle $\Pi E^*$ as $\rho\to \bar \Psi_\mu(t)$, we see that
$\langle \rho, \nabla s\rangle $ corresponds to the fermion bilinear terms in \sqmact.
The curvature produces the four-fermi term. 
Thus the SQM action \sqmact\  coincides exactly with the MQ formula. 
Moreover, the expression \qxtact\ for the action coincides with the 
gauge fermion \secgf.
The differential is 
$$
{\bf Q} = \oint dt~ \left ( \Psi^\mu(t) {\partial \over \partial \phi^\mu(t)}
  + \bar F_\mu(t)  {\partial \over \partial \bar \Psi_\mu(t)} \right )
$$

We now use the results of chapter \sINandTIR\ to compute the partition function, $Z_{SQM}$. 
In the fields-equations-symmetries paradigm the equations are the instanton equations 
$s^\mu(t)=0$.
The localization to the zeroes of  $s$ calculates the Euler character of an appropriate 
bundle over $\CZ(s)$.
For $W=0$, $X$ compact  we have a nongeneric section: $\CZ(s) = X$, the 
constant loops.
Acting on $T^\ast_\phi LX$ at a constant loop $\phi$ the operator $\nabla s$ is
identified with ${d\over dt}$ acting on periodic loops
in $T_\phi X$. 
This operator has index 
$$
\ind~ {d\over dt} =0 \qquad \ker~ (\nabla s) \cong \cok~ (\nabla s) \cong TX
$$
so by \locii\ we have 
$$
Z_{{\rm SQM}}= \int_X\chi ( TX )\qquad .
$$ 
If $W$ is a generic function, we have a generic section and the localization to the zeroes
of $s$ is the localization to constant loops mapping to critical points of $W$.
The localization to $\CZ(s)$ becomes the signed sum: 
$$
\sum_{\{P\in X~ \vert~  dW(P)=0\}}
{\rm sign} \biggl[\det {\partial^2 W\over \partial \phi^\mu \partial \phi^\nu} ( P ) \biggr]
$$
Note that this is just the sum of indices for the vector field $\nabla W$. 
{}From either point of view we see that the partition function is given by
\eqn\sqmpf{
\mathboxit{
Z_{SQM}= \chi(X)}}
We will now verify \sqmpf\  from a more physical point of view using the canonical
quantization of $SQM(X)$. 

\subsec{Canonical Quantization}

To describe the Hilbert space we represent the canonical commutation relations: 
\eqn\clffd{\eqalign{
[\phi^\mu,\pi_\nu]&=i \delta^\mu_\nu\cr
\{ \bar \Psi_\mu, \Psi^\nu\}  &= \delta_\mu^\nu\cr}}
where $\pi_\mu = 2 G_{\mu\nu} \dot\phi^\mu + 2 \partial_\mu W$ is the canonical
momentum to $\phi^\mu$.
The first line is represented in the standard way in terms of wavefunctions in $L^2(X)$.
To represent  the second line, which defines a Clifford algebra, we must choose a
polarization, that is, a fermionic vacuum. 
We can make the choice that $\bar \Psi_\mu$ are annihilation operators:
$$
\bar \Psi_\mu \mid 0\rangle =0
$$
Then wavefunctions are functions $F(\phi,\Psi)$ of $\phi^\mu$ and 
$\Psi^\mu$, that is, functions on  the superspace $\hat X$. 
The Hilbert space of the theory is then identified,  by the  basic tautology of section
\sssSprsce\ above, with $L^2$ sections of the DeRham complex
\eqn\hilbsp{
\HOL_{SQM} = \Omega_{L^2}^\bullet ( X )}
From \intrdque\ and \clffd\ it is clear that the supersymmetry operator may be
written as follows:
\eqn\SSyOperator{\eqalign{
{\bf Q}
=& \oint d \sigma~ \bar F_\mu \Psi^\mu\cr
=& \oint d \sigma ( - 2 i G_{\mu\nu} \dot\phi^\nu - 2i \partial_\mu W ) \Psi^\mu\cr
=& -i \oint d\sigma~ \pi_\mu \Psi^\mu\cr}}
and similarly for ${\bf Q}^\ast$.
Note that the hermitean supersymmetry operators are given by
\eqn\dConj{\eqalign{
{\bf Q} =&i  e^W~ d~ e^{-W}\cr
{\bf Q}^\ast =& -i  e^{-W}~  d^\ast ~ e^W\cr}}
where $d$ and $d^\ast$ are the exterior derivative on $X$ and its adjoint.
(Note that on the wavefunctions $d = \Psi^\mu {\partial\over{\partial \phi^\mu}}$.)

Finally we note that these operators satsify
$$
{\bf Q}^2=({\bf Q}^*)^2=0\quad .
$$
The Hamiltonian is given by 
\eqn\susyham{
H= \{ {\bf Q}, {\bf Q}^*\}.}

\subsubsec{$\tr(-1)^F$}

The motivation for $\tr(-1)^F$ came from the search for a particle theory model of
dynamical supersymmetry breaking.
Witten introduced the index $\tr(-1)^F$ to give a sufficient condition for the presence
of supersymmetric ground states. 
A supersymmetric Hamiltonian satisfies \susyham\ and therefore  boson/fermion states
of nonzero energy are always paired.
On the other hand, the supersymmetric ground states, i.e., the zero energy states need
not be paired. 

Now, small perturbations of the theory may lift $E=0$ states, but bosonic and fermionic
states must always be lifted in pairs.
Thus, while the total number of supersymmetric ground states might be difficult to calculate,
the difference of the number of bosonic $E=0$ states minus the number of fermionic
$E=0$ states is invariant under many perturbations of the theory and might be more
accessible to calculation. 
Thus, for example, it will be a topological invariant of SQM on $X$,  if $X$ is compact.
For this reason Witten introduced the index: 
\eqn\wttndx{
\tr_{E=0}~ (-1)^F \equiv  \# [E=0 \  {\rm bosons}] - \# [E=0 \ {\rm fermions}].}

Let us examine the supersymmetric, $E=0$, states in SQM when $X$ is compact and
$W=0$\foot{Quantization with $W\not=0$ leads to Morse theory. See appendix 4.}.
These are annihilated by both  ${\bf Q}$ and ${\bf Q}^\ast$.
Since, by \susyops, these operators are related to $d$ and $d^\ast$ by conjugation with
an invertible operator,  the supersymmetric states are in 1-1 correspondence with the
harmonic forms on $X$.
By Hodge theory we identify the vector space  of $E=0$ states with the DeRham
cohomology  $H^\bullet (X)$\foot{When $X$ is noncompact there are infinitely many
harmonic forms and we need  $W$ to regulate the behavior at infinity to define a  good
Hodge theory.
See appendix 3 below. }.
Moreover, with our choice of vacuum above, which is  sensible for $X$ compact and
$W=0$,  form degree is identical to fermion number.
So $H^j(X)$ is the space of supersymmetric ground states of  fermion number $j$. 
The Witten index is therefore the Euler character of $X$: 
\eqn\trmnsone{\eqalign{
\tr_{E=0} (-1)^F
& = \sum_j (-1)^j \dim~ H^j ( X; R)\cr
&=\chi ( X ). \cr}}

\subsubsec{The partition function}

We are now in a position to confirm the result \sqmpf. 
Using the pairing of nonzero  energy states the index may be written as 
a trace over the entire Hilbert space: 
\eqn\betanz{\eqalign{
\tr_{E=0} (-1)^F
=& \tr_\cH(-1)^F e^{-\beta H}\cr
=& {\rm str}~ e^{-\beta H}\cr}}
where ${\rm Re}~ \beta > 0$ regulates the supertrace.
Standard manipulations\foot{See J. Fr\"ohlich's contribution to
this volume.} allow us to express the  partition function
${\rm str}~ e^{-\beta H}$  of a theory in terms of a path integral 
with a periodic Euclidean  time of period $\beta$. 
Thus the partition function of  SQM is just the Euler character of $X$: 
\eqn\sqmpf{
\int d \phi \int_{P.B.C.} d \Psi d \bar \Psi~ e^{-I} = \chi(X)}
Note that the partition function is special and topological 
because only the  $E=0$ states contribute to the trace. 
\vskip0.1truein\noindent
{\bf Remark}. The parameter $\beta$ may be identified with the 
parameter $t$ of the MQ formalism. 

\subsec{Appendix 1: The two-dimensional supersymmetric 
nonlinear sigma model}
\subseclab\ssTDSNSM

The theories of chapters \sSQM\ and \sTSM\ are 
closely related to the 2D supersymmetric nonlinear
sigma model. We collect here some basic formulae. 

\subsubsec{Superfields}

The $N=1,2$ supersymmetric sigma models (S$\sigma$M) in $d=2$
are conveniently obtained by dimensional reduction from the $N=1$
S$\sigma$M in $d=4$ \refs{\wittphases}.
The supersymmetry generators of the four dimensional theory are
given by\foot{We adopt the conventions of Wess and Bagger \refs{\WeBa}\ with
a Minkowskian metric, $\eta_{mn} = {\rm diag} ( -1, 1, 1, 1)$ and
with $\sigma^m_{\alpha\dot\alpha}$
$$\eqalign{
\sigma^0 ~=&~ \pmatrix{-1 & 0\cr 0 & -1\cr}\qquad\qquad\qquad
\sigma^1 ~=~ \pmatrix{0 & 1\cr 1 & 0\cr}\cr
\sigma^2 ~=&~ \pmatrix{0 & -i\cr i & 0\cr}\qquad\qquad\qquad
\sigma^3 ~=~ \pmatrix{1 & 0\cr 0 & -1\cr}\cr}
$$
Spinor indices are raised and lowered via
$$
\epsilon_{\alpha\beta} ~=~ \pmatrix{0 & -1\cr 1 & 0\cr}\qquad\qquad\qquad
\epsilon^{\alpha\beta} ~=~ \pmatrix{0 & 1\cr -1 & 0\cr}
$$
Moreover,
$$
\int d^2 \bar\theta d^2 \theta ~  \bar\theta^2 \theta^2 = 1.
$$}
\eqn\FourSSY{
Q_\alpha ~=~ \partial_\alpha
- i \sigma^m_{~\alpha\dot\alpha} \bar \theta^{\dot\alpha} \partial_m
\qquad\qquad
\bar Q_{\dot\alpha} ~=~ - \bar\partial_{\dot\alpha}
+ i \sigma^m_{~\alpha\dot\alpha} \theta^\alpha \partial_m}
where for $\alpha, \dot\alpha = +,-$, we have the two Weyl spinors $\theta^\alpha$
and $\bar \theta_{\dot\alpha}$; moreover,
$\partial_\alpha = {\partial\over{\partial\theta^\alpha}}$
and $\bar\partial_{\dot\alpha} = {\partial\over{\partial\bar\theta^{\dot\alpha}}}$.
The (future) worldsheet coordinates are given by $x^m$, $m = 0, \ldots, 3$.
The covariant superderivatives are
\eqn\FourSD{
D_\alpha ~=~ \partial_\alpha
+ i \sigma^m_{~\alpha\dot\alpha} \bar \theta^{\dot\alpha} \partial_m
\qquad\qquad
\bar D_{\dot\alpha} ~=~ - \bar\partial_{\dot\alpha}
- i \sigma^m_{~\alpha\dot\alpha} \theta^\alpha \partial_m. }

Superspace provides an expedient way to formulate supersymmetric field
theories.
In four dimensions the chiral superfields (satisfying $\bar D_{\dot\alpha} \Phi = 0$)
are given by
\eqn\FourCSF{
\Phi^\mu ( x, \theta, \bar\theta ) ~=~ \phi^\mu ( y ) + \theta^\alpha \psi^{~ \mu}_\alpha ( y )
+ \theta^\alpha \theta_\alpha F^\mu ( y ), }
where
$y^m ~=~ x^m + i \theta^\alpha \sigma_{~\alpha\dot\alpha}^m \bar\theta^{\dot\alpha}$.

\subsubsec{$N=2$, $d=2$ Supersymmetric 
Sigma Models: Field Content}
\subsubseclab\sssSSMi

We dimensionally reduce by dropping the dependencies on $x^1$ and $x^2$.
We also Wick rotate to Euclidean space, by setting $\tau = i x^0$ (and $\sigma = x^3$).
Then
\eqn\TwoTwoSSY{\eqalign{
Q_- ~=&~ \partial_- - \bar \theta^- \bar\partial\qquad\qquad\qquad
Q_+ ~=~ \partial_+ - \bar\theta^+ \partial\cr
\bar Q_- ~=&~ -\bar\partial_- + \theta^- \bar\partial\qquad\qquad\qquad
\bar Q_+ ~=~ -\bar\partial_+ + \theta^+ \partial\cr}}
and
\eqn\TwoTwoSD{\eqalign{
D_- ~=&~ \partial_- + \bar \theta^- \bar\partial\qquad\qquad\qquad
D_+ ~=~ \partial_+ + \bar\theta^+ \partial\cr
\bar D_- ~=&~ -\bar\partial_- - \theta^- \bar\partial\qquad\qquad\qquad
\bar D_+ ~=~ -\bar\partial_+ - \theta^+ \partial .\cr}}
where we have defined local complex coordinates $z = \half ( \tau + i \sigma )$ and $\bar z = \half ( \tau - i \sigma )$;
we have also used the obvious shorthand $\partial = {\partial\over{\partial z}}$ and
$\bar\partial = {\partial\over{\partial\bar z}}$.

The dimensional reduction of the superfields \FourCSF\ leads us to consider the
chiral $N=2$ superfields:
\eqn\TwoTwoSF{
\Phi ( z, \bar z, \theta, \bar\theta )
~=~ \phi ( w_1, w_2 ) + \theta^+ \psi_+ ( w_1, w_2 ) + \theta^- \psi_- ( w_1, w_2 )
+ \theta^- \theta^+ F ( w_1, w_2 )}
where $w_1 = z + \theta^+ \bar\theta^+$ and $w_2 = \bar z + \theta^- \bar\theta^-$
satisfy the conditions: $\bar D_\pm w_{1,2} = 0$.
\foot{
For two dimensional theories, it is also possible to have twisted chiral superfields,
i.e. $\Phi$, such that $\bar D_+ \Phi = D_- \Phi = 0$.
}
Under $N=2$ supersymmetry transformations the component fields transform as follows:
\eqn\TwoTwoSSYT{\eqalign{
\delta \phi^\mu ~=&~ \epsilon^- \psi_-^{~ \mu} + \epsilon^+ \psi_+^{~ \mu}\cr
\delta \psi_+^{~ \mu} ~=&~ \bar\epsilon^+ \p \phi^\mu + \epsilon^- F^\mu\cr
\delta \psi_-^{~ \mu} ~=&~ \bar\epsilon^- \bar{\p} \phi^\mu - \epsilon^+ F^\mu\cr
\delta F^\mu ~=&~ \bar \epsilon^- \bar{\p} \psi_+^{~ \mu}
                                - \bar \epsilon^+ \p \psi_-^{~ \mu}\quad . \cr}}

\subsubsec{$N=1$, $d=2$ Supersymmetric Sigma Models}
\subsubseclab\sssSSMii

We are also interested in 
constructing the $N=1$ S$\sigma$M in $d=2$.
This is easily obtained from 
the $N=2$ S$\sigma$M in $d=2$, by imposing the
condition that $\bar\theta$ is the 
charge conjugate of $\theta$ rather than an independent
spinor\foot{In the conventions that we have adopted, the charge conjugation matrix
is given by
$$
C = \pmatrix{  & -i\cr
                       i &    \cr}
$$
so that $\psi^c_{\dot\alpha} = C_{\alpha\dot\alpha} \psi^{\ast~\alpha}$ and the
self-conjugacy condition is
$\pmatrix{\bar\psi^-\cr \bar\psi^+\cr} = i \pmatrix{\psi^-\cr \psi^+\cr}$.}.
We set
\eqn\SelfConj{
\theta^+ ~=~ -i \bar\theta^+ ~=~ \theta\qquad\qquad\qquad\qquad
\theta^- ~=~ -i \bar\theta^- ~=~ \bar\theta}
and
\eqn\TwoOneSSD{\eqalign{
\bar Q ~=&~ \bar\partial_{\bar\theta} - i \bar\theta \bar\partial\qquad\qquad\qquad\qquad
Q ~=~ \partial_\theta - i \theta\partial\cr
\bar D ~=&~ \bar\partial_{\bar\theta} + i \bar\theta \bar\partial\qquad\qquad\qquad\qquad
D ~=~ \partial_\theta + i \theta\partial . \cr}}
Note that $Q_- = i \bar Q_- = Q$ and $Q_+ = i \bar Q_+ = \bar Q$ and similarly for
$\bar D_\pm$.
The $N=1$ superfields are given by
\eqn\TwoOneSF{
\Phi^\mu ( z, \bar z, \theta, \bar \theta )
= \phi^\mu ( z, \bar z ) + \theta \bar\psi^\mu ( z, \bar z ) + \bar\theta \psi^\mu ( z, \bar z )
+ \bar\theta \theta F^\mu ( z, \bar z ). }
The component fields transform as follows under the supersymmetry
\eqn\TwoOneST{\eqalign{
\delta \phi^\mu ~=&~ \epsilon \bar\psi^\mu + \bar\epsilon \psi^\mu\cr
\delta \bar\psi^\mu ~=&~ i \epsilon \partial \phi^\mu - \bar\epsilon F^\mu\cr
\delta \psi^\mu ~=&~ i \bar\epsilon \bar\partial \phi^\mu + \epsilon F^\mu\cr
\delta F^\mu ~=&~ i \epsilon \partial \bar\psi^\mu - i \bar\epsilon \bar\partial \psi^\mu. \cr}}
The basic action of the $N=1$ S$\sigma$M in $d=2$ is
\eqn\TwoOneAct{
I = \int d^2 z~ d \bar\theta d \theta~ G_{\mu\nu} ( \Phi ) \bar D \Phi^\mu D \Phi^\nu, }
to which we may add a potential term
\eqn\PotTerm{
\Delta I = 2 \int d^2 z~ d \bar\theta d \theta~ W ( \Phi ), }
where the hermiticity of the action requires that $W$ be a real-valued function of $\Phi$.
After integrating out
$F^\mu = \Gamma^\mu_{\lambda\nu} \psi^\lambda \bar\psi^\nu + \partial^\mu W$,
we get the action in components:
\eqn\TwoOneCompAct{\eqalign{
I + \Delta I&\cr
=& \int d^2 z~ \left \{ G_{\mu\nu} ( \partial \phi^\mu \bar\partial \phi^\nu
-i \bar\psi^\mu D_{\bar z}^{{\scriptscriptstyle ( 0 )}} \bar\psi^\nu
- i \psi^\mu D_z^{{\scriptscriptstyle ( 0 )}} \psi^\nu )
+ \half R_{\kappa\lambda\mu\nu} \psi^\kappa \psi^\lambda \bar\psi^\mu \bar\psi^\nu
\right.\cr
&\qquad\qquad\qquad\qquad\left.
+ 2 ( \nabla_\mu \nabla_\nu W ) \psi^\mu \bar\psi^\nu
+ G^{\mu\nu} \partial_\mu W \partial_\nu W \right \}\cr}}
where
$( D_z^{{\scriptscriptstyle ( 0 )}} \psi )^\mu = \partial \psi^{~ \mu}
+ \Gamma^\mu_{\lambda\nu} \partial \phi^\lambda \psi^{~ \nu}$.
Later we will need the coupling to a curved worldsheet with metric $h_{\alpha \beta}$. 
The changes to the action are:
\eqn\CurvedWS{\eqalign{
d^2 z
~\longrightarrow&~ d^2 z \sqrt{h}\cr
D_{\bar z}^{{\scriptscriptstyle ( 0 )}} \bar\psi^{~ \mu}
~\longrightarrow&~ \bar \partial \bar\psi^{~ \mu} - { i \over 2} \omega_{\bar z} \bar\psi^{~ \mu}
+ \Gamma^\mu_{\lambda\nu} \bar\partial \phi^\lambda \bar\psi^{~ \nu}\cr
D_z^{{\scriptscriptstyle ( 0 )}} \psi^{~ \mu}
~\longrightarrow&~ \partial \psi^{~ \mu} + { i \over 2} \omega_z \psi^{~ \mu}
+ \Gamma^\mu_{\lambda\nu} \partial \phi^\lambda \psi^{~ \nu}\cr}}
where $\omega$ is the spin connection.

\subsubsec{Canonical Quantization} 
\subsubseclab\sssCQ

When discussing canonical quantization  we take $\sigma$ periodic with period $L$. 
Canonical quantization leads to 
\eqn\cnqnt{\eqalign{
[  \phi^\mu ( \sigma, \tau ) , \dot\phi^\nu ( \widetilde \sigma, \tau ) ]
=&~ i G^{\mu \nu}  ( \sigma, \tau ) \delta ( \sigma - \widetilde \sigma )\cr
\{ \psi^\mu ( \sigma, \tau ) , \psi^\nu ( \widetilde \sigma, \tau ) \}
=&~ G^{\mu \nu} ( \sigma, \tau ) \delta ( \sigma - \widetilde \sigma )\cr
\{ \bar\psi^\mu ( \sigma, \tau ) , \bar\psi^\nu ( \widetilde \sigma, \tau ) \}
=&~ G^{\mu \nu} ( \sigma, \tau ) \delta ( \sigma - \widetilde \sigma )\cr
\{ \psi^\mu ( \sigma, \tau ), \bar\psi^\nu  ( \widetilde \sigma, \tau )\}
=&~ 0\cr}}
The  hermitean supersymmetry operators are given by 
\eqn\susyops{\eqalign{
Q& = \oint d\sigma \left ( i G_{\mu \nu} \psi^\mu \dot \phi^\nu
           - \psi^\mu \Gamma_{\mu\lambda\nu} \psi^\lambda \bar\psi^\nu
          + \psi^\mu \partial_\mu W \right )\cr
\bar Q&  = \oint d\sigma \left ( i G_{\mu \nu} \bar \psi^\mu  \dot \phi^\nu
          + \bar\psi^\mu \Gamma_{\mu\lambda\nu} \psi^\lambda \bar\psi^\nu
          - \bar\psi^\mu \partial_\mu W \right )\cr}}

\subsubsec{Dimensional Reduction}
\subsubseclab\sssDRSQM

We further dimensionally reduce to $d=1$ SQM by dropping all $\sigma$ dependence
and studying only the zero modes of the fields.
We introduce
$$\eqalign{
\Psi^\mu =& \psi^\mu + i \bar\psi^\mu\cr
\bar\Psi_\mu =& G_{\mu\nu} ( \psi^\nu - i \bar\psi^\nu )\cr
{\bf Q} =& Q + i \bar Q .\cr}
$$
Under ${\bf Q}$ the fields transform as in \intrdque.
Dimensionally reducing \TwoOneCompAct\  we recover the formulae of section \ssRTZO. 

%
%
%
%
%

\subsec{Appendix 2: Index theorems for some elliptic operators}
\subseclab\ssITforSEO

Although somewhat of a digression, SQM yields an elementary proof
of the index theorems \atising\ 
for  various elliptic operators. 
This very beautiful application of SQM was worked out by 
L. Alvarez-Gaum\'e and by D. Friedan and P. Windey \refs{\AGIndex,\FrieWind}.

We have already seen that SQM may be used to compute the index of the
DeRham complex
$$
\ind~ d = \chi ( M )
= \int d \phi \int_{P.B.C.} d \Psi d \bar \Psi~ e^{-I_{SQM} }
$$
The indices of all the other classical elliptic complexes may be computed similarly.
As we will have occasion to use several index theorems below, we will briefly
sketch the main idea.
We refer the interested reader to the literature \refs{\AGIndex,\FrieWind} for
details.

Given an elliptic complex $( E, D )$ one seeks to construct a SQM model whose
supercharge corresponds to $D$.
Then
$$\matrix{
\ind~ ( {\cal H}, Q ) & \eqdef & \sum_F  ( -1 )^F \dim~ \ker~ H_F\cr
 & & \cr
               \parallel                  &    &           \parallel\cr
 & & \cr
\ind~ ( E, D ) & \eqdef& \sum_p ( -1 )^p \dim~ \ker~ \Delta_p ( E, D )\cr}
$$
where $H_F = Q_F Q^\ast_{F+1} +  Q^\ast_F Q_{F-1}$ is the Hamiltonian
on the fermion number $F$ sector of the Hilbert space and $\Delta_p$ is
the Laplacian on $p$-forms in the complex $( E, D )$.

Once one has constructed an appropriate SQM model, $\ind~ ( E, D ) $
may be computed by evaluating the partition function
$$
\ind~ ( E, D ) = \int_{B.C.}  e^{-I_{SQM}}
$$
where {\it B.C.} is schematic for an appropriate set of boundary conditions.

\subsubsec{The Index of the Dirac Complex}

The index of the Dirac operator is obtained by reducing the supersymmetry
to $N = \half$, by imposing the condition
$$
\Psi_+^{~ \mu} = \Psi_-^{~ \mu} = 2^{-1/2} \widetilde \Psi^\mu
$$
{}From the Bianchi identity it follows that the curvature term in $I$ disappears,
leaving
$$
I_{SQM}^{N=\half} = \int dt~ \left \{
-\half G_{\mu\nu} \dot \phi^\mu \dot\phi^\nu
+ \half G_{\mu\nu} \widetilde \Psi^\mu D_t \widetilde \Psi^\nu \right\}
$$
There is one remaining supersymmetry and 
the supercharge $Q$ for this model is related to the Dirac operator, $i \Dsl$, on the
target space.
Evaluating the path integral of this theory (e.g. by expanding about the
constant bosonic background) one finds \refs{\AGIndex}
\eqn\DirInd{\mathboxit{
\ind~ i\Dsl = \int_M \hat A ( M )}}
where $\hat A ( M )$ is the {\it A-roof genus}
\eqn\ARoof{
\hat A ( M )
= \prod_{a=1}^{\half \dim M} {{\half x_a}\over{\sinh \half x_a}}
= 1 - {1\over24} p_1 ( M ) + {1\over5760} ( 7 p_1^{~ 2} - 4 p_2 ) ( M ) + \cdots}
Here $x_a$ label the eigenvalues of the skew-diagonalized form
${1\over{2\pi}}R_{ab}$ and $p_i ( M )$ are the Pontryagin classes.

\subsubsec{The Index of the Twisted Dirac Complex}

Taking the tensor product of the Dirac complex with a vector
bundle, $E$, one obtains the twisted Dirac complex, whose index may be
computed from the following SQM action\refs{\AGIndex}
$$\eqalign{
I =& \int dt~
\left \{ G_{\mu\nu} \left [
\half  \dot\phi^\mu \dot\phi^\nu
- {i\over2}  \widetilde\Psi^\mu D_t \widetilde\Psi^\nu
\right ]
+ i C_i^\ast \left ( \dot C_i
- A_\mu^a ( \phi ) \dot\phi^\mu T^a_{~ij} C_j \right )\right.\cr
&\qquad\qquad\qquad\qquad\left.
- {i\over2} \widetilde\Psi^\mu \widetilde\Psi^\nu F^a_{\mu\nu}~
C_i^\ast T^a_{~ij} C_j \right \}\cr}
$$
where $A_\mu$ is the connection on the associated bundle (viewed here
as an external gauge field) and the $C_i$ form a Clifford algebra in the representation
of $G$ generated by $T^a_{~ij}$.
Evaluating the partition function, one finds
\eqn\TwisDirInd{
\mathboxit{
\ind~ i \Dsl_A = \int_M {\rm ch} ( F ) \wedge \hat A (  M )}}
where ${\rm ch} ( F )$ is the Chern character
\eqn\ChernChar{
{\rm ch} ( F ) = \tr~ \exp~ {i\over{2 \pi}} F
= \rank~ E + c_1 ( F ) + \half ( c_1^{~ 2} - 2 c_2 ) ( F ) + \cdots}
and the $c_i ( F )$ are the Chern classes of $F$.

\subsubsec{The Index of the Dolbeault Complex}

If the target space admits a K\"ahler structure, then the $\sigma$-model
actually has two independent supersymmetries.
The exterior algebra refines to
\eqn\ExtAlg{
\Lambda^\bullet ( M ) = \bigoplus_{p,q=0}^{\dim_{\IC} M} \Lambda^{p,q} ( M ). }
If we restrict the computation of the SQM partition function to the
anti-holomorphic, $( 0, p )$, part of the Hilbert space we find 
\eqn\DolbInd{\mathboxit{
\ind~ \bar \partial = \int_M {\rm td}~ ( M )}}
where ${\rm td} ( M )$ is the {\it Todd class}
\eqn\ToddClass{
{\rm td} ( M ) = \prod_{a=1}^{\half \dim M} {{x_a}\over{1 - e^{-x_a}}}
= 1 + \half c_1 ( M ) + {1\over12} ( c_1^{~ 2} + c_2 ) ( M ) + \cdots}
As with the Dirac operator we may take the tensor product of the
Dolbeault complex with a vector bundle $E$ and compute the index
of the twisted complex:
\eqn\TwisDolbInd{
\mathboxit{
\ind~ \bar \partial_E = \int_M {\rm ch} ( E ) \wedge {\rm td}~ ( M )}}

\subsec{Appendix 3: Noncompact $X$ and runaway vacua}
\subseclab\ssRunvac

We can illustrate one example in which formally 
$Q$-exact perturbations of a topological theory 
in fact change the  physical answers. We will take $X=\IR$ and 
$W$= a polynomial.  The supersymmetric groundstates 
must be annihilated by ${\bf Q}$ and ${\bf Q}^*$.
{}From \dConj\ it is apparent that they must be a linear combination of
\eqn\ncptxi{e^{W(\phi)} \qquad e^{-W(\phi)} d \phi}
which have fermion number ($F =$ form degree) $0$ and $1$, respectively.
These states have norms: 
\eqn\nrmsts{
\eqalign{
\parallel e^{W(\phi)}  \parallel^2 & = \int_{\IR} d \phi ~ e^{2 W}\cr
\parallel e^{-W(\phi)}  d \phi \parallel^2 & = \int_{\IR} d \phi ~ e^{-2 W}\cr}}
which must be finite in order for the state to be admissible. 
It follows that $\tr(-1)^F$ is given by
\eqn\cmpt{
\tr~ ( - 1 )^F ~=~ \cases{
-1 & for $W= a \phi^{2n} + \cdots\qquad$ and $a>0$\cr
& \cr
0 & for $W = a \phi^{2n+1} + \cdots$\cr
& \cr
+1 & for $W= a \phi^{2n} + \cdots\qquad$ and $a<0$\cr}}

To see how a BRST exact perturbation can change a 
physical answer we first reinterpret the system. 
Represent the Clifford algebra 
by $2\times 2$ matrices so that
\eqn\elctrns{\eqalign{
{\bf Q}&= \pmatrix{0&0\cr
1&0\cr} \biggl({d\over d \phi} - W'\biggr)\cr
{\bf Q}^*&= -\pmatrix{0&1\cr
0&0\cr} \biggl({d\over d \phi} + W'\biggr)\cr}}
Computing the Hamiltonian, $H = \{ Q, Q^\ast \}$, one sees that 
we are discussing a theory of electrons in a 
magnetic field with potential $V=(W')^2$. 
\foot{This representation affords a connection 
to the theory of Landau levels as discussed in 
J. Fr\"ohlich's  contribution to this volume.}
If $V$ has an even number of zeroes SUSY is 
broken (nonperturbatively), otherwise it is not. 

Suppose now that $W_0= \phi^{2n}+ \cdots$
and that we add a ``small" perturbation 
$\Delta W = \epsilon \phi^{2n+1} $. 
By  \qxtact\  this leads to a BRST exact change in the 
action. On the other hand, by the above exact 
results we see that it changes the value of $Z$! 
The reason for the failure of BRST decoupling is 
that $\Delta W$ brings in new fixed points 
``from infinity." That is, as $\epsilon\to 0 $ the 
extra zero of $V$ runs out to infinity. 

\ifig\VacuumStates{Vacuum state receding to infinity
for $\epsilon \to 0$.}
{\epsfxsize3.0in\epsfbox{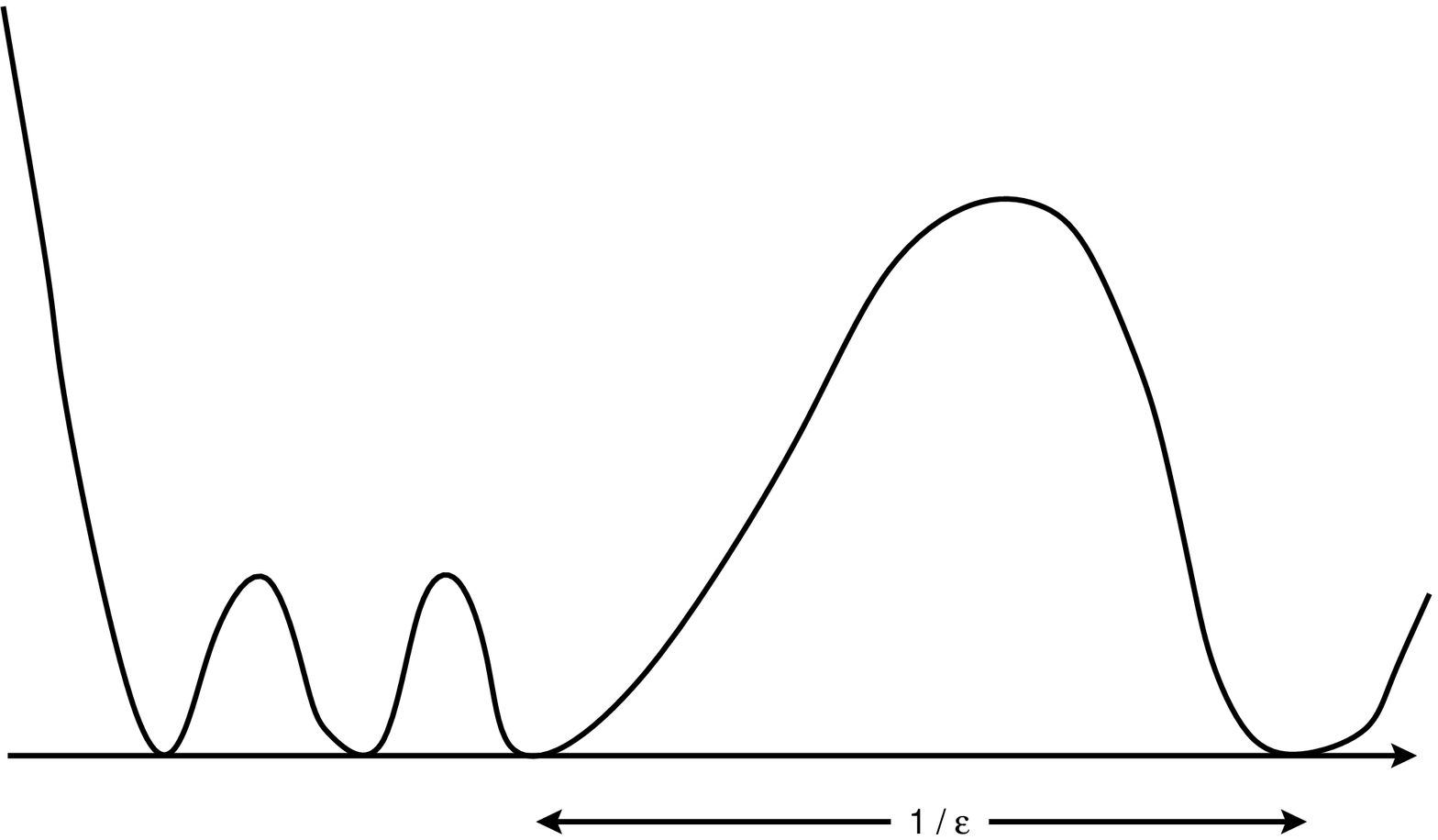}}

\subsec{Appendix 4: Canonical quantization with 
$W\not=0$:  Morse theory}
\subseclab\ssCQMT

In a classic paper \ssymrs\ Witten reformulated 
Morse theory using ideas from supersymmetry. 
Let us take $X$ to be compact and introduce 
a scale $W \to -t W$ so that the supersymmetry 
operators become: 
\eqn\susyops{
Q_t = i e^{-t W}~ d~ e^{t W} \qquad Q_t^* = -i  e^{t W} d^*~ e^{-t W}}
We will search for SUSY groundstates. Since $Q_t$ is 
related to $d$ by conjugation with an invertible 
operator there is a 1-1 correspondence between 
such states and harmonic forms. 

The Hamiltonian of the theory is 
\eqn\hamw{
H_t=\{Q_t, Q_t^*\} =
\{d , d^*\} + t^2 (\nabla W)^2 + t {D^2W\over D \phi^\mu D \phi^\nu}
[\psi^\mu, \bar \psi^\nu] }
where the last term involves a {\it commutator} 
of the Fermi fields. 

Let us look for susy (i.e. $E=0$) ground states at large $t$. 
Because of the potential term these must have 
their wavefunctions localized near the critical points $P$ of 
$W$, i.e.  $\nabla W(P)=0$.  Near such a critical point we can 
choose coordinates which diagonalize the Hessian and write
$$
W=W(P)+ \half \sum_\mu \lambda_\mu (\phi^\mu)^2 + \CO(\phi^3). 
$$ 
We will assume all critical points are nondegenerate. 
This means that $\lambda_\mu\not=0$ for all $\mu = 1, \ldots, \dim~ X$. 
The number of negative directions is called the 
Morse index of $W$ at $P$ and is denoted:
\eqn\mrsindx{
\lambda(P,W) \equiv \# \{ \lambda_\mu<0\}. } 
For $t\to \infty$ the wavefunctions will be 
exponentially damped away from the critical points. 
Therefore, we can find approximate ground states 
by considering the Hamiltonian 
\eqn\appxham{
\bar H_t = \sum_\mu \Biggl\{ \biggl[
-{\partial^2\over {\partial \phi^\mu}^2}
+ t^2 \lambda_\mu^2 (\phi^\mu)^2\biggr]
+ \biggl[
t \lambda_\mu[\psi^\mu, \bar \psi^\mu]\biggr] 
\Biggr\}. }
The operators in square brackets all commute and 
may be simultaneously diagonalized to give the approximate 
spectrum: 
\eqn\appxspct{
t \sum_\mu \Biggl\{  \mid \lambda_\mu\mid (1+2N_\mu) 
+
\lambda_\mu n_\mu\Biggr\}, }
where $N_\mu\in \{0,1,2,\dots\}$ labels the occupation 
number of the  $\mu^{\rm th}$ 
harmonic oscillator and $n_\mu=+1$ if $\psi^\mu(P)$ is 
occupied and $-1$ if not. As $t\to \infty$ the only way to 
get an $E=0$ state is to take: 
$N_\mu=0, n_\mu=+1$ for $\lambda_\mu<0$ and 
$N_\mu=0, n_\mu=-1$ for $\lambda_\mu>0$. 
The fermion number of the ground state at $P$ 
can be taken to be
$F ( P ) = \sum_{\mu = 1}^{\dim~ X} \half( n_\mu + 1) $.
Thus: 
\bigskip

\boxit{
Each critical point contributes one approximate ground 
state and it has fermion number $j$ (is a $j$-form) if $\lambda(P,W)=j$. 
}
\bigskip

Thus  $M_j$, the number of critical points of 
Morse index $j$ is the number of approximate 
ground states of fermion number $j$. In general 
there are strictly more approximate groundstates 
than true groundstates. This is the statement of 
the weak Morse inequalities: 
$$M_j \geq b_j . $$

\ifig\BumpyTorus{Bumpy torus with Morse indices of the
various critical points}
{\epsfxsize3.0in\epsfbox{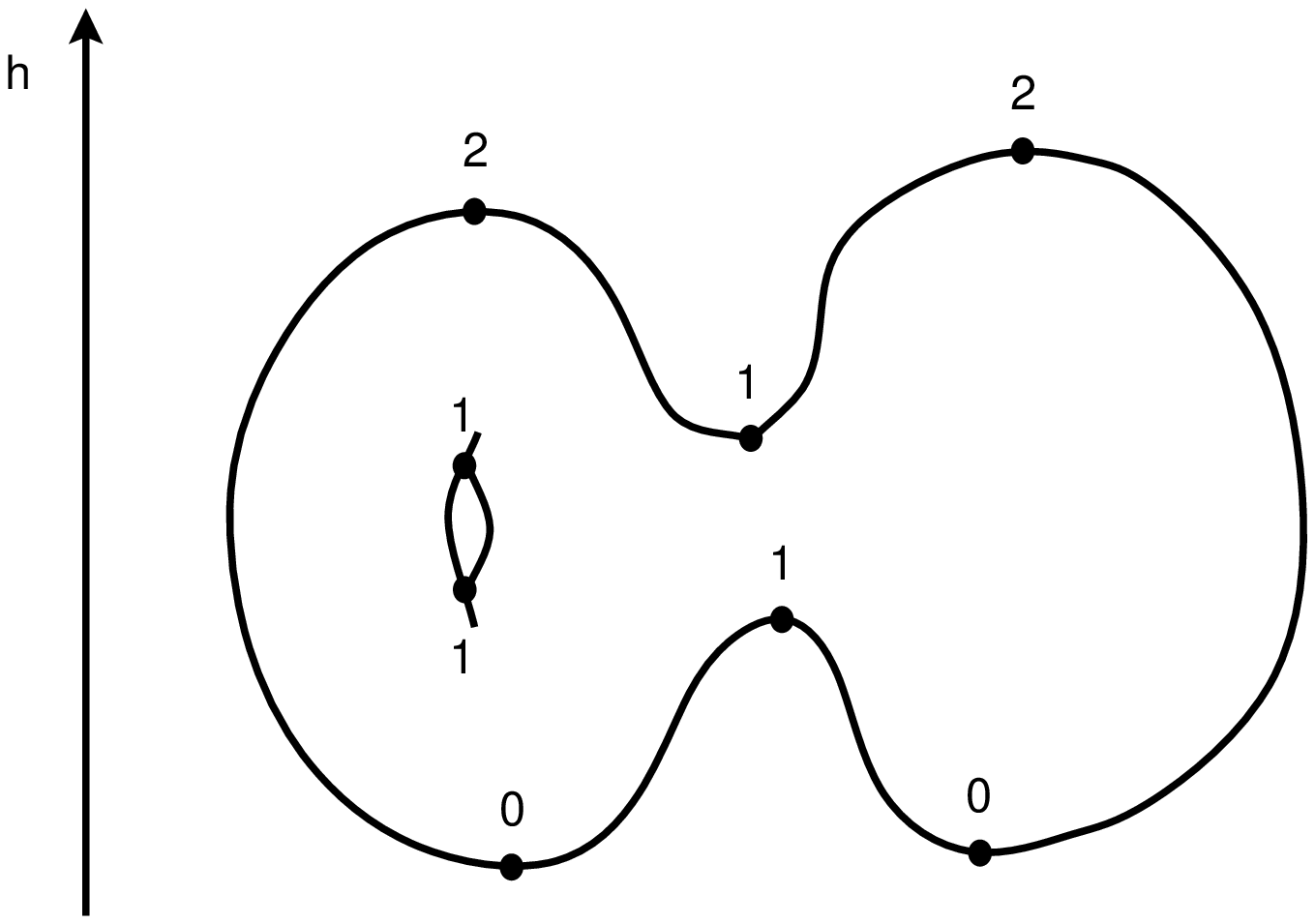}}

\vskip0.1truein\noindent
{\bf Example:} Consider $W$ to be the height function 
in the torus of \BumpyTorus. 
Evidently: 
$$M_2=2 \qquad M_1=4 \qquad M_0 =2 $$
and we have clearly overcounted groundstates, because 
we know that the true counting is given by 
DeRham cohomology: 
$$ b_2= 1 \qquad b_1 =2 \qquad b_0 =1$$

The degeneracy of approximate $E=0$ states
is lifted by instantons - or tunneling 
solutions - between the approximate groundstates. 
Indeed let 
$$
C^j = \{ \mid P; {\rm approx} \rangle \vert \lambda(P,W) = j \}
$$
Then the degeneracy will be lifted if an 
instanton solution gives nonzero matrix elements to 
the supersymmetry operator
\eqn\mtrxele{
\langle P_1 \mid Q_t \mid P_2\rangle
}
this matrix element being zero to all orders in 
perturbation theory. In the one-instanton approximation 
\mtrxele\ is given by a sum over instanton solutions: 
\eqn\instappx{
\langle P_1 \mid Q_t \mid P_2\rangle
= \sum_\Gamma \pm 1
}
where $\Gamma$ runs over solutions to 
\eqn\instants{
\eqalign{ 
s^\mu= {d \phi^\mu\over dt } & + G^{\mu\nu}\partial_\nu W=0 \cr
\phi(t=-\infty) = P_1 \qquad & \phi(t=+\infty) = P_2\cr}
}
and the sign is determined by considerations 
of orientations (see \ssymrs). 

By supersymmetry $Q_t^2=0$ and thus $Q_t$ defines 
a differential on the chain complex $(C^j, Q_t) $. 
Except in extraordinarily symmetric situations
\foot{For example, J. Zinn-Justin suggests 
three wells symmetric about $\phi=0$.} 
one-instanton effects will lift the perturbative 
degeneracy. Mathematically this means that the 
cohomology of the complex $(C^j, Q_t) $
coincides with the DeRham cohomology: 
$$H^\bullet (E^*, Q_t) = H^\bullet _d(X)$$
This last statement is easily seen to be 
equivalent to the strong Morse inequalities
$$\sum t^j M_j - \sum t^j b_j =(1+t)Q(t)$$
where $Q(t)$ is a polynomial with positive coefficients. 

This formulation of Morse theory was generalized 
by A. Floer in two ways known as 
symplectic Floer homology and 
3-manifold Floer homology. 
Each of these Floer homology theories plays a 
role in the canonical formulation of TFT's. 

\newsec{Topological sigma models}
\seclab\sTSM

We will approach the subject of topological sigma models\refs{\Witsm} from the Mathai-Quillen
point of view, generalizing the discussion of SQM in the previous section.

\subsec{Fields and Equations.}

\noindent
We begin with the data:

\item{1.}
$X$ (the ``target space") is an almost K\"ahler manifold. 
This is a symplectic manifold $(X,\omega)$ with symplectic 
form $\omega$ together with an almost complex structure, 
  $J\in \End ( TX )$  with $J^2=-1$ such that
$G_{\mu\nu}=\omega_{\mu\lambda}J^\lambda_\nu$, i.e., 
$\omega(J(\cdot),\cdot) = G(\cdot, \cdot)$ is a positive 
definite metric. The theory simplifies considerably if $J$ 
is an integrable complex structure so that 
$X$  is K\"ahler.
We will make this choice, indicating how the theory generalizes at appropriate
points.
\item{2.}
$\Sigma$ is a 2d surface, with metric $h$ that induces a complex structure,
$\epsilon$.
\vskip0.1truein\noindent
Then in the fields-equations-symmetries paradigm, we have:
\par\noindent
$\bullet$
The {\it fields} will be the space of maps
$$
\MAP ( \Sigma, X) = \{ f \in \CC^\infty( \Sigma, X )  \}
$$
\par\noindent
$\bullet$
The {\it equations} will be the Gromov equations for (pseudo-) holomorphic maps.
As we have seen in section \ssFBAHS, (equation \lcbgfm)
$f \in \MAP ( \Sigma, X )$ is holomorphic iff
\eqn\gromov{
s(f)\equiv df+J df \epsilon=0}
with $J$ the (fixed) complex structure of $X$ and $\epsilon$ the (fixed) complex
structure of $\Sigma$.
We will denote by $\HOL ( \Sigma, X )$ the space of  holomorphic maps from
$\Sigma$ to $X$.
\par\noindent
$\bullet$
The {\it symmetries} in this example are trivial.

\noindent
{\bf Remark}: 

The relationship between $\MAP~ ( \Sigma, X )$
and   $\HOL ( \Sigma, X )$ 
is quite analogous to the relationship
between the moduli space of (anti-) self-dual
 connections and the space
${\cal A} / {\cal G}$ in Yang-Mills theory.
Much of the fundamental information 
about $\MAP ( \Sigma, X )$ is already captured
by $\HOL ( \Sigma, X )$.
In fact, it has been established that the inclusion
$\HOL ( \Sigma, X ) \inclusionmap{i} \MAP( \Sigma, X )$ induces isomorphisms in
homology (and in certain cases also homotopy) groups.
For the degree $k$ components of these two spaces, one finds
$H_i ( \HOL_k ( \Sigma, X )) \cong
H_i ( \MAP_k ( \Sigma, X ))$, for $i < N_k$,
where $N_k$ is an integer growing linearly with $k$.
These results have been established for
$\Sigma$ an arbitrary Riemann surface and $X$ an arbitrary flag
manifold \refs{\Segal,\Kirwan,\Guest,\BoHuMaMi,\Hurt}.

\subsec{Differential Forms on $\MAP ( \Sigma, X )$}

Just as in the example of SQM, we shall study differential geometry on the infinite
dimensional space, $\MAP ( \Sigma, X )$.
In complete analogy to $LX$ we have the natural  isomorphism:
\eqn\tantomap{
T_f ( \MAP ( \Sigma, X ) ) \cong \Gamma( f^\ast ( TX ))}
Consider formally the DeRham complex, $\O^\bullet ( \MAP ( \Sigma, X ) )$, as a
continuous tensor product of DeRham complexes on the target space $X$:
\eqn\mapdrcm{
\O^\bullet (\MAP ( \Sigma, X )) = \otimes_{\sigma \in \Sigma} \O^\bullet ( X )_\sigma}
Formally we may think of $f^\mu ( \sigma )$ as local coordinates in $\MAP ( \Sigma, X )$,
and $\widetilde d f^\mu(\sigma)$ as a basis of one-forms, where $\widetilde d$ is
the exterior derivative on $\MAP ( \Sigma, X )$.
Under the basic tautology \bsctaut\ we identify
$\widehat{\cC^\infty( \MAP ( \Sigma, X ))}$ with forms 
$\O^\ast ( \MAP ( \Sigma, X ))$ using:
\eqn\tautiii{
\widetilde d f^\mu(\sigma) \leftrightarrow \chi^\mu(\sigma)}

\subsec{Vector Bundle}

To apply the MQ formalism we will regard \gromov\  as the equation for the
vanishing of a section of a vector bundle.
Given a map, $f$, we may form the bundle over $\Sigma$, whose fiber at
$\sigma\in \Sigma$ is just 
$$
\biggl[ T^\ast \Sigma \otimes f^\ast ( TX ) \biggr]_\sigma=
T^\ast_\sigma \Sigma \otimes T_{ f  ( \sigma )} X
\cong \Hom ( T_ \sigma  \Sigma ,T_{f ( \sigma )} X )
$$
Sections of this bundle form the fibre of a bundle ${\cal W} \to \MAP ( \Sigma, X )$.
Note that $s ( f )$ satisfies the ``self-duality" constraint:
$$
J s ( f ) \epsilon = s ( f )
$$
whether $f$ is holomorphic or not.
Hence it is natural\foot{In fact, we will see below that it is necessary to make this
restriction.} to consider $\cV$ to be the subbundle of ${\cal W}$ whose
fibre over $f$ is given by:
\eqn\fibatf{
\CV_f:=\Gamma[ T^\ast (\Sw) \otimes f^\ast ( TX ) ]^+}
where the superscript ``+" indicates the ``self-duality" constraint, i.e.
\eqn\sdcnst{
x \in {\cal V}_f ~\Longleftrightarrow~J x \epsilon= x}

The antighosts $\rho_\mu^{~ \alpha}$ used in the integral construction of the MQ form
will live in the dual bundle $\CV^\ast$.
Using the metrics on  $\Sigma$ and $X$ we can write a natural metric on  $\CV^\ast$ :
\eqn\fibmet{
( \rho_{\scriptscriptstyle(1)}, \rho_{\scriptscriptstyle (2)} )_{\CV_f^\ast}
= \int_\Sigma d^2 \sigma~ \sh~ h_{\alpha\beta} G^{\mu \nu} (f (\sigma))~
\rho_{{\scriptscriptstyle (1)} \mu}^{\qquad \alpha}
\rho_{{\scriptscriptstyle (2)} \nu}^{\qquad \beta},}
These formulae simplify in local complex coordinates.
We take these to be $z,\zb$ on the worldsheet and $i, \bar i$ on the target space.
In terms of these the constraint \sdcnst\  reads
$$
\rho_i^{~z }= \rho_{\bar i}^{~ \bar z} = 0
$$

\subsec{Choice of connection}
\subseclab\ssCOC

In order to write a MQ form whose pullback by $s$ is the Poincar\'e dual to
$\HOL(\Sigma,X)$ in $\MAP ( \Sigma, X )$, we need to choose a connection on $\CV$.
We can motivate this connection using the remarks in section \sssGLF.
Recall that, in the MQ formalism, given a connection $\nabla$ on $E=\CV$
and a section, $s$, we form the linear operator
$$
\nabla s: T\ \MAP ( \Sigma, X ) \to \CV
$$
which determines the fermion bilinear terms in the action.
Over the zeroes of $s$, ${\cal Z} ( s )$, we have (from section \sssGLF):
$$
\ker ( \nabla ( s )) \mid_{\CZ(s)} \cong T \CZ ( s )
$$
In the present case we have chosen $s(f)$, so that $\CZ(s)$ is the space of
holomorphic maps.
We therefore examine the tangent space to the space of holomorphic maps.
In the case where $X$ is complex, this tangent space is
defined by the equation \refs{\Gr,\Witsm}:
\eqn\GroTanSpace{
D \delta f + J D \delta f \epsilon = 0}
where $D$ is  the pulled-back connection
\eqn\sgmcnn{
( D_\alpha~ \delta f )^\mu ( \sigma )
= \partial_\alpha \delta f^\mu
+ \Gamma^\mu_{~ \kappa\lambda} \partial_\alpha f^\kappa \delta f^\lambda}
and $\Gamma^\mu{}_{\nu\lambda}$ is the Christoffel connection of the target K\"ahler
metric.
If we consider the case of an almost complex structure, then there is a third term in the
equation involving $\nabla J$.
An important point to note is that $ ( D_\alpha \delta f )^\mu$ does {\it not} satisfy the
self-duality constraint.

\exercise{Variation of the Gromov equation}

Derive \GroTanSpace\ and \sgmcnn.
\vskip0.1truein\noindent
It is important to note that the Gromov equation is non-linear in $f$.
We can make this clear by explicitly indicating that $J$ is evaluated at 
$f ( \sigma )$:
\eqn\Gromov{
d f ( \sigma ) + J [ f ( \sigma )] df ( \sigma ) \epsilon ( \sigma ) = 0}
Now consider a one parameter family of holomorphic maps
$$\eqalign{
F\colon \Sigma \times I \to& X\cr
F ( \sigma; t ) \mapsto& f_t ( \sigma )\cr}
$$
with $f_0 ( \sigma ) = f ( \sigma )$.
This family must also satisfy the Gromov equation
$$
d f_t ( \sigma ) + J [ f_t  ( \sigma )] df_t ( \sigma ) \epsilon ( \sigma ) = 0\qquad
\forall \sigma \in \Sigma~ {\rm and}~ \forall t \in I
$$
Now take the derivative with respect to $t$ and evaluate at $t=0$.
We suppress worldsheet indices where they are obvious.
$$\left [
d \dot f^\mu_t ( \sigma )
+ \partial_\kappa J^\mu_{~ \nu} [ f_t ( \sigma )] \dot f^\kappa_t ( \sigma )
   d f^\nu_t ( \sigma ) \epsilon ( \sigma )
+ J^\mu_{~ \nu} [ f_t ( \sigma )] d \dot f^\nu_t ( \sigma ) \epsilon ( \sigma )
\right ]_{t=0} = 0
$$
Now consider the covariant derivative of $J$
$$
\nabla_\kappa J^\mu_{~ \nu}
= \partial_\kappa J^\mu_{~ \nu}
+ \Gamma^\mu_{~ \kappa\lambda} J^\lambda_{~ \nu}
 - \Gamma^\lambda_{~ \kappa\nu} J^\mu_{~ \lambda}
$$
Then we may write (setting $\delta f^\mu = \dot f^\mu_t \vert_{t=0}$)
$$
\partial_\kappa J^\mu_{~ \nu} \delta f^\kappa d f^\nu \epsilon
= - \Gamma^\mu_{~ \kappa\lambda} J^\lambda_{~ \nu} \delta f^\kappa d f^\nu \epsilon
+  \Gamma^\lambda_{~ \kappa\nu} J^\mu_{~ \lambda} \delta f^\kappa d f^\nu \epsilon
+ \nabla_\kappa J^\mu_{~ \nu} \delta f^\kappa d f^\nu \epsilon
$$
Now since $f_t ( \sigma )$ is, by fiat, a family of holomorphic maps,
$d f_t^\mu \epsilon = J^\mu_{~ \nu} d f_t^\nu$, for all $t$, so that
$$
\partial_\kappa J^\mu_{~ \nu} \delta f^\kappa d f^\nu \epsilon
= \Gamma^\lambda_{~ \kappa\nu} J^\mu_{~ \lambda} \delta f^\kappa d f^\nu \epsilon
-  \Gamma^\mu_{~ \kappa\lambda} \delta f^\kappa d f^\lambda
+ \nabla_\kappa J^\mu_{~ \nu} \delta f^\kappa d f^\nu \epsilon
$$
{}From this deduce \GroTanSpace\ and \sgmcnn.
Determine the equation for the tangent space in the case when $X$ is not
complex.

\endexercise

Equation \GroTanSpace\ suggests 
that we should  define  the 1-form on $\MAP ( \Sigma, X )$:
$$
\bD = D \chi + J D \chi \epsilon
$$
where we identify $\chi^\mu ( \sigma )$ with one-forms as in \tautiii.
A convenient choice of connection will be one for which this is $\nabla s$.
Let $z_\alpha^{~ \mu} [ f ]$ be a local section of $\CV \to \MAP ( \Sigma, X )$ then define:
\eqn\sgmcnn{\eqalign{
\nabla z_\alpha^{~ \mu} &= \widetilde d z_\alpha^{~ \mu}
- \Gamma^\mu_{\kappa\lambda} z_\alpha^{~ \kappa}~ \widetilde d f^\lambda\cr
&=\int d^2 \sigma \sqrt{h} \Biggl(
{\delta z_\alpha^{~ \mu} \over \delta f^\kappa ( \sigma )}~ \widetilde d f^\kappa ( \sigma )
- \Gamma^\mu_{\kappa\lambda} [ f ( \sigma ) ] z_\alpha^{~ \kappa}~
\widetilde d f^\kappa ( \sigma ) \Biggr) \cr}}
In the first line we use condensed notation.
$\widetilde d$ is the exterior derivative on $\MAP ( \Sigma, X )$.
One may easily check that $\nabla s$ satisfies the self-duality constraint.
Thus we will adopt this connection.

\exercise{Connection}

Derive the relation:
\eqn\secconn{
\nabla s = \bD \qquad .}

Remember that $J$ depends on $f$ so the functional derivative acts on it. 
Use the fact that the complex structure is integrable.
When that is not the case, there is an extra term corresponding to the extra
term in the equation for the tangent space.

\endexercise

\par\noindent
{\bf Remark}: The physical reason for imposing the self-duality constraint on
the fields $\rho_\mu^{~ \alpha}$ is that the anti-dual components do not enter into
the action following from the gauge fermion below.
Mathematically the operator $\nabla s$ fails to be Fredholm without this constraint.

\subsec{BRST Complex}

The final ingredient we need in order to apply the formulae 
of chapter \sINandTIR\ is the
differential $Q$.
Having chosen the connection, we have effectively applied the Chern-Weil
homomorphism, so the BRST complex will be made of the basic differential forms,
 $\O(\CV^\ast)$, on the total space $\CV^\ast$.
As usual we identify this with a superspace of functions $\widehat\CF ( \CV^\ast )$.
The coordinates on the base are $f^\mu(\sigma)$ and $\chi^\mu(\sigma)$, as above.
The superspace coordinates on the fiber $\widehat{\cal F} ( \CV^\ast_f )$ will be
even coordinates $\pi_\mu^{~ \alpha}$ and odd coordinates
$\rho_\mu^{~ \alpha} \in \Pi \CV^\ast_f$, the  ``anti-ghosts."
Note that because of the self-duality constraint
$\rho_i^{~ z} = \rho_{\bar i}^{~ \bar z} 
= \pi_i^{~ z} = \pi_{\bar i}^{~ \bar z}=0$.
The grading on the complex is given by the ghost number.
$\chi$ has ghost number $1$
and $\rho$ has ghost number $-1$; the bosonic fields
$f$ and $\pi$ both have ghost number zero.

The BRST differential is given by $Q=\widetilde d \otimes 1 + 1 \otimes \delta $,
where $\delta$ is (implicitly) defined by:
\eqn\sgextdrv{\eqalign{
Q f^\mu &= \chi^\mu\qquad
Q\chi^\mu =0 \cr
Q\rho_\mu^{~ \alpha} & = \pi_\mu^{~ \alpha}\qquad
Q\pi_\mu^{~ \alpha} = 0 . \cr}}
Put more succinctly, $Q$ is the DeRham  exterior derivative on the total space of
$\Pi \CV^*$.

\subsec{Gauge Fermion and Action}

According to sections \ssIRQI, \ssDOCandBRSTD, 
we should take the gauge fermion:
\eqn\tsgf{\eqalign{
\Psi_{t} &= i\langle \rho, s \rangle - t ( \rho, \theta \rho )_{V^\ast} + t  ( \rho, \pi )_{V^\ast}\cr
&=~ \int d^2 \sigma~ \sqrt{h}~ \left \{
\rho_\mu^{~ \alpha} \left ( i s_\alpha^{~ \mu}
- t~ \Gamma^\mu_{\kappa\lambda} G^{\kappa\kappa^\prime} h_{\alpha\alpha^\prime} \rho_{\kappa^\prime}^{~  \alpha^\prime} \chi^\lambda
+ t~ G^{\mu\mu^\prime} h_{\alpha\alpha^\prime} \pi_{\mu^\prime}^{~ \alpha^\prime}
\right ) \right \} \cr}}
where we have used the metric \fibmet\ and the connection \sgmcnn\
to raise and lower indices on $\rho$ and $\pi$.
This formulation of the gauge fermion was first obtained in \refs{\BaSi}.
There are modifications to this action involving $\nabla J$ in the case where
$X$ is not K\"ahler.

Acting with $Q$ and integrating out the Lagrange multipliers,
$ \pi_\mu^{~ \alpha}
= - {i \over {2 t}} G_{\mu\mu^\prime} h^{\alpha\alpha^\prime} s_{\alpha^\prime}^{~ \mu^\prime}
- \rho_\kappa^{~ \alpha} \Gamma_{~ \mu\lambda}^\kappa \chi^\lambda$,
we obtain the action:
\eqn\tsact{\eqalign{
I_{{\rm t}\sigma} ~ &=~ Q ( \Psi_t )\cr
&=\int d^2 \sigma~ \sqrt{h}~ \left \{
{1\over{4t}} G_{\mu\mu^\prime} h^{\alpha\alpha^\prime}
s_\alpha^{~ \mu} s_{\alpha^\prime}^{~ \mu^\prime}
- i  \rho_\mu^{~ \alpha} ( D \chi )_\alpha^{~ \mu}\right.\cr
&\qquad\qquad\qquad\qquad\qquad\left.
-t \rho_\mu^{~ \alpha} R^\mu_{~ \nu\rho\sigma} \chi^\rho \chi^\sigma
G^{\nu\nu^\prime} h_{\alpha\alpha^\prime}
\rho_{\nu^\prime}^{~ \alpha^\prime} \right \}. \cr}}
Passing to local complex coordinates on the target and the worldsheet we may write this as:
\eqn\tsacti{\eqalign{
\int d^2 z& \left ( {2\over t} G_{i \bar j} \partial_{\bar z} f^i \partial_z f^{\bar j}
- i h_{z \bar z} \rho_i^{~ \bar z} D_{\bar z} \chi^i
- i h_{z \bar z} \rho_{\bar i}^{~ z} D_z \chi^{\bar i}\right.\cr
&\qquad\qquad\qquad\qquad\qquad\qquad\left.
- 4 t  G^{j\bar j } \rho_i^{~ \bar z} R^i_{~  j k \bar l} \chi^k \chi^{\bar l} \rho_{\bar j}^{~ z}
(h_{z \bar z} )^2\right ) . \cr}}

\subsec{Observables}
\subseclab\ssObsv

Observables are most elegantly constructed using the ``universal map".
When restricted to the moduli space of holomorphic maps this gives the universal instanton \Witsm.
A very similar construction will appear below in constructing the observables in topological
gauge theory .

Given any point $P\in \Sigma$ there is a canonical map $\Phi_P: \MAP ( \Sigma, X )\to X$
given by evaluation at $P$.
These maps fit together to give the ``universal map:"
\eqn\evlmp{\eqalign{
\Phi\colon \Sigma \times \MAP ( \Sigma, X ) \longrightarrow& X\cr
\Phi ( P, f )\longmapsto& f ( P )\cr}}
Thus, if $A \in \Omega^k ( X )$ is a differential form on the target then
\eqn\tsigobsi{
\Phi_P^*(A) \in \Omega^k (\MAP ( \Sigma, X ))}
is a form on $\MAP ( \Sigma, X )$.
Explicitly:
\eqn\tsigobsii{
\widehat{\CO_A^{(0)}} ( P ) =  \widehat{\Phi_P^\ast ( A )}
= A_{i_1\dots i_k} ( \phi ( P ))~ \chi^{i_1}( P ) \cdots \chi^{i_k} ( P )}
Inspection of \sgextdrv\ shows that $Q$ acts as 
\eqn\tsigobsiii{
\{ Q, \widehat{\CO_A^{(0)}} \} = \widehat{\CO_{dA}^{(0)}}}
In particular we learn that we have explicit representatives of the $Q$-cohomology
in terms of the cohomology of  the target for each $A\in  H^\bullet (X)$ and
$P\in \Sigma$.

To go further we must use the ``descent equations."
If we pull back $A \in H^\bullet ( X )$ by the universal map \evlmp, then we have a
closed form on $\Sigma \times \MAP ( \Sigma, X )$.
Therefore, splitting the exterior derivative into $d+Q$, with $d$ the exterior derivative
on $\Sigma$ we have
\eqn\clsdfrm{
(d+Q) \widehat{\Phi^\ast (A)}=0}
Now $H^\bullet ( \Sigma \times \MAP ( \Sigma, X ))$ is bigraded by the degrees on $\Sigma$ and
$\MAP ( \Sigma, X )$, so we may write:
\eqn\dcmpse{
\widehat{\Phi^\ast ( A )}
= \widehat{\CO^{(0)}_A} + \widehat{\CO^{(1)}_A} + \widehat{\CO^{(2)}_A}}
where the upper index refers to form degree in the $\Sigma$ direction.
Taking into account the grading of $d$ and $Q$ the statement \clsdfrm\ simply becomes
the famous descent equations
\eqn\dscnt{\eqalign{
0 &= \{ Q, \widehat{\CO_A^{( 0 )}} \}\cr
d \widehat{\CO_A^{(0)}}  & = - \{Q, \widehat{\CO_A^{(1)}} \} \cr
d \widehat{\CO_A^{(1)}}  & = - \{Q, \widehat{\CO_A^{(2)}} \} \cr
d \widehat{\CO_A^{(2)}}  & =0 \cr}}

There are two important consequences of  these equations.
First, it might appear that we have infinitely many BRST observables (one for each
$P$) but this is not the case.
The second descent equation shows that
$$
\widehat{\CO^{( 0 )}_A ( P^\prime )} - \widehat{\CO^{( 0 )}_A ( P )}
= -\{ Q, \int_P^{P^\prime} \widehat{\CO^{(1)}} \}
$$
so the class does not depend on $P$.
Thus, the vector space of BRST classes $\widehat{\CO^{(0)}_A}$ is isomorphic to  the
DeRham cohomology of  $X$.
Second, if $\gamma\in H_k ( \Sigma )$ is a homology cycle, then we may form the BRST
invariant observables:
\eqn\grmvcl{
\widehat{W} ( A, \gamma ) \equiv \int_\gamma \widehat{\CO_A^{( k )}}
= \int_\gamma \widehat{\Phi^\ast( A )}}
where the hat indicates translation to superspace.
BRST invariance is proven using the descent equations \dscnt.

\vskip0.1truein\noindent
{\bf Remark}:
In general these are not the only BRST classes in the model.
For example\refs{\DiVVTrieste,\Hott} one can form  ``homotopy observables" when
$X$ is not simply connected. 

\subsec{Correlation Functions}

We now want to study  the correlation functions
$$
\langle \prod_i \widehat{W} ( A_i, \gamma_i ) \rangle
$$

\subsubsec{Localization on moduli space}

Since the action is given by the MQ formula, we know that we will localize to
$\CZ(s) = \HOL ( \Sigma, X )$.
Let us apply the discussion of section \sssGLF\ to the present model.
The operator $\nabla s $ defined above in \sgmcnn\ is, as we have seen,
$\bD: T_f\CM \to \CV_f$.
The sequence \wittseq\ becomes the sequence:
$$
0\rightarrow \Im(\bD) \rightarrow \CV \rightarrow \cok(\bD) \rightarrow 0
$$
The equation \locii\ becomes in this case \refs{\AsMo}:
\eqn\lchli{\eqalign{
\int_{\widehat{\MAP ( \Sigma, X )}} df d \chi~ 
\hat \CO \int_{\Pi\CV_f^\ast} d \rho~
e^{-I_{{\rm t} \sigma}}
&= \int_{\MAP ( \Sigma, X )} \CO~ s^\ast ( \Phi( \CV, \nabla ))\cr
&=\int_{\CZ(s)} i^\ast ( \CO )~  \chi( \cok(\bD))\cr}}
where the BRST observables, $\widehat {\cal O} = \widehat{W} ( A, \gamma )$
of section \ssObsv\ correspond to differential forms, ${\cal O}$ on $\MAP ( \Sigma, X )$,
and hence $i^\ast ( {\cal O} )$ are differential forms on $\HOL ( \Sigma, X ) = {\cal Z} ( s )$,
where $i\colon {\cal Z} ( s ) \hookrightarrow \MAP ( \Sigma, X )$ is the inclusion
map.
We have used the fact that (at least formally)  
\eqn\lchl{
s^\ast ( \Phi ( \CV, \nabla ))= \int_{\Pi \CV_f^\ast} d \rho~ e^{-I_{ts}}}
represents the Euler class of $\chi ( \CV \to \MAP ( \Sigma, X ))$.

Clearly, to get a nonzero integral the form $\CO$ must have degree given by
$$
\Index~ \bD = \dim~ \ker~ \bD - \dim~ \cok~ \bD
$$
In physical terms, $\widehat \CO$ must have ghost number appropriate to the
anomaly in the ghost number current.

\subsubsec{Zeromodes and the Index theorem}
\subsubseclab\sssZandtheIT

Let us now study the kernel and cokernel of $\bD$ more closely.
The $\chi$ zeromodes span the kernel of $\bD$.
The $\rho$ zero-modes are in the bundle $\cok~ \bD \rightarrow \CZ(s)$.
The operator is a direct sum of holomorphic and antiholomorphic operators.
The operator $\bD_{\zb}=D_\zb$ appearing in \tsacti\ is the twisted Dolbeault
operator $\pb_{f^\ast ( TX )}$ discussed in section \ssITforSEO\ above.
Thus the $\chi^i$ zeromodes span  $H^0_{\bar \partial} ( \Sigma; f^\ast T^{1,0} X )$ while
the $\rho_{\bar z}{}^i $ zeromodes span
$H^{0,1}_{\bar\partial} ( \Sigma; f^\ast T^{1,0} X)$.

{}From the Hirzebruch-Riemann-Roch theorem \refs{\Hi}, (cf \TwisDolbInd\
in section \ssITforSEO ) we have
\eqn\DIndex{\eqalign{
\ind{}_{\sst \IC} D_\zb ~=&~ \# \chi~ {\rm zero~ modes}
                     - \# \rho~ {\rm zero~ modes}\cr
~=&~ \int_{\Sigma_W} (1+\half c_1(T^{(1,0)} \Sigma) )(\dim{}_{\IC} X + f^\ast ( c_1 ( T^{(1,0)}X )))\cr
           ~=&~ ( 1 - h ) \dim{}_{\IC} X
              + \int_{\Sigma_W} f^\ast ( c_1 ( T^{(1,0)}X )).\cr}}
In order to find the number of $\chi$ and $\rho$ zeromodes separately we need an
independent argument, typically a vanishing theorem on some cohomology group.

\subsubsec{Reduction to Enumerative Geometry}

Let us return to \lchli.
In terms of  the universal map \evlmp\ we define 
$\omega(A,\gamma)= 
\iota^\ast \int_\gamma \Phi^\ast ( A ) \in H^\bullet ( \CZ ( s ))$.
The correlation functions are intersection 
numbers in the moduli space $\CH ( \Sigma, X ) = \CZ ( s )$: 
\eqn\corrlfn{\eqalign{
\langle \prod W( A_i, \gamma_i ) \rangle
=& \int_{\CZ(s)} \omega( A_1, \gamma_1 ) \wedge - \wedge \omega(A_r,\gamma_r)\cr
&\qquad\wedge \chi(  H^{1,0}_{\bar\partial} ( \Sigma, f^\ast ( T^{0,1} X )) )
\wedge \chi(  H^{0,1}_{\bar\partial} ( \Sigma, f^\ast ( T^{1,0} X )) )\cr}}
The above intersection numbers can be related to classical problems in
enumerative geometry from the following construction
\Witsm. 
Suppose $A$ is Poincar\'e dual to $H_A \subset X$, then we may choose a
representative $A$ with delta function support on $H_A$.
The corresponding representative of the class $\CO^{(0)}_A \in H^\bullet (\MAP ( \Sigma, X ))$
has support on
$$
L(A,P) = \{ f\in\CH(\Sigma,X)\mid f(P) \in H_A \}
$$
We are computing intersection numbers of these cycles.
Roughly speaking, we are ``counting" numbers of curves in $X$ which pass through
specified points (or homology cycles).
This is the initial observation behind the famous curve-counting results provided by
mirror symmetry \refs{\CaDeGrPa}. 
Further discussion of the relation to enumerative 
geometry can be found in \refs{\kontsevichi,\kontsevichii}.  

\subsec{Quantum Cohomology}

In the case of a K\"ahler target space, $X$, the  space of 
local operators of the $N=2$ sigma model
admits a $\IZ \oplus \IZ$ bigrading:
$$
\cO = \bigoplus_{p,q \in \IZ} \cO_{p,q}
$$
which resembles the Dolbeault cohomology of $X$ in certain respects.
Crucial differences are : (i) the degrees of operators range from $-\infty$
to $+\infty$ and (ii) the existence of an anti-unitary involution under which
$\CO_{p,q} \to \CO_{-p,-q}$ (this is related to the CPT invariance of the
physical theory).
The classical cohomology
 corresponds as a vector space to the  $Q$-cohomology classes.
There is a natural way, in the context of string theory, to define an associative and
distributive operator product of these operators.
Under this product, the operator algebra is closed, i.e.
$$
\CO_i \CO_j = C_{ij}^k \CO_k + [ Q, \cdot ]
$$
so that it defines a ring structure called the {\it quantum cohomology
ring} \refs{\Witp,\VaTopMirr}.

The quantum cohomology ring is related to the classical cohomology ring.
Its relation to the classical one can be made apparent if we deform the usual sigma
model action by a purely topological
term\foot{If $\dim H^{{\scriptscriptstyle ( 1, 1)}} > 1$, this deformation generalizes
in the obvious way.
For the sake of simplicity we will consider the case of only one deformation term.} :
$$
I_{\rm top} = t \int_\Sigma  f^\ast ( {\bf K} )
$$
where ${\bf K} \in H^{{\scriptscriptstyle ( 1, 1)}} ( X )$.
Roughly speaking, the role of this term is the following: if $f\colon \IP^1 \to C$ is a map
of degree $n$ into a rigid curve $C \subset X$, then its contribution to the
operator product is weighted by $q^n = e^{-n A t}$, where
$A = \int_{\IP^1} ( f^{(1)} )^\ast ( {\bf K} )$ is the contribution from a map of degree 1,
$f^{(1)}\colon \IP^1 \to C$.

We have seen in \ssObsv\ that the observables of topological sigma models,
${\cal O}_A{}^{(0)}$, are isomorphic to the DeRham cohomology of $X$.
For example, consider a collection of observables corresponding to
$A_{i_a} \in H^{1,1} ( X, \IR )$.
Let $\{ H_{A_{i_a}} \}_{a=1,\ldots,3}$ be homology classes in $X$ dual to the $A_{i_a}$.
Then a three point function in the deformed theory has the following
structure \refs{\DiSeWeWi,\AsMo}:
\eqn\qcinter{\eqalign{
&\left\langle \CO_{A_{i_1}} \CO_{A_{i_2}}  \CO_{A_{i_3}} \right\rangle
= \# \left ( H_{A_{a_1}} \cap H_{A_{a_2}} \cap H_{A_{a_3}} \right )\cr
&\qquad
+ \sum_{\matrix{{\sst C \subset X}\cr {\sst C~ {\rm isolated}}\cr {\sst {\rm rational~ curve}}\cr}}
\sum_{\matrix{{\sst f\colon \IP^1 \to C}\cr {\sst {\rm f~ holo}}\cr}}
\int_{\IP^1} f^\ast ( A_{i_1} ) \int_{\IP^1} f^\ast ( A_{i_2} ) \int_{\IP^1} f^\ast ( A_{i_3} )~
e^{- t \int_{\IP^1} f^\ast ( {\bf K} )}\cr}}
%

Rescaling the K\"ahler class is equivalent to changing $t$.
In the limit $t \to \infty$ only the degree zero (constant) holomorphic maps survive
and we recover the classical cohomology ring.
It is in this sense that the quantum cohomology ring may be considered to be a
deformation of the classical one. Quantum cohomology 
is under intensive investigation by many mathematicians. 
See, for example,  \kontsevichi.

\subsec{Relation to the physical sigma model}

The above presentation of the topological sigma model is somewhat idiosyncratic.
In this section we review the  standard approach to the subject.

\subsubsec{K\"ahler target}

Just as the topological sector of SQM can be related to a larger, nontopological
theory, the topological sigma model can be related to a larger theory.
In this case it is the $N=2$ supersymmetric sigma model.

One way to define the $N=2$ model is to consider
the sigma model of section \ssTDSNSM\ with a K\"ahler target.
The splitting of the target space coordinates into holomorphic and antiholomorphic
coordinates allows us to define {\it two} supersymmetries:
\eqn\twossysi{\eqalign{
\delta_1 \phi^i &= \epsilon^- \psi_-^{~ i} +  \epsilon^+ \psi_+^{~ i}\cr
\delta_1 \phi^{\bar i} &= 0 \cr
\delta_1 \psi_-^{~ i} &=- \epsilon^+ F^i\cr
\delta_1 \psi_-^{~ \bar i} &= -i \epsilon^- \bar\partial \phi^{\bar i}\cr
\delta_1 \psi_+^{~ i} &= \epsilon^- F^i\cr
\delta_1 \psi_+^{~ \bar i} &=i \epsilon^+ \partial \phi^{\bar i}\cr
\delta_1 F^i &= 0\cr
\delta_1 F^{\bar i} &= -i \epsilon^+ \partial \psi_-^{~ \bar i}
                                        -i \epsilon^- \bar\partial \psi_+^{~ \bar i}\cr}}
and
\eqn\twossysi{\eqalign{
\delta_2 \phi^i &= 0 \cr
\delta_2 \phi^{\bar i} &= - \bar \epsilon^- \psi_-^{~ \bar i} - \bar \epsilon^+ \psi_+^{~ \bar i}\cr
\delta_2 \psi_-^{~ i} &= i \bar \epsilon^- \bar\partial \phi^i\cr
\delta_2 \psi_-^{~ \bar i} &= \bar \epsilon^+ F^{\bar i}\cr
\delta_2 \psi_+^{~ i} &=  - i \bar \epsilon^+ \partial \phi^i\cr
\delta_2 \psi_+^{~ \bar i} &= - \bar \epsilon^- F^{\bar i}\cr
\delta_2 F^i &= i \bar \epsilon^+ \partial \psi_-^i
                          + i \bar \epsilon^- \bar\partial \psi_+^i\cr
\delta_2 F^{\bar i} &= 0\cr}}

\subsubsec{Superspace}

More fundamentally, we start with $N=2$ superspace $(z,\zb, \theta^\pm , \bar \theta^\pm)$.
\foot{Our conventions will be those of  \wittphases, hence those of Wess-Bagger
{\it except} that we rescale all $\theta$'s by $\sqrt{2}$. }
The action is
\eqn\ntwoact{
I = \alpha \int d^2 z d^4 \theta~ K  ( \Phi^i, \Phi^{\bar i} )
    +  \beta \left ( \int d^2 z d^2 \theta~ W ( \Phi^i )
                        +  \int d^2 z d^2 \bar \theta~ \bar W ( \Phi^{\bar i} ) \right )}

The simplest kinds of $N=2$ superfields are the chiral and anti-chiral ones.
The chiral superfields obey $\bar D_{\dot\alpha} \Phi^i = 0$ and
$D_\alpha \Phi^{\bar i} = 0$.
They are most simply expressed as functions of the variables 
$z+i \tp\tbp$ and
$\zb-i \tm\tbm$
\eqn\chspfld{
\Phi^i   = \phi^i + \tp \psp^i + \tm \psm^i + \tp\tm F^i}
Similarly, the antichiral superfields
\eqn\acspfld{
\Phi^{\bar i} = \phi^{\bar i} + \tbp \psp^{\bar i}  + \tbm \psm^{\bar i} + \tbp\tbm F^{\bar i}}
are functions of $z-i \tp\tbp$, $\zb+i \tm\tbm$.

Supersymmetry transformations are generated by
\eqn\ntwoqs{\eqalign{
Q_+ &= {\p\over \p \tp} - i \tbp \p_z\cr
Q_- &= {\p\over \p \tm} + i \tbm \p_\zb\cr
\bar Q_+ &= -{\p\over \p \tbp} + i \tp \p_z\cr
\bar Q_- &= -{\p\over \p \tbm} - i \tm \p_\zb\cr}}

The action becomes (with $W=0$):
\eqn\fullntwo{\eqalign{
I = \int_\Sigma d^2 z& \left (
\half G_{i \bar i} ( \partial_z \phi^i \partial_{\bar z} \phi^{\bar i}
                            + \partial_{\bar z} \phi^i \partial_z \phi^{\bar i}
                             - i \psi_-^{~ \bar i} D_z \psi_-^{~ i}
                             - i \psi_+^{~ \bar i} D_z \psi_+^{~ i} )\right.\cr
&\qquad\qquad\qquad\qquad\left.
- R_{i \bar i j \bar j} \psi_+^{~ i} \psi_+^{~ \bar i} \psi_-^{~ j} \psi_-^{~ \bar j} \right )\cr}}
In deriving this one uses the basic formulae of 
 K\"ahler geometry:
\eqn\khlrgeom{
\mathboxit{
\eqalign{
G_{i \jb}                  &= \p_i \p_{\jb} K ( \Phi^i, \Phi^{\bar i} )\cr
\Gamma^i_{jk}       &= G^{i \bar l} \p_j G_{k \bar l}\cr
\Gamma^i_{j\bar k}       &= 0\qquad {\rm etc.} \cr
R_{i \bar j k \bar l} & = - G_{m \bar j} \partial_{\bar l} \Gamma^m_{ik}\cr
R_{i \bar j k \bar l} &= - R_{\bar j i k \bar l}
= - R_{i \bar j \bar l k} = R_{k \bar l  i \bar j}\cr
}}}

\subsubsec{Symmetries}

The symmetries of the action include $N=2$ supersymmetry with
$$
\{ Q_+, \bar Q_+\} = H+P \qquad
\{ Q_-, \bar Q_-\} = H-P
$$
and a $U(1)_L\times U(1)_R$ R-symmetry:
\eqn\rsymm{
\eqalign{
U(1)_L :  \psm^j \to e^{i \alpha}\psm^j
&\qquad  \psm^\jb \to e^{-i \alpha}\psm^\jb\cr
J^{(L)}&= G_{i \jb} \psm^i \psm^{\jb} \cr
U(1)_R :  \psp^j \to e^{i \alpha}\psp^j
&\qquad  \psp^\jb \to e^{-i \alpha}\psp^\jb\cr
J^{(R)}&= G_{i \jb} \psp^i \psp^{\jb} \cr}}
These currents are anomalous, but the vectorlike symmetry $J= J^{(L)}-J^{(R)}$ is
anomaly free.

\subsubsec{Topological twisting}
\subsubseclab\sssTopltw

The topological theory is defined by making a redefinition of the energy momentum
tensor \Witsm:
\eqn\twsttens{
T'_{\alpha \beta}
= T_{\alpha \beta} - \half \epsilon_\alpha{}^\gamma\p_\gamma J_\beta}
This is often called the {\it A-model}  \wttnmirror.
\twsttens\ defines a different coupling to gravity and consistency requires that objects
which are charged under $J$ change their transformation laws.
Before twisting  $Q_+,Q_-,\bar Q_+$, and  $\bar Q_-$
transform under $U(1)_L\times U(1)_R\times SO(2)_{\rm local Lorentz}$ as
\eqn\bftw{
(1, 0, \ha ) \oplus ( 0, 1, -\ha ) \oplus ( -1, 0, \ha ) \oplus ( 0, -1, -\ha )}
%
%
%
After twisting the four supercharges transform as:
\eqn\afttw{
( 0, +1, 0 )\oplus ( 0,-1,+1)\oplus (+1,0,-1)\oplus (-1,0,0)}
giving {\it two} scalar supercharges $Q_+$ and $\bar Q_-$.
The operator $Q= Q_+ + \bar Q_-$ is nilpotent: $Q^2=0$ and the energy momentum
tensor is $Q$-exact, $T'_{\alpha \beta} = \{ Q, \Lambda_{\alpha \beta} \}$.

Another way of understanding the twisting procedure 
is that we couple the theory to
an external $U(1)$ gauge field $A_\mu$, so that correlators 
now depend on $A_\mu$ as well as the spin connection: 
\eqn\physcorr{
\langle \prod \CO\rangle^{\rm N=2\ model}_{\omega_\mu, A_\mu} 
} 
The ``diagonal correlators'' with $A_\mu \sim \half \omega_\mu$ 
of local operators, $\CO$, which are ``chiral primary fields'' 
\lvw\ are the topological correlators. 
To see the equivalence of these viewpoints note that 
coupling to the current adds to the Lagrangian:
\eqn\extcpl{
A_z \psm^i \psi_{-i} + A_\zb \psp^i \psi_{+i}}
Setting the gauge fields to be: 
\eqn\ggespin{
A_z = - {i\over 2} \omega_z \qquad
A_\zb = + {i\over 2} \omega_\zb
}
is equivalent to the redefinition of the 
stress-tensor in \twsttens. 
The extra coupling \extcpl\ modifies 
the covariant derivatives in the Lagrangian
change in a way compatible with \twsttens.

{}From either point of view, the fermions change according to:
\eqn\chgfrm{
\psi_+{}^i \to \chi^i \qquad \psi_+^{\bar i} \to \rho_z{}^{\bar i} \qquad
\psi_-{}^i \to \rho_z{}^{\bar i} \qquad \psi_-^{\bar i} \to \chi^{\bar i}}
%
%

It is important to note that the overall fermion determinant in the $A$ model is an
absolute square, since the $( \chi^i, \rho_z^{~ \bar i} )$ and
$( \chi^{\bar i}, \rho_{\bar z}^{~ i} )$ are complex conjugates of one another.
Hence there is no problem in making sense out of this determinant and actually
this theory is well defined on arbitrary almost complex manifolds.

One might now wonder whether other twistings of the $N=2$ theories lead to
consistent theories.
Since the (worldsheet) spin content of the theory changes linearly with the twisting, it
seems reasonable to restrict attention to integral or half-integral twists, in order that we
end up with integer or half-integer spin particle content.
A half-integral axial twist of the energy-momentum tensor defines what is known
as the B-model \wttnmirror. 
The particle content of this model includes chiral fermions (sections of $T^{{\sst (0,1)}} X$),
for which the fermion determinant is in general afflicted with an anomaly.
As a result, the $B$ model is well-defined only on Ricci-flat K\"ahler target spaces.

\subsubsec{Comparison to the MQ form}
\subsubseclab\sssCtotheMQF

Let us now compare the Lagrangian \fullntwo\ above, after twisting, with \tsact\
and \tsacti.
The purely bosonic terms differ slightly, but we may write:
\eqn\bosprt{
\int \sqrt{h} d^2 z~ h^{\alpha \beta} G_{i \jb} \p_\alpha f^i \p_\beta f^{\bar j}
= \ha \int d^2 z~ G_{i \bar j} \pb f^i \p f^{\bar j} + \ha \int f^\ast {\bf K}}
where ${\bf K} = {i \over 2} G_{i \bar j} dw^i \wedge d w^{\bar j}$ is the K\"ahler form on
the target.
The second term on the RHS of \bosprt\ only depends on the homotopy class of $f$ and
on the cohomology class of ${\bf K}$.
As discussed above, the space of holomorphic maps, $\CH(\Sigma,X)$, has many
components corresponding to the different homotopy classes of  $f$.
The second term becomes a constant on each component of fieldspace and simply provides a weighting factor for these different components.


Finally, putting together \bosprt, \chgfrm, and 
\fullntwo, and identifying  $s_\zb^i = \p_{\zb} f^i $
we obtain precisely the action for the MQ form \tsacti. 

\subsec{Canonical approach}
\subseclab\ssCA

The canonical approach is a very interesting application 
of the supersymmetric quantum mechanics approach to 
Morse theory, and involves
symplectic Floer homology. 
We shall not enter into a thorough discussion of this subject here.
Some references include \refs{\FlSympi,\FlSympii,\Sadov,\Sa,\MDSa}.

From the point of view of the twisted physical 
$\sigma$-model, recall that in the case of 
 SQM we found that if we restrict our attention to the $E=0$ sector
of the Hilbert space (i.e. states annihilated by $Q$ and $Q^\ast$) then the partition
function yields topological information, the index of $Q$.
Analogously in the present case, the restriction to ( chiral, chiral ) primary fields \lvw\ 
(i.e. states in the cohomology of $Q = Q_+ + \bar Q_+$) 
leads to correlation functions
which yield topological information 
about the moduli space, $\CH ( \Sigma, X )$.

\subsec{Appendix: Examples of Moduli spaces of Holomorphic Maps}
\subseclab\ssAEofMSofHM

In this appendix we illustrate the above formal
constructions with some examples
of spaces $\CH(\Sigma,X)$.

\item{1.}
Our first example will be rational maps from
$\IC P^1\to \IC P^1$. These are given by $f(z)=P(z)/Q(z)$
where $P,Q$ are polynomials of degree $n$. The nonnegative
integer $n$ is the degree of $f$. The polynomials may be
factorized so that
$$
f(z) = A {\prod_{i=1}^n (z-a_i)\over \prod_{i=1}^n (z-b_i)}
$$
The moduli space breaks up into components labelled by
$n\in \IZ_+$. The $n^{th}$ component of the
moduli space may be described as:
\eqn\hlexpi{
\{ (A, a_i,b_i) \in \IC^{2n+1}~ \vert~ a_i\not= b_j, A\not=0 \}}
Each component is noncompact and may be parametrized by
the $2n+1$ complex parameters $A,a_i,b_i$.
In this case $H^1(f^*(TX))$ is zero so we may also obtain
the dimension directly from the index theorem.
This example illustrates two important properties of
$\CH(\Sigma, X)$. First there are many components.
Second, the space is noncompact. The intersection
theory is not well-defined until some compactification
is chosen \refs{\MorComp,\RuTi,\MDSa,\morrpless}.

\item{2.}
$X$ is a Riemann surface of genus $G$ and $f$ a map of degree
$n$.
Then \DIndex\ becomes
\eqn\RSIndex{
\ind{}_{\sst \IC}~ D_{\bar z} ~=~ B - 3 ( h - 1 ),}
where we have applied the Riemann-Hurwitz relation to express the result
in terms of the branching number $B = 2 ( h - 1 ) - 2 n ( G - 1 )$.
For $G > 1$ one can show that
$H^0 ( f^\ast ( TX ))=0$, so there are $\rho$ zero modes but no $\chi$
zero modes. The fact that there are no $\chi$ zero modes
means the moduli space is zero dimensional. Thus it can
be empty $\CH(\Sigma,X)= \emptyset$, or
$\CH(\Sigma,X)$ is a discrete set of points. All
this is consistent with the discussion in chapter \sCS\
above. In fact, from the above discussion we see there
are at most a finite set of points.
This is just as well, for $G>1$ the sigma model has constant
negative curvature and its status as a quantum field theory,
before twisting, is rather unclear.

\item{3.} One important example is the case of a topological sigma model
based on a Calabi-Yau manifold of complex dimension $d$.
The first Chern class of a Calabi-Yau vanishes, hence the complex index
is $d ( 1 - h )$.

\newsec{Topological Theories with Local Symmetry}
\seclab\sTTwithLS

When constructing actions for topological theories with gauge invariance one meets
two basic constructions. 
The first, discussed in chapter \sINandTIR, 
 is the MQ construction localizing to 
the set of fields satisfying some basic equations.  
The second, discussed in this chapter, 
 is connected with  gauge invariance. 
The essential principle motivating the second construction is that of {\it spacetime locality}. We will 
 explain the second construction in abstract 
terms. The same construction applies to all topological 
theories with local gauge invariance. We have 
attempted to explain the construction at length and 
in detail since we have found various points quite 
confusing. 

\subsec{Projection and Localization}
\subseclab\ssPandL

Let us return to \fbaspic\ of chapter \sGRTFT. 
The moduli spaces of interest are 
obtained by restricting fields $\Phi$ to 
$D \Phi=0$, and then dividing by $G$. 
These moduli spaces  form a submanifold $\CZ\subset M$, where $M=P/G$ is naturally
viewed as a quotient of a principal bundle\foot{This holds true away from points in $M$
where the action of $G$ is not free.}
The key examples we will discuss are:

\item{1.} 
Topological Yang-Mills theory: 
Here $P$ is $\CA$, the space of all connections and $G$ is $\CG$ the space of gauge transformations. 
$$
\CZ \subset B\CG = \CA/\CG
$$ 
is defined by, for example, 
\eqn\vani{
\CZ= \{ A \in \CA~ \vert~ F(A) = - \ast F(A) \} / \CG}
in the example of $4$-dimensional Donaldson theory. 
\item{2.}
Topological Gravity:
Here $P$ is $\MET(\Sigma)$ and $G$ is $\Diff(\Sigma)$ and we may take:
$$
\CZ \subset \MET ( \Sigma ) / \Diff ( \Sigma )
$$
to be defined by
\eqn\vanii{
\CZ=\{ h \in \MET ( \Sigma )~ \vert~ R(h) = k \} / \Diff ( \Sigma )}
where the Ricci scalar is restricted to be $k = \pm 1$ or $0$, depending on the genus
of $\Sigma$.
\item{3.}
Topological string theory:
Here $P$ is $\MAP ( \Sigma, X ) \times \MET ( \Sigma )$ and $G$ is $\Diff ( \Sigma )$ and
$$
\CZ \subset {\MAP ( \Sigma, X ) \times \MET ( \Sigma ) \over  \Diff(\Sigma)}
$$
is the moduli space of  holomorphic maps defined by 
\eqn\vaniii{ 
\CZ = \{ (f, h ) \in \MAP \times \MET~  \vert~ R(h) = \pm 1, 0~
{\rm and}~ df+J df \epsilon=0)\}/G}

In all these cases, $\CZ$ is obtained by dividing the solutions of a {\it gauge invariant}
equation by the action of the group of gauge transformations\foot{N.B. Although the principal
bundle $P$ and group $G$ are infinite dimensional the spaces $\CZ$ are all finite
dimensional. }.
In the above situations we see that $\CZ$ can be described in terms of the vanishing of
a section of a vector bundle as follows.
The equations \vani, \vanii\ and \vaniii\  define a section, $s$, of a vector bundle 
\eqn\vaniv{
\CV\to P}
The zero set, $\CZ(s)$, is gauge invariant and 
\eqn\vanv{
\CZ=\CZ(s)/G\qquad .}
Since we are writing gauge-covariant equations, the vector bundle, $\CV$,
and section, $s$, are $G$-{\it equivariant}.
Thus if $\CV$ has standard fiber $V$, then $V$ is in a representation $\rho$ of 
$G$ and  we can define a bundle $E=\CV/G\to M= P/G$ as in\foot{In Donaldson
theory $\CV=P\times V$ is naturally trivial over $P$, but  in topological string theory
this is not the case.}:
\eqn\dgrmi{
\matrix{\CV & \mapright{\pi_1} & P \cr
\mapdown{\pi_2} & & \mapdown{\pi_3}\cr
E=\CV/G & \mapright{\pi_4} & M=P/G\cr}}
Moreover, as in \tautequi\ below, if $s\colon P \to {\cal V}$ is a $G$-equivariant section,
then
\eqn\vanvi{
s(p\cdot g) = \rho(g^{-1})\cdot s(p)}
Hence, by \tautequi\ we see that $s$ descends to a section $\bar s$ of $E$.
The situation is summarized by
\eqn\dgrmiii{
\matrix{\CV & \mapleft{s} & P \cr
\mapdown{\pi_2} & & \mapdown{\pi_3}\cr
E & \mapleft{ \bar s} & M=P/G\cr}}
In particular, $\CZ(s)/G = \CZ(\bar s)$. 

The topological field theory path integral essentially constructs a MQ form for
$E$, $\bar s^\ast ( \Phi ( E; \nabla ))$, for some connection $\nabla$ on $E$.
The intersection numbers of homology classes
$H_{{\cal O}_i}\in H_\bullet ( \CZ )$ dual to
$\CO_i \in H^\bullet ( \CZ )$ are, as usual\foot{For simplicity we denote
$\CO_i\in H^\bullet ( M )$ and $i^\ast \CO_i \in H^\bullet (\CZ)$ by the same
thing, where $i$ is the inclusion of $\cZ ( s )$ in $M$.}:
\eqn\tpggcrr{\eqalign{
\# (H_{\CO_1}\cap \cdots \cap H_{\CO_k}\cap D)
& \equiv  \int_{\CZ(\bar s)} \CO_1\wedge -\wedge \CO_k~
\chi( \cok(\nabla \bar s) \to \CZ(\bar s) )\cr
&= \int_{M=P/G} {\bar s}^*(\Phi(E,\nabla)) \wedge \CO_1\wedge - \wedge \CO_k\cr}}
Here $D$ is Poincar\'e dual to $\chi$.

Topological field theory provides expressions for these integrals in terms of {\it local} quantum field theory. 
In order to derive these expressions we must
\item{1.}
Express the integral over $P$, not $M=P/G$
\item{2.}
Choose a connection $\nabla$ on $E$ and write $\Phi ( E, \nabla )$ as a path integral. 

The first issue is quite familiar from ordinary gauge theory.
Conceptually, correlation functions of gauge invariant operators, $\CO$, such as  
Wilson loops, are integrals over gauge inequivalent  fields.
In local field theory we express this as an integral over all gauge fields and divide by 
$\vol~ \CG$. For example, in Yang-Mills theory we write
\eqn\ggecrrs{
\langle \CO \rangle
= \int_{\CA/\CG} d \mu~ \CO
={1\over \vol~ \CG} \int_\CA {dA}~ e^{-S_{Y.M.}} \CO}

We will first address point 1 in secs. 
\ssThePrF\ and \ssABRSTCofPhi\  by trying to lift \tpggcrr\ to an integral over $P$.
After that we address point 2 in sec. \ssAthePNGF.  

\ifig\fibration{Example of a projection.}
{\epsfxsize3.0in\epsfbox{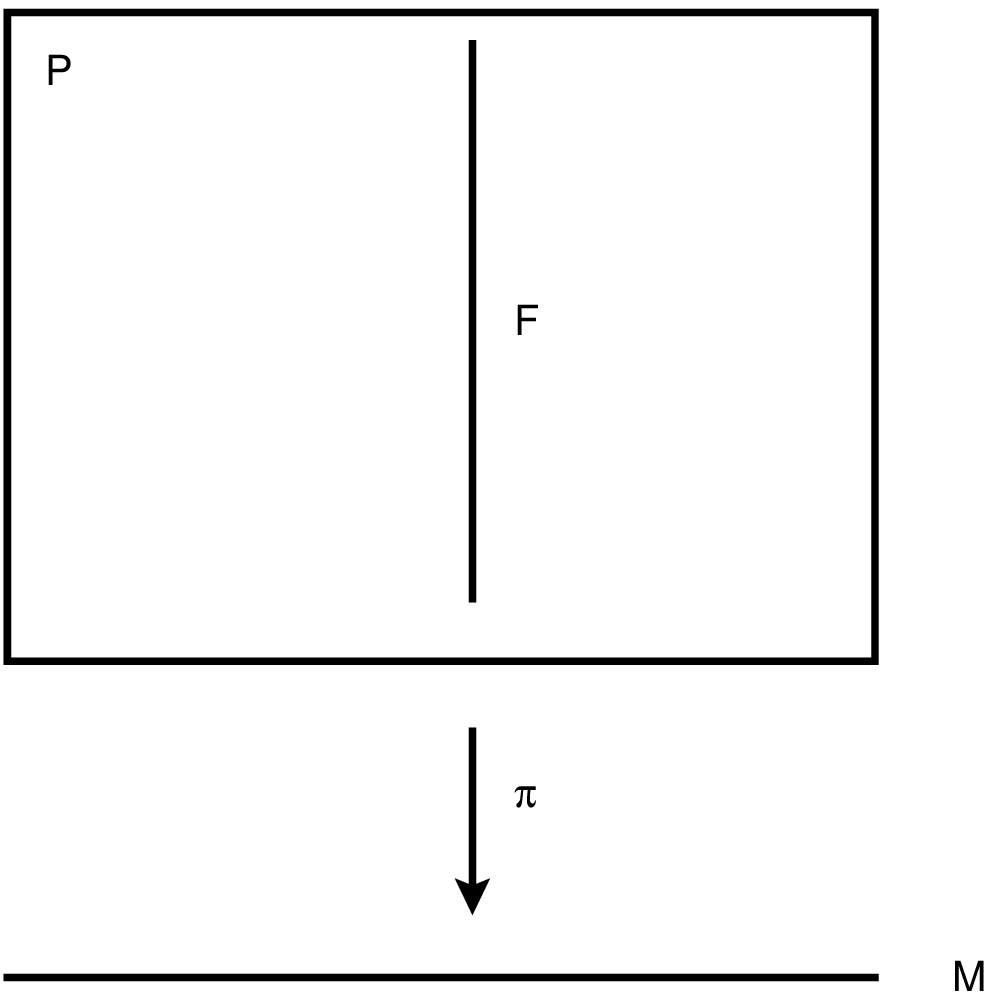}}

\subsec{The Projection Form} 
\subseclab\ssThePrF

Consider the fibration $P\to M$ as shown in  \fibration. 
The fiber might be a vector space (for a vector bundle) or a Lie group (for a principal bundle).
Here the basic problem is: given a form $\xi$ on the base $M$, how can we write $\int_M \xi$
as an integral over $P$?
We will search for a construction of a cohomology class  $\Phi(P\to M)$, such that for $\xi\in H^\bullet ( M )$
\eqn\vertform{
\int_M \xi = \int_P \pi^* \xi \wedge \Phi(P\to M)\quad. 
}

In fact, we will find it much more natural to phrase the problem of constructing a projection 
form in the framework of {\it equivariant cohomology}. 
Thus, we will search for an equivariant differential form
$\Phi ( P \to M ) \in \Omega^\bullet_G ( P )$ such that 
\eqn\verfrmii{
\int_M \xi = \oint_{\epsilon=0,P} \pi^* \xi \wedge \Phi(P\to M)}
where we are using equivariant integration defined in section \sssNL\ above. 

\vskip0.1truein\noindent
{\bf Remarks}
\item{1.}
As noted in \Witdgtr, the limit $\epsilon\to 0$ is  usually a singular limit.
We will find that $\Phi$ is a distribution in $\phi$ so the convergence factor 
is not needed (See, however, remark 
\sssConsPhi.3 below.)
\item{2.}
What we are doing is {\it not} the same as 
Faddeev-Popov (FP) gauge-fixing.
There, one is given a natural measure $dA$ on $P$ and one wishes
to construct a measure $d \mu$ on $\CA/\CG$ so that 
$dA=\pi^*(d \mu) \wedge \mu_{\rm Haar}$.
The FP construction involves (among other things)  an explicit choice of slice,
and a different multiplet of (anti-)ghosts. We describe the FP procedure, from 
the point of view of these lectures, in section \ssFPGF\ below.

\subsec{A BRST  Construction of $\Phi(P\to M$) for Principal Bundles}  
\subseclab\ssABRSTCofPhi

\subsubsec{Characterization of $\Phi(P\to M$)} 

The essential point is that $\Phi(P\to M)$ in \verfrmii\  must contain all the 
vertical directions on the principal bundle.
Recall from sec. \sssVTV\ below that  for any principal bundle, and point $p\in P$ 
we have a canonical mapping 
\eqn\CanIso{
C_p = d R_p: \lieg \to T_p P}
The image of $C_p$ is the space of vertical vectors,  $(T_p P)^{\rm vert}$.

One way to solve the problem posed in the previous 
section is to find an equivariant 
differential  form $\Phi(P\to M)\in \O_G(P)$
that satisfies the three criteria:

\item{(i)}
$\Phi(P\to M)\in \Lambda^{\rm max} (T^*P)^{\rm vert} $
\item{(ii)}
$d_\CC \Phi = 0$, where $d_\CC$ is the Cartan differential 
\CartanDiff, i.e., $\Phi$ is equivariantly closed. 
\item{(iii)}
$\forall p\in P$, $R_p^\ast \Phi = \mu_{\rm Haar}$ where $\mu_{\rm Haar} $ is a
normalized equivariant Haar measure for $G$.
By an equivariant Haar measure we mean an equivariant differential form on
$G$ such that 
\eqn\euihaar{
\oint_{\epsilon,G} \mu_{\rm Haar} = 1}

The three criteria i,ii, and iii  are in close analogy to the three criteria of section \ssUTC\
characterizing the universal Thom class. 
Let us check that these three conditions guarantee that $\Phi$ satisfies \verfrmii. 
Condition $(ii)$ is required since the LHS of \verfrmii\  depends only on 
the cohomology class of $\xi$.
If $F$ is a compactly supported form on $M$ then\foot{Note that pullback commutes with
$d$, so that $\pi^\ast d = d \pi^\ast$; while the image of $\pi^\ast$ is basic, so
that $d \pi^\ast = D \pi^\ast$.}
$\oint_{\epsilon=0,P} \pi^\ast ( dF ) \Phi ( P \to M )
= \oint_{\epsilon=0,P} D\pi^\ast (F)\Phi(P\to M)=0$.
To integrate over the fibers choose a local trivialization $t: ( x, g) \to R_{s(x)} ( g )$
of $\pi^{-1}(U)$ for a patch $U$ in $M$. 
($R_p$ is defined in \rghtgact.)
Using $(i)$ and $(iii)$, one easily checks that the integral over the fiber directions is 1.

In the next two sections we show that one can construct $\Phi$ given a $G$-invariant metric 
on  $P$. 

\ifig\orthcom{Orthogonal complements to gauge orbits define a connection. }{\epsfxsize3.5in\epsfbox{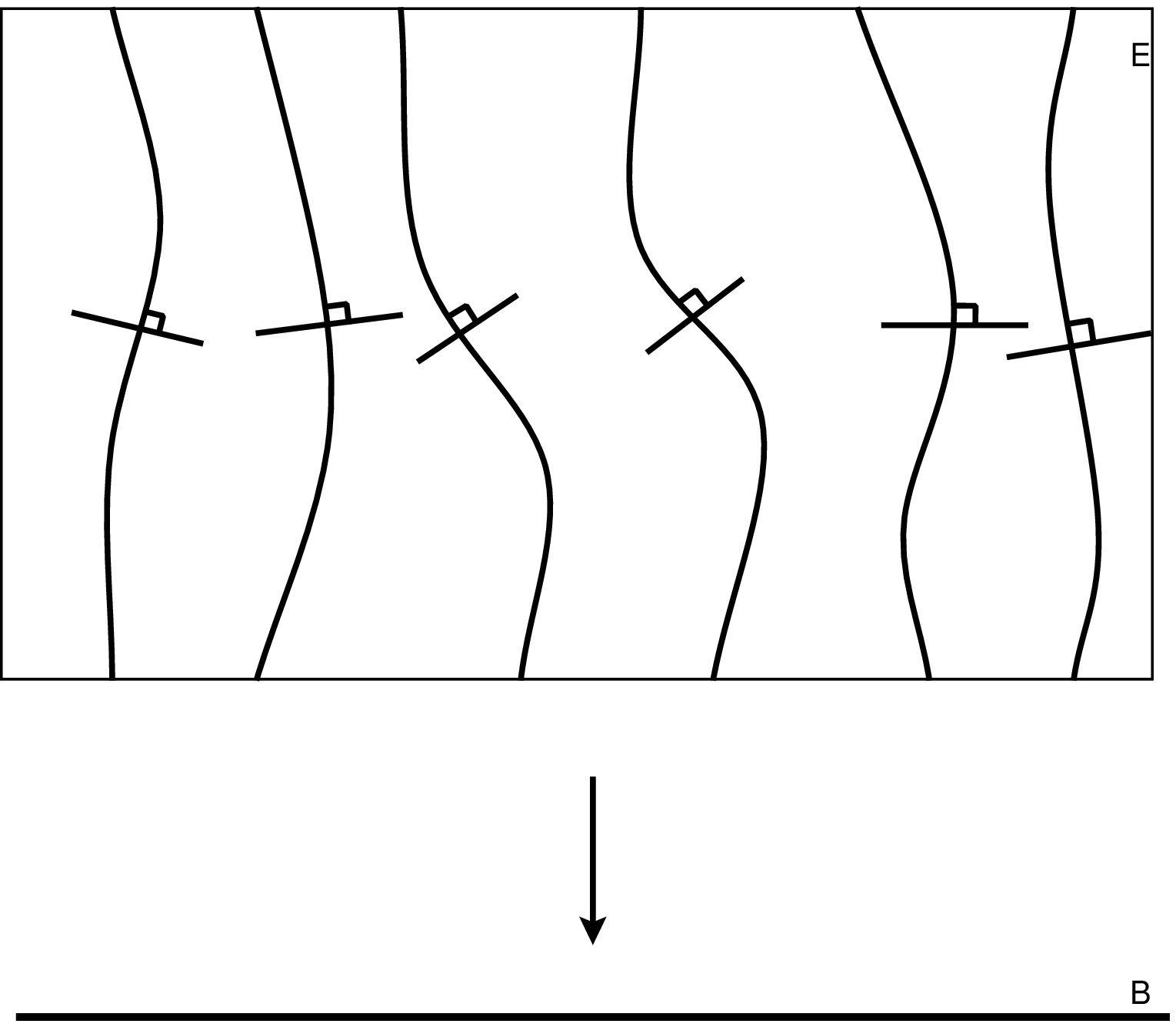}}

\subsubsec{Geometry of Principal Bundles}
\subsubseclab\sssGofPB

We first review some of the differential geometry of principal bundles. 
The essential fact is that, given a $G$-invariant 
metric $( \cdot, \cdot )_{TP}$ on a principal bundle, there is automatically 
associated to it a natural connection. 
As we have seen in \ssCVBFB, a 
connection is simply a $G$-equivariant choice of
horizontal tangent vectors.
Using a $G$-invariant metric one simply defines the connection by declaring the
horizontal subspaces to be  the orthogonal complements to the gauge orbits as
in \orthcom.

As we showed in section \sssVTV\ we always have a map
$
C_p: \lieg\to T_p P
$
as in \defofcee. 
Because we have a metric there is an adjoint,
defining a Lie-algebra valued $1$-form: 
\eqn\cdgger{
C^\dagger_p:  T_p P\to \lieg}
In terms of  $C$ and $C^\dagger$ 
\eqn\genvrt{
\Pi^v  = C {1\over C^\dagger C} C^\dagger}
is the projection onto the vertical tangent space and the connection is 
\eqn\metconn{
\Theta = C^{-1} \Pi^v=  {1\over C^\dagger C} C^\dagger 
\qquad . }
$G$-invariance of the metric guarantees that \pllbckcn\ will be satisfied. 

\exercise{Curvature}

Show that, on horizontal tangent vectors 
$$
d \Theta(X,Y) = {1\over C^\dagger C} d C^\dagger [X,Y]
$$

\endexercise

\subsubsec{Construction of $\Phi$} 
\subsubseclab\sssConsPhi

To motivate the solution to the problem, we return to the 
Lie-algebra-valued 1-form, $C_p^\dagger 
\in T_p^\ast P\otimes \lieg$ defined in \cdgger. 
To be explicit, let $dp^i$ be a basis of one-forms on $P$.
$C^\dagger$ can be written as $T_a (C^\dagger)^a_i dp^i$. 
Moreover, $\ker~ C_p^\dagger=(T_p P)^{\rm horizontal}$. 
Thus we should take a wedge product 
$$
C_p^\dagger\wedge \cdots \wedge C_p^\dagger
$$
with one factor
for each dimension of $(T_pP)^{\rm vertical}$. 
Since $\lieg \cong (T_p P )^{\rm vertical}$ we can use the metric on the Lie algebra,
$\lieg$, $( \cdot ,  \cdot )_\lieg$ to contract the Lie algebra indices to get a vertical 
form\foot{This assumes that $G$ acts freely.
In the case of topological Yang-Mills, the group of gauge transformations does not act freely
on reducible connections.}.
The contraction of Lie algebra indices is most elegantly done by introducing 
odd vectors $\eta_a\in \Pi\lieg$ corresponding to an ON basis of $\lieg$ and forming:
\eqn\inteta{
\int_{\Pi \lieg} d\eta~ e^{( \eta, C_p^\dagger )_\lieg }
\in \Lambda^\bullet (T_p^\ast P)^{\rm vertical}.}
The problem is that this form is not closed, and does not satisfy criterion $(iii)$. 

The above two problems can be fixed by introducing a BRST exact action. 
We now take the complex to be: 
\eqn\projcplx{\CC_{\rm projection}^\ast =
S( \lieg\dual~  )\otimes \O^\bullet ( P ) \otimes ( S ( \lieg ) \otimes \Lambda ( \lieg ))}
The first two factors are identified with the Cartan-model complex of equivariant
cohomology while the last two factors admit an action of $\lieg$ 
from the adjoint representation.
We introduce generators $\lambda_a$ and $\eta_a$ of ghost number $-2$ and $-1$,
respectively, for the last two factors. 
Then we can augment the Cartan differential on the first two factors by 
\foot{
Note that $\phi \in \lieg\dual$, so we must understand
$\CL_\phi \lambda = [ \phi, \lambda ] \in \lieg\dual$ 
as a coadjoint action.
We then must use the metric on $\lieg$ to identify $[ \phi, \lambda ]\in \lieg$. 
In what follows we will use the metric on $\lieg$ to identify $\phi\in \lieg$.
}
\eqn\cardiffii{\eqalign{
Q_\CC&=d_\CC\otimes 1 + 1\otimes \delta\cr
Q_\CC \pmatrix{ \lambda_a \cr \eta_a\cr} &= 
\pmatrix{0 & 1\cr  -\CL_\phi & 0\cr} 
\pmatrix{ \lambda_a \cr \eta_a\cr}  \cr}}
Note that $Q_\CC^2=0$ only on the $G$-invariant subcomplex. 

Using the metric on $\lieg$ we can contract $C^\dagger$ with $\lambda$ to obtain 
an element of $\O ( P ) \otimes S( \lieg )$. 
The contraction is 
\eqn\cntrct{
( \lambda, C^\dagger )_\lieg= \lambda_a ( C^\dagger )^a_i dp^i}
where $\lambda_a$ are coordinates with 
respect to an ON basis. 

\noindent
{\bf Proposition:} The form 
\eqn\projfrm{\eqalign{
\Phi_{\rm projection} ( P \to M) &=  
\bigl( {1\over 2 \pi i} \bigr )^{\dim~ G} \int_{\widehat \lieg} 
d \lambda d \eta~ e^{Q_\CC \Psi_{\rm projection}}\cr
\Psi_{\rm projection} & = i  ( \lambda, C^\dagger )_\lieg \in \O^1 ( P )\cr}}
satisfies criteria $(i)$, $(ii)$ and $(iii)$ above and therefore is a representative of a
projection form, $\Phi(P\to M)$. 
Here
$Q_\CC= 1 \otimes d \otimes 1 - \iota_\phi \otimes1 \otimes 1 + 1 \otimes 1 \otimes \delta$
on \projcplx.

\vskip0.1truein\noindent
{\it Proof}: 
To prove property $(i)$ we expand:
\eqn\expdprj{
Q_\CC ( \lambda, C^\dagger )_\lieg = ( \eta, C^\dagger )_\lieg + ( \lambda, dC^\dagger )_\lieg+
(\lambda, C^\dagger C \phi)_\lieg}
The second and third terms arise from the Cartan differential on $\Omega^\bullet (P)$.  
To justify the third term write: 
$\lambda_a (C^\dagger)^a_i \iota_\phi dp^i = \lambda_a (C^\dagger)^a_i (C\phi)^i
= (\lambda, C^\dagger C \phi)$, where $\phi$ is considered in $\lieg$. 
Thus we have: 
\eqn\prvrtfm{\eqalign{
\int_{\widehat \lieg} 
d \lambda d \eta~ e^{Q_\CC \Psi }
=& (2 \pi)^{\dim~ G} \int_{ \Pi\lieg} d \eta~ e^{i(\eta,C^\dagger)_\lieg}~
\delta(C^\dagger C\phi + d C^\dagger )\cr
=& (2 \pi)^{\dim~ G}~ \delta( C^\dagger C\phi + d C^\dagger ) \int_{ \Pi \lieg} d \eta~ 
e^{i(\eta,C^\dagger)_\lieg }\cr}}
is a top vertical form. 
Hence it is manifest that $(i)$ is satisfied. 

To prove equivariant closure we write, in analogy to \comint: 
\eqn\comintii{\eqalign{
d_\CC \int d \lambda d \eta~ e^{ Q_\CC ( \Psi )} 
=& \int d \lambda d \eta~ [Q_\CC - \delta ] e^{Q_\CC (\Psi)}\cr
&=\int d \lambda d \eta~ [ - \delta ] e^{Q_\CC (\Psi)}\cr}}
where we have used $Q_\CC^2= -\CL_\phi$ which vanishes on $\Psi$.
Moreover, $\delta$ acts as: 
$$
\delta= \eta_a {\p \over \p \lambda_a} 
- f^a_{bc} \phi_b \lambda_c {\p \over \p \eta_a}
$$
so we may drop this operator by the properties of the Berezin integral.  
Thus $\Phi$ is equivariantly closed, $d_\CC \Phi=0$. 

It remains to check that the form is properly normalized, property $(iii)$.
To see this note that $R_p^\ast C^\dagger$ is a left-invariant 1-form on $G$ with values
in the Lie algebra $\lieg$. 
The value of this form on a tangent vector $X_e\in T_e G$ may be computed from 
$$
\langle R^\ast C^\dagger, X_e \rangle
= \langle C^\dagger, R_\ast X_e \rangle
= \langle C^\dagger, C X_e\rangle
= C^\dagger C X
$$ 
where in the last equality we have used the isomorphism $T_e G \cong \lieg$.
It follows that, if $\theta^a$ is an ON basis of left-invariant 1-forms on $G$ dual to a 
basis $T_a$ for $\lieg$, then 
$$
R_p^\ast C^\dagger = \sum_a C^\dagger C (T_a) \theta^a \in \Omega^1(G;\lieg)
$$
and 
\eqn\gpvol{\eqalign{
R_p^\ast \int_{ \Pi \lieg} d \eta~ e^{i(\eta,C^\dagger)_\lieg }
=& \int_{ \Pi \lieg} d \eta~  e^{i(\eta,R_p^\ast C^\dagger)_\lieg } \cr
=& i^{\dim~ G} \det (C^\dagger C)~ \theta^1\wedge - \wedge 
\theta^{\dim~ G} \cr}}
where $\theta^1\wedge - \wedge \theta^{\dim~ G}$ is the normalized Haar measure on $G$. 
Combining this with \prvrtfm\ shows that $\Phi$ pulls back to a correctly normalized
equivariant Haar measure, thus completing the proof. $\spadesuit$

\ifig\varradius{A circle bundle with metric.}
{\epsfxsize3.0in\epsfbox{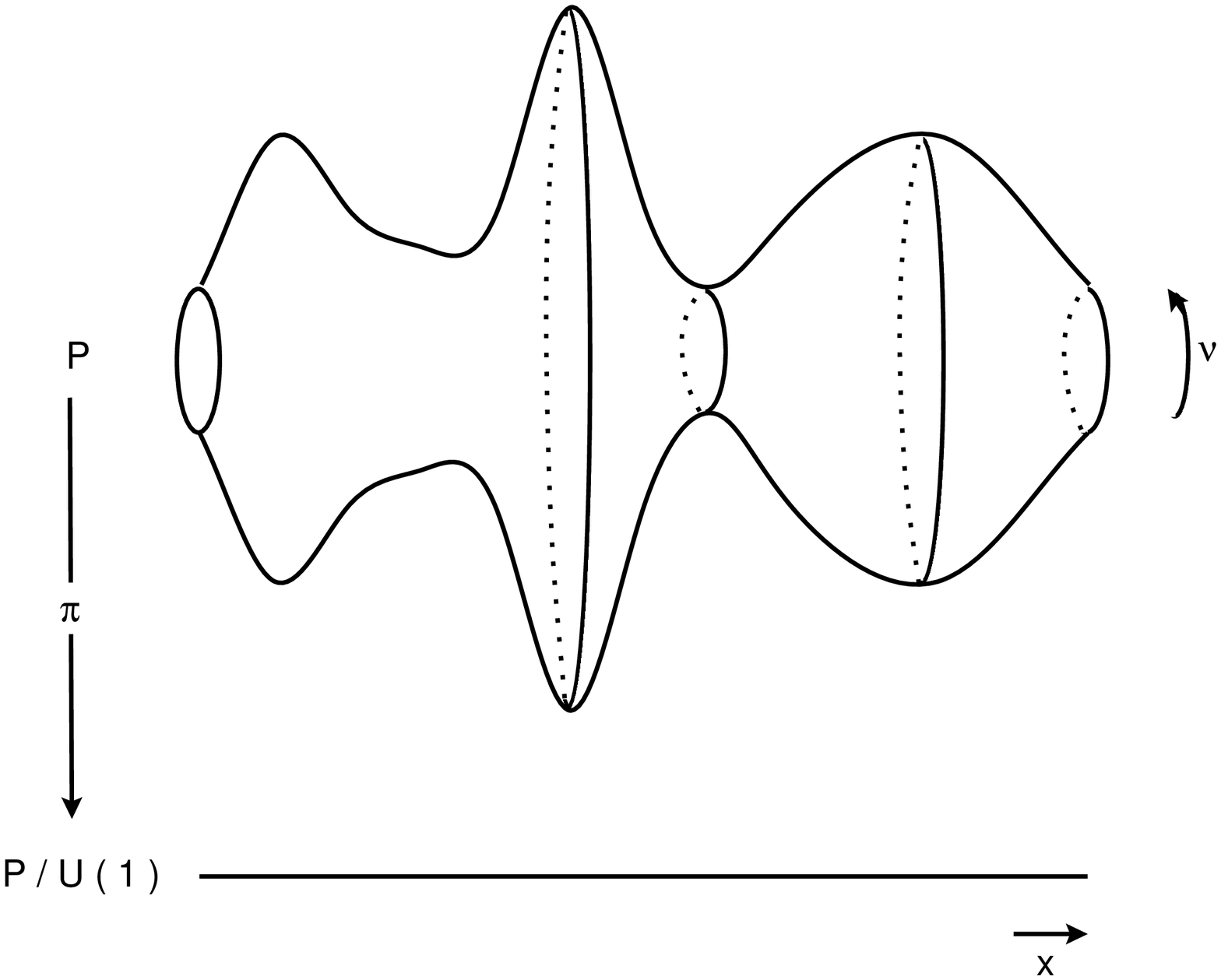}}

\noindent
{\bf Example:} Consider a principal $U(1)$ bundle, $P$, with coordinates $(x,\nu)$.
$x$ is a coordinate on the base and $0\leq \nu < 2 \pi$ is a periodic fiber coordinate.
$P$ has an invariant metric $dx^2 + g(x) d \nu^2$, so the circles have a variable radius
as in \varradius.

The standard coordinate on the group $U(1)$ will be called $\theta$. 
The Lie algebra is $\IR\cdot {\p \over \p \theta}$. 
Then we may easily compute\foot{The normalization for the dual pairing has been
chosen to be $\left\langle d \nu, {\partial \over {\partial \nu}} \right\rangle = 2 \pi$}: 
\eqn\expli{\eqalign{
R_{x,\nu} (e^{i \theta} )=& ( x,\nu+\theta)\cr
C\colon {\p \over \p \theta} & \rightarrow  {\p \over \p \nu}\cr
C^\dagger & = g(x) {d \nu \over 2 \pi} \otimes {\p \over \p \theta}\cr
C^\dagger C & = g(x) \cr}}
It is easily seen that 
\eqn\explii{
\int d \eta e^{i (\eta {\p \over \p \theta},
C^\dagger)_{u(1)}} = i g(x) d \nu}
does not satisfy good properties.
For example, if $F(x)$ is compactly supported there is no reason for
$\int_P \pi^\ast (dF(x)) g(x) d\nu$ to vanish. 
On the other hand, one easily checks: 
\eqn\expliii{
{1\over 2 \pi i} \int  d \eta d \lambda~ e^{ i Q(\lambda, C^\dagger)_{u(1)} }
= {d \nu \over 2 \pi} g(x) \delta [ g(x) \phi - g'(x) dx d\nu ]}
which satisfies 
\eqn\expliv{
\oint_{\epsilon=0,U(1)} {d \nu \over 2 \pi}~
g(x) \delta [ g(x) \phi - g'(x) dx d\nu) ] =1 . }

\vskip0.1truein\noindent
{\bf Remark \sssConsPhi.1:} 
Note that one by-product of the above discussion is that the integral over $\phi$ is
delta-function supported on 
$$
\phi\rightarrow - {1\over C^\dagger C} d C^\dagger
$$
which is the horizontal part of the curvature of $P\to M$

\vskip0.1truein\noindent
{\bf Remark \sssConsPhi.2:}
In 2D gravity the metric on  the diffeomorphism group is {\it not} Weyl invariant.
This leads to extra complications when trying to interpret the Weyl group as a gauge
group. 

\ifig\volcollap{A circle bundle with metric.}
{\epsfxsize3.0in\epsfbox{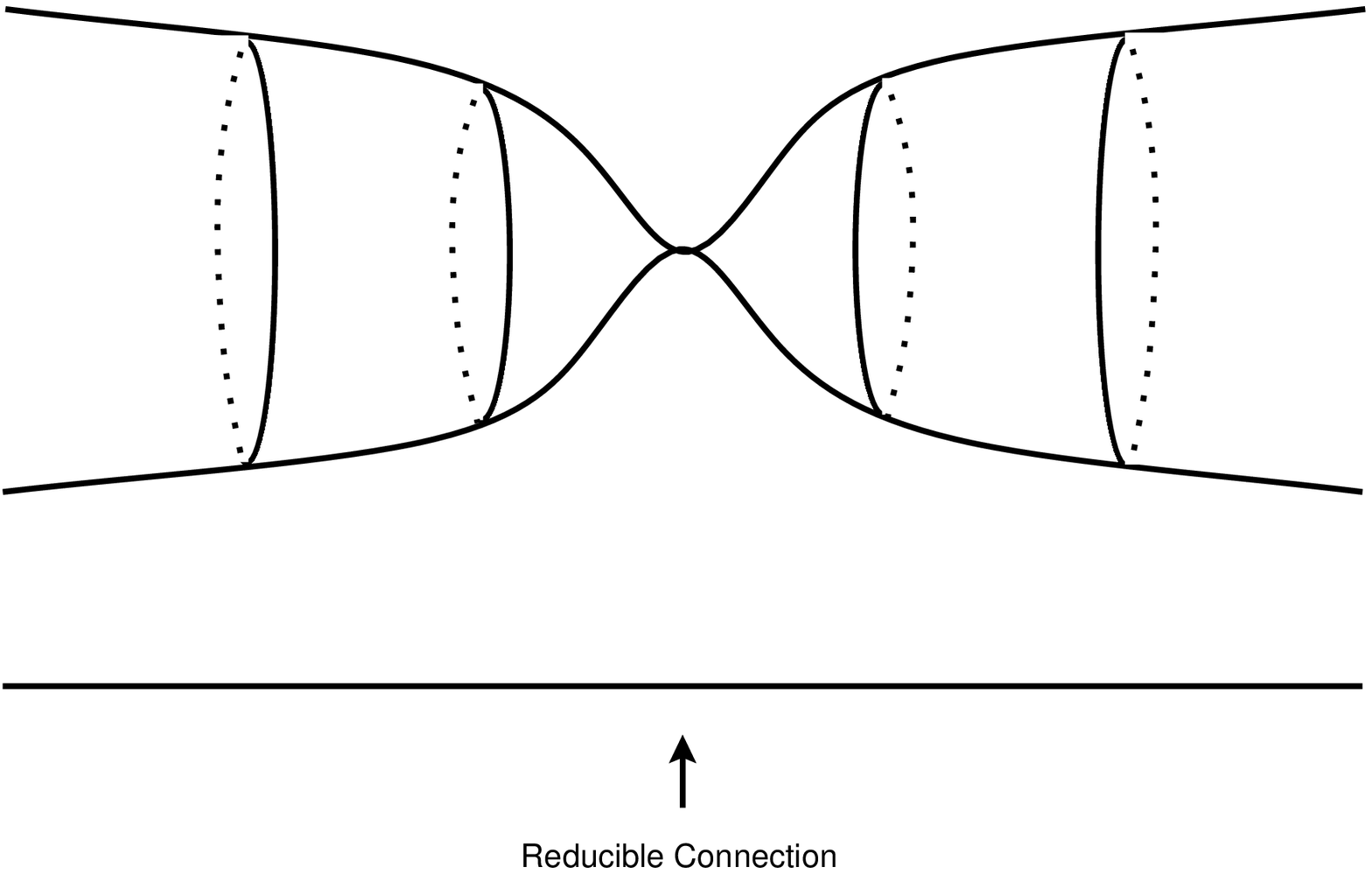}}

\vskip0.1truein\noindent
{\bf Remark \sssConsPhi.3}. 

One of the main technical obstacles in Donaldson theory is the problem that on subspaces
of  ``reducible gauge connections'' the operator $C^\dagger C$ has zero modes, rendering
the above and subsequent discussions more complicated. 
A picture of what happens over a reducible connection is the following.  
We lose some dimensions in the fiber and 
the volume collapses to zero as in \volcollap. 
It might be of interest to regulate these 
singularities using $\epsilon>0$ in \verfrmii. In topological 
Yang-Mills a closely related idea of adding 
a mass term to the topological Lagrangian 
 has led to some success \wittsusygt.

\subsec{Assembling the Pieces: No Gauge Fixing. }
\subseclab\ssAthePNGF

Now that we have constructed the projection form let us return to the problem of writing 
\tpggcrr\ as a field theory correlator with $Q$-exact action. 
Consider again \dgrmi\ and \dgrmiii\ and denote the group $G$ by $G_{\rm gauge}$. 
We use the projection form  to write \tpggcrr\ as an integral over $P$:
\eqn\assmbli{\eqalign{
\int_{M=P/G_{\rm gauge}}& {\bar s}^*(\Phi(E,\nabla))  \wedge \bigwedge_i \CO_i\cr
&= \oint _{P } \pi_3^\ast ( {\bar s}^\ast (\Phi(E,\nabla) )
\wedge \Phi_{\rm projection} (P \to M ) \wedge \bigwedge_i \pi_3^\ast ( \CO_i )\cr}}
where the projections are defined in \dgrmi\ above. 
The classes $\CO_i$ are more naturally expressed in terms of their pullbacks: 
$\pi_3^\ast (\CO_i)$.
These pullbacks will be local operators in the field theory. 
We will denote these by $\tilde \CO_i$. 

\subsubsec{Equivariant Thom Class}

Finally we must choose a connection $\nabla$ on $E$ and build the MQ form. 
Suppose first that we were trying to construct the MQ form for $\CV\to P$, and were trying
to localize to $\CZ(s)$.
In this case we would follow the logic of chapter \sINandTIR. 
Let $\CV$ have standard fiber $V$ with a metric $( \cdot , \cdot )_V$, defining an orthogonal 
group $SO(V)$. 
The bundle $\CV\to P$ is itself associated to a  principal $SO(V)$ bundle $\tilde P\to P$:
\eqn\dgrmiv{
\matrix{\tilde P\times V & \rightarrow & \tilde P \cr
\downarrow \tilde \pi_2 & & \downarrow \tilde\pi_3\cr
\CV =\tilde P\times_{SO(V)} V&  {\buildrel \pi_1 \over \rightarrow } & 
 P=\tilde P/SO(V) . \cr}}
Choose an $SO(V)$ connection $\nabla_s$.
Then corresponding to a universal Thom class, $U_s$, in
$SO(V)$-equivariant cohomology we use
$$w_s ( U_s ) = \tilde \pi_2^\ast ( \Phi ( \CV, \nabla_s )), $$
to obtain
$\Phi ( \CV, \nabla_s ) \in \Omega^\bullet_{SO ( V )} ( \widetilde P \times V )$. 
However, we are {\it not} trying to localize to $\CZ(s)$ but to $\CZ(\bar s)$.
Moreover, $\nabla_s$ does {\it not} simply descend to a connection on $E$.
Even if  $\CV$ were trivial, i.e.  $\CV=P\times V$,  then $E=\CV/G_{\rm gauge}$ would 
still have a nontrivial connection $\nabla_g$,
since it  is associated to the principal bundle $P\to P/G_{\rm gauge}$. 
As we saw in section \sssGofPB\  above $P$ has a nontrivial connection.
We must account for both $SO(V)$ and $G_{\rm gauge}$. 

We see that there are {\it two groups} involved in the construction of the MQ class
of $E$ and that the relevant universal Thom class $U_{\rm total}$ is constructed using  
$G_{\rm total} =SO(V)\times G_{\rm gauge}$-equivariant cohomology. 
The fiber $V$ is a representation of $G_{\rm total}$. 
The formulae above continue to hold, but 
we must remember that a Weil homomorphism 
involves a sum of connections and curvatures.
That is, for instance
\eqn\nabtot{\eqalign{
\nabla=& \nabla_s \oplus \nabla_g\cr
R =& R_s + R_g\cr}}
$\bar w_s \bar w_g (U_{\rm total})$ descends to a differential form $\Phi(E,\nabla_s \oplus \nabla_g)$
in $\O ( E )$. 
The whole thing fits together as:
\eqn\dgrmii{\matrix{
\tilde P \times V & \mapright{} & \tilde P \cr
\mapdown{\tilde \pi_2} & & \mapdown{\tilde\pi_3}\cr
\CV =\tilde P \times_{SO(V)} V &  \mapright{\pi_1} & P=\tilde P/SO(V)\cr
\mapdown{\pi_2} & & \mapdown{\pi_3}\cr
E =\CV/G_{\rm gauge} & \mapright{\pi_4} & M= P/G_{\rm gauge}=\cr
&  &  \tilde P / ( SO(V) \times G_{\rm gauge})\cr}}

Although we have a sum of connections, from the point of view of field theory
the nature of the two connections is very different. $\nabla_s$ involves local expressions 
in spacetime, while $\nabla_g$ is {\it nonlocal in spacetime}. 
The nonlocality of $\nabla_g$ is eliminated by lifting the form to $P$ in 
the way we describe next.

Referring back to \dgrmiii,  we use the trivial observation: $\bar{s} \pi_3= \pi_2 s$, to write:
$$
\pi_3^\ast {\bar s}^\ast  \Phi( E, \nabla_s \oplus \nabla_g )
= s^\ast  \pi_2^\ast ( \Phi ( E, \nabla_s \oplus \nabla_g ))
$$
The form $ \pi_2^\ast ( \Phi ( E, \nabla_s \oplus \nabla_g ))$ is in $\O^\bullet ( \CV )$ and
arises from a Thom class for $SO ( V )\oplus  G_{\rm gauge}$-equivariant cohomology. 
Because it is inconvenient to make explicit horizontal projections, we use the {\it Weil
model} for $SO ( V )$-cohomology. 
Thus we choose a connection $\nabla_s$ and corresponding Weil homomorphism $w_s$. 
On the other hand, because of the nonlocality of $\nabla_g$ it is convenient to use the
{\it Cartan model} for $G_{\rm gauge}$-equivariant cohomology. 
Thus, the universal Thom class will take values in the complex
\eqn\mxcplx{
U_{\rm total} \in \cW( so (V) ) \otimes S ( \lieg_{\rm gauge}\dual )\otimes \O^\bullet (V)}
By applying $w_s$ we have a form 
\eqn\carcplx{\eqalign{
U_{\rm gauge} & \in S( \lieg_{\rm gauge}\dual~ )\otimes\O^\bullet ( \CV )\cr
w_s (U_{\rm total} )&=\tilde \pi_2^\ast ( U_{\rm gauge}) \cr}}
In fact, $U_{\rm gauge}$ will be $\lieg_{\rm gauge} $-invariant and hence is 
a $G_{\rm gauge}$-equivariant form in $\O_{G_{\rm gauge}}(\CV)$. 

As we saw in section \ssCartMod, when applying the Cartan model for the MQ form,
we must take a horizontal projection.
Thus strictly speaking we must choose a connection, $\nabla_g$, corresponding to the
Weil homomorphism $w_g$ and consider
$w_g (U_{\rm gauge})^{\rm horizontal} \Phi(P\to M)$ in \assmbli. 

At this point the projection form comes to the rescue. 
We need not apply $w_g$, nor need we enforce the horizontal projection.
All of this is done automatically for us by the projection form $\Phi ( P \to M)$.
Being fully vertical we have: 
\eqn\keyrep{
w_g ( U_{\rm gauge} )^{\rm horizontal} \Phi ( P \to M) 
= w_g ( U_{\rm gauge} ) \Phi ( P \to M )}
Moreover, as we have seen in the previous section, $\Phi$ is $\delta$-function supported
on values of  $\phi$ given by the curvature of $P\to P/G_{\rm gauge}$, hence we needn't
apply $w_g$. 

Thus, we have finally found a home for $\pi_2^\ast (\Phi (E, \nabla_s \oplus \nabla_g))$. 
It should lie in the equivariant cohomology $\O_{G_{\rm gauge}}(\CV)$ and can be 
identified with $U_{\rm gauge}$ which is related to $U_{\rm total}$ by  \mxcplx\ and \carcplx. 
The pullback by $s$, $s^\ast (U_{\rm gauge})$, will be an equivariant differential form on $P$. 

\subsubsec{$Q$-exact actions} 
\subsubseclab\sssQexct

We are now finally ready to express the intersection numbers \tpggcrr\  as correlators with
a $Q$-exact action. 
We have managed to write the intersection numbers on $\CZ$ in terms of integration of 
equivariant differential forms on $P$, i.e. \tpggcrr\ can be written as: 
\eqn\finfrmi{
\oint_{\epsilon=0,P} 
s^\ast ( U_{\rm gauge}) \wedge \Phi ( P \to M) \wedge \bigwedge_{i=1}^k
\tilde \CO_i}
By the results of the above sections, each of the first two factors can be written in QFT
language in terms of a $Q$-exact action. 
Thus we have the general integral formula for the intersection
 numbers in 
$\CZ(\bar s)\subset M$ generalizing \locii:
\eqn\ajwfrm{\mathboxit{
\langle \hat \CO_1  \cdots  \hat \CO_k  \rangle
=  \int_{\CZ(\bar s)}
\widetilde \CO_1\wedge -\wedge \widetilde \CO_k~ \chi(\cok(\nabla \bar s))}}
where $\hat \CO_k$ are superspace representatives of the forms $\tilde \CO_k$ and
the measure for the correlation functions is: 
\eqn\ajwfrmi{
\mathboxit{
\langle \hat \CO \rangle \equiv  
{1\over \vol~ \CG} \int_{\lieg\times \hat \CS } d \phi \hat \mu~
e^{Q(\Psi )}~  \tilde \CO}}
where the
superspace 
 $\hat \CS= \widehat  \CV^\ast  \times \hat \lieg$  has 
functions
generated by coordinates $A$ and $\psi$ for the base, $P$, of $\CV^\ast$, coordinates
$\rho$ and $\pi$ for the fibers of $\CV^\ast$, and coordinates $\lambda$ and $\eta$ 
for $\lieg$.
The superspace measure is standard: 
$$
\hat \mu = 
(dA d \psi)(d \pi d\rho )( d \lambda   d\eta) 
$$
The action of Q will be the Cartan-model action for $G_{\rm gauge}$ -equivariant 
cohomology.
We explicitly apply the Chern-Weil homomorphism to obtain the connection $\nabla_s$
and need not talk about $SO(V)$-equivariant cohomology.
Thus $Q$ is defined by \brsii\ and \cardiffii.
Explicitly
$$
Q=(d-\iota_\phi)\otimes 1\otimes 1 
+ 1\otimes Q_\CC\otimes 1 + 1\otimes 1\otimes \delta$$
on $(S( \lieg_{\rm gauge}\dual~ )\otimes \hat\CF( P )) \otimes \hat\CF(V^\ast ) \otimes \hat\CF(\lieg_{\rm gauge} )$. 
(We oversimplify somewhat assuming that $\CV^\ast =P\times V^\ast$ 
is a trivial bundle. )
The gauge fermion has the form:
\eqn\ggefrm{
\mathboxit{
\Psi  = \Psi_{\rm localization} + \Psi_{\rm projection} }}
Here we may take from \projfrm: 
\eqn\anthrfrm{
\Psi_{\rm projection}  = i  (\lambda,C^\dagger)_{\lieg_{\rm gauge}}}
for the projection gauge fermion. 
The localization gauge fermion can be represented in many ways, as we saw above.
In accord with \gfunv, \mxcplx\  we take:
\eqn\projnlc{\eqalign{
\Psi_{\rm localization}
=& \rho_a ( i s^a -t \theta^{ab} \rho_b + t \pi_a)\cr
=&  i\langle \rho,s\rangle - t(\rho,\theta \cdot \rho)_{V^\ast} + t ( \rho, \pi)_{V^\ast } \cr}}
where $\theta$ refers to the $SO ( V )$ connection $\nabla_s$. 

This representation will strike many readers as utterly bizarre, but we will see that correlation 
functions of topological field theories of cohomological type all fit into this scheme. 

\vskip0.1truein\noindent
{\bf Remarks:}
\item{1.}
The measure $d \phi \hat \mu$ in 
\ajwfrmi\  
may look a little strange, especially since it appears that 
the number of fermionic and bosonic degrees of freedom differ.
This is related to the use of the Cartan differential which does {\it not} square to zero,
$Q^2= - \CL_\phi$, but is only zero on the invariant subcomplex. 
Since $Q$ does not square to zero there need not be equality of degrees of freedom. 
If we work with the Weil model or the BRST model of equivariant cohomology, then we
also introduce $c^a$, thus restoring the balance between commuting and anticommuting 
degrees of freedom. 
\item{2.}
The vector bundle,  $\cok~ \nabla \bar s \to \cZ ( \bar s )$, is crucial in describing the
above integrals, but the operator is somewhat awkward to work with.
As usual, it is better to speak about  $\nabla s: T_p P\to T_{s ( p )} \CV$.
Since $s$ is gauge-covariant this has $\infty$-dimensional kernel and cokernel.
However, the operator
\eqn\genbgo{
\mathboxit{
\bO = \pmatrix{ \nabla s\cr C^\dagger\cr}: 
T_p P \to T_{s ( p )} \CV \oplus \lieg}}
defines  equivariant vector bundles, $\ker~ \bO$ and $\cok~ \bO$, with {\it finite} dimensional
fibers which descend to $\ker~ \nabla \bar s$ and $\cok~ \nabla \bar s$. 
The operator $\bO$ is the operator appearing the the fermion kinetic terms in the complete 
lagrangian, and plays a crucial role in the following chapters. 
\item{3.}
As we will discuss in detail later the moduli spaces are in fact not smooth.
This leads to extra complications.
If there are orbifold singularities then we compute an orbifold Euler class in \ajwfrm.
For worse singularities we do not know a general prescription. 

\subsec{Faddeev-Popov Gauge Fixing}
\subseclab\ssFPGF 

We now describe how the standard FP gauge-fixing procedure fits into the formalism of
this paper. 
We will use the language of Yang-Mills theory although the same considerations 
apply to gravity. 

What one would like to do is construct a measure $d \bar \mu$ on $\CA/\CG$ such that 
we can identify 
\eqn\sfpi{
\pi^\ast (d \bar \mu ) d \mu_{H} = dA~ e^{-I_{\rm Y.M.}}}
where $d \mu_{H}$ is an $A$-independent standard Haar measure on the group.

In order to separate the gauge degrees of  freedom from gauge inequivalent degrees 
of freedom one trivializes the principal bundle $\CA\to \CA/\CG$.
This can be done by choosing a local cross section,  or, equivalently, by choosing
a local slice $\CZ\subset \CA$. \foot{There are topological obstructions to 
choosing global sections of $\CA$ associated with the ``Gribov ambiguity'' \singgrib.
These do not affect our calculations since we are constructing a local measure.}

A local cross section can be defined by considering a gauge noninvariant function 
$\CF$ on $\CA$.
We will assume that $\CF$ takes its values in a vector space $V$, that is: 
$\CF\colon \CA\to  V$.
The statement that this defines a good gauge is the statement that
\eqn\sfpii{
\forall A\in \CA \qquad \exists ! ~~ \bar g_A \in \CG~ {\rm such}~ {\rm that}~
\CF[A^{\bar g_A} ] =0}
For example, in nonabelian gauge theory one often takes $\CF[A] = \p_\mu A^\mu$, 
which fixes the gauge, once suitable boundary conditions are imposed. 

Let us consider the trivial $V$ bundle over the gauge group
\eqn\sfpiii{
E= \CG \times V}
For any $A\in \CA$ we can define a section of this bundle:
\eqn\sfpiv{
s_A\colon g \longmapsto s_A ( g ) = (g, \CF [A^g] )}
Since $\CF$ defines a good slice, this section will have a unique zero.
Hence, if  $\eta[\CZ(s_A)\hookrightarrow \CG]$ is the Poincar\'e dual to the zero section, we
have:
\eqn\sfpv{
\int_{\CG}  \eta[\CZ(s_A)\hookrightarrow \CG] = 1}
We will now rewrite this formula using the MQ formalism. 
We will choose a metric $( \cdot, \cdot )_V$ and the trivial connection on $E$ in order to 
construct the MQ representative of the Thom class $\Phi(E)$ in the standard way: 
\eqn\ThomFP{
\Phi ( E  ) = \left ( {1\over{ 2 \pi}} \right )^{\dim~ V}
\int_{\widehat{ V}^\ast} d \bar c d \bar{\pi} \exp \left [ Q ( \Psi_{\rm g.f.}) \right ]}
where $Q= d_E + \delta$  is the sum of the exterior derivatives on $E$ and $V$: 
\eqn\CAEA{\eqalign{
\delta \bar c_a =& \bar{\pi}_a\cr
\delta \bar{\pi}_a =& 0\cr}}
And, in the standard way: 
\eqn\FPGF{\eqalign{
Q \Psi_{\rm g.f.}
=& Q \left ( -i \langle  \bar c, x \rangle -  ( \bar c, \bar{\pi} )_{V^*} \right )\cr
=& -i \langle \bar{\pi}, x \rangle + i \langle \bar c, d_E x \rangle - ( \bar{\pi}, \bar{\pi })_{V^\ast} \cr}}
where $x$ is a fiber coordinate.

The pullback, $s_A^\ast ( \Phi )$, defines a top-form on the group $\CG$.
Using the basic tautology \bsctaut:
\eqn\sfpvi{
\hat \CF( \CG ) \qquad \longleftrightarrow \qquad \O^\bullet ( \CG )}
we may write instead:
\eqn\ThomFPi{
s_A^\ast  (\Phi ( E \to \CG )) = \left ( {1\over{ 2 \pi}} \right )^{\dim~ V}
\int_{\widehat{ V}^\ast} d \bar c d \bar{\pi}~
\exp \biggl( Q\bigl[ \langle \bar c, \CF[A^g]\rangle - (\bar c, \bar{\pi})_{V^\ast} \bigr] \biggr)}
where now $Q=Q_\CG + \delta$, and $Q_\CG$, the DeRham differential reads, under 
the correspondence \sfpvi: 
\eqn\sspcque{\eqalign{
c & \longleftrightarrow g^{-1} d g\cr
Q g& = g c\cr
Q g^{-1} & = -c g^{-1} \cr}}
The standard superspace measure \supmeas\ is $\hat \mu_\CG =  \mu_H(g) dc$. 
Thus equation \sfpv\ can be written 
\eqn\sfpvii{
1= \int dc~ \mu_H ( g )~ \widehat \eta [ \cZ ( s_A ) \hookrightarrow \cG ] = 
\int_\CG \mu_H(g) \int dc d \bar c d \bar{\pi}~ e^{Q \Psi_{\rm g.f.}}}

We now - following the standard derivation - insert ``1'' into the path integral over gauge
fields: 
\eqn\sfpvii{\eqalign{
Z & \equiv \int_{\CA} dA e^{ -I_{\rm Y.M.} [A]}\cr
&= \int_{\CA} dA e^{-I_{\rm Y.M.}[A]} 
\int_\CG \mu_H(g) \int dc d \bar c d \bar{\pi}~ 
\exp \biggl( Q\bigl[ \langle \bar c, \CF[A^g]\rangle - (\bar c, \bar{\pi})_{V^\ast}
\bigr]\biggr)  \cr}}

Now, we would like to change variables $A \to A^g$ and factor out the $A$-independent 
Haar measure, but there is one important conceptual point we must first address.
If we simply set $A \to A^g$ then $Q$, as defined, would not act on $\CF$: making the
change of variables in the argument of $Q$ is wrong.
However, one easily checks that 
\eqn\sfpviii{
Q A^g = -d_x c - [A^g,c]_+}
That is, the action of $Q$ {\it coincides with the differential of Lie algebra cohomology.}: 
$Q_{\rm L.A.} A = - D_A c$. 
Thus, we can rewrite the RHS of \sfpvii\ as: 
\eqn\sfpviv{
\int_\CG \mu_H(g)
\int_{\CA} dA^g~ e^{-I_{\rm Y.M.}[A^g]} \int dc d \bar c d \bar{\pi}~ 
\exp \biggl( \tilde Q \bigl[ \langle \bar c, \CF[ A^g ] \rangle - ( \bar c, \bar{\pi})_{V^\ast}
\bigr] \biggr)}
where $\tilde Q$ is now expressed in terms of Lie algebra cohomology: 
\eqn\lacoho{
\tilde Q = Q_{\rm L.A.} + \delta}
and hence, with
$\tilde \Psi_{\rm g.f.}= \langle \bar c, \CF  [ A ] \rangle -  (\bar c, \bar{\pi})_{V^\ast}$ we have
\eqn\sfpvv{
\mathboxit{
{1 \over \vol \CG} \int_{\CA}  dA e^{-I_{\rm Y.M.}}
= \int dA dc d\bar c d \bar{\pi}~  e^{-I_{\rm Y.M.}+ \tilde Q ( \tilde \Psi_{\rm g.f.})}}}

\vskip0.1truein\noindent
{\bf Remarks}: 
\item{1.}
In standard treatments of perturbative Yang-Mills theory, with the choice $\CF=\p\cdot A$,
the overall scale of the metric $(\cdot, \cdot)_V$ 
is the parameter $\xi$  of ``$R_\xi$-gauge.'' The 
$\xi$-independence of correlators of gauge-invariant
observables is conceptually the same as topological 
invariance. 
\item{2.}
It is interesting to compare the actions from the FP procedure with the projection form
$\Phi(P\to M)$.
When  constructing   the latter  we make no choice of gauge fixing term, nor do we choose
a slice for the gauge orbit. In the FP procedure we have an integral over two odd fields
$\bar c$ and $c$ (and one even field $\bar{\pi}$, if we introduce the Lagrange multiplier).
In the projection form we integrate over two even fields $\lambda$ and $\phi$ and one
odd field $\eta$. 
\item{3.}
A simple extension of the  above argument shows that  if there are other fields in the
theory, transforming in representations of the gauge group then $\tilde Q$ becomes
the differential for Lie algebra cohomology in the representation defined by those
fields. 
\item{4.}
When there are other fields in the theory, it can happen that the last innocent-looking
step $dA^g d \psi^g = d A d \psi $ is wrong.
This is the phenomenon of anomalies. 
\item{5.}
If $\CG$ does not act freely then there is an extra factor on the RHS of \sfpvv\ 
of the volume of the subgroup that fixes $\CA$. 
This factor is important in comparing \ymt\ to topological results \Witdgt. 

\subsec{The general construction of 
cohomological field theory. }
\subseclab\sstheGCOFGL

We now summarize what we have learned in 
chapters \sEQCO,\sINandTIR,\sTTwithLS. 

\subsubsec{Fields}
\subsubseclab\sssFlds

We  introduce fields with ghost numbers 
\eqn\gnsghst{\eqalign{ 
A,\psi  \qquad & \qquad U=0,1\cr
\rho,\pi  \qquad & \qquad U=-1,0\cr
\lambda_a, \eta_a  \qquad & \qquad U=-2,-1\cr
\bar c,\bar \pi  \qquad & \qquad U=-1,0\cr
\phi^a, c^a  \qquad & \qquad U=2,1 . \cr}}
These should be interpreted as generating the DeRham complex of the total space of a
certain vector bundle $\CE\to \CC$ over field space which is a sum of three factors: 
\eqn\sumthrf{
\CE = \Pi\CE_{\rm loc} \oplus \CE_{\rm proj} \oplus \Pi\CE_{\rm g.f. }}
The first line in \gnsghst\ 
gives the DeRham complex of the base. 
The next three lines give the DeRham complex of the three kinds of fibers.
Thus we may regard $\psi=\tilde d A$, $\pi=\tilde d \rho$, $\eta=\tilde d \lambda$ and
$\bar \pi=\tilde d \bar c$.
The last line generates the Weil complex. 
The commutation properties are dictated by the total grading (form degree plus ghost number):
all fields with odd total grading are anti-commuting; those
with even total grading are commuting.

\subsubsec{Observables}
\subsubseclab\sssObservables

The observables and action are formulated using the ``BRST model'' of $\CG$-equivariant 
cohomology described in section  
\ssECvLAC\ \refs{\stora,\Kal,\zuckerman}.
Recall that 
to any Lie algebra, $\lieg$, there is an associated differential graded Lie algebra (DGLA)
$\lieg [ \theta ] \equiv \lieg \otimes \Lambda^\bullet \theta$; $\theta^2=0$, $\deg~ \theta=-1$,
$\deg~ \lieg =0$ and   $\p \theta=1$.
Moreover, if $M$ is a superspace with a $\lieg$-action then $\O^\bullet (M)$ is a 
differential graded $\lieg [ \theta ]$ module, with $X\in \lieg \to \CL_X$ and
$X\otimes\theta\to \iota_X$. 
In our case $\lieg \to \liebg$ and $M$ is the total space of $\CE$. 
The BRST complex is
$\hat\CE\equiv \Lambda^\bullet \Sigma ( \liebg [ \theta ] )^\ast \otimes \O^\bullet ( \CE )$
where $\Sigma$ is the suspension, increasing grading by $1$.
The differential on the complex is 
$Q=( d_\CE+ \p\dual ) + Q_{\rm L.A.} $ where $\p\dual$ is dual to $\p$ and 
$Q_{\rm L.A.}$ is the BRST differential for the DGLA $\liebg [ \theta ]$ acting
on $\O^\bullet ( \CE )$.
Physical  observables $\hat\CO_i$ are $Q$-cohomology classes of the ``basic'' ($\liebg$-relative) subcomplex and correspond to basic forms $\CO_i \in \O^\bullet (\CC)$
which descend and restrict to cohomology classes $\omega_i \in H^\bullet (\CM)$. 

\subsubsec{Action}
\subsubseclab\sssAction

The basic data needed to construct the action are: 
\item{1.}
$\CG$-invariant metrics on $\CC$ and $\CE$. 
\item{2.}
A  $\CG$-equivariant section $s: \CC\to \CE_{\rm loc}$, and a $\CG$-equivariant connection
$\nabla s= d s + \theta s \in \O^1(\CC;\CE_{\rm loc}) $.
\item{3.}  
A $\CG$-{\it nonequivariant}  section $\CF\colon\CC\to \CE_{\rm g.f.}$, whose zeros
determine local cross-sections for $\CC\to \CC/\CG$. 

The action is then $I=Q\Psi$ where 
\eqn\gaugefermis{\eqalign{
\Psi  &= \Psi_{\rm localization} + \Psi_{\rm projection} + \Psi_{\rm gauge\ fixing} \cr
\Psi_{\rm loc}
&=  -i\langle \rho,s\rangle - (\rho,\theta \cdot \rho)_{\CE_{\rm localisation}^\ast}
+ (\rho,\pi)_{\CE_{\rm localisation}^\ast}\cr 
\Psi_{\rm proj}  & =i  (\lambda,C^\dagger)_{\liebg}\cr 
\Psi_{\rm gauge\ fixing}
&= \langle \bar c, \CF [A] \rangle -  (\bar c, \bar \pi)_{\CE_{\rm g.f.} }\cr}}

\subsubsec{Correlators}
\subsubseclab\sssCorrelators

Correlation functions are intersection numbers according to the localization formula: 
\eqn\ggefxd{
\mathboxit{\int_{\CE\times \widehat \liebg} \hat \mu e^{-I} 
\hat \CO_1  \cdots  \hat \CO_k 
= \int_{\CZ( s)/\CG} \omega_1\wedge -\wedge \omega_k~ \chi( \cok~ \bO / \CG )}}
where $\bO$ is the operator defined in \genbgo. 

\newsec{Topological Yang-Mills Theory}
\seclab\sTYMT

\subsec{Basic Data}
\subseclab\ssDGonA

\par\noindent
The data are:
\item{$\bullet$} a Riemannian spacetime, $M$, with metric, $g_{\mu\nu}$,
of Euclidean signature.
\item{$\bullet$} a compact finite dimensional Lie group, $G$.
\item{$\bullet$}  a principal $G$-bundle, $P\to M$.
\item{$\bullet$} the group, $\CG$, of gauge transformations for $P$.
(In the case where $P$ is the trivial bundle, $P=M\times G$, this is just
$\CG= \MAP (M, G)$.)
\item{$\bullet$} the space $\CA$ of connections on $P$.

\par\noindent
In analogy to the case of SQM and the topological sigma model we have 
\eqn\tanconn{
T_A \CA = \Omega^1 ( M; \lieg)\quad . }
Indeed, $\CA$ is an affine space, and the difference of two connections
lies in $\Omega^1 ( M; \lieg)$. 
The Lie algebra of the group of gauge transformations is
$Lie( \CG ) = \Omega^0 ( M; \lieg)$. 

The spaces $\Omega^k ( M; \lieg)$ all inherit metrics:
\eqn\mtrcs{
( a, b) = \int_M \tr ( a \wedge \ast b)}
where $\ast$ is the Hodge-$\ast$ operation.
The trace $\tr$ is in the fundamental representation of $SU(N)$\foot{In general we follow the normalization of  \Witdgt\ so that $-{1\over 8 \pi^2} \Tr F^2$ generates $H^4(B\tilde G;\IZ)$,
where $\tilde G$ is the universal cover of $G$. }.

\subsec{Equations}

We will take the $D=4$ Donaldson theory as an example and indicate the generalization to
other TYM theories below.
In $D=4$, $\ast\colon \Omega^2 ( M ) \to \Omega^2 ( M )$ and
$\ast^2 = 1\vert_{\Omega^2 ( M )}$, so that we may define the eigenspaces
$\Omega^{2,+} ( M )$ and $\Omega^{2,-} ( M )$ with eigenvalues under $\ast$
of $+1$ and $-1$, respectively:
$$
\Omega^2 ( M ) = \Omega^{2, -} ( M ) \oplus \Omega^{2, +} ( M )
$$
Note that the $\ast$-operator on the degree two forms
 depends only on the
conformal class of the Riemannian metric.
Conversely the subspaces $\Omega^{2,+} ( M )$ define a conformal structure
on $M$.

The anti-self-dual (ASD) Yang-Mills connections satisfy:
\eqn\asd{ F+ * F = 0}
The solution space $\CA_+ \subset \CA$ admits a $\CG$-action and we wish
to localize to $\CM_+=\CA_+/ \CG$.

\subsec{Vector Bundles}

We choose 
\eqn\tymvb{
\CV_+ = \CA \times \Omega^{2,+}(M,\lieg)}
which has section $s(A) = F+ * F$. 
The fibers have an action of the gauge group $\CG$, so we may form the
 vector bundle 
\eqn\tymvbii{
\CE_+= \CA \times_{\CG}  \Omega^{2,+}(M,\lieg)}
associated to the principal bundle $\CA \to \CA / \CG$. 
Since $s ( A )$ transforms equivariantly it defines a section $\bar s$ of $\CE_+$.
We have 
$$
\CZ( s ) / \CG = \CZ( \bar s )
$$
so we are in the situation of \ssPandL. 

\subsec{Connection on $\CV_+$ and $\CE_+$}

$\CV_+$ is trivial when viewed as a bundle over $\CA$, so we can choose the trivial
connection for $\nabla_s$ in \nabtot. 
On the other hand,  $\CE_+$ has a nontrivial connection. 
Since $\CA$ has a $\CG$-invariant metric, it  inherits a connection $\nabla$
from the natural connection defined in section \sssGofPB\ above. 
The operator $C$ becomes, in this case, $C\colon \Omega^0 (M, \lieg) \to T_A \CA$
defined by 
$$
C \epsilon = D_A \epsilon = d \epsilon +[A,\epsilon]\in \Omega^1 ( M; \lieg )
$$
The image of $C$ defines the vertical tangent space.
Therefore, the horizontal subspace of $T_A \CA$ is orthogonal to the gauge
orbit and therefore horizontal tangent vectors $\tau\in \Omega^1( M, \lieg)$ 
satisfy: 
\eqn\horspace{\eqalign{
D_A^\ast \tau &=( D^\mu_A \tau_\mu) \cr
&=g^{\mu\nu} \bigl( \partial_\nu \tau_\mu+[A_\nu,\tau_\mu] + \Gamma^\lambda_{\nu\mu}\tau_\lambda\bigr) = 0 \cr}}
where $\Gamma^\lambda_{\mu\nu}$ is the Levi-Civita
 connection on $M$ associated
to $g_{\mu\nu}$.

To be explicit, if $\tau\in T_A \CA$ is a tangent vector, so $\tau\in \Omega^1 ( M; \lieg )$,
then the connection evaluated on $\tau$ is given by \genvrt: 
\eqn\valtua{ 
{1\over D^\dagger D} D^\dagger \tau =
\int_M \sqrt{g} d^4 y~ G ( x, y)^a_b (D^\mu \tau_\mu)^b ( y )~
\in~  \Omega^0 (M;\lieg)}
where $G ( x, y )^a_b$ is the Green function of the Laplacian
$D^\dagger D \colon \Omega^0 ( M, \lieg ) \to \Omega^0 ( M, \lieg )$.
Since this expression involves $1/{D^\dagger D}$, it is manifestly non-local
in spacetime.

\subsec{BRST Complex}
\subseclab\ssBRSTCPLX

The BRST complex of  fields is a superspace realization of a complex for computing the
$\CG$-equivariant cohomology of $\CA$ tensored  with a $\CG$-module.
The differential forms $\Omega^\bullet ( \CA )$ are represented by functions on
superspace:  $\hat \CF(\CA)$ generated by\foot{Strictly speaking, $A$ should be
evaluated at a point $p\in P$, since $A\in \Omega^1 ( P; \lieg )$.} 
\eqn\tymfldi{
A_\mu^a (x)~ \in~ \Omega^0 ( {\cal A} ) \qquad
\psi_\mu^a (x) \leftrightarrow \widetilde d A_\mu^a (x)~ \in~ \Omega^1 ( {\cal A} )}
of ghost numbers $0$ and $1$, respectively. 
The remaining fields and transformation
 properties depend on what model
we use for $\CG$-equivariant cohomology of $\CA$. 

\subsubsec{Weil Algebra}
\subsubseclab\sssDWA

The Weil algebra is constructed from the dual of the algebra
of gauge transformations $\liebg\dual \cong \Omega^4 ( M; \lieg)$. 
We define
 $c$ and $\phi  \in \Omega^0( M; \lieg)$ 
which can be expanded: 
$$
\eqalign{
c= c^a(x)T_a  \in \Omega^0 ( M; \lieg )\cr 
\phi = \phi^a ( x ) T_a \in \Omega^0 ( M; \lieg )\cr}
$$
The coordinate functions $c^a(x)$ or $\phi^a(x)$ on the Lie algebra of $\CG$
can be taken as generators of the dual Lie algebra, $\liebg\dual$
(corresponding to $\delta$-function supported measures in $\Omega^4 ( M; \lieg)$). 
These are the analogs of $\theta^i$ and  $\phi^i$ in the general Weil algebra
discussed in chapter \sEQCO, section \ssWA. 
As always $c^a(x)$ and $\phi^a(x)$ have grading (ghost number) $1$ and $2$,
respectively. 

\subsubsec{BRST Model}
\subsubseclab\sssDBRSTM

According to chapter \sEQCO\ we should take the complex:
$$
\CW( \liebg ) \otimes \Omega^\bullet ( \CA )
$$
using the ``BRST'' differential of \sssBRSTM. 
The generators of the complex are $A,\psi,c$ and $\phi$ as above.
Translating \BRSTOp\  into the present example we find: 
\eqn\dondi{\eqalign{
d_B A &=\psi - D_A c \cr
d_B \psi & = [\psi,c] - D_A \phi \cr
d_B c & = \phi-\half [c,c]\cr
d_B\phi &= -[c,\phi] \cr}}

\exercise{}

Show how the above transformations can be considered as BRST transformations 
plus a ``supersymmetry'' transformation,  as in  \ssECvLAC. 

\endexercise

\ifig\assmghfr{Assembling fields according to ghost number and form degree.}
{\epsfxsize2.0in\epsfbox{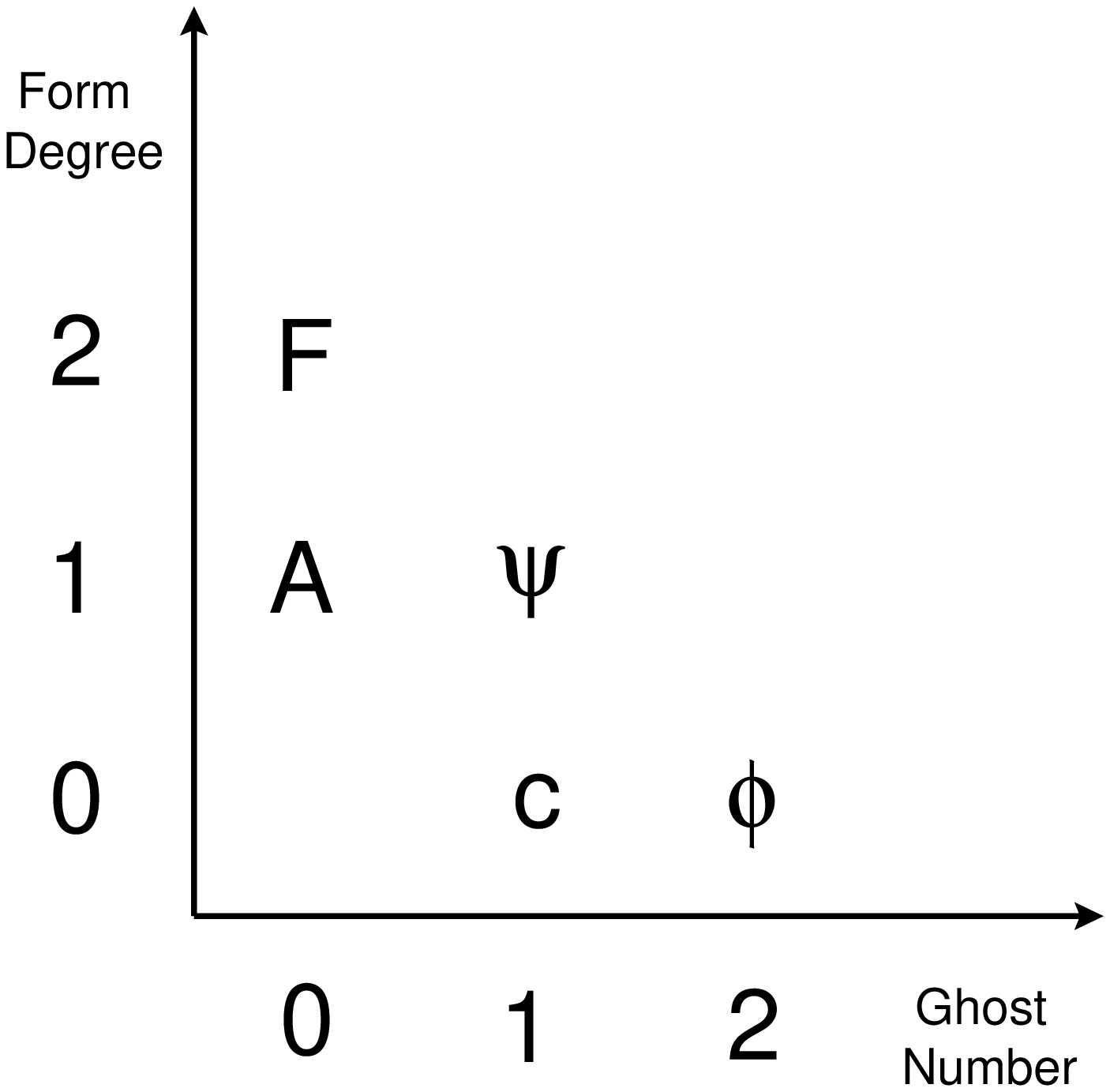}}

The equations defining the BRST model of $\CG$-equivariant cohomology of
$\CA$ were given a very elegant formulation by Baulieu and Singer \BaSiii.
Let us consider, formally, the sums
\eqn\baulsgi{
\eqalign{
\bA & = A+c\cr
\bF & =  F + \psi + \phi \cr}}
These linear combinations are homogeneous in the total grading: ghost number
plus form degree (see \assmghfr).

One defines the differential $d + d_B$ of total degree $1$. 
The relations of the Weil algebra
\eqn\baulsgii{\eqalign{
(d+d_B) \bA   & = \bF - \half [\bA, \bA]\cr
(d+d_B) \bF   & = [\bA, \bF] \cr}}
are equivalent to equation \dondi, the Bianchi identity, and the equation for the curvature.

\subsubsec{Cartan Model}

This is the model originally used in \donaldson. 
Generators of the complex $S( \lieg\dual )\otimes \O^\bullet ( \CA )$ are $A$, $\psi$
and $\phi$. 
Translating \CartanDiff\  into the present example, we find: 
\eqn\dondii{\eqalign{
d_\CC A &=\psi\cr
d_\CC \psi & = - D_A \phi \cr
d_\CC \phi  &= 0 \cr}}

Thus, $d_\CC^2$ is a gauge transformation by $\phi$, and thus as expected vanishes
on gauge invariant operators, i.e. the {\it $\CG$-invariant subcomplex}. 

\subsubsec{Antighosts}

The antighosts for the construction of
$\Phi_{{\rm localization}} ( {\cal A} \to {\cal A}_+ )$ are 
\eqn\tymii{\eqalign{
\rho & \longrightarrow \chi_{\alpha \beta}~ \in~ \Omega^{2,+} ( M, \lieg )\cr
\pi & \longrightarrow H_{\alpha \beta}~ \in~ \Omega^{2,+} ( M, \lieg)\cr}}
These have ghost numbers $-1$ and $0$, respectively.
The antighosts for the construction of
$\Phi_{\rm projection} (\CA\to \CA/\CG  )$ are: 
\eqn\tymiii{\eqalign{
\lambda & \longrightarrow \lambda^a \in \Omega^0 ( M, \lieg )\cr
\eta & \longrightarrow \eta^a \in \Omega^0 ( M, \lieg )\cr}}
of ghost numbers $-2$ and $-1$. 

Finally, for gauge fixing we choose antighosts
\eqn\tymiv{\eqalign{
\bar c~ \in~ \Omega^0 (M, V^\ast )\cr
\bar \pi~ \in~ \Omega^0 (M, V^\ast )\cr}}
of ghost numbers $-1$ and $0$ as is standard in FP gauge fixing.
Here $V$ is the vector space in which the noninvariant function $\CF[A]$
of section \ssFPGF\ takes values.

The action of  $Q_\CC$ is that described in previous chapters
\eqn\tymiv{\eqalign{
Q_\CC \pmatrix{\chi  \cr H\cr} &= 
\pmatrix{0 & 1\cr  -\CL_\phi & 0\cr} 
\pmatrix{\chi  \cr H\cr}= \pmatrix{H \cr   - [\phi, \chi]  \cr}  \cr
Q_\CC \pmatrix{ \lambda \cr \eta\cr} &= 
\pmatrix{0 & 1\cr  -\CL_\phi & 0\cr} 
\pmatrix{ \lambda \cr \eta\cr} =
\pmatrix{\eta\cr - [\phi,\lambda]\cr}\cr
Q_\CC \pmatrix{ \bar c \cr \bar \pi\cr} &= 
\pmatrix{0 & 1\cr  -\CL_\phi & 0\cr} 
\pmatrix{ \bar c \cr \bar \pi\cr} =
\pmatrix{ \bar \pi \cr - \phi\cdot \bar c \cr}\cr}}

\subsec{Lagrangian}

\subsubsec{Localization Lagrangian}

The Cartan representation of the MQ form in \mqii\ becomes :
\eqn\loclagi{
\Psi_{{\rm Localization}} ={1\over e^2} \int_M d^4 x \sqrt{g}~ \tr~ \chi(i F_+ - t H)}
and
\eqn\loclagii{
Q_\CC \Psi_{{\rm Localization}} =
{1\over e^2} \int d^4 x \sqrt{g}~ \left \{ \tr~ H (i F_+ - t H) 
- \tr~ \chi( i (D_A\psi)_+ + t~ [\phi,\chi] ) \right \}. }
Integrating out $H$ gives 
\eqn\loclagiii{
H= {i \over 2 t} F_+. }
yielding the Lagrangian: 
\eqn\loclagiv{
{1\over e^2} \int d^4 x \sqrt{g}~ \Biggl\{ -{1\over 4 t} \tr~ F_+^2
- i~ \tr~ \chi^{\mu\nu} (D_A\psi)^+_{\mu\nu}  
+ t~ \chi_{\mu\nu}[ \phi, \chi]^{\mu\nu} ) \Biggr\}. }
As we remarked in section \ssAthePNGF\  there is no need for an explicit
projection on the horizontal components of this form since all of the vertical
directions on $\CA$ are taken up by the projection part of the path integral. 

\subsubsec{Projection Lagrangian}

We now apply \projfrm\ to the present case.
$C^\dagger$ is a one-form with values in the Lie algebra $Lie ( \CG )$.
From \horspace\ we see that it may be identified with 
\eqn\tymcdgg{
C^\dagger \longrightarrow  - \ast D_A  \ast \psi = - ( D^\mu \psi_\mu )^a
\in \Omega^1 ( \CA; Lie ( \CG ))}
so that:
\eqn\tymprj{\eqalign{
\Psi_{{\rm Projection}}
&= -{i\over e^2} ( \lambda, C^\dagger )
= {i \over {e^2}} \int_M \tr~ \lambda D_A \ast \psi\cr
&= -{i\over e^2} \int_M d^4 x \sqrt{g}~
\tr~ \biggl(  \lambda \ast D_A \ast \psi \biggr)\cr}}
and hence 
\eqn\tymprji{
Q_\CC \Psi_{{\scriptscriptstyle \rm Projection}}  =
-{i \over {e^2}} \int_M
\tr~ \biggl ( \eta D_A \ast \psi + \lambda \{ \psi, \ast \psi \} 
+ \lambda D_A \ast D_A \phi \biggr )}
corresponding to the three terms in \expdprj. 

Combining the actions for projection and localization we recover 
Witten's celebrated Lagrangian\refs{\donaldson} for Donaldson
theory\refs{\DoApp,\Don,\DoKro}:
\eqn\donalduck{\eqalign{
I_{{\scriptscriptstyle \rm Donaldson}} = {1 \over e^2} \int d^4 x \sqrt{g}~ &\Biggl[
i~ \tr~ \biggl( \eta D_A^\mu \psi_\mu + \lambda \{ \psi_\mu,\psi^\mu \} 
+ \lambda D_A \ast D_A \phi \biggr)\cr
& +\biggl(  -{1\over 4 t} \tr~ F_+^2 
- \tr~ \chi_{\mu\nu} [ i ( D_A \psi)_+^{\mu\nu}
+ t [\phi,\chi^{\mu\nu}] ] \biggr) \Biggr ] \cr}}

\subsubsec{Gauge-Fixing Lagrangian}

The Donaldson Lagrangian of \donaldson\ is gauge invariant. 
Therefore, if one were to attempt to evaluate it with standard methods of local
quantum field theory the Lagrangian would have to be gauge-fixed, 
as emphasized by Baulieu and Singer \refs{\BaSiii}. 
As discussed in \ssFPGF\ this can be done by adding the action:
\eqn\ggefx{
\Psi_{{\rm gauge\ fixing}}
=\int_M d^4 x \sqrt{g}~ \tr~ \bar c~ ( \CF[A] - \xi  \bar \pi)}
where $\CF[A]$ is gauge noninvariant,
and we have chosen $V=\lieg$, for simplicity.
$\xi$ is an arbitrary (bosonic) constant gauge-fixing parameter.

\subsubsec{MQ form on $\CE_+\to \CM$}

We have been writing forms ``upstairs'' on $\CA$.
It is nice to see how integrating out various fields\foot{
This discussion is subject to difficulties related to reducible connections.}
produces the standard MQ formula for a section $\bar s$ of $\CE_+$. 
Of course, the resulting Lagrangian is no longer local. 

Integrating out $\lambda$ produces 
\eqn\idenphi{
\phi = - {1\over D_A^\dagger D_A} \ast \{ \psi, \ast \psi \}}
in accord with the identification of $\phi$ with the curvature of $\CE_+$. 

Integrating out $\eta$ enforces 
$$ 
D_A^\mu \psi_\mu =0 
$$
and \horspace\ shows that this means $\psi$ is horizontal :
$\langle D \epsilon, \psi\rangle =0 $. 
The connection on the section $\nabla s$ in the MQ formula is very simple 
on $\CA$, since the connection on $\CV_+$ is trivial: 
$$
\nabla  s = \tilde d  s = (D \psi)_+ 
$$
where $( \quad )_+$ denotes projection onto the self-dual part.
For horizontal $\psi$ this descends to the same expression on $\CA/\CG$. 
Substituting back \idenphi\  we get the Lagrangian: 
\eqn\loclagv{
I= {1\over e^2} \int d^4 x \Tr~ ( -{1\over 4 t}  F_+^2 + 
i~ \chi_{\mu\nu}(D_A\psi)_+^{\mu\nu}
- t~ [\chi,\chi] {1\over D^\dagger D} [ \psi_\mu, \psi^\mu] )}
which coincides with the form \nrmunv\ interpreted as a form on $\CE_+$.  
 
\subsec{$D=2$: Flat Connections}
\subseclab\ssDtwoFLC

Another possible moduli space is the space of flat connections
\refs{\bbrt}.
The deformation complex which is relevant to this case is the following
\eqn\FlatComplex{
0 \mapright{} \Omega^0 ( M, \lieg ) \mapright{D_A^{(0)}} \Omega^1 ( M, \lieg )
   \mapright{D_A^{(1)}} \Omega^2 ( M, \lieg ) \mapright{D_A^{(2)}}
   \cdots \mapright{D_A^{(\scriptscriptstyle \dim~ M)}}
   \Omega^{{\scriptscriptstyle \dim\ M}} ( M, \lieg ) \mapright{} 0}
The tangent space is given by the first cohomology of this complex.
We recognize this as the twisted deRham complex, whose index, in the case of
flat connections, is just
\eqn\TwistedDRIndex{\eqalign{
\ind~ D_A
~=&~ \sum_{i=0}^{\dim~ G} ( -1 )^i h_i\cr
~=&~ \dim~ G~ \chi ( M )\cr}}
The case of two dimensions is particularly interesting.
In the case of a Riemann surface of genus $g \ge 2$, one
finds the dimension of the moduli space to be
\eqn\DimTwo{\eqalign{
\dim~ {\cal M} ( M, G )
~=&~ 2 h_0 - \dim~ G~ \chi ( M )\cr
~=&~ 2 h_0 + 2 ( g - 1) \dim~ G\cr}}
For irreducible connections, $h_0 = 0$, so that
$\dim~ {\cal M} ( M, G ) = 2~ \dim~ G ( g - 1 )$, 
in accord with section \sssetzfc.
Reducible connections increase the dimension of the moduli space.

For $d=4$, we have
$$
h_1 = h_0 + {1 \over 2} \left ( h_2 - \dim~ G  \right )~ \chi ( M )
$$
For irreducible connections $h_0 = 0$.
In the case that both $h_0$ and $h_2$ vanish, non-trivial moduli spaces exist
only for manifolds with $\chi ( M ) < 0$.
However, moduli spaces of {\it reducible} connections can continue to be
non-trivial, even if $\chi  ( M ) > 0$.

To define a topological field theory that studies the moduli space of flat connections
in $d=2$, we choose following section of  $\CE_0=\CA\times_\CG V$ with
$V= \Omega^2 ( M; \lieg )$:
$$
s ( A ) = F
$$ 
The only point that changes from our discussion above is that now
$\chi$ and $H \in \CE_0^\ast $ lie in $\Omega^0 ( M; \lieg )$.
All other manipulations leading to the Lagrangian are identical. 
Thus we simply set $\chi_{\mu\nu} = \epsilon_{\mu\nu}\chi$ in the above Lagrangian.
The theory localizes to the space $\CM_0(\Sigma)$ of {\it flat} connections on a
surface $\Sigma$.

\subsec{Observables} 

The observables, i.e. the BRST 
cohomology of the complex described 
in section \ssBRSTCPLX\  above, are
quantum field theory representatives 
of certain cohomology classes as we shall
now describe. 
Mathematically, these  can be obtained by a ``universal'' construction analogous 
to that used for topological sigma models. 

\subsubsec{The Big Bundle}

Consider the product space ${\cal A} \times P$.
Naively it seems we could regard this as a ${\cal G} \times G$ bundle
over a  base space, ${\cal A} / {\cal G} \times M$.
However, there are difficulties which ensue from the fact that the action
of ${\cal G}$ on ${\cal A}$ is not free {\it even} at points $A \in {\cal A}$
where the connection is irreducible.
The stabilizer, $\Gamma_A$, contains at least the center of the group, $C ( G)$.
For this reason (and others) mathematicians work with {\it framed connections}
( See, for example, \refs{\DoKro}. )

Let $( M, m_0 )$ be a four manifold with base point $m_0 \in M$.
Then a framed connection is a pair $( A, \varphi )$ where $A$ is a connection
and $\varphi\colon G \to P_{m_0}$ is an isomorphism from the gauge group into
the fibre over $m_0$.
The gauge group acts naturally (and without fixed points) on (irreducible) framed
connections.
So consider
$$
\widetilde {\cal B} ~=~ ( {\cal A} \times \Hom ( G, P_{m_0} ) ) / {\cal G}
$$
There is a natural map which forgets the framing
$$
\beta\colon \widetilde {\cal B} \longrightarrow {\cal B} = {\cal A} / {\cal G}
$$
Equivalently we may think of the framed connections as the quotient
$$
\widetilde {\cal B} = {\cal A} / {\cal G}_0
$$
where ${\cal G}_0 = \{ g \in {\cal G}~ \vert~ g ( m_0 ) = 1 \}$.
The ``forgetful" map, $\beta$, takes the quotient with the remainder of the
gauge group:
${\cal G} / {\cal G}_0 \equiv \Aut ( P_{m_0} ) \equiv G$.
The inverse image under $\beta$ of $[ A ] \in {\cal A} / {\cal G}$ is isomorphic
to $G / \Gamma_A$.

Finally if $\widetilde {\cal B}^\ast \subset \widetilde {\cal B}$ is the space of framed
irreducible connections, then
$$
\beta\colon \widetilde {\cal B}^\ast \longrightarrow {\cal B}^\ast
$$
is a principal $G_0 = G / C ( G )$ bundle, called the {\it base point} fibration.

Next we introduce the universal family of connections parametrized by
$\widetilde {\cal B}^\ast$.
$$\matrix{
 & {\cal A}^\ast \times P & \cr
\mapsw{{{\cal G}_0}} & & \mapse{{G_0}} \cr
\widetilde {\cal B}^\ast \times P & &  {\cal A}^\ast \times M \cr
\mapse{{G_0}} & & \mapsw{{{\cal G}_0}} \cr
 & \widetilde {\cal B}^\ast \times M & \cr}
$$
It is apparent that
$$\eqalign{
\widetilde {\cal B}^\ast \times P & \leftarrow G_0\cr
\downarrow{} & \cr
\widetilde {\cal B}^\ast \times M & \cr}
$$
is a principal $G_0$-bundle and as such has a classifying map
$$
\Phi\colon \widetilde {\cal B}^\ast \times M \longrightarrow B G_0
$$
So we obtain characteristic classes via pullback
$$
c_r ( \widetilde {\cal B}^\ast \times M )
= \Phi^\ast ( c_r ( B G_0 ))
\in H^{2r} ( \widetilde {\cal B}^\ast \times M )
$$
or more generally
$$
\prod_{i=1}^n c_r ( \widetilde {\cal B}^\ast \times M )^{d_i}
= \Phi^\ast ( \prod_{i=1}^n c_r ( B G_0 )^{d_i} )
\in H^s ( \widetilde {\cal B}^\ast \times M )
$$
where $s = 2 \sum_{i=1}^n r_i d_i$.

To construct observables in topological Yang-Mills theories (= cohomology classes on
$\widetilde {\cal B}^\ast$)
one defines the slant product pairing.
For $c \in H^d ( \widetilde {\cal B}^\ast \times M )$
and $[ \alpha ] \in H_i ( M )$,
$$\eqalign{
/ \colon H^d ( \widetilde {\cal B}^\ast \times M ) \times H_i ( M )
~\longrightarrow&~ H^{d-i} ( \widetilde {\cal B}^\ast )\cr
c / [ \alpha ] ~=&~ \int_\alpha c\cr}
$$
We shall refer to the associated map on $H_\bullet (M)$
 as the {\it Donaldson map}.

For the case of $G = SU(2)$, $c = c_2 ( B G_0 )$ and $\Sigma \in H_2 ( M )$,
this construction yields:
\eqn\DonMuMap{\eqalign{
\mu&\colon H_2 ( M ) \longrightarrow H^2 ( \widetilde {\cal B}^\ast )\cr
\mu& [ \Sigma ] ~=~ \int_\Sigma \left [
\Phi^\ast ( c_2 ( B G_0 ) \right ]_{{\scriptscriptstyle ( 2, 2 )}}\cr}}
where $[ \cdots ]_{{\scriptscriptstyle ( p, q )}}$ is the projection of $\cdots$ to
the form degree $( p, q )$ part on $\widetilde {\cal B}^\ast \times M$.

If $M$ is connected and simply-connected, then the only non-trivial
homology groups are $H_0 ( M ), H_2 ( M )$, and $H_4 ( M )$.
We may then consider the following cohomology classes on $\widetilde {\cal B}^\ast$:
\item{1.}
By Poincar\'e duality $h^4 = h^0 = 1$ for connected $M$.
In this case the slant product is equivalent to the pushdown via the
projection operator
$$
\pi_1\colon \widetilde {\cal B}^\ast \times M \longrightarrow \widetilde {\cal B}^\ast
$$
We may take any characteristic class (of degree $\ge 4$) and integrate
it over $M$:
\eqn\FourForm{
( \pi_1 )_\ast \Phi^\ast\colon H^d ( BG_0 ) \times H_4 ( M )
\longrightarrow H^{d-4} ( \widetilde {\cal B}^\ast )}
\item{2.}
If $h^2 \not= 0$, then given a basis 
$\{ [ \Sigma_i ] \}_{i=1,\cdots, h^2}$ of $H_2 ( M )$
and a characteristic class 
$c ( \widetilde {\bf P} ) \in H^d ( \widetilde {\bf P} , \IZ )$ we may
construct the observables
$$
\mu_c ([ \Sigma_i ] ) = \int_{\Sigma_i} \left [
\Phi^\ast ( c ( \widetilde {\bf P} )) \right ]_{{\scriptscriptstyle ( d-2, 2 )}}
$$
\item{3.}
$h^0 = 1$ for connected $M$.
Then we may evaluate the $0$-form part of any characteristic class on
$H^d ( \widetilde {\bf P} )$ on any $m \in M$:
$$
\nu_c ( m ) = \left [ \Phi^\ast ( c ( \widetilde {\bf P} )) \right ]_{{\scriptscriptstyle ( d, 0 )}} (m)
$$

All of the above classes are defined on ${\widetilde {\cal B}}^\ast$.
By pushing down via the forgetful map $\beta$, there is a natural way to
obtain cohomology classes on ${\cal B}^\ast$.

\vskip0.1truein\noindent
{\bf Remarks:}
\item{1.} If $P$ is an $SU ( 2 )$ bundle over a connected and simply-connected
four manifold and $\Sigma_1, \ldots, \Sigma_{h^2}$ are a basis for $H_2 ( X; {\bf Z} )$,
then the rational cohomology ring, $H^\ast ( {\cal B}^\ast; {\bf Q} )$, is a polynomial
algebra on the generators: $\mu ( \Sigma_1 ), \ldots, \mu ( \Sigma_{h^2} )$, and $\nu$,
where  $\nu$ is an extra generator in dimension four\refs{\DoKro}.
\item{2.} There is a construction of the cohomology classes, $\mu ( \Sigma )$,
as the first Chern classes of determinant line bundles of a family of Dirac operators
on $\Sigma$ \refs{\DoKro}.

\subsubsec{Universal Connection}

Finding an explicit representative of the characteristic classes on
$\widetilde {\cal B}^\ast \times M$ requires choosing an explicit connection.
We follow the general construction used in  chapter
\sTTwithLS. 
We have ${\cal G} \times G$ invariant metrics on ${\cal A}^\ast \times P$.
For $\tau_i \in T_A \widetilde {\cal A}^\ast$ and $X_i \in T_p P$ we have
\eqn\NatConnMet{\eqalign{
&\left\langle ( X_1, \tau_1 ), ( X_2, \tau_2 ) \right\rangle_{( p, A )}\cr
&\qquad\qquad
= g_{\mu\nu} X_1^\mu X_2^\nu ( p ) + \tr A ( X_1 ) A ( X_2 )
+ \int_M \sqrt{g} d^4 x~ g^{\mu\nu} \tr~ \tau_{1\mu} \tau_{2\nu}\cr}}

The connection has a curvature $\CF\in \Omega^2 (\CA^\ast \times P; \liebg \oplus \lieg)$
which is horizontal (but not invariant).
Decomposing according to the natural bigrading we have
\eqn\curvcpts{\eqalign{
\CF_{2,0} ( \tau_1,\tau_2 )
 & = - {1\over D_A^\dagger D_A} [ \tau_1, \ast \tau_2] \cr
\CF_{1,1} ( X, \tau )& = \tau ( X )\cr
\CF_{0,2} ( X_1, X_2 )& = F_A ( X_1, X_2 )\cr}}
where $\pi\colon P \to M$ is the projection.

This connection descends to a $G$-connection on $\tilde {\bf P}$ called the universal
connection. 
Using \curvcpts\  we can then find explicit representatives for
$c_r \in H^{2r}  ( \widetilde {\cal B}^\ast \times M)$ from
\eqn\explreps{
c_r\equiv {1\over r! (2 \pi i)^{r} } \Tr~ \CF^{r}}

\vskip0.1truein\noindent
{\bf Remarks}: 
\item{1.}
The universal connection may be described in a way  very similar to 
\evlmp.
Recall that $\widetilde {\cal B}^\ast$ parametrizes classes of $G$-connections.
Denote by $[ A ( p, y)]$ the class parametrized by a coordinate $y \in\widetilde {\cal B}^\ast$.
A $G$-connection on $\widetilde {\cal B}^\ast \times P$ is a $\lieg$-valued one-form.
The components of this one-form along $P$ form a $G$-connection on $P$, hence
define a point in $\widetilde {\cal B}^\ast$. 
The universal connection has the property that  the components along $P$ at $( p,y)$
define the class $[ A ( p, y) ]$. 
\item{2.}
The universal connection descends to a conection on moduli space which 
can   be described in more down-to-earth terms as follows. 
Suppose $A_\mu(x,Z)$ is a family of ASD instantons parametrized by some 
coordinates $Z^i$. 
Tangent vectors, ${\partial\over \partial Z^i}$, to a horizontal slice for $\CM_+$ correspond
to elements of $\Omega^1 ( M; \lieg)$ given by 
$$
\delta_i A_\mu= {\partial \over \partial Z^i } A_\mu - D_\mu \epsilon_i
$$
The $D_\mu \epsilon_i$ corresponds to a gauge transformation which we perform in order
to keep the vector tangent to the moduli space (i.e. we require $D_A^\mu ( \delta_i A_\mu)=0$).
Now we introduce $s_i = \partial_i  + [ \epsilon_i, \cdot ]$ which defines a $G$-connection on
$\CM^+\times M$ through the covariant derivative $\CD= dZ^i s_i + dx^\mu D_\mu$. 
One checks that $\CD^2$ has the components given in \curvcpts. 
This description appears in an  intriguing interpretation of the Donaldson map and polynomial
in the context of the theory of heterotic string fivebranes,  discussed by
J. Harvey and A. Strominger in \refs{\harstrom}. 

\subsubsec{QFT Representative}

Let us return to the BRST complex of 
section \ssBRSTCPLX. 
The relations \baulsgi\ and \baulsgii\
 are the standard Weil algebra relations, so if we consider
\eqn\baulsgv{
\widehat \CO_r\equiv {1\over {r! (2 \pi i)^{r}}} \tr~ \bF^{r}}
This will be $d+d_B$-closed by standard arguments. 
Given the bigrading into ghost number and form degree, we may decompose this
according to its spacetime form degree as in equation \dcmpse\ for the topological 
sigma model
\eqn\baulsgvi{
\widehat \CO_r \equiv \widehat \CO_r^{(0)} + \cdots + \widehat \CO_r^{(4)}}
where ${\cal O}_r^{(i)}$ is a spacetime $i$-form.
Now
\eqn\baulsgvii{
( d + d_B) \widehat \CO_r=0}
leads to the descent equations
\eqn\DescentEqs{
\eqalign{
d \widehat {\cal O}_r^{(4)} ~=&~ 0\cr
d \widehat {\cal O}_r^{(3)} ~=&~ -\{ Q, \widehat {\cal O}_r^{(4)} \}\cr
d \widehat {\cal O}_r^{(2)} ~=&~ -\{ Q, \widehat {\cal O}_r^{(3)} \}\cr
d \widehat {\cal O}_r^{(1)} ~=&~ -\{ Q, \widehat {\cal O}_r^{(2)} \}\cr
d \widehat {\cal O}_r^{(0)} ~=&~ -\{ Q, \widehat {\cal O}_r^{(1)} \}\cr
0 ~=&~ \{ Q, \widehat {\cal O}_r^{(0)} \} \cr}}

We can therefore form the BRST invariant observables 
\eqn\baulsgv{
\widehat W_r(\gamma) \equiv  \int_\gamma \widehat{\CO}_r^{(k)}}
by integration over a cycle
$\gamma\in H_k(M)$, thus providing a map
\eqn\descentii{
H_k(M)\to H_Q^{2r-k} ( \hat \CF ( \CA ) )}

Clearly the parallels between the mathematical and QFT construction are 
very close.
The precise  relation is provided by the natural connection on $\CA$,
induced by $\CG$-invariant metric, \NatConnMet.
We may think of  this connection as a Chern-Weil homomorphism,
$w_\natural$. 
Then given $A \in \CA$ we may regard:
\eqn\baulsgiii{
A+w_\natural (c) \in \Omega^1(\CA \times  P; \liebg \oplus \lieg)}
as a ``universal connection'' on $\CA \times  P$. 
The curvature of this connection is 
\eqn\baulsgiiii{
\CF= F + \psi +w_\natural(\phi)
\in \Omega^2(\CA \times  P; \liebg \oplus \lieg)}
The expression  $\psi = \tilde d A_\mu \otimes dx^\mu$ may be regarded as an
element of $\Omega^{1,1}( \CA\times P; \lieg)$.
Its value on a tangent vector is given by: 
$$
\psi_{A,p}(\tau,X_p) = \tau_p(\pi_\ast X_p) \in \lieg
$$
where $(\tau,X_p)\in T_{( A, p )}( \CA \times P) =\Omega^1 (M;\lieg) \oplus T_p P$. 
Thus, $\bF$ in \baulsgi\ 
becomes the curvature of  the universal connection and $\hat \CO_r$ 
is a basic form which descends to a characteristic class of $\CQ$ in 
$H^{2r} ( \CA / \CG \times M)$.
We thus recognize the observables 
$\widehat W_r(\gamma) $ as those given by 
the Donaldson map, $\mu ( \gamma )$. 
\foot{Related discussions appear in \refs{\bbrt, \storaii}.}

\vskip0.1truein\noindent
{\bf Remark}: 
It is interesting to compare classes in $H^\bullet (BG)$ for 
$G$ compact with the above classes. 
Let us take, for example, $G=SU(N)$.
In this case
 $\tr~ \phi^{j+1}$ generate the cohomology ring for $j=1,\dots, N-1$.
The analog of $\phi$ for $\liebg$ is $\phi(x)$, but  the analog of $\tr~ \phi^{j+1}$ is not
just $\tr~ \phi^{j+1}(x)$.
While these do generate BRST classes, by the descent equations, the
$x$-dependence does {\it not} lead to independent cohomology classes.
Moreover, forming invariant polynomials on the Lie algebra, we miss the cohomology classes associated to other elements of  $H_\bullet ( M )$. 

\subsec{Correlation Functions}

We are now set up to present Witten's path-integral formulation of the Donaldson
invariants. 
Using the basic result \ggefxd\  we have
\eqn\pathdnld{
{1\over \vol~ \CG}
\int d \phi \hat \mu ~ e^{-I_{\rm Donaldson}} 
\prod \widehat W^{r_i}(\gamma_i)
=\int_{\CM_+} \prod \omega_{r_i}(\gamma_i)~
\chi(\cok(\bO)/\CG) }
where
$
\omega_{r_i} ( \gamma_i ) = w_\natural \left ( 
\widehat W^{r_i} ( \gamma_i ) \right )
$. 
$\nabla s$ may be identified with the operator: 
\eqn\idnbls{
p_+ D_A: \Omega^1 ( M; \lieg) \to \Omega^{2,+} ( M; \lieg)}
and the operator $\bO$ of sec. \sssQexct\ is 
\eqn\dnldo{
\bO = \pmatrix{ p_+ D_A^{(1)} \cr  (D_A^{(0)})^\dagger \cr}}
These operators fit into the well-known Atiyah-Hitchin-Singer instanton
deformation complex \refs{\AHS}:
\eqn\ahscplx{\matrix{
0 & \mapright{} & \Omega^0 ( M; \lieg ) & \mapright{D_A^{(0)}} & \Omega^1 ( M; \lieg) &
\mapright{p_+ D_A^{(1)}} & \Omega^{2,+} (M;\lieg) & \mapright{} & 0}}

The complex \ahscplx\ is used to study the 
tangent space to the moduli space $\CM_+$. 
The index of the complex 
 \ahscplx\ was computed to be\foot{Note that we have adopted the
following convention for the sign of $\tau ( M )$: If $\dim~ M = 4n$, then
\eqn\SignConvTau{
\tau ( M ) = b_{2n}^+ - b_{2n}^-}
}
$$\eqalign{
\ind ( \CC, D^\ast_A )
~=&~ p_1 ( \lieg ) - {1 \over 2} \dim~ G~ ( \chi ( M ) + \tau ( M ) )\cr
~=&~ 4 h_G k - {1 \over 2} \dim~ G~ ( \chi ( M ) + \tau ( M ) )\cr}
$$
where $h_G$ is the dual Coxeter number of $G$ and $k$ is the instanton number
of the connection.

For generic situations 
\foot{See \refs{\Freed,\DoKro,\Hitchin} for a detailed 
discussion of these points.}
 the connection is irreducible ($h^0 = 0$) and there are no further obstructions ($h^2 = 0$). In this case
\eqn\indxasd{
\dim~ {\cal M}_+ ( M, G; k ) = 
4 h_G k - {1 \over 2} \dim~ G~ ( \chi ( M ) + \tau ( M ) )
}
and at the smooth points of $\CM^+$, we have
$ T\CM^+ \cong \ker (p_+D_A)/\Im D_A$.  
Moreover 
$$\ker \bO \cong \ker \nabla \bar s\cong \ker (p_+ D_A^1 \oplus (D_A^{(0)})^\dagger)
\cong \ker (p_+ D_A^1) / \Im D_A $$ 
so we recover  \zrmd.  
Moreover, generically  $\cok~ p_+D_A =0$, so 
 $\chi(\cok \nabla s)$ is a constant (= $\pm 1$) on 
each component of the moduli
space. Therefore, 
on the instanton number $k$ component of 
$\CM_+$ the correlator \pathdnld\ must satisfy the 
 the ghost number selection rule 
\eqn\ghstslct{
\sum_i {\rm ghost} \#~ ( \widehat W^{r_i} (\gamma_i)) = \sum_i (2r_i - k_i) = 4 h_G k - {1 \over 2} \dim~ G~ ( \chi ( M ) + \tau ( M ) ) }
where $\gamma_i \in H_{k_i} ( M )$
in order to obtain an nonzero answer. 

%
%

\subsubsec{Reducible connections}

The above discussion has ignored an important subtlety: the moduli space 
$\CM_+$ has singularities.
One source of these singularities are reducible connections.
These correspond to connections $A \in {\cal A}$, where the isotropy group,
$\Gamma_A \subset {\cal G}$, is non-minimal\foot{Equivalently, for these connections
the holonomy group is smaller than $G$.}, i.e. there exist non-trivial $\phi \in {\cal G}$,
such that $D_A \phi = 0$.
Correspondingly there exist nontrivial zero modes of  $C^\dagger C$ , so that  this operator
is not invertible. 
A glance at \gpvol\ and \idenphi\ shows that such connections lead to singularities
in the path integral measure.
One cannot neglect these singularities in calculations of  topological invariants. 

One, more physical,  way to express the problem of reducible connections is the
following \refs{\Witr}. 
The path integral should localize to the $Q$-fixed points; these follow from the RHS of
\tymiv\ and \dondi,  and are given by 
\eqn\qfxdpt{\eqalign{
H ={i\over 2 t} F^+ &=0\cr
D_A \phi&=0\cr
[ \phi, \lambda ] &=0\cr}}
The branch of solutions to these equations containing irreducible connections has
$\phi=0$. 
$\lambda$ is unconstrained, but its integral leads to a delta function as we have seen. 
However, at the reducible connections there are nonzero solutions for $\phi$.
Thus, the space of $Q$-fixed points is larger than the space we wish to localize to.
Worse, this space is noncompact and the definition of the integration over this
noncompact region has been problematic. 

\subsec{The canonical formulation: Floer Homology}

The canonical formulation of Donaldson theory involves Floer homology for 3-folds.
Indeed Floer theory played a crucial role in the 
origin of cohomological field theory.  For further 
information see  
\refs{\FlThreeD, \atiythrfr, \Braam,\donaldson} . 

\subsec{Relation to ``physical'' Yang-Mills: $D=4$}

\subsubsec{The $N=2$ Lagrangian}

This is discussed in 
\refs{\BrSchSch,\WeBa,\sohnius,\wittsusygt}, 
as well as in many other references. 
We shall obtain $N=2$ super Yang-Mills theory in $d=4$ from the dimensional
reduction of $N=1$ SYM in $d=6$.
The field content of the six dimensional theory is simply a vector field, ${\bf A}_\mu$,
and a fermion, $\Lambda$.
The theory has a simple action:
\eqn\sixSYM{
I_{{\scriptscriptstyle d=6\ N=1}} = \int d^6 x~ \sqrt{g}~ \left \{
-{1\over 4} \tr~ F_{\mu\nu} F^{\mu\nu}
+ i \tr~ \bar \Lambda \Gamma^\mu ( D_{\bf A} )_\mu \Lambda \right \}}
where $D_{\bf A}$ is the gauge covariant derivative.
The action is invariant under the supersymmetry transformations:
\eqn\DSixSSY{\eqalign{
\delta {\bf A}_\mu
=& i \bar E \Gamma_\mu \Lambda  - i \bar \Lambda \Gamma_\mu~ E\cr
\delta \Lambda =& \Sigma^{\mu\nu}  F_{\mu\nu}~ E\cr
\delta \bar\Lambda =& -\bar E~ \Sigma^{\mu\nu}  F_{\mu\nu}\cr}}
where $\Sigma^{\mu\nu}
= {1 \over 4} \left ( \Gamma^\mu \Gamma^\nu - \Gamma^\nu \Gamma^\mu \right )$
and $E$ is a Grassmann supersymmetry transformation parameter.

We make the following choice for the representation of the $\Gamma$ matrices
in six dimensional Euclidean space, which is convenient in what follows
\eqn\SixDirac{\eqalign{
\Gamma^m =& \pmatrix{ \gamma^m & 0 \cr 0 & \gamma^m\cr}\qquad m = 0,\ldots, 3\cr
& \cr
\Gamma^4 = \pmatrix{ 0 & i \gamma_5\cr i \gamma_5 & 0\cr}\qquad
\Gamma^5 =& \pmatrix{ 0 & \gamma_5\cr - \gamma_5 & 0\cr}\qquad
\Gamma^7 = \pmatrix{ \gamma_5 & 0 \cr 0 & - \gamma_5\cr}\cr}}
where $\Lambda = \pmatrix{\Lambda_L\cr \bar\Lambda_R\cr}$ is a left-handed six dimensional
Weyl spinor.
Separately $\Lambda_L$ and $\Lambda_R$ are (in our choice of representations of the
Dirac matrices) left-handed and right-handed four dimensional Dirac matrices.
We may write
$$
\Lambda_L = \pmatrix{\lambda_L\cr 0\cr}\qquad\qquad
\Lambda_R = \pmatrix{ 0 \cr \bar \lambda_R\cr}
$$
These may be combined into a single four-dimensional Dirac spinor
$$
\lambda = \pmatrix{ \lambda_L\cr \bar\lambda_R\cr}
$$
Now introduce
\eqn\FourComps{\eqalign{
A_m =& {\bf A}_m\qquad m = 0,\ldots, 3\cr
B = {1 \over \sqrt{2}} ( {\bf A}_4 - i {\bf A}_5 )\qquad&
\bar B = {1 \over \sqrt{2}} ( {\bf A}_4 + i {\bf A}_5 )\cr}}
We dimensionally reduce by eliminating  $x^4,x^5$ dependence
of the fields. 
It is a simple matter to obtain the dimensionally reduced action expressed in
terms of the  fields above \refs{\BrSchSch,\WeBa,\sohnius,\wittsusygt}:
\eqn\NTwoDFour{\eqalign{
I_{{\rm N=2\ SYM}} =& \int \sqrt{g} d^4x~ \Tr \left \{
-{1\over 4} F_{mn} F^{mn} - ( D_A \bar B )_m ( D_A B )^m -
 {1 \over 2} [ B, \bar B ]^2\right.\cr
&\qquad\left.
-i \bar\lambda_{\dot\alpha i} ( \bar\sigma \cdot D_A )^{\dot\alpha \alpha} \lambda_\alpha^{~ i}
+ {i \over \sqrt{2}}  B \epsilon^{ij} [ \bar\lambda_{\dot\alpha i}, \bar\lambda^{\dot\alpha}_{~ j} ]
-{i \over {\sqrt{2}}} \bar 
B \epsilon_{ij} [ \lambda^{\alpha i}, \lambda_\alpha^{~ j} ]
\right\}\cr}}
From \DSixSSY\ we find that the supersymmetry transformations in $d=4$ are
given by
\eqn\DFourSSY{\eqalign{
\delta A_m
=& -i \bar\lambda_{\dot\alpha i}~ \bar\sigma_m{}^{\dot\alpha\alpha}~ \epsilon_\alpha{}^i
+ i \bar\epsilon_{\dot\alpha i}~ \bar\sigma_m{}^{\dot\alpha\alpha}~ \lambda_\alpha{}^i\cr
\delta \lambda_\alpha{}^i
=& \sigma^{mn}{}_{\alpha\beta}~ \epsilon^{\beta i}~ F_{mn} + i \epsilon_\alpha{}^i~ [ B, \bar B]
+i \sqrt{2}~ \sigma^m{}_{\alpha\dot\alpha}~ D_m B~ \epsilon^{ij}~ \bar\epsilon^{\dot\alpha}{}_j\cr
\delta\bar\lambda_{\dot\alpha i}
=& \bar\sigma^{mn}{}_{\dot\alpha\dot\beta}~ \bar\epsilon^{\dot\beta}_{~i}~ F_{mn}
-i\bar\epsilon_{\dot\alpha i}~ [ B, \bar B ]
+ i\sqrt{2}~ \sigma^m{}_{\alpha\dot\alpha}~ D_m \bar B~ \epsilon_{ij}~ \epsilon^{\alpha j}\cr
\delta B =& \sqrt 2\epsilon^{\alpha i}~ \lambda_{\alpha i}\cr
\delta \bar B =& \sqrt{2} \bar\epsilon^{\dot\alpha}{}_i~ \bar\lambda_{\dot\alpha}{}^i,\cr}}
The classical theory possesses a global $SU ( 2 )_L \times SU ( 2 )_R 
\times U (2 )_I$
symmetry group.
 $SU ( 2 )_L \times SU ( 2 )_R$ is the local Lorentz 
group and the $U ( 2 )_I$  is the ``$R$-symmetry'' 
 of $N=2$ supersymmetry.
The various fields of the action transform as follows under this symmetry:
$$\eqalign{
\lambda_\alpha^{~ i }&\qquad\qquad ( 2, 1, 2^{+1} )\cr
\bar\lambda_{\dot\alpha i}&\qquad\qquad ( 1, 2, 2^{-1} )\cr
B&\qquad\qquad ( 1, 1, 1^2 )\cr
\bar B&\qquad\qquad ( 1,1, 1^{-2} ), \cr}
$$
and the supercharges transform as $(2,1,2^{-1}) \oplus (1,2,2^{+1})$
(the superscript in the third entry denoting the 
$U(1)$ charge in $U(2)_I$). 
These symmetries will play a role in making a topological theory from the physical
$N=2$ theory.

\subsubsec{Topological Twisting}

Historically, Donaldson theory provided the first example of a topological twisting of an 
$N=2$ theory\refs{\donaldson}.
The procedure is analogous to that of sec. \sssTopltw.

Since the local Lorentz group is
$SO ( 4 ) \equil SU ( 2 )_L \times SU (2 )_R$, we may
 decompose the spin
connection: $\omega_m = ( ( \omega_L )_m,  ( \omega_R )_m )$.
To twist the physical theory, we now gauge the $SU(2)_I$ with external gauge fields, $B_m$. As in the topological sigma model, 
we next consider ``diagonal'' correlators with $B_m$ 
related to $(\omega_R)_m$.  The remaining symmetry 
group is 
$SU ( 2 )_L \times SU ( 2 )_R^\prime \times U ( 1 )_G$ 
(the $U(1)_G$ now being interpreted as ``ghost number'') 
where  $SU(2)_R'$ is diagonally embedded in 
$SU(2)_R\times SU(2)_I$. 
In particular, the supercharges now transform as: 
\eqn\nwchrges{
( 2, 2, -1 ) \oplus ( 1, 3, 1 ) \oplus ( 1, 1, 1 )
}
under
$SU ( 2 )_L \times SU ( 2 )_R^\prime \times U ( 1 )_G$. 
The first two are vectors and self-dual tensors, respectively.
Since a general 4-manifold does not admit nonvanishing vectors or self-dual forms
these charges will not, in general, exist on arbitrary 4-manifolds.  
The last is identified with the BRST operator. 
Being a scalar it can be defined on a general 4-manifold.
If $M$ is a K\"ahler manifold there is a second globally defined BRST operator. 
This has recently been exploited in \refs{\wittsusygt}\  to derive some explicit 
formulae for the Donaldson invariants. 

It is convenient to relabel the fields of the twisted theory in a way that reflects
their transformation behaviour under the new embedding of the local Lorentz
group\refs{\wittsusygt}.
For the supersymmetry generators:
$$\eqalign{
\epsilon^{\alpha i}\quad&\longrightarrow\quad 0\cr
\bar\epsilon^{\dot\alpha i}\quad&\longrightarrow\quad - \epsilon^{\dot\alpha \dot\beta} \rho\cr}
$$
while for the fields
$$\eqalign{
\psi_m =&~ {1\over2}\bar\sigma_m{}^{i\alpha} \lambda_{\alpha i}\cr
\chi_{mn} =&~ {1\over2}\bar\lambda^{\dot\alpha}_{~ i} \bar\sigma_{mn}{}^i{}_{\dot\alpha}\cr
\eta =&~ \bar\lambda^{\dot\alpha}{}_{\dot\alpha}\cr
\phi =&~ {i\over{\sqrt{2}}}~ B\cr
\lambda =&~ \bar B\cr}
$$
In terms of these fields and supersymmetry generators, the topological supersymmetry
transformations are quite familiar
\eqn\TopSSY{\eqalign{
\delta A_m =& i \rho \psi_m\cr
\delta \psi_m =& - \rho D_m \phi\cr
\delta \phi =& 0\cr}}
%
%

\subsubsec{Comparison to the MQ form}

As with the sigma model,  when comparing the MQ form to the twisted $N=2$ theory 
there is a difference.
In this case the difference consists of a 
topological term for
 the gauge fields and a term involving 
 the potential energy for  the scalar fields. 
The topological term is simply the Pontryagin index. 
The other term is  $Q$-exact and is given by:
\eqn\dffdldact{
\eqalign{
\Delta I  = Q \Delta \Psi
&= \int d^4 x \sqrt{g}~ \Tr~ \left (- \eta [ \phi, \eta ]+ [\phi,\lambda]^2 \right ) \cr
\Delta \Psi &= \int d^4 x \sqrt{g}~ \Tr~ ( \eta [ \phi, \lambda ])\cr}}
We must also rotate $\phi\to i \phi$ to compare 
to the physical theory. Note  from \FourComps\ 
that in the physical theory 
 $ i \phi $ and $  \lambda^\ast$ are proportional
complex fields,  while in the topological theory 
it is more natural to take $\phi$ and $\lambda$ to be real. 

\subsubsec{The Energy-Momentum Tensor}

As we mentioned in chapter \sINandTIR\ in our general discussion of topological actions,
the cohomology class defined by the path integral is independent of metric on the bundle $E_{\rm localization}$. 
Thus, even though a metric is used to construct the Donaldson Lagrangian, 
the correlation functions, being topological intersection numbers, cannot depend on the
choice of metric on spacetime. 
Physically this arises from the key fact that the energy-momentum tensor is a BRST commutator. 
That in turn follows immediately from the BRST exactness of the action 
\eqn\exctenmom{\eqalign{
T_{\mu\nu} & = \{ Q, \lambda_{\mu\nu} \}  \cr
\lambda_{\mu \nu} & ={1\over \sqrt{g}}  {\delta\over \delta g^{\mu\nu} } \Psi\cr}}

\subsec{Relation to ``physical'' Yang-Mills: $D=2$}
\subseclab\ssRelphysym

The relation of the physical and cohomological Yang-Mills theories in two dimensions is 
very intriguing and was described in \refs{\Witdgtr}. 
As we have seen, the cohomological theory localizes to the space of {\it flat connections}. 
Let us re-examine the physical theory. 

As discussed in section \sssWkcpl, the path integral of the physical theory can be written as in \ymtss. 
%
%
{}From this equation we see that the action is
 a sum of two BRST invariant operators in a cohomological
gauge theory on $\Sigma$ with basic multiplet $A$, $\psi$ and $\phi$.
Indeed, in a cohomological gauge theory the BRST class of  $\Tr~ \phi^2$ is independent
of $P$, hence the action \ymtss\ is just
\eqn\ymtssi{
I = i \CO_2^{(2)} ( \Sigma ) + \epsilon a \CO_2^{(0)} ( P )}
up to a BRST commutator. 
Moreover, as discussed in section \sssWkcpl, 
 $\CO_2^{(2)}(\Sigma)$ is just the equivariantly closed 
symplectic form 
of section \ssECSGA, and \ymtss\ is 
 the equivariant integration of the exponential of the 
equivariant symplectic form. 

These two observations lead to a  relation between the physical
and cohomological gauge theories.
Recall that to discuss localization of integrals of equivariant differential forms we must
add a $G$-invariant one-form $\lambda$, and perturb by $t D \lambda$. 
This perturbation can  then be interpreted as a $Q$-exact action in a cohomological
gauge theory. 
In this case we take \ymtwgf\ from section \sssWkcpl. 

In \Witdgtr\ Witten shows that the cohomological theory 
with gauge fermion \ymtwgf\  is equivalent to 
a theory with gauge fermion: 
\eqn\ssymtiii{
\Psi = \Psi_{\rm Donaldson} + {1\over t e^2} \int_\Sigma \Tr~ \chi~ \lambda}
He then shows that the theory 
with gauge fermion \ssymtiii\  is equivalent to $D=2$
Donaldson theory with the standard  gauge 
fermion  \tymprj\ + \loclagi, 
up to terms of order  $\sim \CO (e^{-c/te^2})$
for $t\to 0^+$. The difference comes about 
because the second term in \ssymtiii\  
introduces new $Q$-fixed points. 

The result is that  the generator of 
intersection numbers on the moduli space
of flat connections is related to the physical partition function by
\eqn\fltcnns{\eqalign{
{1 \over \vol(\CG)} \int DA~& \exp\Biggl[ {2 \pi^2\over \epsilon}
\int_\Sigma d \mu\Tr f^2\Biggr]\cr
&= e^{\alpha_1 (2-2p)}
\sum_R (\dim~ R)^{2-2p} e^{- e^2 a (C_2(R)+ \alpha_2) }\cr
&=\langle e^{\omega + \epsilon a \CO^{(0)} }\rangle_{\CM_N}
+\CO(e^{-c/(\epsilon a)})
\cr}}
where $\CM_N$ is the moduli spaceof flat connections
on $\Sigma$  for $G=SU(N)$.
 Formulae for the
renormalization constants  $\alpha_1,\alpha_2$ are given in \refs{\mooricm}. 

\vskip0.1truein\noindent
{\bf Remarks:}
\item{1.}
As emphasized in \Witdgtr, \fltcnns\ shows that the leading term for 
$e^2\to 0$ only has polynomial dependence, 
of order ${1\over 4} \dim~ \CM_N=\ha (G-1) (N^2-1)$. 
\item{2.}
Reference  \park\  makes the interesting suggestion that 
one can apply the ideas of nonabelian localization to
$D=4$ topological Yang-Mills theory. 

\subsubsec{Holomorphic maps and flat connections}
\subsubseclab\sssHolMpsF

We can  combine the results of the 
present section with those from chapter \sEC\  to
discover a  relation between the moduli spaces of holomorphic maps of
Riemann surfaces $\Sw \to \ST$ (Hurwitz space) 
and those of the moduli of flat $SU(N)$
bundles over $\ST$. 

Consider  \fltcnns\ for $N\to \infty$.
In order to define the large $N$ limit we must hold $N \epsilon = \lambda$ fixed.
Thus, the contributions of the unstable solutions are exponentially
 small for $N\to \infty$,
that is, {\it physical \ymt\ and $D=2$ Donaldson theory are identical at large $N$.}
Combining this with the results of part I, we obtain a relation between intersection 
theory of the moduli space of flat connections on $\ST$ and intersection theory for
the moduli space of holomorphic maps into $\ST$\foot{To make this statement rigorous
one must (a.) take care of the singularities in $\CM$ and (b.) ensure that the corrections
$\sim \CO(e^{-2 N c/\lambda})$ from \wkcpl\ are not overwhelmed by the ``entropy
of unstable solutions'' \refs{\grssmatyt}.
The absence of phase transitions as a function of $\lambda$ for $G>1$ suggests that,
for $G>1$, these terms are indeed $\sim \CO(e^{- N c'})$ for some constant $c'$. }:
\eqn\dualmdli{
\eqalign{
\biggl \langle exp\biggl[ \omega + 
{8 \pi^2  a \over N} \CO^{(0)} \biggr]\biggr \rangle_{\CM(F=0,\ST,SU(N) )} &\cr
{\buildrel N\to \infty \over \sim}\qquad
\sum_{\chi } N^\chi \sum_{d\geq 0} & e^{-\ha d a } P_d(a) 
\chi_{\rm orb} (\CC\CH(\Sw,\ST,d))\cr}}
where $\CC\CH(\Sw,\ST,d)$ is the coupled Hurwitz moduli space for maps of degree $d$, 
which discussed in chapter \sEC.  
$P_d(a) $ is a polynomial  with $P_d(0)=1$.
We will return to this formula in the concluding chapter. 

\newsec{2D Topological Gravity}
\seclab\sTDTG

At the fundamental level, 2D topological gravity is the study of intersection theory
on the moduli space $\CM_{h,n}$ of Riemann surfaces of genus $h$ with 
$n$ punctures through field-theoretic methods. 
There are many ways of describing the moduli $\CM_{h,n}$
and correspondingly many different field theoretic formulations. 
How one describes the theory by equations, fields, and symmetries depends on which formulation of moduli space one uses. 
This has led to a bewildering jungle of  formulations in the literature. 
In our opinion a satisfactory field theoretic formulation of topological gravity
remains to be completed.

\subsec{Formulation 1: $\CG=\Diff(\Sigma)\sdtimes \Weyl(\Sigma)$}

Ideally we would like to have no equations at all. 
For the above group we can take universal bundle and classifying space to be
\eqn\dffwyli{\eqalign{ 
E\CG &=\MET(\Sigma) \cr
B\CG  &=\CM_{h,0}.  \cr }}

\subsubsec{BRST Complex}

Algebraically we can formulate the complexes for equivariant cohomology as: 

\noindent
A. BRST Model

The generators of the complex are: 
\eqn\dgeni{\eqalign{
h_{\alpha\beta} \in \O^0 ( \MET )  & \qquad
\psi_{\alpha\beta} = \widetilde dh_{\alpha \beta} \in \O^1 ( \MET )\cr
c^\alpha \in \CW^1( \diff ) & \qquad
\gamma^\alpha \in \CW^2( \diff )\cr
\rho \in \CW^1 ( \weylg )& \qquad \tau \in \CW^2( \weylg ) \cr}}
Translating \BRSTOp\  into the present example we find: 
\eqn\dondi{\eqalign{
d_B h_{\alpha\beta} &=\psi_{\alpha\beta} + \cL_c h_{\alpha\beta} - \rho h_{\alpha\beta} \cr
d_B \psi_{\alpha\beta} & = \cL_c \psi_{\alpha\beta} - \rho \psi_{\alpha\beta}
- \cL_\gamma h_{\alpha\beta} + \tau h_{\alpha\beta}\cr
d_B c^\alpha & =\gamma^\alpha -\half [c,c]^\alpha\cr
d_B\gamma^\alpha &= \cL_c \gamma^\alpha  \cr
d_B\rho &=\tau + \cL_c \rho \cr
d_B\tau &= \cL_c \tau - \cL_\gamma \rho. \cr}}
The occurrence of Lie derivatives with respect to $c^\alpha$ in the last two formulae arises
because we actually have a {\it semi-direct} product of $\Diff$ and $\Weyl$. 
Exactly this multiplet of fields was used in the formulation of 2D topological gravity
in \refs{\LaPeWi}. 

\noindent
B. Cartan Model

This is easily obtained from the BRST model by setting the odd (degree one) generators to
zero.
In this case the generators of the complex are: 
\eqn\dgeni{\eqalign{
h_{\alpha\beta} \in \O^0( \MET )  & \qquad
\psi_{\alpha\beta} = \widetilde dh_{\alpha\beta} \in \O^1( \MET ) \cr
\gamma^\alpha \in \CW^2( \diff )   &  \qquad
\tau \in \CW^2( \weylg )\cr}}
Translating \CartanDiff\  into the present example we find: 
\eqn\dondii{\eqalign{
d_\CC h_{\alpha\beta} &= \psi_{\alpha\beta} \cr
d_\CC \psi_{\alpha\beta} & = -\cL_\gamma h_{\alpha\beta} + \tau h_{\alpha\beta}\cr
d_\CC \gamma^\alpha & = 0\cr
d_\CC \tau & = 0\cr}}

\subsubsec{Lagrangian}

There is no localization Lagrangian so we must only construct projection
and gauge-fixing Lagrangians. 
Here we meet a new element.
The construction of the projection Lagrangian in Chapter \sTTwithLS\ requires a
$\CG$-connection on the principal $E\CG$ bundle $\MET$.
In Chapter \sTTwithLS\ this was constructed using a $\CG$-invariant metric. 
The natural metric on $\MET$ is
\eqn\metonmet{
( \delta h^{{\sst ( 1 )}},  \delta h^{{\sst ( 2 )}} )
= \int_\Sigma d^2 z \sqrt{h}~ \left (
h^{\alpha\gamma} h^{\beta\delta} + t~ h^{\alpha\beta} h^{\gamma\delta} \right )
\delta h^{{\sst ( 1 )}}_{\alpha\beta}~ \delta h^{{\sst ( 2 )}}_{\gamma\delta}}
where $t$ is an arbitary positive real number\foot{The restriction of $t$ to positive
real numbers ensures that the metric is positive definite.}.
Unfortunately,  the general construction of section \ssABRSTCofPhi\  cannot be
applied because the metric \metonmet\  is {\it not} Weyl invariant, nor is the
corresponding metric on $\Diff(\Sigma)$. 

Nevertheless, the general reasoning of section \ssABRSTCofPhi\ can be applied.
All we really need to construct  $\Phi ( P \to M )$ is a vertical 1-form, valued in a
vector bundle, that is, an analogue of $C^\dagger$. 
In the present case we can construct a one-form whose kernel is the space of
solutions to 
\eqn\qdffrls{
P^\dagger \delta h_{\alpha\beta} = D^\alpha \delta h_{\alpha \beta} =0.}
where
\eqn\orlpi{
( P \xi )_{\alpha \beta} = D_{(\alpha}\xi_{\beta)} - 
h_{\alpha\beta} D\cdot \xi,}
(The notation $P$ and  $P^\dagger$ is standard in string theory.)
In conformal coordinates such zero modes are given by $\delta h_{zz} (dz)^2$ which
are holomorphic quadratic differentials $\in H^0(\Sigma; K^2)$.
Let
$\Pi_{\alpha \beta}{}^{\gamma \delta}$ be the projection onto $H^0 ( \Sigma; K^2 )$
\foot{In terms of the $P$, this is
$$
\Pi_{\alpha \beta}^{~~ \gamma \delta}
= ( {\bf 1} - P {1\over{P^\dagger P}}  P^\dagger )_{\alpha \beta}^{~~ \gamma \delta}
$$
modulo some complications which occur in low genus due to the nontriviality of the
kernel of $P^\dagger P$.}. 
Then $C^\dagger = \Pi_{\alpha \beta}^{~~\gamma \delta} \psi_{\gamma \delta}$

For antighosts we introduce $\lambda^{\alpha \beta}$ and  $\eta^{\alpha \beta}$;  we
form\foot{The antighosts do not live in the Lie algebra because we are not using 
the canonical connection employed in chapter \sTTwithLS.}
\eqn\dffwylii{
\Psi_{{\rm projection}} = i (\lambda, C^\dagger)
= \int_\Sigma d^2 z \sqrt{h}~ \lambda^{\alpha \beta} \Pi_{\alpha \beta}{}^{\gamma \delta} 
\psi_{\gamma \delta}}
We must also introduce a gauge-fixing term in the standard way by using the BRST 
model, choosing a slice $ h_{\alpha \beta}^{(0)}$ for $\CG$ and introducing:
\eqn\dffwyliii{
\Psi_{{\rm gauge-fixing}} = \int d^2 z \sqrt{h}~ 
b^{\alpha \beta} ( h_{\alpha \beta} - h_{\alpha \beta}^{{\scriptscriptstyle (0)}} )}
where $Q b^{\alpha\beta} = d^{\alpha\beta}$ and 
$b^{\alpha\beta}, d^{\alpha\beta}$ are 
symmetric tensors of ghost number
$-1$ and $0$, respectively. 

The Lagrangian can now be easily worked out.
The slice is $6h-6$ dimensional.
That is, in $\MET$ almost all ( $\infty - (6h-6)$ ) directions are vertical.
For the moment let us assume that, in fact, {\it all} directions are vertical. 
Then we could simply take $C^\dagger = \psi$ and drop the
$\Pi_{\alpha \beta}{}^{\gamma \delta}$ in \dffwylii.
Applying the BRST model differential we arrive at the Lagrangian: 
\eqn\dffwyliv{\eqalign{
d_B&  ( \Psi_{{\rm projection}} + \Psi_{{\rm gauge\ fixing}} )\cr
&\quad = \int d^2 z \sqrt{h}~ \Biggl\{
\eta^{\alpha \beta} \psi_{\alpha \beta} 
+ \lambda^{\alpha \beta} ( \tau h_{\alpha \beta} - D_{(\alpha} \gamma_{\beta)} ) \Biggr\}\cr
&\qquad +\int \sqrt{h} d^2 z~ \Biggl\{ 
d^{\alpha \beta} (h_{\alpha \beta} -h_{\alpha \beta}^{(0)})
- b^{\alpha \beta}  \psi_{\alpha\beta} - b^{\alpha\beta} (D_{(\alpha} c_{\beta)} - \rho~ h_{\alpha \beta}) \Biggr\}\cr
&\qquad
+ \int \sqrt{h} d^2 z~ \half h^{\alpha\beta} ( {\cal L}_c h_{\alpha\beta}
- h_{\alpha\beta} \rho )\Biggl\{ \lambda^{\gamma\delta} \psi_{\gamma\delta} 
+ b^{\gamma\delta} ( h_{\gamma\delta} - h_{\gamma\delta}^{{\scriptscriptstyle (0)}} )
\Biggr\}\cr}}
%
%

Physically, we may integrate out fields to simplify the Lagrangian:
Integrating out $\eta^{\alpha\beta}$ sets $\psi_{\alpha\beta}=0$. 
Integrating out $d^{\alpha\beta}$ sets $h_{\alpha\beta}=h^{(0)}_{\alpha\beta}$.
Integrating out $\tau$ and $\rho$, makes $\lambda$ and $b$ traceless. 
We are then left with: 
\eqn\dffwyliv{
\int d^2 z \sqrt{h}~ \Biggl[ 
b^{\alpha \beta}  D_{(\alpha} c_{\beta)} -
\lambda^{\alpha \beta} D_{(\alpha} \gamma_{\beta)} 
\Biggr]}
Geometrically the integral on $\eta$ produces a totally vertical form, so we may drop
$\psi$ in the remainder of the Lagrangian while the integral over $d^{\alpha\beta}$
produces a Poincar\'e dual in the space of metrics to the slice $h^{(0)}_{\alpha\beta}$. 

The field $\lambda^{\alpha\beta}$ is usually denoted $\beta$ and equation \dffwyliv\ is
the ``$bc \beta \gamma$''  formulation of 2D gravity obtained in \refs{\LaPeWi, \Di}.
This field theory is an example of a conformal field theory with total conformal 
anomaly $c=-26+26=0$.
As we saw in chapter \sEQCO, we expect such systems to arise in discussions
of equivariant cohomology on very general grounds. 

Finally, let us return to the complications of $\Pi_{\alpha \beta}^{~~ \gamma \delta}$. 
There will be extra nonlocal terms from the variation of this projector.
However, by introducing a finite set of quantum mechanical degrees of freedom 
(in correspondence with the superspace $\Pi T \CM_{h,0}$) this nonlocality can be 
eliminated \refs{\LaPeWi, \becchi}. 

\subsec{Formulation 2: $\CG=\Diff(\Sigma)$ } 
\subseclab\ssFormtwo

We study $\Diff( \Sigma )$-equivariant cohomology of $\MET ( \Sigma )_k$, the 
space of metrics of constant scalar curvature $R = k$, where
$$
k = \cases{+1 & if $g = 0$\cr
                     0  & if $g = 1$\cr
                     -1 & if $g \ge 2$\cr}
$$

\subsubsec{Equations} 

The fundamental field variable will be $h_{\alpha \beta} \in \MET ( \Sigma )$. 
Therefore, we must localize to $\MET ( \Sigma )_k$.
The equations will be 
\eqn\vanivi{
\CZ = \{ h ~\vert~ R (h) = k \} / \CG}
so $\CV=\MET \times \IR$.
The corresponding antighosts are $\rho,\pi\in \Omega^2 ( \Sigma )$. 
We take the trivial connection on $\CV \to \MET$. 

\subsubsec{BRST Complex}

The BRST complex is easily obtained from the complex \dgeni\ and \dondi\  by
dropping Weyl degrees of freedom: $\rho,\tau\to 0$. 
As usual there are BRST and Cartan models. 
The Cartan model is especially simple, with generators:
\eqn\dgeni{
h_{\alpha\beta} \in \Omega^0 ( \MET )\qquad
\psi_{\alpha\beta} = \widetilde d h_{\alpha\beta} \in \Omega^1 ( \MET )\qquad
\gamma^\alpha \in \CW^2( \diff )}
and Cartan differential, according to  \CartanDiff: 
\eqn\dondii{\eqalign{
d_\CC h_{\alpha\beta} &=\psi_{\alpha\beta} \cr
d_\CC \psi_{\alpha\beta} & = -\cL_\gamma h_{\alpha\beta}
= - ( D_\alpha \gamma_\beta + D_\beta \gamma_\alpha) \cr
d_\CC \gamma^\alpha & =0. \cr}}
This multiplet of fields is very popular and was used in \refs{\Witp, \Winm}, among
other places. 

The antighosts for the projection form are vector fields 
\eqn\cartfri{
\lambda^\alpha, \eta^\alpha \in \liebg= {\rm diff}~ (\Sigma)}
of ghost numbers $-2$ and $-1$, respectively.
In the Cartan model the transformation laws are: 
\eqn\cartfrii{
d_\CC \lambda^\alpha = \eta^\alpha\qquad
d_\CC \eta^\alpha = - \CL_\gamma \lambda^\alpha . }

\subsubsec{Lagrangian}

\noindent
A. Localization Lagrangian

For the localization Lagrangian we introduce a scalar antighost $\rho$ and its
Lagrange multiplier $\pi$. 
Then the gauge fermion for localizing to $\MET ( \Sigma )_k$ is:
\eqn\wylloci{
\Psi_{\rm weyl\ loc} = \int d^2 z \sqrt{h}~ \rho~ ( R - k )}

Using the following two relations:
\eqn\UseFull{\eqalign{
Q \Gamma^\alpha_{\beta\gamma}
=& \half h^{\alpha\delta} \left (
D_\beta Q h_{\gamma\delta} + D_\gamma Q h_{\beta\delta} - D_\delta Q h_{\beta\gamma}
\right )\cr
Q R
=& - \half D_\alpha D^\alpha ( h^{\gamma\beta}~ Q h_{\beta\gamma} )
+ D^\alpha D^\beta~ Q h_{\alpha\beta}
- \half R~ h^{\alpha\beta}~ Q h_{\alpha\beta}
\cr}}
where $Q = d + d_\CC$, we may write this action as 
\eqn\wyllocii{
I_{\rm weyl\ loc}= 
\int d^2 z \sqrt{h}~ \left \{ \pi~ ( R - k ) - \rho~ L^{\alpha\beta} \psi_{\alpha\beta}
\right \}}
where 
\eqn\LOpDef{
L^{\alpha\beta} = D^\alpha D^\beta - \half h^{\alpha\beta} D^2 - \half k h^{\alpha\beta}}

\noindent
B. Projection Lagrangian

The metric \metonmet\  {\it is} $\Diff$ invariant. 
Therefore we can directly transcribe the construction of chapter 
\sTTwithLS. 
For the projection Lagrangian we  use the formulae in section 
\sssQexct\  above. 
The operator $C$ is  
\eqn\CGrav{\eqalign{
C\colon \gamma &\to \CL_\gamma h_{\alpha\beta}\cr
\gamma^\alpha &\to (P \gamma )_{\alpha\beta} = D_{(\alpha} \gamma_{\beta)} \cr}}
so that the operator $C^\dagger$ is
\eqn\CDagGrav{\eqalign{
C^\dagger\colon T_h~ \MET &= \Gamma ( \Sym~ ( T^\ast \Sigma^{\otimes 2} ))
\to {\rm Vect} ( \Sigma )\cr
C^\dagger &= P^\dagger \psi\cr}}
where the last expression represents $C^\dagger$ in superspace.

Thus we take the gauge fermion:
\eqn\rdgf{
\Psi_{\rm projection} ~=~ i  \int d^2 z \sqrt{h}~
\left \{ \lambda^\alpha D^\beta \psi_{\alpha\beta} \right \}}
with corresponding action 
\eqn\rdact{\eqalign{
I_{\rm projection} =& ~ i \int d^2 z \sqrt{h}~ \left \{
\lambda^\alpha [ ( D_\alpha \psi_{\beta\gamma} ) \psi^{\beta\gamma}
+ ( D_\beta \psi_{\alpha\gamma} ) \psi^{\beta\gamma} ]
\right.\cr
&\qquad\qquad\qquad\qquad\left.
- \lambda^\alpha ( D^\beta D_\beta \gamma_\alpha + D^\beta D_\alpha \gamma_\beta ) 
+ \eta^\alpha D^\beta \psi_{\alpha\beta} \right \}\cr}}

The combined action $I_{\rm weyl\ loc} + I_{\rm proj}$ is still $\Diff$ invariant so that,
just as in the case of Donaldson theory, it remains for us to fix the gauge symmetry.
We fix the $\Diff$ symmetry as in \dffwyliii,  introducing the
standard $b$ and $c$ ghosts.
Altogether, the Lagrangian is
\eqn\totallii{\eqalign{
I_{\rm tot}
=& I_{\rm weyl\ loc} + I_{\rm projection} + I_{\rm gauge-fixing}\cr}
}
\vskip0.1truein\noindent
{\bf Remarks}
\item{1.}
This Lagrangian is not manifestly conformally invariant, although the $d_\CC$-invariance
suggests it should have a conformally invariant quantization.
It does, formally, have a cancelling conformal anomaly $52-52 =0$ as it must by
$d_\CC$-symmetry. 
\item{2.}
This formulation of 2D gravity is essentially that given in \refs{\LaPeWi}, eq. 4.8, and is closest
to our general formulation in chapter \sTTwithLS.
We will use it in formulating topological string theory and the \ymt\ theory in the next
two chapters. 

\subsec{Formulation 3: $\CG=\Diff(\Sigma) \times L.L.(\Sigma)$}

Another formulation makes 2d topological gravity seem very similar to Donaldson theory.
The physical fields in this formulation are the zweibeine, local sections of the frame
bundle, $F \to \Sigma$, and the spin connection, an $SO ( 2 )$ connection on $F$.
We shall denote the space of zweibeine and spin connections by $\FRAME$.

\subsubsec{Equations}

As before, we fix the Weyl symmetry by imposing a curvature constraint on the spin
connection.
In defining topological gravity, we must ensure that our connection is Riemannian,
hence we must localize to the subspace of $( e^a, \omega_a^{~ b} )$, which
satisfy a set of  torsion constraints in addition to the curvature constraint:
\eqn\tdgeqs{
\CZ = \{ e^a , \omega_{~ b}^a \in \FRAME~ \vert~
R_{~ b}^a = d \omega_{~ b}^a + \omega_{~ c}^a \omega_{~ b}^c = k \epsilon^a_{~ b}~
{\rm and}~ T^a = d e^a + \omega^a_{~ b} e^b = 0 \}}
where $k = 0, \pm 1$, depending on the genus of $\Sigma$ and
$\epsilon = \pmatrix{0 & 1\cr -1 & 0\cr}$.
The obvious choice for $\CV$ is the trivial bundle $\CV = \FRAME \times \IR^3$.
The corresponding anti-ghosts and Lagrange multipliers are
$r \in \Omega^0 ( \Sigma; so(2))$, $\rho_a \in \Omega^0 ( \FRAME^\ast )$ 
and $p \in \Omega^0 ( \Sigma; so(2))$, $\pi_a \in \Omega^0 ( \FRAME^\ast )$.

\subsubsec{BRST Complex}

Generators of the complex:

$$\eqalign{
e^a \in& \O^0 ( \FRAME )\qquad\qquad
\psi^a = \widetilde d e^a \in \O^1 ( \FRAME )\cr
\omega^a{}_b \in& \O^0 ( \FRAME, so ( 2 ))\qquad
\psi_0{}^a{}_b = \widetilde d \omega \in \O^1 ( \FRAME, so ( 2 ))\cr
c^\alpha \in& \CW^1 (\diff )\qquad\qquad\qquad
\gamma^\alpha \in \CW^2 ( \diff )\cr
c_0{}^a{}_b \in& \CW^1 ( \loclor )\qquad\qquad
\gamma_0{}^a{}_b \in \CW^2 ( \loclor )\cr}
$$

\vskip0.1truein\noindent
A. BRST Model

In the BRST model the differential is given by

\eqn\totbrst{\eqalign{
 Q~e^a   ~=&~ \CL_c e^a - (  e \cdot c_0 )^a + \psi^a, \cr
 Q~\psi^a   ~=&~ \CL_c \psi^a + (  e \cdot \gamma_0 )^a
	    -\CL_\gamma e^a - ( \psi \cdot c_0 )^a, \cr
 Q~\omega_{~ b}^a  ~=&~ i_c R_{~ b}^a  - D c_0{}^a{}_b + \psi_0{}^a{}_b, \cr
 Q~\psi^a_{0~ b} ~=&~ i_c ( D \psi_0 )_{~ b}^a - [ c_0, \psi_0 ]_{~ b}^a
	    - i_\gamma R_{~ b}^a  - D \gamma_0{}^a{}_b, \cr
 Q~c^\alpha   ~=&~ \half \CL_c c^\alpha + \gamma^\alpha, \cr
 Q~\gamma^\alpha   ~=&~       \CL_c \gamma^\alpha, \cr
 Q~c_{0~b}^a ~=&~ \half i_c i_c  R_{~ b}^a
   -\half \{ c_0, c_0 \}_{~ b}^a + \gamma_{0~ b}^a, \cr
 Q~\gamma_{0 b}^a ~=&~ \half i_c i_c ( D \psi_0 )_{~ b}^a
   + [ c_0, \gamma_0  ]_{~ b}^a. \cr}}

\vskip0.1truein\noindent
B. Cartan Model

As always the Cartan model is readily obtained from the BRST model by
setting the gauge ghosts (in this case, 
 the $\diff$ ghosts $c^\alpha$ and
the local Lorentz ghosts $c_0^{~ a}$,  to zero.)
\eqn\totCart{\eqalign{
 Q_\CC~e^a   ~=&~ \psi^a, \cr
 Q_\CC~\psi^a   ~=&~ (  e \cdot \gamma_0 )^a -\CL_\gamma e^a, \cr
 Q_\CC~\omega_{~ b}^a  ~=&~\psi_0{}^a{}_b, \cr
 Q_\CC~\psi_0{}^a{}_b ~=&~ - i_\gamma R_{~ b}^a - D \gamma_0{}^a{}_b, \cr
 Q_\CC~\gamma^\alpha   ~=&~ 0, \cr
 Q_\CC~\gamma_0{}^a{}_b ~=&~ 0.\cr}}

\subsubsec{Lagrangians}

The only novel feature of this model is the localisation
Lagrangian:
\eqn\tdgadl{
I_{\rm loc}= Q
\int d^2 z \sqrt{h}~ \left \{
r~ \epsilon^b_{~ a} ( d \omega_{~ b}^a + \omega_{~ c}^a \omega_{~ b}^c - k \epsilon^a_{~ b} )
+ \rho_a~ ( d e^a + \omega^a_{~ b} e^b ) \right \}}
This is sometimes described as ``Donaldson theory of the spin connection''.

\subsec{Formulation 4: $\CG=\widehat{\Diff(\Sigma)} $} 

This formulation 
is an illustration of the general remark of section \ssECvLAC.
There we saw that $\CG$-equivariant cohomology can be formulated as ordinary 
Lie algebra cohomology for a ``supersymmetrized''  Lie algebra.
Thus we can approach the subject as an ordinary gauge theory, but for a
supersymmetrized version of $\Diff( \Sigma )$. 

To begin, we localize to the space of constant curvature metrics using 
\wylloci\ and \wyllocii,  but this time we gauge fix $h_{\alpha \beta}$ and
$\psi_{\alpha \beta}$ to 
\eqn\wyllociii{
h_{\alpha \beta} = e^\phi~ \delta_{\alpha \beta} 
\qquad 
\psi_{\alpha \beta} = e^\phi \psi~  \delta_{\alpha \beta}}
The gauge fixing introduces the $b,c,\beta,\gamma$ ghost systems. 
The theory breaks up into two sectors: a $b,c,\beta,\gamma$ and a ``topological Liouville'' 
sector.
Both have twisted $N=2$ supersymmetry. 
This formulation  was discussed in \refs{\DiVVTrieste}.

\subsec{Observables}

Next we turn to the matter of observables.
Further discussion of observables in topological gravity can be found 
in \refs{\BaSii, \Witdgit, \storaii,\becchi}

Let $X$ be an orientable, $d$-dimensional Riemannian manifold and let $F$
be  the frame bundle.
$F$ is a principal $GL(d,\IR)$ bundle over $X$.
Let ${\cal A}$ be the space of connections on $F$.
Two local symmetries act on ${\cal A} \times F$: (i) the diffeomorphisms of $X$,
$\Diff ( X )$, and (ii) the local linear frame transformations, ${\cal G}$.
We shall denote ${\cal H} = {\cal G} \times \Diff ( X )$.
In contrast to the ordinary gauge theory case, the automorphisms of $F$ that we
need to consider  here may be both horizontal (Diff) and vertical ($\CG$).
As in the case of Yang-Mills moduli space, the naive moduli spaces,
${\cal A} / {\cal H}$, are problematical.
The resolution to (some of) these problems is to choose base points
$x_1, \ldots, x_n \in X $ at which we specify trivialisations, $\varphi_i$,
of $F$.
We then consider the $n+1$-tuples $( \omega; \varphi_1, \ldots, \varphi_n )$,
where $\omega$ is a spin connection of $F$, modded out by automorphisms
(which now include diffeomorphisms) that act trivially at the base points.
We shall denote
${\cal H}_0 = \{ h \in {\cal H}~ \vert~ h ( x_i ) = ( 1, x_i ),~ \forall i = 1, \ldots, n \}$.

Then consider
\eqn\UnivBundGravi{\matrix{
 & {\cal A}^\ast \times F & \cr
\swarrow{{\cal H}_0} & & \searrow{H_0} \cr
&&\cr
 \widetilde{\cal B}^\ast \times F\quad & & {\cal A}^\ast \times X \cr
&&\cr
\searrow{H_0} & & \swarrow{{\cal H}_0} \cr
 & \widetilde {\cal B}^\ast \times X & \cr}}
where $H_0 = GL(d,\IR) / C ( GL(d,\IR) )$ and where
$\widetilde {\cal B}^\ast = \widetilde{\cal A}^\ast / {\cal H}_0$.
So $\widetilde {\cal B}^\ast \times F$ is a principal $H_0$ bundle over
$\widetilde {\cal B}^\ast  \times X$:
\eqn\UnivBundGravii{\matrix{
\widetilde {\cal B}^\ast \times F & \leftarrow & H_0\cr
\downarrow{} &&\cr
\widetilde {\cal B}^\ast \times X & &\cr}}
Consider now  the case of $d=2$.
Let $X = \Sigma_{h,n}$ be some Riemann surface of genus
$h$ with $n$ marked points.
Then we study the moduli spaces, ${\cal M}_{h,n}$ of  stable genus $h$ Riemann
surfaces with $n$ marked points.
The Deligne-Mumford compactification of these moduli spaces will be denoted
by $\bar {\cal M}_{h,n}$.
There is the ``forgetful" map $\beta_n\colon \bar{\cal M}_{h,n} \to \bar{\cal M}_{h,n-1}$
which forgets about the first trivialisation \refs{\Witdgit}.
This map does not exist for certain low values of $h$ and $n$
(e.g. $( h, n ) = ( 0,3 )$ or $( h, n ) = ( 1,1)$.)

We have a classifying map:
\eqn\ClassMap{
\Phi\colon \bar{\cal M}_{h,n} \times \Sigma_{h,n} \to BGL(2,\IR) }
which we may use to obtain cohomology classes
$c \in H^d ( \widetilde {\cal B}^\ast \times X )$ from characteristic classes
associated with the maximal compact subgroup $SO(2)$.

The cohomology classes on $\bar {\cal M}_{h,n}$ may be constructed in a manner
similar to the case of topological Yang-Mills theory \refs{\BaSii,\Witdgit}:
\item{1.}
There are the Mumford-Morita-Miller classes\refs{\Mu,\Morita,\miller} which are
constructed as follows:
Consider the moduli space $\bar {\cal M}_{h,1}$.
The location of the marked point at $x_1 \in X$, may be viewed as a section
of the universal curve: $\bar {\cal M}_{h,1} \to \bar {\cal M}_{h,0}$.
The cotangent along the fibre defines the relative cotangent bundle:
$$\matrix{
{\cal L}_{{\scriptscriptstyle (1)}} = K_{{\cal C}/{\cal M}}\cr
\mapdown{} \cr
\bar {\cal M}_{h,1}\cr}
$$
The first Chern class of ${\cal L}_{{\scriptscriptstyle (1)}}$ (and its powers) define a set
of cohomology classes on $\bar{\cal M}_{h,1}$, which we may pushdown via the
``forgetful" map: $\beta\colon\bar{\cal M}_{h,1} \to \bar{\cal M}_{h,0}$, which forgets
about the first point.
We then define
\eqn\MumfordObs{
\kappa_n = \beta_\ast ( c_1 ( {\cal L}_{{\scriptscriptstyle (1)}} )^{n+1} )}
The resulting objects are degree $2n$ cohomology classes on $\bar {\cal M}_{h,0}$.
\item{2.}
There is a set of cohomology classes on $\bar{\cal M}_{h,n}$ which is
very natural from the point of view of topological Yang-Mills theory.
There is an obvious generalization of the line bundle ${\cal L}_{{\scriptscriptstyle (1)}}$
to a collection of line bundles ${\cal L}_{{\scriptscriptstyle (i)}} \to \bar {\cal M}_{h,n}$
for $i = 1,\ldots, n$.
Just as in the case of topological Yang-Mills theory, we define observables via the slant
product pairing:
$$\eqalign{
/\colon H_i ( X ) \times H^r ( \bar{\cal M}_{h, n} \times X )&\to H^{r-i} ( \bar{\cal M}_{h,n} )\cr
{\cal O}_c~ [ \alpha ] & \mapsto c / [ \alpha ]\cr}
$$
In particular, for any $x_i \in \Sigma_{h,n}$, this yields the
classes:
$$
\sigma_n ( x_i ) = ( c_1 ( {\cal L}_{{\scriptscriptstyle (i)}} ))^n ( x_i )
$$
which live in $H^{2n+2} ( \bar {\cal M}_{h,n} )$.

\subsec{Hamiltonian Approach} 

The Hamiltonian approach to the construction of operators in 2D gravity within the
framework of equivariant cohomology has been discussed in \refs{\getzler}.
The cohomology can be identified with $S^1$-equivariant cohomology, 
and this continues to hold when coupled to matter.
The result is somewhat surprising since one might well have expected to compute 
$\Diff(S^1)$-equivariant cohomology. 
Conceivably,  we can replace $\Diff(S^1) \to S^1$ in homotopy theory. 
(Many other papers, too numerous to list here, have examined 2D topological
gravity from the canonical point of view.) 

\subsec{Correlation Functions}

According to general principles we know that the correlation functions of the 
field theory will be related to intersection theory on moduli space by
\eqn\twdgrvcr{\eqalign{
\langle \prod_i \hat \sigma_{n_i}  \rangle
&=\int_{\CM_{h,0}} \prod_i \sigma_{n_i}~ \chi (\cok \nabla \bar s)\cr
&=\int_{\CM_{h,0}} \prod \sigma_{n_i} \cr}}
The second equality arises since, in 2d topological gravity, the appropriate operators have 
$\cok \nabla \bar s=\{ 0\} $ for $h > 1$.
This can be seen in different ways from the different formulations as follows: 

\vskip0.1truein\noindent
Formulation 1:  Here there are no equations, hence no section $\bar s$! Thus it is trivially true that $\chi=1$. 
\vskip0.1truein\noindent
Formulation 2: Here we have
\eqn\twdgrvcri{
\bO = \pmatrix{L^{\alpha \beta} \psi_{\alpha \beta}\cr
\nabla^\beta \psi_{\alpha \beta} \cr}}
where $L^{\alpha\beta}$ is defined in \LOpDef. 
It is well-known  that $P^\dagger$ has a trivial cokernel when the genus is greater than one.
In other words, any vector field may be expressed as $P^\dagger$ acting on a {\it traceless}
quadratic differential.
The question is now whether any function on $\Sigma$ can be expressed via
as $L^{\alpha\beta}$ acting on the trace part of a quadratic differential.
From \LOpDef, it is apparent that the diferential part of $L$ does not act on
the trace part of $\psi$.
This leaves only the algebraic piece of $L$ to act on it.
This may be solved without obstruction for genus greater than one.
It follows that
\eqn\twdgrvcri{
\ker~ \bO \cong H^0(\Sigma; K^2) \qquad\qquad 
\coker~ \bO \cong \{0\} }

\newsec{Topological string theory}
\seclab\sTST

\subsec{Basic Data}
\subseclab\ssTSTBD

We continue our examination of $\CG$-equivariant cohomology for
$\CG=\Diff(\Sigma),\- \Diff(\Sigma)\sdtimes \Weyl(\Sigma)$, etc.
A natural $\CG$-space is the space of ${\cal C}^\infty$ maps
from a worldsheet $\Sigma$ to a target $X$, 
$\MAP~ ( \Sigma, X ) $, of
chapter \sTSM.  
We will take $X$ to be compact and K\"ahler.
\foot{More generally, $X$ can be almost K\"ahler. 
The formal construction of Lagrangians discussed 
in this paper works equally well for $X$ 
noncompact. More work is required to formulate 
precisely the observables and correlators in the 
noncompact case.}
Thus, 
the fields in the theory will  be pairs
$$\bF= (h, f)\in \tcM\equiv \MET\times \MAP \quad . $$ 
These are the fields of topological string theory\refs{\Witsm,\DiVVTrieste,\MoSo}.

\subsec{Equations}
\subseclab\ssTSTeqs

The quotient space $\MAP \times_\CG \MET$ is 
again infinite dimensional, and again
we  must localize on an interesting finite-dimensional  subspace thereof.
The space of holomorphic maps, $\HOL ( \Sigma, X )$,  is one very interesting subspace, and we will focus on this. 
We localize by the same section as in chapter \sTSM:
$s (h, f ) = d f + J df \epsilon (h)$.
The complex structure on the 
worldsheet is related to the (conformal class of the)
worldsheet metric, $h$, via
$\epsilon_\alpha^{~ \beta} ( h ) = \sh \hat\epsilon_{\alpha\gamma} h^{\gamma\beta}$,
as in section \ssHSHM.
$J$ is the complex structure of the target space, $X$.
Note that, in contrast to the topological sigma model of
chapter \sTSM,  in topological string theory, 
we allow the complex structure of
$\Sigma$ to vary.

Depending on how we formulate topological gravity we 
might also have equations for the metric $h_{\alpha \beta} $. 
To fix ideas we will use formulation 2 of chapter \sTDTG, hence
the equations are the constant curvature equation 
$R[h]=-1$ (for genus $>1$). $\tcM_{-1}$ denotes the 
subspace of $\tcM$ satisfying this equation. We 
also let 
 $\tilde \CH = \{ \bF\in \tcM_{-1}: s(\bF)=0\} $. 

\subsec{Antighost bundle}

As in the topological sigma model \fibatf\ we can take:
$\tcV\to  \tcM$
whose fiber at $f$ is given by:
$$
\tcV_f :=\Gamma [ T^\ast (\Sigma)\otimes f^\ast ( TX ) ]^+
$$
The superscript $+$ indicates the ``self-duality" constraint
$
\rho \in \widetilde {\cal V}_f 
\leftrightarrow 
J~ \rho~ \epsilon = \rho
$. 
Clearly this constraint introduces $\MET$ dependence to the fibre.
Correspondingly the fibre metric, \fibmet, varies in the metric directions.

\subsec{Connection on antighost bundle}

Referring back to the general discussion in  chapter \sTTwithLS, we recall that
there are two
connections used to construct the Lagrangian of a topological field theory with local
symmetry: $\nabla = \nabla_s \oplus \nabla_g$.
In the case of  topological string theory both connections are nontrivial.
Essentially, the connection $\nabla_s$ is that used in the construction of the topological 
sigma model, while that for $\nabla_g$ follows from the $\Diff( \Sigma )$ gauge symmetry. 

The connection on $\tcV$ will be a $\diff(\Sigma) \oplus so(V)$ connection. 
The $so( V )$ part of the connection can be deduced following the 
same reasoning as in section \ssCOC.
The tangent space to $\CZ(s)$, determined by the Gromov equation,
\eqn\gromov{
\CZ ( s )
= \widetilde \HOL
= \{ ( h,  f) \in \widetilde {\cal M}_{-1}~ \vert~ s ( h, f ) = d f + J df \epsilon(h) = 0 \}}
is defined by 
\eqn\taneq{
T_{{\scriptscriptstyle f,h}} \widetilde \HOL =\{ (\delta f,\delta h)~ \vert~
D (\delta f)+ J D (\delta f) \epsilon +J df k(\delta h)=0\}.}
where $D$ is the pulled-back connection
$( D_\alpha \delta f )^\mu = \partial_\alpha \delta f^\mu
+ \Gamma^\mu_{\kappa\lambda} \partial_\alpha f^\kappa \delta f^\lambda$
and where $k(\delta h)$ is the variation of the complex structure:
$$ 
\delta \epsilon_\alpha^{~ \beta} = k ( \delta h ) _\alpha^{~ \beta}
= - \epsilon_\alpha^{~ \gamma} \widehat{( \widetilde d h )}_\gamma{}^\beta
$$
where $\widehat{( \widetilde d h )}_\alpha{}^\beta
= ( \delta_\alpha^{~ \gamma} h^{\beta\delta} - \half \delta_\alpha^{~ \beta} h^{\gamma\delta} )
( \widetilde d h )_{\gamma\delta}$ is the traceless part of $\widetilde d h$.
Equation \taneq\  suggests that we introduce an operator:
$$
\bD_\tcV \equiv D  \chi + J ( D \chi ) \epsilon - J df \hat \psi \epsilon\colon T\tcM \to \tcV
$$
where $\chi^\mu$ represents $1$-forms in $T^\ast \MAP$ and $\psi$ represent 1-forms
in $T^\ast \MET$.
The circumflex over the $\psi$ indicates projection to the traceless part.
One easily checks that the same formulae \sgmcnn\ (with $\tilde d$ now the 
exterior on $\tilde \CM$) defines a connection $\nabla_\tcV$ on
$\tcV$.  
Let $x_\alpha^{~ \mu} \in \Gamma ( \tcV )$ be a section of $\tcV$ over
$\tcM$, then in local coordinates $\{ f^\mu , h_{\alpha\beta} \}$ on
$\tcM$, the covariant derivative of $x_\alpha^{~ \mu}$ is given by
$$
\nabla_\tcV x_\alpha^{~ \mu}
=~ \widetilde d x_\alpha^{~ \mu}
- \Gamma^\mu_{\kappa\lambda}~ x_\alpha^{~ \kappa} \widetilde d f^\lambda
$$
where $\widetilde d$ is the exterior derivative on $\tcM$,
and
$\Gamma^\mu_{\kappa\lambda}$ is the 
Christoffel connection on $X$. Moreover, 
 $\nabla_\tcV (s) = \bD_\tcV$.

\subsec{BRST Complex}
\subseclab\ssBRSTCts

We can think of the generators of the complex in terms of 
$$\matrix{
{\rm Fields:}&\quad  \bF ~=~ \pmatrix{f^\mu\cr h_{\alpha\beta}\cr}\cr
& \cr
{\rm Ghosts:}&\quad \bG ~=~ \pmatrix{ \chi^\mu\cr \psi_{\alpha\beta}\cr}\cr
& \cr
{\rm Antighosts:}&\quad \rho^{\alpha}_{~ \mu}\cr
& \cr
{\rm Lagrange\ multipliers:}&\quad \pi^{\alpha}_{~ \mu}\cr}
$$

The Cartan model differential for $\Diff$ acts as: 
$$
Q_\CC \pmatrix{\bF\cr \bG\cr}
=~ \pmatrix{       0               & 1\cr
                        -\cL_\gamma & 0\cr} \pmatrix{\bF\cr\bG\cr}\qquad 
\qquad Q_\CC \pmatrix{\rho\cr \pi\cr}
=~ \pmatrix{       0               & 1\cr
                        -\cL_\gamma & 0\cr} \pmatrix{\rho\cr \pi\cr}
$$
How we handle remaining ghosts/antighosts depends on how we represent
2D gravity. As mentioned above, we will choose formulation 2 of 
chapter \sTDTG. 

\subsec{Lagrangian}

\subsubsec{Localization}

The localization 
$\tcM_{-1}\hookrightarrow \tcM$ 
eliminating 
 the Weyl degrees of freedom proceeds
exactly as in the discussion of 2D gravity in section \ssFormtwo.
The localization $\widetilde \CH\hookrightarrow \tcM_{-1}$ 
proceeds as in chapter \sTSM. 
%
Therefore 
the gauge fermion for localization is that of the topological 
sigma model: 
\eqn\tsgf{
\Psi_{\rm localization} ~=~  \int d^2 z~ \sqrt{h}~ \left \{
\rho^{\alpha}_{~ \mu} [ i s_{\alpha}^{~ \mu} 
- {1 \over 4} \Gamma^\mu_{\kappa\lambda} \rho_{\alpha}^{~ \kappa} \chi^\lambda
+ {1 \over 4} \pi_{\alpha}^{~ \mu}] \right \}}
\vskip0.1truein

\subsubsec{$\Psi_{\rm projection} $}

Next we turn to the gauge fermion for projecting out  the gauge degrees of freedom.
We follow the procedure of section \ssAthePNGF, taking as gauge group just the
diffeomorphism group.
Thus, following the discussion of section \sssQexct, we replace 
$\phi \to \gamma^\alpha$.
To find $C$ we note that there is a canonical isomorphism between the Lie algebra
$\diff ( \Sigma )$ of  $\Diff^+ ( \Sigma )$ and the vertical tangent space,
$T^{{\rm  vert}} \tcM_{-1} $, given by:
\eqn\isoC{\eqalign{
C \colon \diff~ ( \Sigma ) ~&\to~ T^{{\rm  vert}} \tcM_{-1}\cr
C_{{\scriptscriptstyle (f,h)}}  \colon \gamma
{}&~\longmapsto~ \pmatrix{ \cL_\gamma f \cr \cL_\gamma h\cr}
{}~=~ \pmatrix{ i_\gamma d f \cr P \gamma,\cr}\cr}}
where $P$ is defined in \orlpi\ above. 
Thus we have 
$C^\dagger = D^\alpha \psi_{\alpha \beta} + \p_\beta f^\mu \chi^\nu G_{\mu\nu}$. 

To form the projection gauge fermion we introduce the antighosts 
$\lambda^\alpha$ and $\eta^\alpha$ of ghost numbers $-2$ and $-1$, as in the 
previous chapter. 
The gauge fermion for projection in the gauge directions is $(\lambda, C^\dagger)$
which in this case reads:
\eqn\rdgfi{
\Psi_{\rm projection} ~=~ i  \int d^2 z~ \sh
\left \{ ( P \lambda )^{\alpha\beta}  \psi_{\alpha\beta}
+ i_\lambda  d f^\mu \chi^\nu G_{\mu\nu} \right \}}

\subsubsec{The Full Action}
\subsubseclab\sssfullaction

Combining all of the above gauge fermions, expanding out the $Q$ actions
and integrating out the Lagrange multipliers, we obtain the general action
for coupling topological sigma models to topological gravity.
As we have seen repeatedly, in formulating the correct intersection formula the 
key point is to examine the fermion kinetic term, which may be written in the
compact form
\eqn\tpstrkin{
\pmatrix {\rho & \eta} \bO {\bG}    = \bA \bO \bG, } 
 where
\eqn\defO{
\bO ~=~
\pmatrix{
\CD & J df k \cr
\p f & P^\dagger\cr}
\colon T_{{\scriptscriptstyle f,h}}\tcM_{-1}  \to
\Gamma[T \Sigma \otimes f^\ast (TX)] \oplus \Gamma[T\Sigma].}
and  the components are given by:
\eqn\dfcomps{\eqalign{
( \CD \chi )_\alpha^{~ \mu} &= \nabla_\alpha \chi^\mu
+ J^\mu_{~ \nu} ( \nabla_\beta \chi^\nu ) \epsilon_\alpha^{~ \beta}\cr
( \partial f \chi )_\alpha & = G_{\mu\nu} (\partial_\alpha f^\mu ) \chi^\nu \cr
[ J df k ( \psi ) ]_\alpha^{~ \mu}
& = -J^\mu_{~ \nu} \partial_\gamma f^\nu \epsilon_\alpha^{~ \beta} \psi^\gamma{}_\beta\cr}}

\subsec{Index} 
\subseclab\sstopstindx

Although we do not have space to examine correlation functions in topological string theory 
we will give a formula for the most important quantity needed when considering correlators, 
namely, the index of $\bO$. 

Consider deformations $(\delta f,\delta h)$ in the kernel of  $\bO$.
The first line of \defO\ ensures that $(\delta f,\delta h)\in T 
\tilde \HOL$
and the second ensures that 
$( \delta f, \delta h ) \not\in T^{\rm vert} \tilde \HOL$.
In conformal gauge the operator $\IO\colon T \tcM_{-1} \to \tcV \oplus \diff$
splits into the direct sum of two operators $O \oplus {\bar O}$, which are,
of course, related by complex conjugation.
\eqn\ODef{
O ~=~ \pmatrix{ 2 D_{\bar z} & - \partial_z f^w\cr
\partial^z f_w & D^{\bar z}\cr}}
Consider the operators
$$
O_0
~=~ \pmatrix{\cD& 0\cr
0 & P^\dagger\cr}
\qquad\qquad
\delta O
~=~ \pmatrix{0 & -\partial_z f^w\cr
\partial^z f_w & 0\cr}
$$
The operators $O_0$ and $O$ share the same leading symbols
and hence have equal indices.
It is easily seen from the Hirzebruch-Riemann-Roch theorem (see section
\sssZandtheIT) that 
\eqn\indxbo{
\mathboxit{
\ind~ (\bO ) = (h-1)(3-\dim~ X ) +  \int_{\Sigma_W} f^\ast ( c_1 ( T X ))}}
In sufficiently high genus ($h \ge 2$), $\dim~ \ker~ O_0 = 3 h - 3$,
where the kernel is spanned by the holomorphic quadratic differentials
$\left \{ \pmatrix{ 0 \cr \psi^{\sst ( m )} \cr} \right \}_{\sst m = 1,\ldots, 3h-3}$;
furthermore, $\dim~ \coker~ O_0 = 3 h - 3 - B$, where the cokernel is
spanned by $\left \{ \pmatrix{\rho^{\sst ( a )} \cr 0\cr} \right \}_{a = 1,\ldots,3h-3-B}$.
Though the indices of $O$ and $O_0$ are equal, their kernels
and cokernels are markedly different, for $\p_{z}f^{w} \ne 0$. 
In fact, $O$ generically does not possess a cokernel.

\subsec{Example: Riemann surface target}
\subseclab\ssExplRS

Adapting the general formula \indxbo\  above to the  case of a Riemann surface
target, $X=\ST$,  we get: 
$$
\dim{}_{\IC}~ \ker~ \bO - \dim{}_{\IC}~  \coker~\bO=B=2h-2-n(2p-2)
$$
which is the total branching number.

On the other hand, we have seen that the complex dimension of Hurwitz space is $B$.
Since the 
$$
\pmatrix{\chi \cr \psi\cr}
$$ 
zeromodes span the cotangent space to $\HOL$ we see that
$\dim_{\IC}~  {\rm coker }~\bO=0$. 
The result $\dim_{\IC}~ \HOL =B$ can also be seen from a generalisation of
Kodaira-Spencer theory described in appendix A of  \CMROLD. 
In the generic case ($G, h > 2$),  $\dim_{\IC}~  \coker~ \bO=0$.
Thus, in strong contrast to the case at fixed metric we have 
a positive number of $\pmatrix{ \chi & \psi\cr}$ zero modes. 

\newsec{\ymt\   as a topological string theory}
\seclab\sYMTasaTST

Let us return to the problem of part I of this review.
The main conclusion there was that the $1/N$ expansion
of the partition functions of \ymt\ generate
Euler characters of moduli spaces of holomorphic maps. 
We can now apply the machinery of 
topological field theory to find an action principle
for the underlying string theory of \ymt.

\subsec{Extending the fieldspace}

We are searching for a string theory whose connected partition function is
\eqn\extfldsi{
Z_{\rm string} \sim \chi_{\rm orb} ( \CH ( \Sw, \ST ))
= \int_{\CH(\Sw,\ST)} \chi [T \CH ( \Sw, \ST )\to \CH ( \Sw, \ST )]}
Clearly, in order to localize to $\CH ( \Sw, \ST )$, we should introduce the basic fields
of topological string theory
$$
\bF = \pmatrix{ f^\mu\cr h_{\alpha \beta}\cr}\in \tcM
$$
together with the standard section $s(\bF)= (df + J df \epsilon,R[h]+1)$.

We must also produce the correct density on the moduli space $\CH ( \Sw, \ST )$.
As we have seen, this is given by the basic formula \ajwfrm.
In particular, for the partition function, the density on moduli space is
$\chi ( \cok~ \nabla \bar s ) = \chi ( \ker~ \bO / \CG)$.
Recall from section \ssExplRS\  that, for the topological string theory
describing holomorphic maps $\Sw \to \ST$ we have a fermion kinetic operator
$\bO$ satisfying:
\eqn\propsofo{\eqalign{
\ker~ \bO_{f,h} & \cong T_{f,h} \CZ(s)\cr
\cok~ \bO_{f,h} & = \{ 0 \} \cr}}
In the usual way,
$\ker~ \bO$ is a $\Diff ( \Sigma_W)$-equivariant bundle over $\CZ(s)$, so that
$\ker~ \bO/\Diff ( \Sigma_W) \cong T_{[(f,h)]} \CH$.

Evidently, one way to obtain the required 
density $\chi(T\CH\to \CH)$ is to use the elementary
fact that
\eqn\propsofoi{\eqalign{
\ker~ \bO_{f,h}^\dagger & = \{ 0 \} \cr
\cok~ \bO_{f,h}^\dagger & \cong T_{f,h} \CZ(s)\cr}}
so that, if the fermion kinetic operator were
\eqn\propsofoi{
\bO \oplus \bO^\dagger}
then we would obtain the desired answer.
Note that, since $\Index ( \bO \oplus \bO^\dagger ) = 0$,
no insertions of operators are necessary, the
total ghost number anomaly is cancelled.

\subsubsec{``Cofields''}

In order to obtain
\propsofoi\  as the fermion kinetic operator,
we must extend the field  space relative to that
of the standard topological string theory.
The  new fields are
completely determined by the requirement that
$\bO^\dagger$ maps ghosts to antighosts, and
by $Q$-symmetry.
In describing the new  fields it is easiest
to begin with the ``dual'' set of ghosts $\tbG$.
These take values in the domain of
$\bO^\dagger$ and hence have the index
structure:
\eqn\coghosts{
\tbG ~=~ \pmatrix{\hat \chi^\alpha_\mu\cr
\hat \psi^\alpha\cr}
}
or,
\eqn\coghstsp{
\tbG \in \Gamma( T \Sw \otimes f^\ast ( T^\ast \ST ))^+
     \oplus \Gamma( T \Sw )
}
As for $\rho^\alpha_\mu$ the selfduality constraint
must be imposed, so
\eqn\coghstsd{
\hat \chi^z_w= 0
}
Since the ghosts represent differential forms
on  fieldspace, (according to \bsctaut), 
we must replace the original fieldspace $\tcM$ of
topological string theory by the total space of
a vector bundle
$\widehat E \to \tcM$, 
where the fiber directions are spanned by fields
$$\tbF\in \Gamma( T \Sw \otimes f^\ast ( T^\ast \ST ))^+
     \oplus \Gamma( T \Sw )$$
that is:
\eqn\cofields{
\tbF
{}~=~
\pmatrix{\hat f^\alpha_\nu \cr
\hat h^\alpha\cr}
}
where $\hat f^z_w=0$.
We refer to the hatted fields as 
``cofields.''

\subsec{Equations}

Our choice of section will be
\eqn\cosction{
s: (\bF, \tbF) \rightarrow (df + J df \epsilon, \bO^\dagger \tbF) =(s_1,s_2)
}
and the zero set is just
$\CZ(s) = (\CZ(s_1),0) $
since $\bO^\dagger$ has no zero modes. The appropriate
antighost bundle is therefore dual to
$\tcV \oplus \tcV_{\rm cf} $
where
\eqn\cfldblde{
\tcV_{\rm cf} = \Gamma( f^\ast (T  \ST )) 
\oplus \Gamma( \Sym ( T \Sw^{\otimes 2} ))\quad . 
}
In indices we have fields
\eqn\coantighst{
\tbA
{}~=~ \pmatrix{\hat \rho^\mu\cr
\hat \eta_{\alpha \beta} \cr} \quad . 
}
The bundle $\tcV_{\rm cf}$ has a natural metric, and we may take the trivial connection
on $\tcV_{\rm cf}$ plus the usual $\Diff(\Sw)$ connection.

\subsec{BRST Complex}

Since we use formulation 2 of 2D
topological gravity we take the
BRST complex of fields to be that given in
section \ssBRSTCts. Similarly, for the cofields
we
choose the Cartan model for $\Diff$-equivariant cohomology:
$$Q \pmatrix{\tbF\cr \tbG\cr} = \pmatrix{0& 1\cr -\CL_\gamma & 0\cr}
\pmatrix{\tbF\cr \tbG\cr} \qquad 
Q \pmatrix{\tbA\cr \tbPi\cr} = \pmatrix{0& 1\cr -\CL_\gamma & 0\cr}
\pmatrix{\tbA\cr \tbPi\cr}$$

The addition of the cofields does not  change the $Q$-cohomology, so we expect 
to have the same observables as in topological string theory. 

\subsec{Lagrangian}
\subseclab\ssECLAG\

The Lagrangian for the \ymt\ string will
be a sum of a Lagrangian for the topological
string theory $\Sw\to \ST$ plus a Lagrangian
for localizing to $\tbF=0$:
\eqn\lrgymt{
I_{YM_2 } = I_{\rm top~ string} + I_{\rm ``cofield"}}
Following chapter \sINandTIR\ we 
write down the gauge fermion
for the co-fields:
\eqn\pgf{\eqalign{
\Psi_{\rm ``cofield"}&=
\langle  \tbA,  \bO^\dagger \tbF\rangle - t( \tbA, \tbPi) \cr
{}~ & =~ \ofp \int d^2 z \sqrt{h}~ \tbA^T ( \bO^\dagger \tbF - t \tbPi )\cr}}
where
\eqn\Odaggdef{
\bO^\dagger ~=~ \Odagger}

\subsec{Localization}

With our choice of section, $s=(s_1,s_2)$ in
\cosction, and connection, 
we can analyze \wittseq\ in this case. When restricted
to the zeroset $\tbF=0$ we find $\nabla s_2 =\tilde ds_2$
is a one-form with values in $\tcV_{\rm cf}$ given by
$$\bO^\dagger \pmatrix{\chi\cr \psi\cr} \quad . $$
The analog of \wittseq\ in this case becomes:
$$
0\rightarrow \Im ( \bO \oplus \bO^\dagger )
\rightarrow \tcV\oplus \tcV_{\rm cf}
\rightarrow \cok(\bO \oplus \bO^\dagger)
\rightarrow 0
$$
as a sequence of bundles over $\HOL$.

By the general principles we have explained at length
we see, by combining \propsofo\ and \propsofoi,
with \ggefxd, 
 that the path integral computes the Euler character of the cokernel (=obstruction)
bundle, $T\HOL(\Sw,\ST)$. 
Therefore, the Lagrangian \lrgymt\ solves problem
$\beta$ of \sssWhtlk\ for chiral \ymt.

\subsec{Nonchiral case}
\subseclab\ssNonchir

The nonchiral analog of the theory of sec. \ssECLAG\ must localize on
both the space of holomorphic {\it and} antiholomorphic maps.
\foot{Some of the text in this section has been 
cut and pasted from \CMROLD.}
When we regard the topological string path integral as an
infinite dimensional version of an equivariant Thom class it becomes
clear that we need a section $\tilde w$ of some bundle
which localizes on the submanifolds $\widetilde \HOL^\pm$
of $\tcM$ defined by $df \pm J df \epsilon=0$.
It is therefore natural  to choose a section of the form:
\eqn\ncsect{
\widetilde w~ ( f, h ) ~\mapsto ~\dot F=
[ df + J df \epsilon ] \otimes [ df - J df \epsilon ]}
which has the index structure $\dot F_{\alpha \beta}^{\mu\nu }$.
We have  fields, ghosts, antighosts, and Lagrange-multipliers:
\vskip0.1truein
\hbox{\hfill
\divide\hsize by 2{
\vbox{
$$\eqalign{
\bF ~=&~ \pmatrix{f^\mu\cr h_{\alpha\beta}\cr}\cr
  A ~=&~ \rho_{\alpha \beta}^{\mu\nu }\cr}
$$}
\hfill
\vbox{
$$\eqalign{
\bG ~=&~ \pmatrix{ \chi^\mu\cr \psi_{\alpha\beta}\cr}\cr
  \Pi ~=&~ \pi_{\alpha \beta}^{\mu\nu }. \cr}
$$}
\hfill}}
Only the anti-ghosts and Lagrange multipliers of the sigma model
have changed relative to the chiral theory.
In particular, the appropriate bundle for the antighosts $\rho$
has fiber:
\eqn\fibreVpm{
\tcV^{nc}_{{\scriptscriptstyle ( f, h)}}
{}~=~ \Gamma \left [
( T ^\ast \Sw )^{\otimes 2} \otimes ( f^\ast  ( T \ST ))^{\otimes 2} \right ]_\pm}
where the subscript $\pm$ indicates that the sections must
satisfy ``self-duality" constraints:
\eqn\GenSDCon{
\rho ~\in~ \tcV^{nc}_{{\scriptscriptstyle ( f, h)}}
\qquad\Longleftrightarrow\qquad
\cases{
\rho - ( J \otimes {\bf 1} )~ \rho~ ( \epsilon \otimes {\bf 1} ) ~=~ 0 & \cr
                                                                     & \cr
\rho + ( {\bf 1} \otimes J )~ \rho~ ( {\bf 1} \otimes \epsilon ) ~=~ 0 & \cr}}

The  BRST transformations are the same as 
above. 
The nonchiral theory has an action
\eqn\ncact{
I_{\rm YM_2  string} ~=~ I_{\rm tg} + 
I^{nc}_{\rm t \sigma} +I^{nc}_{\rm cofield}}
The gravity part of the action is the same as before.  The topological
sigma model part is
\eqn\tsgfnc{
I^{nc}_{t\sigma}
{}~=~ Q \int d^2 z~ \sh \left \{
\rho^{\alpha \beta}_{\mu \nu } \bigl[i  \dot F^{\mu\nu }_{\alpha \beta}
 -\Gamma^\mu_{\lambda\rho} \chi^\lambda \rho^{\rho \nu }_{\alpha \beta}
-\Gamma^\nu_{\lambda\rho} \chi^\lambda \rho^{\mu \rho }_{\alpha \beta}
+ \half \pi_{\alpha \beta}^{\mu\nu } \bigr]  \right \}}
where the indices on $\rho, \pi$ are raised and lowered with the metrics on the
worldsheet ($h$), and target space ($G$) .

If we expand \tsgfnc\ and integrate out the Lagrange
multiplier then the bosonic term becomes (in local
conformal coordinates)
\eqn\tsgfnci{
I^{nc}_{\rm t \sigma}
{}~=~ \int h^{z\zb} G_{w\wb}^2~ \vert\p_z f^w\vert^2 \vert\p_{\zb} f^w\vert^2 +
\cdots}
thus clearly localizing on both holomorphic and antiholomorphic maps.
Moreover, when we work out the quadratic terms in the fermions
we find that many components of $\rho$ do not enter the
Lagrangian.
These components are eliminated by the  constraints \GenSDCon.
In locally conformal coordinates the only non-trivial components of
$\rho \in \tcV_{{\scriptscriptstyle ( f, h)}}^{nc}$ are
$\rho_{zz}^{\bar w w}, \rho_{z \bar z}^{\bar w \bar w},
\rho_{\bar z z}^{ww}$, and $\rho_{\bar z \bar z}^{w \bar w}$.
(Note that $\rho_{\alpha\beta}^{\mu \nu }$ is not symmetric in interchanging
$\{(\alpha\beta)(\mu\nu)\} \leftrightarrow \{(\beta\alpha)(\nu\mu)\}$.).
The kinetic term for the fermions is given by
\eqn\tsgfncii{
I^{nc}_{\rm t \sigma}
{}~=~ i \int d^2 z~ \sh~ \pmatrix{ \rho & \eta\cr} \bO^{nc} \pmatrix{\chi\cr
\psi\cr} +\cdots
}
where $\bO_{nc}$ is a $2\times 2$ matrix operator
with entries:
\eqn\NCO{\eqalign{
\bO^{{\scriptscriptstyle nc}}_{11}
{}~=&~ \CD^+ \otimes [ df - J df \epsilon ]
   + [ df + J df \epsilon ] \otimes \CD^-\cr
\bO^{{\scriptscriptstyle nc}}_{12} ~=&~ J~  d f~ k \otimes [ df - J df \epsilon
]
   - [ df + J df \epsilon ] \otimes J~ d f~ k\cr
\bO^{{\scriptscriptstyle nc}}_{21} ~=&~ \partial f\cr
\bO^{{\scriptscriptstyle nc}}_{22} ~=&~ P^\dagger\cr}}
Here $\CD^\pm \chi^\mu  =  D \chi^\mu \pm J ( D \chi^\mu) \epsilon$
and, as usual, $k[ \delta h ]$ is the variation of the complex structure
on $\Sw$ induced from a variation of the metric $\delta h$.

The co-model is introduced using the same principles as
before.
The contact terms have not been carefully analyzed in this
nonchiral theory. They should be interesting. Heuristically
the contributions of coupled maps can be pictured as
boundary contributions.
For further discussion see \CMROLD. 

\ifig\arcrvcoll{An area operator collides with a 4 fermi
curvature operator.}
{\epsfxsize3.5in\epsfbox{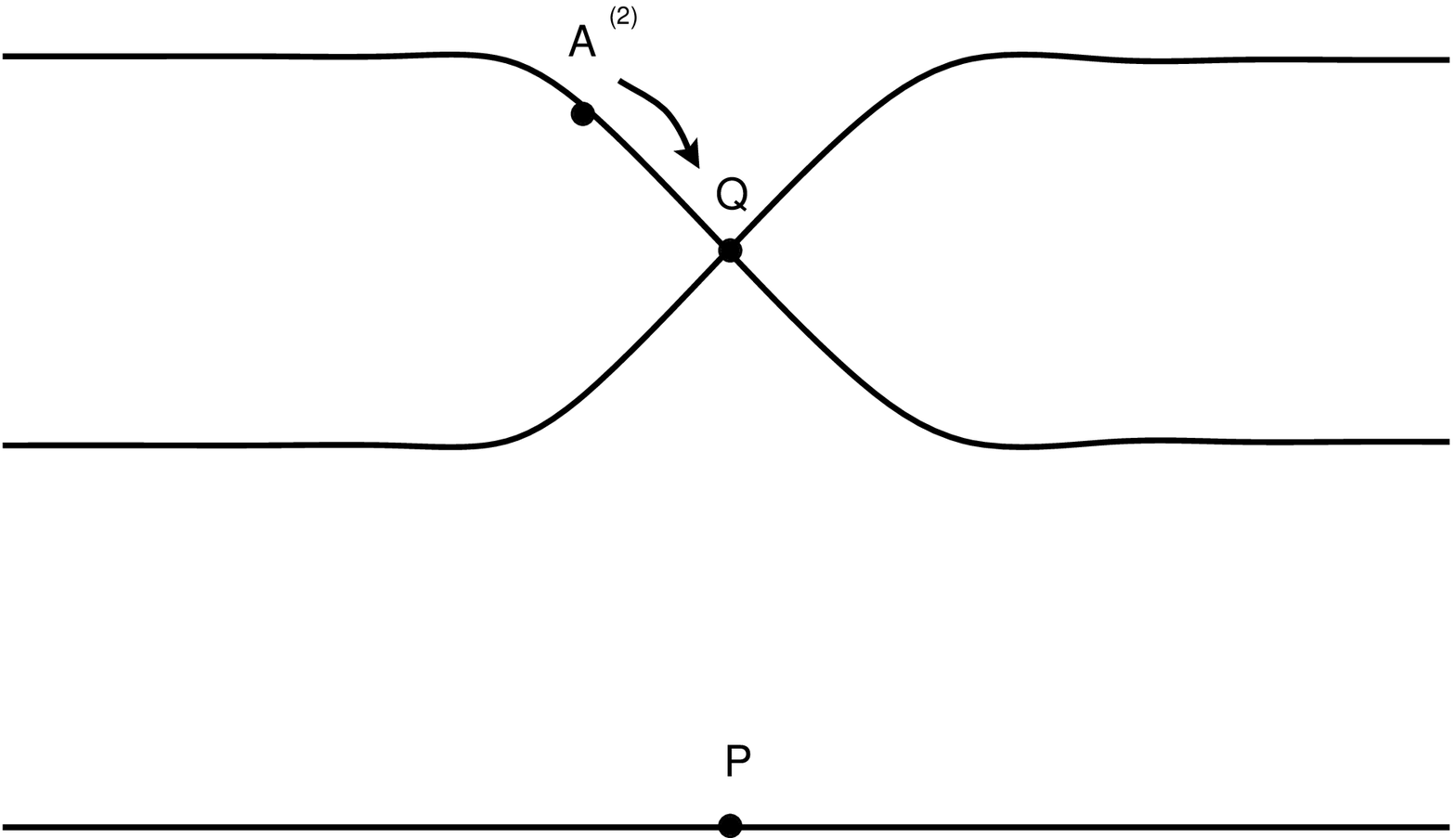}}

\subsec{Perturbing by the area}
\subseclab\ssPBtheA

In order to formulate a string  picture of the
physical phase of \ymt\  we must perturb by the area
operator:
\eqn\defdact{
I_0 ~\longrightarrow~ I_0 + \half \int {\cal A}^{{\scriptscriptstyle (2)}}.  }
Here $\CA^{(2)}$ fits into the area operator descent multiplet:
\eqn\areamult{\eqalign{
\CA^{(0)}&=\sigma _0\bigl(\omega_{ij }(f(x)) \chi^i \chi^j\bigr)\cr
\CA^{(1)}&=\sigma _0\bigl(dx^\alpha \omega_{ij }(f(x))\p_\alpha f^i
\chi^j\bigr)\cr
\CA^{(2)}&=\sigma_0\bigl(dx^\alpha\wedge dx^\beta \omega_{ij}(f(x))\p_\alpha
f^i
\p_\beta f^j\bigr), \cr}
}
where $\omega$ is the  K\"ahler two-form from the target space.
Naively, the contribution of $\half \int \CA^{(2)}$ in
a path integral over maps $f$ of index $n$ is
$e^{-\half n A}$. This accounts nicely for the
 exponential  factors in \arpol\ 
but fails to account for the
polynomial dependence on $A$, $Z_{n,h,p}$ of 
\arpol. 
The latter may be accounted for by contact interactions
between area operators and the curvature operators, illustrated in
\arcrvcoll.

An analysis of contact terms modelled on that
used in \DiVeVe\  leads to recursion relations
of the form:

\eqn\RecursePrimitive{\eqalign{
\langle\!\langle
&\overbrace{
{\cal A}^{{\scriptscriptstyle (0)}}\cdots
{\cal A}^{{\scriptscriptstyle (0)}}
}^{B-r}
\overbrace{
{\cal A}^{{\scriptscriptstyle (2)}}
 \cdots
{\cal A}^{{\scriptscriptstyle (2)}}
}^k
\rangle\!\rangle_{\HOL(B,k; r  )}\cr
&~=~ n A \langle\!\langle
\overbrace{
{\cal A}^{{\scriptscriptstyle (0)}}\cdots
{\cal A}^{{\scriptscriptstyle (0)}}
}^{B-r}
\overbrace{
{\cal A}^{{\scriptscriptstyle (2)}} \cdots
{\cal A}^{{\scriptscriptstyle (2)}}
}^{k-1}
\rangle\!\rangle_{\HOL(B,k-1;r )}\cr
&\qquad -2r  \langle\!\langle
\overbrace{
{\cal A}^{{\scriptscriptstyle (0)}}
\cdots
{\cal A}^{{\scriptscriptstyle (0)}}
}^{B-r+1}
\overbrace{
{\cal A}^{{\scriptscriptstyle (2)}} \cdots
{\cal A}^{{\scriptscriptstyle (2)}}
}^{k-1}
\rangle\!\rangle_{\HOL(B,k-1 ;r-1)}\cr}}
The subscripts refer to a stratification of 
Hurwitz space described in 
\CMROLD. 
The first term represents the bulk contribution.
In the second
term there is one extra insertion of
 ${\cal A}^{{\scriptscriptstyle (0)}}$ which
has replaced a curvature operator
at a ramification point, and there is one  fewer
${\cal A}^{{\scriptscriptstyle (2)}}$
operator.  The coefficient $r$ in the second term comes from the fact
that for each area integral there are $r$ collisions with  curvature insertions
at ramification points. The factor of $-2$ comes from the
normalization of the contact term.  From \RecursePrimitive\ 
 one can derive the area polynomial
associated with the  portion of Hurwitz space corresponding 
to simple branched covers \CMROLD. 

\subsec{Horava's Theory}

In \Hora\  P. Horava made an independent and
apparently different proposal for a topological
string theory underlying \ymt.  As in the
above discussion he begins by trying to
reproduce the $A=0$ theory of \ymt.
The $A\not= 0$ will be obtained as a perturbation
by a BRST invariant operator.

Horava begins with a
matter sector generalizing the topological sigma
model.  Instead of localizing onto holomorphic
maps his matter localizes onto {\it harmonic maps}
$\Sw\to \ST$.  This theory is then coupled to
2D topological gravity. It is argued in
\Hora\  that the resulting theory localizes on the
space of {\it minimal area maps}.

It would be very interesting to develop the
proposal in \Hora\ further. We have seen
that the {\it chiral} theory of \ymt\ involves
holomorphic maps in a very natural way.
On the other hand, the {\it nonchiral}
theory involves  the ``degenerated coupled covers''
of \CMROLD. These are maps
which are holomorphic or antiholomorphic
on different components of the normalized
curve $\Sw$.  Perhaps an approach using
 minimal area maps would lead to
a more natural description of the
degenerated coupled covers. It is curious
that both the nonchiral theory \ncact\ and
Horava's theory  involve higher numbers of
derivatives than two.

\newsec{Euler character theories}
\seclab\sECT

In the standard topological theories one
localizes to an interesting moduli subspace,
and computes intersection numbers within
that subspace.
It is also of interest to  find a theory
which calculates the Euler character of
the moduli space
$\CZ(s)$ itself. The ``cofield construction''
of the previous chapter  can be generalized.
Whenever a topological theory has
$\cok \nabla \bar s = \cok~ \bO = \{0\}$ we can
introduce a second set of fields with
fermion kinetic operator $\bO^\dagger$, by
 reversing the roles of
the ghosts and the antighosts.
As in the previous chapter, the change of
operator $\bO\to \bO^\dagger$
involves adding new fields:
$(\tbF, \tbG, \tbPi, \tbA)$. In the context of
Witten's  fields/equations/symmetries
paradigm the new ``cofield'' multiplet
$(\tbF, \tbG)$ has the ``quantum numbers of
the equations,'' (and of the Lie algebra, in
the presence of gauge symmetry), while
the antighosts $\tbA$ have the
``quantum numbers of the fields.''

Recently, C. Vafa and E. Witten have applied
the cofield construction to explain how the partition
function of a topological twisting of D=4, N=4
supersymmetric Yang-Mills theory computes
Euler characters of the moduli spaces of
ASD instantons \refs{\VaWi}.
In the notation of \VaWi, one begins with a topological
field theory with BRST field multiplets
$(u^i,\psi^i)$ ($i$ running over a set of
coordinates for field space), equation
antighost multiplets $(\rho^a,\pi^a)$,
($a$ running over a basis for the fiber of
$\CV$), and projection antighost multiplets
$(\lambda^x,\eta^x)$, ($x$ running over a basis
for the Lie algebra of the gauge symmetry group).
The fermion kinetic term of the original topological model
is given by:
\eqn\kts{
\pmatrix{ \rho^a  & \eta^x \cr}  \pmatrix{\nabla s\cr  C^\dagger\cr}=
\langle \pmatrix{ \rho^a  & \eta^x \cr} , \bO \psi \rangle =
 \pmatrix{ \rho^a & \eta^x \cr}  \pmatrix{D_{ai} \cr  D_{xi} \cr} \psi^i}
where the  $D$'s  differential operators.
One then adds fields such that BRST symmetry
is preserved and the new fermions have kinetic
operator $\bO^\dagger$. Thus the ``cofields'' have
indices:
\eqn\cfld{
\hat\bF = \pmatrix{ \hat u^a \cr  \hat u^x\cr}}
and so forth.

It is pointed out in \VaWi\ that, quite generally, the
cofield construction involves {\it two} BRST charges and there is an
underlying $sl(2)$ symmetry in the construction.
Indeed the fields and cofields fit together according to:
\eqn\fes{
\matrix{
 2&    &    & & & & & & & & & & &            \phi^x  & &  \cr
   &&&&&&&&&&&&&&\searrow& \cr
 1&   &  & \psi^i &    & & & & & \hat \psi^a & & & & & & \hat \psi^x\cr
   &   & \nearrow & & \searrow & &&&  \nearrow & & \searrow & & & & \nearrow &
\cr
  0 &   u^i &  &    & & \hat \pi^i & & \hat u^a & & & & \pi^a  & & \hat u^x & &
\cr
   & & \searrow & & \nearrow & & & & \searrow & & \nearrow & & & & \searrow &
\cr
-1& & & \hat \rho^i & & & & & &  \rho^a  & & & & &  & \eta^x \cr
 & & & & & & & & & & & & & & \nearrow & \cr
-2 & &&&&&&&&&&&& \lambda^x & & \cr}}
Where  $\nearrow$ indicates action of $Q$, and
$\searrow$ the action of the second symmetry $Q'$.
$\phi^x$ is the generator of the polynomials on the
Lie algebra of the gauge symmetry, as appropriate to
the Cartan model of equivariant cohomology.

\vskip0.1truein\noindent
{\bf Remarks}
\item{1.}
As opposed to the \ymt\ example, the crucial
vanishing theorem for  $\ker~ \bO^\dagger$ can fail
for twisted $N=4,D=4$ SYM. This necessitates the
choice of a more elaborate section $s(\bF, \tbF)$
in \VaWi.
\item{2.}
The presence of two topological charges
$Q$ and $Q'$ and of the underlying $sl(2)$ symmetry is related
to the original $N=4$ supersymmetry in the Yang-Mills
case. It can be shown that the chiral \ymt\ string
is similarly related to a topological string arising from
an $N=4$ string \refs{\DiInMoPl}.
\item{3.}
Another construction of Euler character
theories appears in the work of Blau and Thompson
\blauthom.

\newsec{Four Dimensions: A Conjecture}
\seclab\sFDAC

The original motivation for the program of 
Gross described in section \ssTheCofTD\ was to find a 
string interpretation of $YM_4$. Gross
proposed that, once \ymt\  was interpreted as 
a string theory that theory, suitably generalized 
to four-dimensional targets, might be the 
string theory of $YM_4$.  In closing, let us 
describe a somewhat refined version of 
Gross' conjecture. 

Interestingly enough, the string theory of \ymt\ 
{\it does} have a natural extension to four-dimensional 
target spaces, these are the ``Euler sigma models'' $\CE \sigma(X)$.
They are obtained by applying the ``cofield construction''  of 
chapter \sECT\ to the topological string associated 
to a target space $X$ which  is compact and K\"ahler.
\foot{As noted in chapter \sTST, as wider class 
of target spaces can be considered.}
If $f$ is a holomorphic map we may generalize the degree of the map 
by choosing a basis $e^\alpha$ of $H_2(X,\IZ)$ 
and writing 
$f(\Sigma) = \sum d_\alpha e^\alpha\in H_2(X,\IZ)$. 
In analogy with $A>0$ in \ymt, the sum over the 
degrees may be regulated by perturbing the 
theory by $\Delta I = t^\alpha\int f^* \omega_\alpha$ 
where $\omega_\alpha$ is Poincar\'e dual to
$e^\alpha$. By analogy with two dimensions 
we may expect that the partition function of  
$\CE \sigma(X)$ should have the form: 
\eqn\frdesm{
Z= \sum_{\chi } N^{ \chi } 
\sum_{d_\alpha\geq 0} \prod_\alpha e^{-t^\alpha d_\alpha} 
P_{d_\alpha}(t^\alpha) \chi_{\rm orb}(\CC\CH(\Sw, X; d_\alpha) )
}
where experience from \ymt\ suggests that we must 
allow $\Sw$ to be singular and we must allow 
``coupled maps'' which are holomorphic and 
anti-holomorphic on different components of 
the normalization of $\Sigma_W$. By  analogy 
we conjecture that $P_{d_\alpha}(t^\alpha)$ is a 
polynomial whose value at zero is one.  

The natural question then arises: what is the 
spacetime physics of $\CE \sigma(X)$? 
Experience with known topological string theories
strongly suggests that the spacetime physics 
should be that of a 4-dimensional topological 
field theory, or an almost topological field theory, 
which depends on the choice of a K\"ahler class. 
We do not expect it to describe ordinary 
Yang-Mills theory. 

Thinking back to the case of $D=2$ we may 
recall that at large $N$, Donaldson theory and 
``physical'' \ymt\ are the same. We may then 
take a leap and guess that the correct relation 
which generalizes to $D=4$ is the relation between 
$\CE \sigma(X)$ and topological Yang-Mills theory 
on $X$. This would mean, for example, that, 
in analogy to \dualmdli\  the large $N$ 
asymptotics of the intersection number 
generator on instanton moduli space: 
\eqn\dnldlgn{
\biggl\langle e^{r^\alpha \CO_2^{(2)}(e_\alpha)}
\biggr\rangle_{\CM_+(X;SU(N))}
}
would be closely related to the series
\frdesm\  with $g_{\rm string} = 1/N$
with some transformation between $t^\alpha$ and 
$r^\alpha$. 

\newsec{Conclusion}

$\qquad\qquad$
{\it It is all true. Or  it ought to be; and 
more and better besides. 

- Winston Churchill}

\bigskip
\centerline{\bf Acknowledgements}

We would like to thank M. Henningson, 
M. Marino, R. Plesser, R. Stora, W. Taylor and
R. von Unge for remarks on preliminary 
drafts of this paper.  Many students 
at the Les Houches school also made useful 
comments on the notes. 
We would also like to thank V.I. Arnold, 
S. Axelrod, T. Banks, M. Bershadsky, M. Douglas,
S. Cecotti, R. Dijkgraaf, I. Frenkel, E. Getzler, D. Gross, 
M. Guest, J. Harris, K. Intrilligator, 
M. Khovanov,
V. Kazakov, I. Kostov,  H. Ooguri, R. Plesser, 
R. Rudd,  V. Sadov, J. Segert, S. Shenker, 
R. Stora, W. Taylor, C. Vafa, 
and G. Zuckerman for discussions and correspondence.
S.C. thanks the department of physics at Harvard University
for hospitality.
G. M. thanks the department of physics at Rutgers 
University for hospitality while  part of these 
lectures were being written. He also thanks 
L. Alvarez-Gaum\'e and E. Verlinde for their 
hospitality at CERN. Finally G.M. thanks 
R. Dijkgraaf and I. Klebanov for the opportunity to 
present some of this material at the 1994
Trieste Spring school and he thanks 
F. David and P. Ginsparg for the opportunity to 
present some of this material at the 1994
Les Houches school on fluctuating geometries. 
This work is supported by DOE grant DE-AC02-76ER03075,
DOE grant DE-FG02-92ER25121, 
and by a Presidential Young Investigator
Award.  

\appendix{A}{Background from Differential Geometry}
\applab\aBfromDG

We begin by reviewing some important background 
material on the theory of bundles and connections 
on bundles.  This is intended for lightning-review
and for establishing notation. More leisurely treatments 
can be found in\refs{\bbrt\BoTu, \ccd, \egh, \danvia, \isham, \steenrod, \lagpg, \kobnom}.

\subsec{Differential forms}

We assume a knowledge of metrics, differential forms, tangent vectors etc.
In particular we assume a knowledge of the exterior differential $d$
\eqn\extrdiff{\eqalign{
\cdots \mapright{d} &  \O^k \mapright{d} \O^{k+1} \mapright{d} \cdots \cr
d^2 & =0 \cr}}
Moreover, given a vector field $\xi$ there are two natural operations on forms 
\eqn\twonatop{\eqalign{
\iota_\xi\colon \O^k &\longrightarrow \O^{k-1}\cr
\cL_\xi : \O^k &\longrightarrow \O^k \cr}}
These satisfy the basic relation:
$
\cL_\xi = [d , \iota_\xi]_+
$. 

\subsubsec{Hodge dual}

Given a metric define $\ast\colon \O^k ( M ) \to \O^{n-k} ( M )$ as follows.
Let  $\epsilon_{\mu_1\dots \mu_n} = \pm 1$ be the signature of the permutation. 
In a local coordinate basis take: 
$$
\ast (dx^{\mu_1}\wedge \cdots \wedge dx^{\mu_p}) 
\equiv {\sqrt{\mid \det g \mid} \over (n-p)!} 
{\epsilon^{\mu_1\cdots \mu_p}}_{\mu_{p+1} 
\cdots \mu_n}  dx^{\mu_{p+1} }\wedge \cdots\wedge dx^{\mu_n } . 
$$
\vskip0.1truein\noindent
Two key properties of $*$ are: 
\item{1.}
$\ast^2 = {\mid \det g\mid \over \det g} (-1)^{p(n-p)} $ on $\O^p ( M )$
\item{2.}
$*$ defines an inner product on $\O^p ( M )$: 
$
\langle \alpha, \beta\rangle = \int \alpha \wedge \ast \beta
$.

\exercise{Adjoint} 

Compute the adjoint of the exterior derivative
\eqn\deead{
d^\dagger = (-1)^{n(p+1)+1} \ast d \ast}

\endexercise

\subsec{Bundles}

\subsubsec{Fiber Bundles}

Locally a fiber bundle is a product space: $E= F\times B$. $F$ is called the fiber,
$B$ the base, and $E$ the total space.
When the base is topologically nontrivial we can have ``twisted'' bundles: 

\vskip0.1truein\noindent
{\bf Example 1:} A band is a trivial bundle, 
$[0,1]\times S^1$. The M\"obius strip is a nontrivial 
bundle with $F= [0,1]$. 
\vskip0.1truein\noindent
{\bf Example 2:} Consider the map of unit circles $S^1\to S^1$ given by $z\to z^n$.
This exhibits the unit circle as an $n$-fold covering of itself.
We have a fiber bundle with discrete fiber $\IZ/n\IZ$. 
Compare $S^1\times \IZ_n$.
\vskip0.1truein\noindent
The formal definition is

\vskip0.1truein\noindent
{\bf Definition:}  A  {\it fiber bundle}, $E$, with fiber $F$, over the basespace $M$ is a
topological space $E$ with a continuous projection 
$$
\pi\colon E\to M
$$
such that, $\forall x\in M$ $\exists$ neighborhood $U$, $x\in U$ with a homeomorphism:
$$
\Phi\colon U \times F \cong \pi^{-1}(U)
$$
with $\pi ( \Phi ( x, f )) =x$. 
\vskip0.1truein\noindent
The bundle is called a trivial bundle if it is globally a product space $E=F\times M$. 

\vskip0.1truein\noindent
{\bf Definition:} {\it Transition functions}:
On overlaps $U_1 \cap U_2$, we have fiber-preserving functions,
$\Phi_{12} = \Phi_2^{-1} \Phi_1\colon (U_1\cap U_2)\times F\to (U_1\cap U_2)\times F$
are the transition functions.
They have the form $( x, f) \rightarrow ( x, \Phi_{12} ( x )( f ))$ where $\Phi_{12}(x)\in \Diff(F)$.

\exercise{Cocycle condition}

Show that 

\item{a.} $\Phi_{ii}$ is the  identity map.
\item{b.} $\Phi_{ij}\Phi_{jk}= \Phi_{ik}$ on triple overlaps.

\endexercise

\ifig\Section{Picture of a section}
{\epsfxsize3.0in\epsfbox{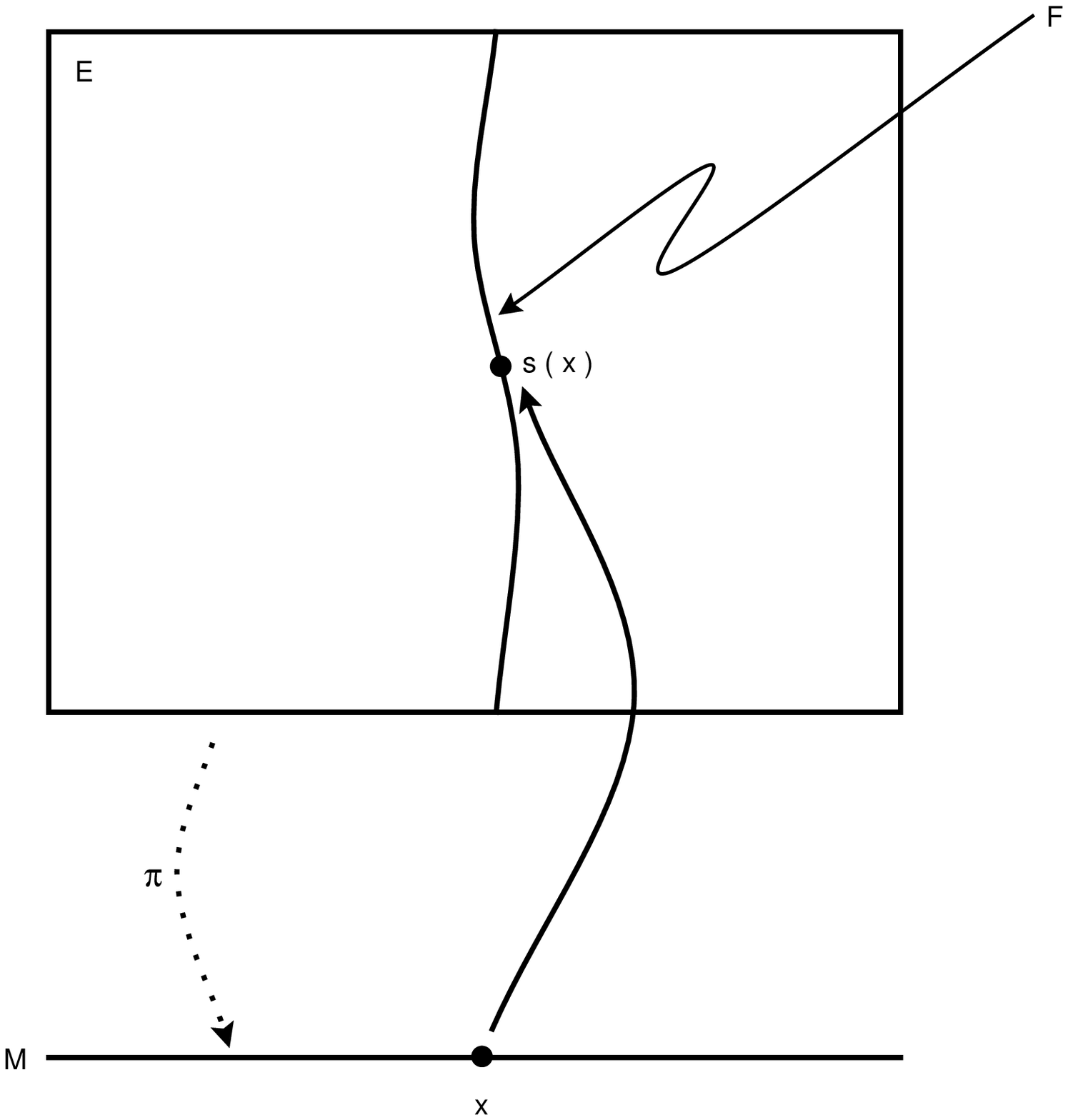}}

\vskip0.1truein\noindent
{\bf Definition:}  A {\it section} of $E$ is a map $s\colon M\to E$, such that
$\pi(s(x))=x$.
The picture is: $s$ takes a point $x$ to a point in the fiber above $x$
as shown in \Section. 
If $s$ is only defined locally it is called a local section. 

\subsubsec{Pulled back bundles}

There are many formal constructions with bundles.
One which will be often used is the pullback.
If $f\colon B_1 \to B_2$ is a map, then given a bundle $\pi_2\colon E_2\to B_2$ we can
construct the pullback $f^\ast E_2\to B_1$. 
This is the bundle over $B_1$ whose fiber at $x$ is the fiber of $E_2$ over $f(x)$. 
The total space is just:
\eqn\pllbck{
f^\ast E_2 = \{ (p,x)\in E_2\times B_1~ \vert~ \pi_2(p)=f(x) \}}
and the projection $\pi: f^\ast E_2 \to B_1$ is $\pi(p,x)=x$. 

\exercise{}
What is the dimension of the total space of $f^\ast E_2$ ?
\endexercise

\subsubsec{Vector Bundles}

\vskip0.1truein\noindent
{\bf Definition:} A vector bundle is a fiber bundle $E$ where the fiber $F$  is a
{\it vector space} and the transition functions are {\it linear transformations}. 

\exercise{Sections}

Show that a vector bundle always has a canonical global section

\endexercise

The extra structure of a vector space opens the way for many new constructions. 
The basic point is that we can do all of linear algebra point by point and glue together
globally.
For example, we can carry out globally the basic definitions of  direct sum, product,
linear transformation and dual.
Let $E$ and $F$ be any vector bundles.
We can define: 
\eqn\vcti{\eqalign{
(E\oplus F)_x & = E_x \oplus F_x\cr
 (E\otimes F)_x & = E_x \otimes F_x\cr
\Hom(E,F)_x &= \Hom(E_x ,F_x) \cr
E^\ast_x &= \Hom(E_x, \IR)\cr}}
\vskip0.1truein\noindent
{\bf Examples:}
\item{1.}
Trivial bundles $V\times M$. $V$ a vector space. 
\item{2.}
Tangent space: $T_x M= $ vector space of directional derivatives. 
A coordinate system defines a natural basis:
 $\{ { \p\over \p x^\mu}\mid_x  \}$. 
Tangent bundle:  $TM = \{ (x,v)~ \vert~ v \in T_x M \}$. 
Explicitly a local section is 
$$
( x^\mu, v^\mu(x) {\p\over \p x^\mu} )
$$
\item{3.}
Cotangent bundle: $T_x^\ast M =$ dual space to $T_xM$. 
Explicitly a local section is 
$$
(x^\mu, \xi_\mu(x) dx^\mu )
$$
\item{4.}
Bundle of Differential $j$- forms. 
Let:  $\Omega^j (M) =\Gamma[\Lambda^j T^\ast M]$ . 
Local sections are: 
$$
(x^\mu, \xi_{\mu_1 \cdots \mu_j} (x)  dx^{\mu_1}\wedge \cdots \wedge dx^{\mu_j} )
$$
\item{5.}
If $E\to M$ is a vector bundle then $\O^k ( M; E )$ denotes the space of 
$k$-forms on $M$ with values in $E$:
$\O^k  ( M; E ) \equiv \Gamma(\Lambda^k(T^\ast M)\otimes E)$.  

\subsubsec{Principal Bundles} 

A {\it principal bundle} is a  fiber bundle $E$,  where the fiber $F$ 
is a Lie group $G$ and the transition 
functions are right-multiplication by elements 
of $G$. More formally: 

\vskip0.1truein\noindent
{\bf Definition}:
A {\it principal $G$-bundle} for a Lie group $G$ is a manifold $P$ with a 
continuous projection $\pi:P\to M$ such that $G$ acts smoothly and freely on the right on $P$ and 
if $s_U$ is a smooth section over $U$ then 
$$\eqalign{
\Phi\colon U\times G & \to \pi^{-1}(U)\cr
\Phi\colon (x,g) & \to s_U(x)g \cr}
$$
is a diffeomorphism. 
\vskip0.1truein

The transition functions across patch boundaries are
local gauge transformations: $s_U(x) g_{UV}(x) = s_V(x)$, giving the 
local transition functions: 
$$
\Phi_U^{-1} \Phi_V ( x, g ) = ( x, g_{UV} ( x ) \cdot g)
$$

\exercise{Cross sections}

Show that a principal bundle is trivial iff it has a global cross-section

\endexercise 

\vskip0.1truein\noindent
{\bf Examples:} 
\item{1.}
Consider the $n$-fold cover, $S^1\to S^1$,  above.
This is a principal $\IZ_n$ bundle. 
\item{2.}
{\it Frame bundle}:  The set of frames on $TM$ form
 a principal $GL(n,\IR)$ bundle. 
Points are $(x;(e_1, \dots , e_n))$ where $(e_1, \dots,  e_n)$ is a linear basis
for $T_x M$. 
\item{3.}
{ \it Orthogonal frame bundle}:  
If $M$ has a metric we can restrict  the basis to an ON basis
Then we have a principal $O(N)$ bundle. 
If $M$ is orientable, then we can further restrict to an oriented ON basis 
and we get an $SO(N)$ bundle. 
\item{4.}
{ \it Homogeneous spaces }:  
$G$= Lie group. $H$ = subgroup.  $\pi:G\to G/H$ is a principal $H$ bundle. 
\item{5.}
{ \it The magnetic monopole bundle} : This is 
obtained from the homogeneous space bundle with
$G=SU(2)$, $H=U(1)$, $G/H=S^2$. 
\item{6.}
{ \it Instanton bundle} : 
Let $\IH$= quaternions, i.e., 
$\IH=\{ x_0 + i \vec x \cdot \vec \sigma= q~ \vert~ (x_0,\vec x)\in\IR^4\}$.
The unit quaternions form the group $SU(2)$. 
$S^7=$ unit sphere in $\IH^2$. $\IH P^1=S^4$, and the quotient of the unit sphere by the 
unit quaternions gives the $SU(2)$ instanton bundle $S^7 \to S^4$. 

\subsubsec{Associated vector bundles.}

Let $V$ be in a representation $\rho$ of $G$.
Then consider the trivial bundle $P\times V$.
This admits a $G$ action: 
$$
g ( p, v ) = ( p\cdot g^{-1}, \rho ( g ) \cdot v )
$$
and the quotient by this $G$-action
$$
{P \times V \over G} \equiv P \times_G V
$$ 
is a vector bundle over $P/G$ with fiber $V$. 
This is called the {\it associated bundle} to $P$ for the representation $\rho$. 
\vskip0.1truein\noindent
{\bf Examples:} 
\item{1.}
$\rho$ = the trivial representation. 
$P\times_G V = P/G \times V$. 
\item{2.}
$V=\IC$, the basic representation of $U(1)$. 
$SU(2)\times_{U(1)} V$ is a complex line bundle over $S^2$.
Wavefunctions in the presence of a magnetic monopole will take values in this bundle. 

Later we will use a basic tautology: 
\bigskip

\boxit{
The sections $\Gamma(P\times_G V) $ may be naturally identified with the equivariant
functions $f\colon P\to V$: 
\eqn\tautequi{
\Gamma ( P \times_G V) = \MAP_G ( P \rightarrow V )
\equiv \{ f\colon P\to V~ \vert~ f(p\cdot g) = \rho ( g^{-1} ) f ( p ) \}}}

\exercise{From trivial bundles to associated bundles, and back} 

\item{a.}
Show that $P\times V\to P\times_G V$ is itself a principal $G$-bundle.
\item{b.}
Show that the definition $P\times_G F$ extends to any space $F$ with a left $G$-action. Show, moreover, that $P\times_G G=P$. 
\item{c.}
Show that if $\pi: P\to M$ is the projection,  then the pullback bundle
$\pi^\ast ( P \times_G F) = P \times F$

\endexercise

\subsec{Connections on Vector Bundles and Principal Bundles}
\subseclab\ssCVBFB

\subsubsec{Connections on Vector Bundles}

\vskip0.1truein\noindent
{\bf Definition:}
A connection,  $\nabla$,  on a vector bundle $E$ is a linear map
\eqn\dfncnn{
\nabla : \Gamma(E)  \to \O^1 ( M; E )}
That is, if $s$ is a section then $\nabla s$ is a one-form with values in the bundle. 
$\nabla$ must  satisfy: 
\eqn\dfncnni{\eqalign{
\nabla (s_1+s_2) & =  \nabla s_1+ \nabla s_2 \cr
\nabla (f s) & = s \otimes df + f \nabla s\cr}}
where $f$ is a function on $M$.

\vskip0.1truein\noindent
{\bf Remark:}
$\nabla$ extends naturally to $\nabla: \O^k ( M; E ) \to \O^{k+1} ( M; E )$
\eqn\dfncnnii{\eqalign{
\nabla s(X_1,\dots , X_{p+1}) &= \sum_j (-1)^{j+1} 
\nabla_{X_j} s(X_1, \dots, \hat X_j , \dots X_{p+1}) \cr
&+\sum_{i<j} (-1)^{i+j} s([X_i,X_j],\dots 
\hat X_i \dots \hat X_j\dots), \cr}}
where $\nabla_{X} = i_X \circ \nabla$. 
There are three immediate constructions obtained from a connection on $E$: 
\item{1.}
$\underline{\rm Covariant\  derivative}$: If $s$ is a section, 
$X$ a tangent vector, we can form the 
covariant derivative,  $\nabla_X s$. In local coordinates
it is given by: 
$$
(\nabla_Xs)^\alpha = X^\mu \biggl(\p_\mu s^\alpha + ({A_\mu})^\alpha_\beta s^\beta\biggr)
$$
where $(A_\mu(x))_\alpha^\beta dx^\mu$ is a one-form with values in $\End (E_x)$. 
\item{2.}
$\underline{\rm Parallel\  transport}$: $\nabla_X $ is a covariant derivative. 
If $Z(\tau)$ is a path, $\dot Z(\tau)$ a  tangent vector along the path, then 
\eqn\ptrans{
\nabla_{\dot Z(\tau)} s =0}
is a first order differential equation.
This equation determines parallel transport, allowing us to ``connect'' different fibers. 
The solution to \ptrans\ in local coordinates reads: 
\eqn\liftlocal{
s(\tau) = P \exp \biggl[-\int_0^\tau  A  \biggr] s(0)}
\item{3.}
$\underline{\rm Horizontal \  tangent\ vectors}$ 

Recall  that $T V = V\oplus V$ for a vector space $V$. 
Consider $TE$. 
At $(x,v)\in E$ there is a natural {\it vertical tangent space} along the fiber of $E$.
In general there is no natural choice for a complementary subspace such that 
\eqn\complsbspc{
T_e E = V_e \oplus H_e . }
Such a subspace is called a horizontal subspace.

A smooth 
choice of horizontal subspaces gives another way to define parallel transport of
vectors. 
If a curve, $\gamma$,  is in the base 
and $\pi(e(t))=\gamma(t)$,  then we may unambiguously 
define a lift of the tangent vector $\dot \gamma(t)$.
Choosing an initial point $\tilde\gamma(0)$,  we then define a lifted path $\tilde\gamma(t)$.
Moreover, if we locally trivialize,  so that the lifted path may be described by
$\tilde\gamma(t)= 
(\gamma(t),s(t))$, then  ``$s$ is parallel transported''
 means that the tangent to
$\tilde \gamma$ is 
$$
\dot \gamma^\mu ( t ) {\p\over\p x^\mu} + {\dot s}^\alpha ( t ) {\p\over \p v^\alpha}
= \dot \gamma^\mu ( t )\biggl[ {\p\over \p x^\mu} - (A_\mu)^\alpha_\beta v^\beta
{\p \over \p v^\alpha} \biggr]
$$
Hence the horizontal subspace is given by:
\eqn\horsubsp{
H_{(x,v)} = \Span \Biggl\{ {\p\over \p x^\mu}-
 (A_\mu)^\alpha_\beta v^\beta
{\p \over \p v^\alpha}  \Biggr\}}

\subsubsec{Vertical Tangent Vectors on Principal Bundles}
\subsubseclab\sssVTV

As preparation for discussing connections on principal bundles we first define 
the vertical vector fields.
Let $P$ be a principal bundle. 
Through any point $p\in P$,  we define a right $G$-action: $R_p: G\to P$
\eqn\rghtgact{
R_p: g \mapsto p\cdot g}
embedding a copy of $G$ inside $P$. 
The differential of this mapping
\eqn\defofcee{
C_p = dR_p : \lieg \to T_pP}
defines tangent vectors to $P$ for every element of the Lie algebra $X\in \lieg$. 
These are called the {\it fundamental vector fields} on $P$: 
\eqn\fndvf{
\forall X\in \lieg \qquad \xi(X)_p = C_p (X)}

They are globally defined vector fields and $X\to \xi(X)$ is a homomorphism of 
Lie algebras:
$$
[ \xi( X ), \xi( Y )] = \xi( [ X, Y ]) \quad . 
$$

The image of $C_p$ is called the {\it vertical tangent space at $p$}.
A figure of a vertical tangent vector for a principal $U(1)$ bundle 
appears in \varradius.

\subsubsec{Connections on Principal Bundles} 

A connection on a principal bundle $P$ is  an equivariant splitting of the tangent 
space of $P$ into horizontal and vertical vectors.
Recall from section \sssVTV\  that there is a canonical vertical tangent space 
$(T_p P)^{\rm vert} \cong \lieg$, {\bf } given by the right $G$-action.
Indeed, 
the differential defines an isomorphism: 
\eqn\vrrtgsp{
C_p = dR_p : \lieg \cong (T_p P)^{\rm vert}}
There is, in general,  no natural way to find a complementary (=horizontal) subspace
- a choice of horizontal subspace is called a connection.
Formally, we have:

\vskip0.1truein\noindent
{\bf Definition A:} 
A connection is a $G$-equivariant splitting of the tangent space $TP$ into horizontal
and vertical spaces, i.e., a choice
\eqn\spltting{
T_p P = H_p \oplus (T_p P)^{\rm vert}}
such that  
\eqn\horequiv{
(R_g)_\ast H_p = H_{pg}.  }

A more concrete way of defining the connection is to define its associated 
connection one-form: 
\vskip0.1truein\noindent
{\bf Definition B}.  The connection form is a {\it globally defined} 1-form $\Theta$ with values
in $\lieg$, i.e., $\Theta\in\O^1 ( P, \lieg )$ such that 
\item{1.}
$\iota_{\xi(X)} \Theta = X$ 
\item{2.}
$\CL_{\xi(X)} \Theta = - \Ad(X) \Theta= -[X, \Theta] $

Condition 2 can be exponentiated to give the right action of the group on $\Theta$: 
\eqn\pllbckcn{
R_p^\ast ( g ) \Theta_p = \Ad ( g^{-1}) \Theta = g^{-1} \Theta g}

\vskip0.1truein\noindent
{\bf Equivalence of Definitions A and B:}
\vskip0.075truein\noindent
$A \Rightarrow B$: If $X_p\in T_pP$, then we have a splitting
$X_p = X_p^{\rm hor} + X_p^{\rm vert} $. 
Define 
\eqn\eqvatob{
\Theta_p(X_p) = C^{-1}\biggl[ X_p^{\rm vert}\biggr]}
Identifying $(T_p P)^{\rm vert} \cong \lieg$ using the isomorphism in \vrrtgsp\ shows that
$\Theta$ is a Lie algebra-valued 1-form $\in \O^1 ( P; \lieg )$. 
\vskip0.075truein\noindent
$B\Rightarrow A$: $H_p = \ker~\Theta$. 

\vskip0.1truein\noindent
{\bf Examples:} 
\item{1.}
Let $G$ be a matrix group, $SU(N),SO(N), \ldots$.
Regard matrix elements of $g\in G$ as functions on $G$. 
Regard $G\to pt$ as a principal $G$-bundle. 
Then we can form the matrix of one-forms: 
\eqn\maurcart{
\Theta = g^{-1} d g}
One can easily check that this satisfies properties 1 and 2 of definition B.
\maurcart\ is called the Maurer-Cartan form.
It is the unique solution to these conditions for a Lie group.
\item{2.}
Trivial bundle $G\times M$.
A connection on a trivial bundle must look like: 
\eqn\trivblde{
\Theta_{(g,x)} =  g^{-1} d g + g^{-1} A_\mu g dx^\mu}
where $A_\mu dx^\mu$ is a Lie-algebra-valued one-form on $M$, choosing a basis $T_a$
for $\lieg$ we can write
\eqn\gaugefield{
A= A_\mu dx^\mu = A_\mu^a T_a \otimes dx^\mu}
\item{3.}
Patch-by-patch. 
When the bundle is not trivial we can describe it
 by gluing together trivial bundles using 
transition functions.
In a local trivialization 
\eqn\loctriv{\eqalign{
\Phi_U\colon U \times G &\to \pi^{-1} ( U ) \cr
\Phi_U\colon ( x, g) &\rightarrow  s_U ( x ) \cdot g \cr}}
we have
\eqn\ptchcnn{
\Phi_U^\ast ( \Theta ) = g^{-1} d g + g^{-1} A_\mu^{( U )} g dx^\mu}
$\Theta$ is globally defined but will have different descriptions in different patches. 
Using the transition functions $s_U( x ) g_{UV}( x )=s_V( x )$ we get:
$$
\Phi_U^{-1} \Phi_V( x, g )=( x, g_{UV} ( x )\cdot g )
$$
We calculate:
\eqn\ptchtoptch{\eqalign{
g^{-1} d g + g^{-1} A^{(V)} g &=\Phi_V^\ast ( \Theta )\cr
&=(\Phi_U^{-1} \Phi_V )^\ast ( g^{-1} d g + g^{-1} A^{(U)} g )\cr}}
Leading to the transformation law across patch boundaries:
\eqn\ptchtrn{\mathboxit{
A^{(V)}=  g_{UV}^{-1} A^{(U)} g_{UV} + g_{UV}^{-1} d g_{UV}(x)}}

\exercise{Connections}

Consider the principal $H$- bundle $G\to G/H$. 
Using an invariant metric define a natural connection. 

\endexercise

\subsubsec{Connection on an Associated Vector Bundle} 

A connection on a principal bundle induces a connection on an associated bundle
$E=P\times_G V$.
There are two ways to see this, corresponding to the two definitions above: 
\item{A.}
We need to choose horizontal subspaces of the tangent space $TE$.
Think of $E$ as the basespace of the $G$-
bundle $P\times V \to E$.
The choice of connection of $P$ defines a splitting of the tangent space 
$T_{(p,v)} ( P \times V) = H_{p,v} \oplus \lieg \oplus V$. 
These horizontal subspaces descend to $TE$. 
\item{B.}
Use the basic tautology \tautequi\ to identify $s\in \Gamma[E] $ with an equivariant
function $f\colon P \to V$.
Then $\nabla s$ corresponds to the equivariant $1$-form: $\rho(\Theta) \cdot f$, 
where $\rho$ is the representation of the Lie algebra.

\vskip0.1truein\noindent
In down-to-earth local coordinates, the associated bundle has a connection defining
a covariant derivative: 
If $s$ is a local section then 
\eqn\dwnerth{
\nabla_Z s = Z^\mu( \p_\mu (s(x) )^\alpha + 
{A}^a_\mu(x) \rho(T_a)^\alpha_\beta s(x)^\beta )}
which transforms across patch boundaries as
\eqn\dwnerthii{
\nabla_Z s_U = \rho(g_{UV}) \cdot \nabla_Z s_V}

\subsec{Curvature and Holonomy}
\subseclab\ssCandH

\subsubsec{Holonomy}

Now let $\gamma(\tau)$ be a  path in $M$ starting at $x$. 
If $P\to M$ is a principal bundle with connection then we can define the lifting of the
vector fields $\dot \gamma$.
We choose a point $p$ over $x$ and the lift is the horizontal vector field such that 
$d \pi \dot{\tilde \gamma} = \dot \gamma$. 
The lifted curve $\tilde \gamma$ has this as tangent vector. 

Now consider a closed curve $\gamma$. 
We have $\pi(\tilde \gamma) = \gamma$. 
The lifted curve need not be a closed  curve, the endpoint
of  $\tilde \gamma$ need only lie in the same fiber above 
as the initial point.  When $\tilde \gamma$ is not 
closed we have holonomy. 
For a principal bundle we can write
\eqn\hologpi{
\tilde \gamma(1) = \tilde \gamma(0)\cdot h(p,\gamma)
=p\cdot h(p,\gamma)}
for some group element $h(p,\gamma)$. 

\ifig\Holonomy{Holonomy}
{\epsfxsize3.0in\epsfbox{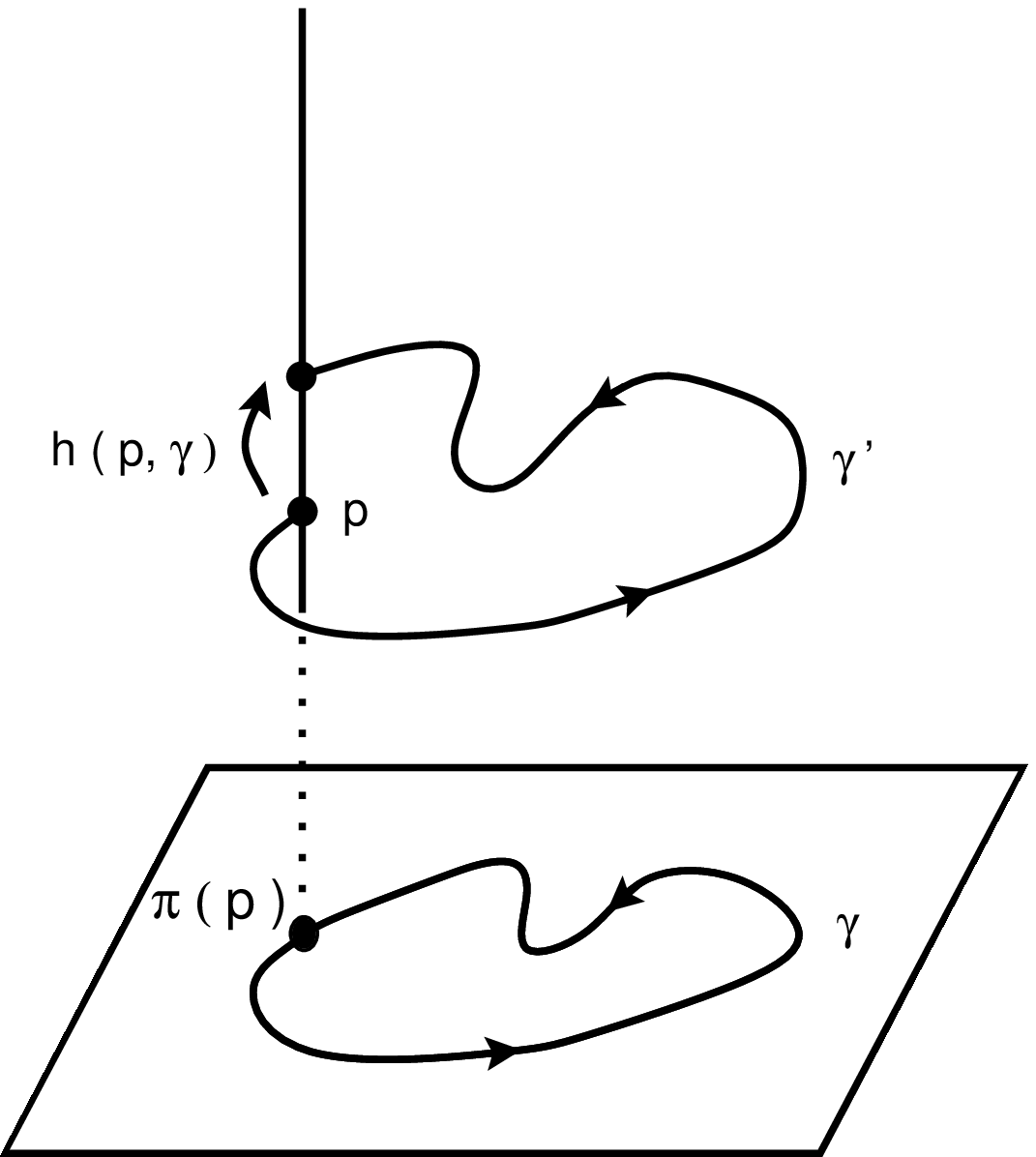}}

\exercise{Holonomy elements}

\item{a.} Show that
$$
h(p,\gamma_1) h(h(p,\gamma_1), \gamma_2) = h(p,\gamma_1 \gamma_2)
$$
\item{b.} 
$$
h(pg,\gamma) = g^{-1} h(p,\gamma) g
$$

\endexercise

\vskip0.1truein\noindent
{\bf Definition:} The {\it holonomy group} $\CH(p,\Theta)$ of the connection 
on $P$ at $p$ is the subgroup of $G$ generated by $h(p,\gamma)$,  for all closed
curves $\gamma\in \O_x(M)$. 

For an associated vector bundle we can also lift curves.
In this case the holonomy is a multiplication of the initial vector by the group element in the 
appropriate representation. 
In a local patch we have: 
\eqn\liftlocali{\eqalign{
\tilde \gamma(1) &= \tilde \gamma(0)
P \exp \biggl[\oint A  \biggr] \cr
\tilde v(1) &= \rho(P \exp \biggl[-\oint A  \biggr] ) 
\tilde v(0). \cr}}

\vskip0.1truein\noindent
{\bf Examples:}
\item{1.}
$P=U(1)\times( \IR^2-\{0\})$, $A=ia d\theta$, $h=\exp(2\pi i n a)$. The holonomy group is
$\IZ$. 
\item{2.}
Consider the tangent bundle to the  sphere $S^2$ 
with the round metric.
The holonomy around a geodesic 
triangle is proportional to the area. 

\vskip0.1truein\noindent
These two examples represent the two ways one can get holonomy: nontrivial $\pi_1$
and nontrivial curvature. 

\ifig\SmallLoop{}
{\epsfxsize3.0in\epsfbox{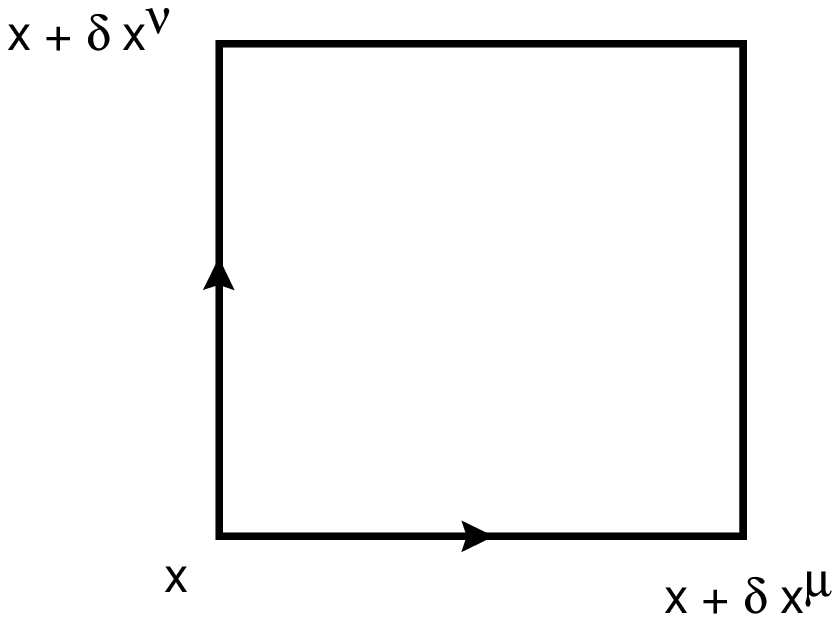}}

\subsubsec{Curvature} 

Consider \liftlocali\ for an infinitesimal loop surrounding an area element 
$\sigma^{\mu\nu}\p_\mu \wedge \p_\nu$ as in \SmallLoop. 
We can trivialize the bundle and introduce $A$. 
An elementary, but extremely important calculation shows that the holonomy is 
\eqn\infholo{\eqalign{
h(\gamma) &= 1 + \sigma^{\mu\nu} F_{\mu\nu} +\CO(\sigma^2)\cr
F_{\mu\nu}&=\p_\mu A_\nu - \p_\nu A_\mu +[A_\mu, A_\nu].  \cr}}

Formally we define the curvature of the connection to be 

\vskip0.1truein\noindent
{\bf Definition} The curvature of a connection $\Theta$ on $P$ is 
\eqn\defncurv{
\bF = d \Theta + \Theta^2 = d \Theta +\half [ \Theta, \Theta] 
\in \O^2 ( P; \lieg ). }

\par\noindent
One easily shows: 
\item{1.}
$\bF$ is horizontal $\iota_{\xi(X)} \bF=0$. 
\item{2.}
$\bF$ transforms in adjoint representation. 

Indeed, in a local trivialization,  
\eqn\crvloc{\eqalign{
\bF&=g^{-1} F_{\mu\nu}(x) g ~dx^\mu \wedge dx^\nu\cr
&=(g^{-1}T_a g) F^a_{\mu\nu}(x)~ dx^\mu \wedge dx^\nu . \cr}}

Once again, by the basic tautology \tautequi\ we have 
\eqn\drnfcrv{
F \in \O^2 ( M; \Ad ~ P )}
where $\Ad ~P$ is the vector bundle associated to $P$ by 
the adjoint representation.

\exercise{Pure gauge}

Show that for the principal bundle $G\to pt$ the Maurer-Cartan form has curvature zero. 

\endexercise

For the case of connections on vector bundles we have:

\vskip0.1truein\noindent
{\bf Definition:}
The curvature of a connection on a vector bundle
$E\to M$ 
is the form in $\O^2( \End(E))$ defined as follows. 
Let $X,Y$ be tangent vectors to $M$. 
Then  $F(X,Y): \Gamma(E) \to \Gamma(E)$ is defined by 
$$
F(X,Y)s= \nabla_X(\nabla_Y s) - 
\nabla_Y(\nabla_X s) -\nabla_{[X,Y]} s . 
$$

When the vector bundle is associated to the principal bundle, then 
$F(X,Y)= \rho(F_{P.B.} (X,Y))$.

\exercise{Curvature as the obstruction to a complex}

We saw above that $\nabla: \O^k ( M; E ) \to \O^{k+1} ( M ;E )$. 
Consider the sequence of spaces 
$$
\cdots {\buildrel \nabla\over\rightarrow}
\O^k(M;E) {\buildrel \nabla\over\rightarrow} \O^{k+1}(M;E) 
{\buildrel \nabla\over\rightarrow}\cdots
$$
Show that $\nabla^2$ = multiplication by the curvature.
The failure of $\nabla$ 
to define a ``complex'' (e.g. a ``BRST complex'') is measured by the curvature. 

\endexercise

\subsec{Yang-Mills  equations and action }
\subseclab\ssYMEQA

Let $\ST$ be a spacetime of dimension $n$, $G_{\mu\nu}$ a metric on it.
$P\to \ST$ a principal $G$ bundle. 
Let $\CA(P)$ be the space of all connections on $P$. 
This space is infinite-dimensional, and is the space of gauge fields in nonabelian gauge
theory. 
We would like to write an action on $\CA(P)$ which is gauge invariant. 

To get an action consider $\ast F$.
This is an $(n-2)$ -form with values in the Lie algebra. 
Let ``$\Tr$''  be an invariant form on the Lie algebra -- for example 
the ordinary trace in the
fundamental representation for $SU(N)$. 
The gauge-invariant action is 
\eqn\ymaction{\eqalign{
I_{\rm YM}[A] &= {1\over 4 e^2}\int_{\ST}  \Tr~ { \tilde F}\wedge \ast F\cr
&={1\over 4 e^2}\int_{\ST} d^D x \sqrt {det G_{\mu\nu}} 
~G^{\mu\lambda} G^{\nu\rho}~ \Tr  ~ F_{\mu\nu}F_{\lambda\rho} \cr}}

The equations of motion and Bianchi identities are: 
\eqn\eombian{\eqalign{
D_A F = d F + [A,F] & = 0  \qquad {\rm Bianchi\ identity} \cr
D_A \ast F & = 0  \qquad  \hbox{{\rm Equations of motion}} . \cr}}

In local coordinates \eombian\ is: 
\eqn\eomii{\eqalign{
D_{[\mu} F_{\nu\lambda]} & = 0 \cr
D^\mu F_{\mu\nu}& =0. \cr}}

\vskip0.1truein\noindent
{\bf Remarks:} 
\item{1.}
\eombian\  are a nonabelian generalization of the equations for a 
harmonic differential form: $d \omega = d * \omega =0$. 
\item{2.}
$D_\mu$ involves the Levi-Civita connection of the metric.

\exercise{Simple solutions}

Show that $A= \lambda U^{-1}(x) d U(x)$ solves the equations for $\lambda = 1/2,1$.
The first has infinite action (it is called a ``meron'') and the second is pure gauge. 

\endexercise 

\exercise{Self-Duality}

\item{a.}
Suppose $D=4$. 
Show that either of the first order equations $F=\pm \ast F$ implies the Yang-Mills
equations. 
These are called the (anti-) self dual equations. 
\item{b.}
Show that the 't Hooft ansatz: 
\eqn\instntn{\eqalign{
A_\mu &= i \bar{\Sigma}_{\mu\nu}\nabla^\nu \log f\cr
f(x) & = 1 + \sum_{i=1}^N {\rho_i^2\over (x-a_i)^2}\cr
 \bar{\Sigma}_{ij}&= \half \epsilon_{ijk}\sigma^k\cr
\bar{\Sigma}_{4i}=- \bar{\Sigma}_{i4}&=\half \sigma^i, \cr}}
where $i$ runs over $\{ 1,2,3\} $ ,  satisfies the ASD instanton equations.

\endexercise

\subsubsec{Infinite Dimensional Principal Bundles}

In part II of these lectures we will often be discussing infinite dimensional principal bundles.
Two fundamental examples:

1. As above, let $\CA(P)=$ the set of all connections on a principal bundle $P$.
This is an infinite-dimensional manifold.
In fact, it is - itself- a principal bundle for the infinite-dimensional group $\CG(P)$
of gauge transformations. 
Formally 

\vskip0.1truein\noindent
{\bf Definition:} $\CG(P)$ is the group of automorphisms of $P$.
These may be regarded as patchwise defined functions $g_U\colon U\to G$,
acting on the local sections as $s_U\to s_U g_U$.
If $P$ is trivial then $\CG ( P ) = \MAP ( \ST \to G)$. 

Let us assume $P$ is trivial.
The action of $\CG(P)$ on $\CA(P)$ is easily seen to be that of gauge transformation:
\eqn\ggetmn{
A\to g^{-1} d g + g^{-1} A g}

This doesn't quite act freely on $\CA(P)$, because of global gauge transformations in the
center $C(G)$ of $G$. 
However, $C(G)$ is a normal subgroup so we can consider the group $\CG(P)/C(G)$.
This acts freely except at the ``reducible connections.'' 
Reducible connections are connections for which the holonomy group
is a proper subgroup of $G$.
For example if $A$ has the block decomposition
$$
A=\pmatrix{ A_1 & 0 \cr  0 & A_2 \cr}
$$
then nontrivial global gauge transformations will fix $A$.
Thus, at reducible connections the dimension of the fiber collapses and the quotient space
is singular.
However 
\eqn\gaugebdle{
\CA^{irr}  \to \CA^{irr} /(\CG/C(G))}
is a well-defined principal fiber bundle. 
The fiber and the base are infinite-dimensional. 

2. $P=\Met(\Sigma)$, the space of metrics on a surface 
$\Sigma$ of genus $p$. $G=\Diff(\Sigma)\sdtimes \Weyl(\Sigma)$, the group of 
diffeomorphisms of $\Sigma$ and Weyl rescalings acting as 
$$\eqalign{
h & \to f^\ast h \cr
h & \to e^\phi h \cr}
$$
The base space is $\CM_{p,0}$, the moduli space of Riemann surfaces of genus $p$. 
We'll see below that it has dimension $6p-6$. 
We will see another  example, Hurwitz space, in chapter \sCS.

\listrefs

\bye